\documentclass{article}

\usepackage{cite}
\usepackage[utf8]{inputenc}
\usepackage{geometry}
\usepackage{authblk}
\usepackage{subcaption}
\usepackage[utf8]{inputenc} 
\usepackage[T1]{fontenc}
\usepackage{url}
\usepackage{ifthen}
\usepackage{cite}
\usepackage[cmex10]{amsmath} 
\usepackage{amssymb,lipsum} 
\usepackage{ bm}
\allowdisplaybreaks
\usepackage{amsthm}                       
\usepackage{tikz}
\usetikzlibrary{patterns}
\usepackage{pgfplots}
\pgfplotsset{compat=1.9}

\usepgfplotslibrary{fillbetween}
\usepackage{graphicx}
\usepackage{pgfplots}
\usepackage{caption}
\usepackage{float, stfloats}
\usepackage{latexsym}
\usepackage{nicefrac}
\usepackage{Notations}

\usepackage{sidecap}
\sidecaptionvpos{figure}{t}

\usepackage{booktabs}       
\usepackage{amsfonts}       
\usepackage{microtype}      
\usepackage{xcolor}         

\title{\bf{Bayesian Extensive-Rank Matrix Factorization with Rotational Invariant Priors}}

\author{Farzad Pourkamali,
Nicolas Macris}
\affil{\emph{School of Computer and Communication Science, Ecole Polytechnique F{\'e}d{\'e}rale de Lausanne}}
\date{}

\begin{document}

\maketitle

\begin{abstract}
    We consider a statistical model for matrix factorization in a regime where the rank of the two hidden matrix factors grows linearly with their dimension and their product is corrupted by additive noise. Despite various approaches, statistical and algorithmic limits of such problems have remained elusive. We study a Bayesian setting with the assumptions that (a) one of the matrix factors is symmetric, (b) both factors as well as the additive noise have rotational invariant priors, (c) the priors are known to the statistician. We derive analytical formulas for \textit{Rotation Invariant Estimators} to reconstruct the two matrix factors, and conjecture that these are optimal in the large-dimension limit, in the sense that they minimize the average mean-square-error. We provide numerical checks which confirm the optimality conjecture when confronted to {\it Oracle Estimators} which are optimal by definition, but involve the ground-truth. Our derivation relies on a combination of tools, namely random matrix theory transforms, spherical integral formulas, and the replica method from statistical mechanics.
\end{abstract}

\section{Introduction}
Matrix factorization (MF) is the problem of reconstructing two matrices $\bX$ and $\bY$ from the noisy observations of their product. Applications in signal processing and machine learning abound, such as for example dimensionality reduction \cite{tovsic2011dictionary,mairal2008supervised}, sparse coding \cite{olshausen1996emergence,olshausen1997sparse,kreutz2003dictionary}, representation learning \cite{bengio2013representation}, robust principal components analysis \cite{candes2011robust, perry2018optimality}, blind source separation \cite{belouchrani1997blind}, or matrix completion \cite{candes2010power,candes2012exact}. 

In this work we approach the problem from a Bayesian perspective and assume that an observation or data matrix $\bS = \sqrt\kappa \bX\bY + \bW$ is given to a statistician who knows the prior distributions of $\bX$ and $\bY$ as well as the prior of the additive noise matrix $\bW$ and the signal-to-noise ratio $\kappa>0$. The task of the statistician is to construct estimators $\bm{\Xi}_{X}(\cdot)$, $\bm{\Xi}_{Y}(\cdot)$ for the matrix factors $\bX$, $\bY$, that ideally, minimize the average mean-square-error (MSE) $\mathbb{E}\Vert \bX - \bm{\Xi}_{X}(\bS)\Vert_{\rm F}^2$ and $\mathbb{E}\Vert \bY - \bm{\Xi}_{Y}(\bS)\Vert_{\rm F}^2$ ($\Vert . \Vert_{\rm F}$ the Frobenius norm and $\mathbb{E}$ the expectation w.r.t $\bX,\bY,\bW$).
We consider priors which are rotation invariant for all three matrices $\bX$, $\bY$, $\bW$ and for $\bX$ we furthermore impose that it is square and symmetric. These matrix ensembles are defined precisely in  section \ref{MF-model-sec}, but the reader can keep in mind the examples of Wigner or Wishart matrices for $\bX$, and general Gaussian $\bY$ and $\bW$ with i.i.d elements. We look at the asymptotic regime where all matrix dimensions and ranks  tend to infinity at the same speed. We remark that the usual "rotation ambiguity" occuring in MF is not present because we impose that at least one of the two matrix factors is symmetric. We also remark that MF is different (and more difficult) than matrix denoising which would consist in constructing an estimator $\bm{\Xi}_{\bX\bY}(\bS)$ for the signal as a whole by minimizing $\mathbb{E}\Vert \bX\bY - \bm{\Xi}_{XY}(\bS)\Vert_{\rm F}^2$.

The rotation invariance of the model implies that the estimators minimizing the MSE belong to the class of rotation invariant estimators (RIE). RIEs are matrix estimators which have the same singular vectors (or eigenvectors) as the observation or data matrix. These estimators have been proposed for matrix {\it denoising} problems (see references \cite{stein1975estimation,takemura1984orthogonally, bun2016rotational, bun2017cleaning,Pourkamali2023Matrix} for covariance estimation, \cite{benaych2019optimal} for cross-covariance estimation, and \cite{troiani2022optimal},  \cite{pourkamali2023rectangular, Landau2023SingularVO} for extensions to rectangular matrices). For the present MF model, we derive optimal estimators (minimizing the MSE) that belong to the RIE class and can be computed explicitly in the large dimensional limit from the observation matrix and the knowledge of the priors. We propose:
\begin{enumerate}
    \item an explicit RIE to estimate $\bX$, which requires the knowledge of the priors of \textit{both} $\bX, \bY$ and of the noise $\bW$. Moreover, under the assumption that $\bX$ is positive-semi-definite, a \textit{sub-optimal} RIE can be derived which {\it does not} require any prior on $\bX$. 
    \item  an explicit RIE to estimate $\bY$, which requires the knowledge of the priors of the noise $\bW$ and $\bX$ {\it only} (the prior of $\bY$ is not required).
    \item combined with the singular value decomposition (SVD) of the observation matrix, our explicit RIEs provide a spectral algorithm to reconstruct both factors $\bX$ and $\bY$.
\end{enumerate}

The derivation of the proposed estimators relies on the replica method from statistical mechanics combined with techniques from random matrix theory and finite-rank spherical integrals \cite{guionnet2005fourier, benaych2011rectangular}. Although the replica method is not rigorous and involves concentration assumptions, the derivation is entirely analytical and suggests that the estimators are optimal in the limit of large dimensions. This is corroborated by numerical calculations comparing our explicit RIEs with Oracle Estimators which are optimal by definition and involve the ground-truth matrices. 

\subsection{Related literature and discussion}
When the matrices $\bX$ and $\bY$ are assumed to have {\it low-rank} compared to their dimension, the mathematical theory of MF has enjoyed much progress under various settings (Bayesian, spectral, algorithmic) and fundamental information theoretical and algorithmic limits have been 
rigorously derived \cite{baik2005phase, benaych2011eigenvalues,benaych2012singular,
dia2016mutual, miolane2017fundamental, lelarge2019fundamental, luneau2020high}.

In extensive-rank regimes, when the rank grows like the matrix dimensions, despite various attempts there is no solid theory of MF. 
One approach is based on Approximate Message Passing (AMP) methods developed in \cite{kabashima2016phase, parker2014bilinearD, zou2021multi}. Despite acceptable performance in practical settings \cite{parker2014bilinearA}, as pointed out in \cite{maillard2022perturbative} the AMP algorithms developed in these works are (theoretically) sub-optimal. Other approaches rooted in statistical physics have been considered in \cite{maillard2022perturbative,barbier2022statistical} but have not led to explicit reconstructions of matrix factors or algorithms.  A practical probabilistic approach to MF problem is based on variational Bayesian  approximations \cite{bishop1999variational, lim2007variational, ilin2010practical}, in which one tries to approximate the posterior distribution with proper distribution. 
In \cite{nakajima2013global} it is shown that under Gaussian priors, the solution to the MF problem is a reweighted SVD of the observation matrix. We point out here that these estimators can be seen as a RIE and therefore there seems to be a rather close relation between the RIE studied here and the variational Bayesian approach. This also suggests that adapting RIEs to real data is an interesting direction for future research. Finally, let us also mention optimization approaches where one constructs estimators by following a gradient flow (or gradient descent) trajectory of a training loss of the type $\Vert \bS - \bX\bY\Vert_{\rm F}^2 +  \textrm{reg. term}$ (see \cite{tarmoun2021understanding}, \cite{bodin2023gradient} for analysis in rotation invariant models). 
Benchmarking these various other algorithmic approaches against our explicit RIEs (conjectured to be optimal) is outside the scope of this work and is left for future work.

Constraints such as sparsity or non-negativity of the matrix entries which have important applications \cite{lee1999learning} are not covered by our theory. Despite this drawback, we believe that the proposed estimators  are important both for theoretical and practical purposes. Even in non-rotation invariant problems our explicit RIEs may serve as sub-optimal estimators, and as we show in an example  they can be used as a "warmed-up" spectral initialization for more efficient algorithms (see for example \cite{mondelli2021approximate, montanari2021estimation} for related ideas in other contexts). The methodology developed here may open up the way to further analysis in inference and learning problems perhaps also in the context of neural networks  where extensive rank weight matrices must be estimated.
\subsection{Organization and notations} 
In section \ref{model-RIE}, we introduce the precise MF model, general class of RIEs, and the Oracle estimators. In section \ref{propsoed-RIE}, we present the explicit RIEs (and algorithm) to estimate $\bX$ and $\bY$. We provide the numerical examples and calculations in section \ref{numerical}. In section \ref{Derivation}, we sketch the derivation of RIE for $\bX$, while the one for $\bY$ is similar and deferred to the appendices.

The following notations are used throughout. For a vector $\bm{\gamma}\in \bR^N$ we denote by $\bm{\Gamma} \in \bR^{N \times M}$ a matrix constructed as 
$\bm{\Gamma}  = \left[
\begin{array}{c|c}
\bm{\Gamma}_N & \mathbf{0}_{N \times (M-N)}
\end{array}
\right]
$ with $\bm{\Gamma} _N \in \bR^{N \times N}$ a diagonal matrix with diagonal $\bm{\gamma}$. 
The same notations will also be used for the vector $\bm{\sigma}$ and the corresponding matrix $\bm{\Sigma}$ and . 
For a sequence of non-symmetric matrices $\bA$ of growing size, we denote the limiting empirical singular value distribution (ESD) by $\mu_A$, and the limiting empirical eigenvalue distribution of $\bA \bA^\intercal$ by $\rho_A$. For a sequence of symmetric matrices $\bB$ of growing size, we denote the limiting empirical eigenvalue distribution by $\rho_B$, and the limiting eigenvalue distribution of $\bB^2$ by $\rho_{B^2}$.

\section{Matrix factorization model and  rotation invariant estimators } \label{model-RIE}
\subsection{Matrix factorization model}\label{MF-model-sec}
Let $\bX = \bX^\intercal \in \bR^{N \times N}$ a symmetric matrix distributed according to a rotationally invariant prior $P_X(\bX)$, i.e.,  for any orthogonal matrix $\bO \in \bR^{N \times N}$ we have $P_X(\bO \bX \bO^\intercal) = P_X(\bX)$. Let also $\bY \in \bR^{N \times M}$ be distributed according to a bi-rotationally invariant prior $P_Y(\bY)$, i.e. for any orthogonal matrices $\bU \in \bR^{N \times N}, \bV \in \bR^{M \times M}$ we have $P_Y(\bU \bY \bV^\intercal) = P_Y(\bY)$. 
We observe the data matrix $\bS \in \bR^{N \times M}$,
\begin{equation}
    \bS = \sqrt{\kappa} \bX \bY + \bW
    \label{MF-model}
\end{equation}
where $\bW \in \bR^{N \times M}$ is also bi-rotationally invariant distributed, and $\kappa \in \bR_+$ is proportional to the signal-to-noise-ratio (SNR). The goal is to recover {\it both factors} $\bX$ and $\bY$ from the data matrix $\bS$. For definiteness, we consider the regime $M \geq N$ with  aspect ratio $N/M \to \alpha\in (0,1]$ as $N \to \infty$. The case of $\alpha > 1$ can be analyzed in the same manner and is presented in section \ref{alpha>1}. 
Furthermore, we assume that the entries of $\bX, \bY$ and $\bW$ are of the order $O(\nicefrac{1}{\sqrt{N}})$. This scaling is such that the eigenvalues of $\bX$ and singular values of $\bY, \bW$ and $\bS$ are of the order $O(1)$ as $N \to \infty$. 

\begin{assumption}\label{assumption}
The empirical eigenvalue distribution of $\bX$ converge weakly to measure $\rho_X$, and the ESD of $\bY, \bW$ converge weakly to measures $\mu_Y, \mu_W$ with bounded support on the real line. Moreover, these measures are known to the statistician. He can deduce (in principle) these measures from the priors on $\bX,\bY,\bW$.
\end{assumption}


\begin{remark}\label{general MF}
In a general formulation of matrix factorization the hidden matrices have dimensions $\bX \in \bR^{N \times H}, \bY \in \bR^{H \times M}$, and in the Bayesian framework with bi-rotational invariant priors for both factors, the optimal estimators are trivially the zero matrix. Indeed, from bi-rotational invariance we have $P_X(-\bX) = P_X(\bX)$, $P_Y(-\bY) = P_Y(\bY)$, which implies that the Bayesian estimate is zero. Here, by imposing that $\bX\in \mathbb{R}^{N\times N}$ is symmetric and $P_X(\bO \bX \bO^\intercal) = P_X(\bX)$, we can break this symmetry and find non-trivial estimators. This is due to the fact that the map $\bX \to - \bX$ cannot be realized as a (real) orthogonal transformation, so 
$P_X(-\bX) = P_X(\bX)$ does not hold in general (various examples are given in section \ref{numerical} and appendices). Of course, if the prior is even, e.g. Wigner ensemble, again the Bayesian posterior estimate is trivially zero for both factors. As we will see our RIEs are consistent with these observations.

\end{remark}
\subsection{Rotation invariant estimators}
To recover matrices $\bX, \bY$ from $\bS$, we consider two denoising problems. One is recovering $\bX$ by treating both $\bY, \bW$ as "noise" matrices, and the other is estimating $\bY$ by treating $\bX, \bW$ as "noise". As will become clear the procedure is not iterative, and the two denoising problems are solved independently and simultaneously. In the following, for each of these two problems, we introduce two rotation invariant classes of estimators and discuss their optimum {\it Oracle} estimators. We then provide an explicit construction and algorithm for RIEs which we conjecture have the optimum performance of Oracle estimators in the large $N$ limit.

\subsubsection{RIE class for $\bX$}
Consider the SVD of $\bS = \bU_S \bGam \bV_S^\intercal$, where $\bU_S \in \bR^{N \times N}$, $\bV_S \in \bR^{M \times M}$ are orthogonal, and $\bGam \in \bR^{N \times M}$ is a diagonal matrix with singular values of $\bS$ on its diagonal, $\big( \gamma_i \big)_{1 \leq i \leq N}$. A rotational invaraint estimator for $\bX$ is denoted $\bm{\Xi}_X(\bS)$, and is constructed as:
\begin{equation}\label{X-RIE-class}
    \bm{\Xi}_X(\bS) = \bU_S \, 
{\rm diag}( {\xi_x}_1, \hdots, {\xi_x}_N) \, \bU_S^\intercal
\end{equation}
where ${\xi_x}_1, \hdots, {\xi_x}_N$ are the eigenvalues  of the estimator. 

First, we derive an {\it Oracle estimator} by minimizing the squared error $\frac{1}{N} \big\| \bX - \bm{\Xi}_X(\bS) \big\|_{\rm F}^2 $ for a given instance, over the RIE class or equivalently over the choice of the eigenvalues $\big( {\xi_x}_i \big)_{1 \leq i \leq N}$. Let the eigen-decomposition of $\bX$ be $\bX = \sum_{i =1}^{N} \lambda_i \,  \bx_i \bx_i^\intercal$ with $\bx_i \in \bR^N$ eigenvectors of $\bX$. The error can be expanded as:
\begin{equation*}
    \begin{split}
        \frac{1}{N} \big\| \bX - \bm{\Xi}_X(\bS) \big\|_{\rm F}^2 &= \frac{1}{N} \sum_{i=1}^N \lambda_i^2 + \frac{1}{N} \sum_{i=1}^N {\xi_x}_i^2 - \frac{2}{N} \sum_{i=1}^N {\xi_x}_i \sum_{j=1}^N \lambda_j \, \big(  \bu_i^\intercal \bx_j \big)^2
    \end{split} 
\end{equation*}
where $\bu_i$'s are columns of $\bU_S$. Minimizing over ${\xi_x}_i$'s, we find the optimum among the RIE class:
\begin{equation}
\bm{\Xi}^*_X(\bS) = \sum_{i = 1}^N {\xi_x^*}_i \, \bu_i \bu_i^\intercal, \quad
    {\xi_x^*}_i = \sum_{j=1}^N \lambda_j \, \big(  \bu_i^\intercal \bx_j \big)^2 = \bu_i^\intercal \bX \bu_i
    \label{X-oracle-estimator}
\end{equation}
Expression \eqref{X-oracle-estimator} defines the Oracle estimator which requires the knowledge of signal matrix $\bX$. Surprisingly, in the large $N$ limit, the optimal eigenvalues $\big( {\xi_x^*}_i \big)_{1 \leq i \leq N}$ can be computed from the observation matrix and knowledge of the measures $\rho_X, \mu_Y, \mu_W$. In the next section, we show that this leads to an {\it explicitly computable} (or algorithmic) RIE, which we conjecture to be optimal as $N \to \infty$, in the sense that its performance matches the one of the Oracle estimator.

Now we remark that the Oracle estimator is not only optimal within the rotation invariant class but is also Bayesian optimal.
From the Bayesian estimation point of view, one wishes to minimize the average mean squared error (MSE) 
$
{\rm MSE}_{\hat{\bX}}\equiv\frac{1}{N} \bE \big\| \bX - \hat{\bX}(\bS) \big\|_{\rm F}^2
$,
where the expectation is over $\bX, \bY, \bW$, and $\hat{\bX}(\bS)$ is an estimator of $\bX$. The MSE is minimized for $\hat{\bX}^*(\bS) = \bE [ \bX | \bS]$ which is the posterior mean. Therefore, the posterior mean estimator has the minimum MSE (MMSE) among all possible estimators, in particular ${\rm MSE}_{\hat{\bX}^*} \leq {\rm MSE}_{\bm{\Xi}^*_X}$ for any $N$. In section \ref{X-posterior-RIE}, we show that, for rotational invariant priors, the posterior mean estimator is inside the RIE class. Thus, since $\bm{\Xi}^*_X(\bS)$ is optimum among the RIE class ${\rm MSE}_{\bm{\Xi}^*_X} \leq {\rm MSE}_{\hat{\bX}^*}$. Therefore, we conclude that the Oracle estimator \eqref{X-oracle-estimator} is Bayesian optimal in the sense that ${\rm MSE}_{\bm{\Xi}^*_X} = {\rm MSE}_{\hat{\bX}^*} = {\rm MMSE}$.

\subsubsection{RIE class for $\bY$}
Estimators for $\bY$ from the rotation invariant class are denoted $\bm{\Xi}_Y(\bS)$, and are constructed as:
\begin{equation}\label{Y-RIE-class}
    \bm{\Xi}_Y(\bS) = \bU_S
\left[
\begin{array}{c|c}
{\rm diag}( {\xi_y}_1, \hdots, {\xi_y}_N) & \mathbf{0}_{N \times (M-N)}
\end{array}
\right] \bV_S^\intercal
\end{equation}
where ${\xi_y}_1, \hdots, {\xi_y}_N$ are the singular values of the estimator. 

Let the SVD of $\bY$ be $\bY = \sum_{i =1}^{N} \sigma_i \, \by_i^{(l)} {\by_i^{(r)}}^\intercal$ with $\by_i^{(l)} \in \bR^N, \by_i^{(r)} \in \bR^M$ the left and right singular vectors of $\bY$. To derive an {\it Oracle estimator}, we proceed as above. Expanding the error, we have:
\begin{equation*}
    \begin{split}
        \frac{1}{N} \big\| \bY - \bm{\Xi}_Y(\bS) \big\|_{\rm F}^2 &= \frac{1}{N} \sum_{i=1}^N \sigma_i^2 + \frac{1}{N} \sum_{i=1}^N {\xi_y}_i^2 - \frac{2}{N} \sum_{i=1}^N {\xi_y}_i \sum_{j=1}^N \sigma_j \, \big(  \bu_i^\intercal \by_j^{(l)}\big) \big(  \bv_i^\intercal \by_j^{(r)} \big)
    \end{split} 
\end{equation*}
where $\bv_i$'s are columns of $\bV_S$. Minimizing over ${\xi_y}_i$'s, we find the optimum among the RIE class:
\begin{equation}
\bm{\Xi}^*_Y(\bS) = \sum_{i = 1}^N {\xi_y^*}_i \, \bu_i \bv_i^\intercal, \quad
    {\xi_y^*}_i = \sum_{j=1}^N \sigma_j \, \big(  \bu_i^\intercal \by_j^{(l)}\big) \big(  \bv_i^\intercal \by_j^{(r)} \big) = \bu_i^\intercal \bY \bv_i
    \label{Y-oracle-estimator}
\end{equation}
Expression \eqref{Y-oracle-estimator} defines the Oracle estimator which requires the knowledge of signal matrix $\bY$.
Like for the case of $\bX$, in the large $N$ limit we can derive  the optimal singular values $\big( {\xi_y^*}_i \big)_{1 \leq i \leq N}$ in terms of the singular values of observation matrix and knowledge of the measures $\rho_X, \mu_W$. 
This leads to an {\it explicitly computable} (or algorithmic) RIE, which is conjectured to be optimal as $N \to \infty$, in the sense that it has the same performance as the Oracle estimator. Note that unlike the estimator for $\bX$, we do not need the knowledge of $\mu_Y$. 

In section \ref{Y-posterior-RIE}, we show that for bi-rotationally invariant priors the posterior mean estimator $\hat{\bY}^*(\bS) = \bE [ \bY | \bS]$ belongs to the RIE class, which (by similar arguments to the case of $\bX$) implies that the Oracle estimator \eqref{Y-oracle-estimator} is Bayesian optimal.
\section{Algorithmic RIEs for the matrix factors}\label{propsoed-RIE}
In this section, we present our explicit RIEs for $\bX, \bY$ and the corresponding algorithm. We conjecture that their performance matches the one of Oracles estimators in the large $N$ limit and they are therefore Bayesian optimal in this limit. Let us first give a brief reminder on useful transforms in random matrix theory.

\subsection{Preliminaries on transforms in random matrix theory}
For a probability density function $\rho(x)$ on $\bR$, the \textit{Stieltjes}  (or \textit{Cauchy}) transform  is defined as 
\begin{equation*}
   \mathcal{G}_\rho (z) = \int_{\bR} \frac{1}{z - x} \rho(x) \, dx \hspace{10 pt} \text{for } z \in \mathbb{C} \backslash {\rm supp}(\rho)
\end{equation*}
By  Plemelj formulae we have for $x\in \mathbb{R}$,
\begin{equation}
    \lim_{\epsilon \to 0^+} \mathcal{G}_\rho(x - \ci \epsilon) = \pi \sH [\rho](x) + \pi \ci \rho(x) 
    \label{Plemelj formulae}
\end{equation}
with $\sH [\rho](x) = {\rm p.v.} \frac{1}{\pi} \int_{\bR} \frac{\rho(t)}{x - t}  d t$ the \textit{Hilbert} transform of $\rho$ (here ${\rm p.v.}$ stands for "principal value"). Denoting the inverse of $\mathcal{G}_\rho (z)$ by $\mathcal{G}_\rho^{-1} (z)$, the \textit{R-transform} of $\rho$ is defined as \cite{mingo2017free}:
\begin{equation*}
    \mathcal{R}_{\rho}(z) = \mathcal{G}_\rho^{-1} (z) - \frac{1}{z}
\end{equation*}
For a probability density function $\mu$ with support contained in$[-K, K ]$ with $K>0$, we define a generating function of (even) moments $\mathcal{M}_{\mu} : [0,K^{-2}] \to \bR_+$
as
$
    \mathcal{M}_{\mu} (z) = \int \frac{1}{1-t^2 z} \mu(t) \, d t - 1
$. For $\alpha \in (0,1]$, define $T^{(\alpha)}(z) = (\alpha z +1)(z+1)$, and $\mathcal{H}_{\mu}^{(\alpha)}(z) = z T^{(\alpha)}\big(\mathcal{M}_{\mu} (z)\big)$. The \textit{rectangular R-transform} with aspect ratio $\alpha$ is defined as \cite{benaych2009rectangular}:
\begin{equation*}
    \mathcal{C}_{\mu}^{(\alpha)}(z) = {T^{(\alpha)}}^{-1}\Big(\frac{z}{{\mathcal{H}_{\mu}^{(\alpha)}}^{-1}(z)} \Big)
\end{equation*}

\subsection{Explicit RIE for $\bX$}\label{X-RIE-explicit}
The RIE for $\bX$ is constructed as $\widehat{\bm{\Xi}^*_X}(\bS) = \sum_{i = 1}^N \widehat{{\xi}_x^*}_i \bu_i \bu_i^\intercal$ with eigenvalues $\big( \widehat{{\xi}_x^*}_i \big)_{1 \leq i \leq N}$ :
\begin{equation}
    \widehat{{\xi}_x^*}_i = \frac{1}{2 \kappa \pi \bar{\mu}_{S}(\gamma_i) } \, {\rm Im}  \, \lim_{z \to \gamma_i - \ci 0^+}  \,  \bigg\{  \frac{1}{\zeta_3}  \Big[\mathcal{G}_{\rho_X}\Big(\sqrt{\frac{z-\zeta_1}{\kappa \zeta_3}}\Big) +  \mathcal{G}_{\rho_{X}}\Big(-\sqrt{\frac{z-\zeta_1}{\kappa \zeta_3}}\Big) \Big] \bigg\}
\label{X-optimal-ev}
\end{equation}
where $\gamma_i$ is the $i$-th singular value of $\bS$, $\bar{\mu}_{S}$ is the symmetrized limiting ESD of $\bS$, and 
\begin{equation}
    \zeta_1 = \frac{1}{\mathcal{G}_{\bar{\mu}_S} (z)} \mathcal{C}_{\mu_W}^{(\alpha)} \Big( \mathcal{G}_{\bar{\mu}_S} (z) \big[\alpha \mathcal{G}_{\bar{\mu}_S} (z) + \frac{1-\alpha}{z} \big] \Big)
    \label{zeta1-X-est}
\end{equation}
and $\zeta_3$ satisfies \footnote{$\zeta_1, \zeta_3$ are the only parameters which appear in the final estimator. However,
in derivation of the RIE, we have defined other parameters which do not appear in the final estimator and we omit them here.}:
\begin{equation}
    (z - \zeta_1) \mathcal{G}_{\bar{\mu}_S} (z) - 1 = \mathcal{C}_{\mu_Y}^{(\alpha)}\Big( \frac{1}{\zeta_3} \big[\alpha \mathcal{G}_{\bar{\mu}_S} (z) + \frac{1-\alpha}{z} \big] \big[(z - \zeta_1) \mathcal{G}_{\bar{\mu}_S} (z) - 1 \big] \Big)
    \label{zeta3-X-est}
\end{equation}

\begin{remark}\label{symmetrix-rho}
If $\rho_X$ is a symmetric measure, $\rho_X(x) = \rho_X(-x)$, then $\mathcal{G}_{\rho_{X}}(-z) = - \mathcal{G}_{\rho_{X}}(z)$. This implies that the optimal eigenvalues $\big( \widehat{{\xi}_x^*}_i \big)_{1 \leq i \leq N}$ in \eqref{X-optimal-ev} are all zero, and $\widehat{\bm{\Xi}^*_X}(\bS) = \bm{0}$, see figure \ref{fig:X-Overlap-Wigner X}.
\end{remark}

\subsubsection{An estimator for $\bX^2$}
It is interesting to note that we can construct a RIE for $\bX^2$ as $\widehat{\bm{\Xi}^*_{X^2}}(\bS) = \sum_{i = 1}^N \widehat{{\xi}_{x^2}^*}_i \bu_i \bu_i^\intercal$ with eigenvalues $\big( \widehat{{\xi}_{x^2}^*}_i \big)_{1 \leq i \leq N}$:
\begin{equation}
    \widehat{\xi_{x^2}^*}_i = \frac{1}{\kappa} \frac{1}{\pi \bar{\mu}_{S}(\gamma_i) } \, {\rm Im} \, \lim_{z \to \gamma_i - \ci 0^+}  \, \frac{z - \zeta_1}{\zeta_3} \mathcal{G}_{\bar{\mu}_S} (z) - \frac{1}{\zeta_3} 
    \label{X2-optimal-ev}
\end{equation}
with $\zeta_1,\zeta_3$ as in \eqref{zeta1-X-est}, \eqref{zeta3-X-est}. Note that, $\zeta_1,\zeta_3$ can be evaluated using the observation matrix and the knowledge of $\mu_Y, \mu_W$, and therefore this time the statistician {\it does not need to know the prior of} $\bX$. Furthermore, assuming that $\bX$ is positive semi-definite (PSD), we can construct a sub-optimal RIE for $\bX$ by using  $ \sqrt{\widehat{\xi_{x^2}^*}_i} $ for the eigenvalues of the estimator.

\subsubsection{Case of Gaussian $\bY, \bW$}
If $\bY, \bW$ have i.i.d. Gaussian entries with variance $\nicefrac{1}{N}$, then $\mathcal{C}_{\mu_Y}^{(\alpha)}(z) = \mathcal{C}_{\mu_W}^{(\alpha)}(z) = \nicefrac{z}{\alpha}$. Consequently, $\zeta_1,\zeta_3$ can easily be computed to be $\zeta_1=\zeta_3=\mathcal{G}_{\bar{\mu}_S} (z) + \nicefrac{(1 - \alpha)}{(\alpha z)}$, thus the estimator \eqref{X-optimal-ev} can be evaluated from the observation matrix. In particular, the estimator \eqref{X2-optimal-ev}  simplifies to:
\begin{equation}
    \widehat{\xi_{x^2}^*}_i= \frac{1}{\kappa} \, \Bigg[ -1 + \frac{1}{\alpha \Big( \pi^2 \bar{\mu}_{S}(\gamma_i)^2 + \big(\pi \sH [\bar{\mu}_S](\gamma_i) + \frac{1 - \alpha}{\alpha\gamma_i} \big)^2 \Big)   } \Bigg]
    \label{X2-optimal-ev-Y,W-G-main}
\end{equation}

\subsection{Explicit RIE for $\bY$}\label{Y-RIE-explicit}
 Our explicit RIE for $\bY$ is constructed as $\widehat{\bm{\Xi}^*_Y}(\bS) = \sum_{i = 1}^N \widehat{{\xi}_y^*}_i \bu_i \bv_i^\intercal$ with singular values $\big( \widehat{{\xi}_y^*}_i \big)_{1 \leq i \leq N}$:
\begin{equation}
    \widehat{{\xi}_y^*}_i = \frac{1}{\sqrt{\kappa}} \frac{1}{\pi \bar{\mu}_{S}(\gamma_i)} \,  {\rm Im} \,  \lim_{z \to \gamma_i - \ci 0^+} \,  q_4 
    \label{Y-optimal-sv}
\end{equation}
where $\gamma_i$ is the $i$-th singular value of $\bS$, and $q_4$ is the solution to the following system of equations \footnote{Like the case for $\bX$, we omit some of the parameters which do not appear in the final estimator.}:
\begin{equation}
    \begin{cases}
    \beta_1 = \frac{\mathcal{C}_{\mu_W}^{(\alpha)}(q_1 q_2)}{q_1} + \frac{1}{2} \sqrt{\frac{q_3}{q_1}} \Big( \mathcal{R}_{\rho_X} \big( q_4 + \sqrt{q_1 q_3} \big) - \mathcal{R}_{\rho_X} \big( q_4 - \sqrt{q_1 q_3} \big) \Big)\\
    \beta_4 = \frac{1}{2}  \Big( \mathcal{R}_{\rho_X} \big( q_4 + \sqrt{q_1 q_3} \big) + \mathcal{R}_{\rho_X} \big( q_4 - \sqrt{q_1 q_3} \big) \Big)\\
    q_1 = \mathcal{G}_{\bar{\mu}_S}(z), \quad q_2 = \alpha \mathcal{G}_{\bar{\mu}_S}(z) + (1- \alpha ) \frac{1}{z} \\
    q_3 = \frac{(z - \beta_1)^2}{\beta_4^2} \mathcal{G}_{\bar{\mu}_S}(z) - \frac{z - \beta_1}{\beta_4^2}, \quad q_4 = \frac{z - \beta_1}{\beta_4} \mathcal{G}_{\bar{\mu}_S}(z) - \frac{1}{\beta_4}
    \end{cases} 
    \label{Y-sol}
\end{equation}
Similarly to the estimator derived for $\bX$, if $\rho_X$ is a symmetric measure then the optimal singular values for the estimator of $\bY$ are all zero, see remark \ref{symmetric-rho-X-Y-RIE}.

If $\bX$ is a shifted Wigner matrix, i.e. $\bX = \bF + c \bI$ with $\bF = \bF^\intercal \in \bR^{N \times N}$ having i.i.d. Gaussian entries with variance $\nicefrac{1}{N}$ and $c \neq 0$  a real number, then $\mathcal{R}_{\rho_X}(z) = z +c$. Moreover, if $\bW$ is Gaussian matrix with variance $\nicefrac{1}{N}$, then the set of equations \eqref{Y-sol} simplifies to a great extent, and we can compute $q_4$ analytically in terms of $\mathcal{G}_{\bar{\mu}_S}(z)$, see section \ref{Y-examples}.




\subsection{Algorithmic nature of the RIEs}
The explicit RIEs \eqref{X-optimal-ev} and \eqref{Y-optimal-sv} proposed in this section, provide  spectral algorithms to estimate the matrix factors from the data matrix (and the priors).
An essential ingredient that must be extracted from the data matrix is $\mathcal{G}_{\bar{\mu}_S}(z)$. This quantity can be approximated from the observation matrix using Cauchy kernel method introduced in \cite{potters2020first}(see section 19.5.2), from which $\bar{\mu}_S(.)$ can be approximated using \eqref{Plemelj formulae}. Therefore, given an observation matrix $\bS$, the spectral algorithm proceeds as follows:
\begin{enumerate}
    \item Compute the SVD of $\bS$.
    \item Approximate $\mathcal{G}_{\bar{\mu}_S}(z)$ from the singular values of $\bS$.
    \item Construct the RIEs for $\bX, \bY$ as proposed in paragraphs \ref{X-RIE-explicit}, \ref{Y-RIE-explicit}.
\end{enumerate}
\section{Numerical results}\label{numerical}


\subsection{Performance of RIE for $\bX$}
We consider the case where $\bY, \bW$ both have i.i.d. Gaussian entries of variance $\nicefrac{1}{N}$, and $\bX$ is a Wishart matrix, $\bX = \bH \bH^\intercal$ with $\bH \in \bR^{N \times 4N}$ having i.i.d. Gaussian entries of variance $\nicefrac{1}{N}$. For various SNRs, we examine the performance of two proposed estimators, the RIE \eqref{X-optimal-ev}, and the square-root of the estimator \eqref{X2-optimal-ev} (since $\bX$ is PSD), which is sub-optimal.
In figure \ref{fig:X-RIE-Wishart}, the MSEs of these algorithmic estimators are compared with the one of Oracle estimator \eqref{X-oracle-estimator}. We see that the average performance of the algorithmic RIE $\widehat{{\bm{\Xi}_X^*}}(\bS)$ is very close to the (optimal) Oracle estimator ${\bm{\Xi}_X^*}(\bS)$ (relative errors are small and provided in the appendices) and we believe that the slight mismatch is due to the numerical approximations and finite-size effects. Note that, although the estimator \small $\sqrt{\widehat{{\bm{\Xi}_{X^2}^*}}(\bS)}$ \normalsize is sub-optimal, it does not use any prior knowledge of $\bX$. For more examples, details of the numerical experiments and the relative error of the estimators, we refer to section \ref{X-examples}.

\begin{figure}
\centering
\begin{minipage}[t]{.48\textwidth}
  \centering
\begin{tikzpicture}[scale = 0.55]

\definecolor{darkgray176}{RGB}{176,176,176}
\definecolor{lightcoral}{RGB}{240,128,128}
\definecolor{lightgreen}{RGB}{144,238,144}
\definecolor{lightslategray}{RGB}{119,136,153}
\definecolor{royalblue}{RGB}{65,105,225}
\definecolor{seagreen}{RGB}{46,139,87}

\begin{axis}[
legend cell align={left},
legend style={fill opacity=0.8, draw opacity=1, text opacity=1, draw=white!80!black},
tick align=outside,
tick pos=left,
x grid style={darkgray176},
xlabel={$ \kappa $},
xmin=-0.145, xmax=5.245,
xtick style={color=black},
y grid style={darkgray176},
ylabel={MSE},
yticklabel style={
  /pgf/number format/precision=3,
  /pgf/number format/fixed},
ymin=0.0608968192932828, ymax=0.131350065473814,
ytick style={color=black}
]
\path [draw=lightcoral, semithick]
(axis cs:0.1,0.11597352696668)
--(axis cs:0.1,0.116226286174364);

\path [draw=lightcoral, semithick]
(axis cs:0.3,0.0858004833813655)
--(axis cs:0.3,0.085934401854847);

\path [draw=lightcoral, semithick]
(axis cs:0.6,0.0751090320297627)
--(axis cs:0.6,0.0752794051870071);

\path [draw=lightcoral, semithick]
(axis cs:1,0.0702933183594271)
--(axis cs:1,0.0704626460099733);

\path [draw=lightcoral, semithick]
(axis cs:2,0.0665407707982209)
--(axis cs:2,0.0666836105777739);

\path [draw=lightcoral, semithick]
(axis cs:3,0.0651400173923888)
--(axis cs:3,0.0653269798137931);

\path [draw=lightcoral, semithick]
(axis cs:4,0.0644517949655576)
--(axis cs:4,0.0645970281938319);

\path [draw=lightcoral, semithick]
(axis cs:5,0.0640992395742161)
--(axis cs:5,0.064272661692868);

\path [draw=lightslategray, semithick]
(axis cs:0.1,0.117182008302904)
--(axis cs:0.1,0.117473960098773);

\path [draw=lightslategray, semithick]
(axis cs:0.3,0.0864583860418636)
--(axis cs:0.3,0.0866594082830361);

\path [draw=lightslategray, semithick]
(axis cs:0.6,0.0758041347484414)
--(axis cs:0.6,0.0760078225706573);

\path [draw=lightslategray, semithick]
(axis cs:1,0.0709352696450034)
--(axis cs:1,0.0711644066024661);

\path [draw=lightslategray, semithick]
(axis cs:2,0.0671669121333483)
--(axis cs:2,0.0673852905425038);

\path [draw=lightslategray, semithick]
(axis cs:3,0.0658746868555549)
--(axis cs:3,0.0660754618122143);

\path [draw=lightslategray, semithick]
(axis cs:4,0.0652096219585044)
--(axis cs:4,0.0654422904068857);

\path [draw=lightslategray, semithick]
(axis cs:5,0.0650395395714377)
--(axis cs:5,0.0652097566820946);

\path [draw=lightgreen, semithick]
(axis cs:0.1,0.127857616228765)
--(axis cs:0.1,0.128147645192881);

\path [draw=lightgreen, semithick]
(axis cs:0.3,0.0906377780377246)
--(axis cs:0.3,0.0907596986850807);

\path [draw=lightgreen, semithick]
(axis cs:0.6,0.078360828879579)
--(axis cs:0.6,0.078517069849001);

\path [draw=lightgreen, semithick]
(axis cs:1,0.0731690751296091)
--(axis cs:1,0.0733393839785486);

\path [draw=lightgreen, semithick]
(axis cs:2,0.0690622773897066)
--(axis cs:2,0.0691960380512968);

\path [draw=lightgreen, semithick]
(axis cs:3,0.0676585519485961)
--(axis cs:3,0.0678269439007362);

\path [draw=lightgreen, semithick]
(axis cs:4,0.0671052684798546)
--(axis cs:4,0.0672516785190221);

\path [draw=lightgreen, semithick]
(axis cs:5,0.067000845616663)
--(axis cs:5,0.0671602188686105);

\addplot [semithick, red, mark=triangle*, mark size=3, mark options={solid,rotate=180}, only marks]
table {%
0.1 0.116099906570522
0.3 0.0858674426181063
0.6 0.0751942186083849
1 0.0703779821847002
2 0.0666121906879974
3 0.0652334986030909
4 0.0645244115796947
5 0.064185950633542
};
\addlegendentry{Oracle estimator, ${\bm{\Xi}_X^*}(\bS)$}
\addplot [semithick, royalblue, mark=triangle*, mark size=3, mark options={solid}, only marks]
table {%
0.1 0.117327984200839
0.3 0.0865588971624499
0.6 0.0759059786595493
1 0.0710498381237347
2 0.067276101337926
3 0.0659750743338846
4 0.0653259561826951
5 0.0651246481267661
};
\addlegendentry{RIE, $\widehat{{\bm{\Xi}_X^*}}(\bS)$}
\addplot [semithick, seagreen, mark=x, mark size=3, mark options={solid}, only marks]
table {%
0.1 0.128002630710823
0.3 0.0906987383614027
0.6 0.07843894936429
1 0.0732542295540789
2 0.0691291577205017
3 0.0677427479246661
4 0.0671784734994383
5 0.0670805322426368
};
\addlegendentry{$\sqrt{\widehat{{\bm{\Xi}_{X^2}^*}}(\bS)}$}
\end{axis}

\end{tikzpicture}
   \caption{\small  MSE of estimating $\bX$. MSE is normalized by the norm of the signal, $\| \bX \|_{\rm F}^2$. $\bX$ is a Wishart matrix with aspect ratio $\nicefrac{1}{4}$, $\bX = \bH \bH^\intercal$ with $\bH \in \bR^{N \times 4N}$ having i.i.d. Gaussian entries of variance $\nicefrac{1}{N}$. Both $\bY$ and $\bW$ are $N \times M$ matrices with i.i.d. Gaussian entries of variance $\nicefrac{1}{N}$. RIE is applied to $N=2000, M =4000$, and the results are averaged over 10 runs (error bars are invisible).}
   \label{fig:X-RIE-Wishart}
\end{minipage}%
\hfill
\begin{minipage}[t]{.48\textwidth}
  \centering
\begin{tikzpicture}[scale = 0.55]

\definecolor{darkgray176}{RGB}{176,176,176}
\definecolor{lightcoral}{RGB}{240,128,128}
\definecolor{lightslategray}{RGB}{119,136,153}
\definecolor{royalblue}{RGB}{65,105,225}

\begin{axis}[
legend cell align={left},
legend style={fill opacity=0.8, draw opacity=1, text opacity=1, draw=white!80!black},
tick align=outside,
tick pos=left,
x grid style={darkgray176},
xlabel={$\kappa$},
xmin=-0.145, xmax=5.245,
xtick style={color=black},
y grid style={darkgray176},
ylabel={${\rm MSE}$},
scaled y ticks=false,
yticklabel style={
  /pgf/number format/precision=3,
  /pgf/number format/fixed},
ymin=0.0432528431531933, ymax=0.39457092906823,
ytick style={color=black}
]
\path [draw=lightcoral, semithick]
(axis cs:0.1,0.373512909726238)
--(axis cs:0.1,0.378437895486202);

\path [draw=lightcoral, semithick]
(axis cs:0.3,0.195568222689053)
--(axis cs:0.3,0.198455746381194);

\path [draw=lightcoral, semithick]
(axis cs:0.6,0.129719083279093)
--(axis cs:0.6,0.130959578976959);

\path [draw=lightcoral, semithick]
(axis cs:1,0.0999373844438707)
--(axis cs:1,0.101667574902742);

\path [draw=lightcoral, semithick]
(axis cs:2,0.0749482719476051)
--(axis cs:2,0.0762414479090259);

\path [draw=lightcoral, semithick]
(axis cs:3,0.0662698130081426)
--(axis cs:3,0.0682014561186748);

\path [draw=lightcoral, semithick]
(axis cs:4,0.062168909249415)
--(axis cs:4,0.0630242993297389);

\path [draw=lightcoral, semithick]
(axis cs:5,0.0592218470584223)
--(axis cs:5,0.0604937979516368);

\path [draw=lightslategray, semithick]
(axis cs:0.1,0.373668314553029)
--(axis cs:0.1,0.378601925163001);

\path [draw=lightslategray, semithick]
(axis cs:0.3,0.195673773538471)
--(axis cs:0.3,0.198554358375894);

\path [draw=lightslategray, semithick]
(axis cs:0.6,0.129836791528522)
--(axis cs:0.6,0.131071783500439);

\path [draw=lightslategray, semithick]
(axis cs:1,0.10008161988262)
--(axis cs:1,0.101814811169172);

\path [draw=lightslategray, semithick]
(axis cs:2,0.0751747889504974)
--(axis cs:2,0.0764695398504073);

\path [draw=lightslategray, semithick]
(axis cs:3,0.0665717243180533)
--(axis cs:3,0.0685106206536402);

\path [draw=lightslategray, semithick]
(axis cs:4,0.0625469464427706)
--(axis cs:4,0.0634091398549601);

\path [draw=lightslategray, semithick]
(axis cs:5,0.0596922287297291)
--(axis cs:5,0.0609637801713104);

\addplot [semithick, red, mark=triangle*, mark size=3, mark options={solid,rotate=180}, only marks]
table {%
0.1 0.37597540260622
0.3 0.197011984535123
0.6 0.130339331128026
1 0.100802479673306
2 0.0755948599283155
3 0.0672356345634087
4 0.062596604289577
5 0.0598578225050295
};
\addlegendentry{Oracle estimator, ${\bm{\Xi}_Y^*}(\bS)$}
\addplot [semithick, royalblue, mark=triangle*, mark size=3, mark options={solid}, only marks]
table {%
0.1 0.376135119858015
0.3 0.197114065957182
0.6 0.13045428751448
1 0.100948215525896
2 0.0758221644004523
3 0.0675411724858467
4 0.0629780431488653
5 0.0603280044505198
};
\addlegendentry{RIE, $\widehat{{\bm{\Xi}_Y^*}}(\bS)$}
\end{axis}

\end{tikzpicture}
   \caption{\small  MSE of estimating $\bY$. MSE is normalized by the norm of the signal, $\| \bY \|_{\rm F}^2$. $\bY$ has uniform spectral density, $\mathcal{U}\big([1,3] \big)$. $\bX$ is a shifted Wigner matrix with $c = 3$, and  $\bW$ is a $N \times M$ matrix with i.i.d. Gaussian entries of variance $\nicefrac{1}{N}$. RIE is applied to $N=2000, M =4000$, and the results are averaged over 10 runs (error bars are invisible).}
   \label{fig:Y-RIE-Uniform}
\end{minipage}
\end{figure}

\subsection{Performance of RIE for $\bY$}
We consider the case where $\bW$ has i.i.d. Gaussian entries of variance $\nicefrac{1}{N}$, and $\bX$ is a shifted Wigner matrix with $c=3$. Matrix $\bY$ is constructed as $\bY = \bU_Y \bSig \bV_Y^\intercal$ with $\bU_Y \in \bR^{N \times N}, \bV_Y \in \bR^{M \times M}$ are Haar distributed, and the singular values are generated independently from the uniform distribution on $[1,3]$. MSEs of the RIE \eqref{Y-optimal-sv} and the Oracle estimator \eqref{Y-oracle-estimator} are illustrated in figure \ref{fig:Y-RIE-Uniform}. We see that the performance of the algorithmic RIE $\widehat{{\bm{\Xi}_Y^*}}(\bS)$  is very close to the optimal estimator ${\bm{\Xi}_Y^*}(\bS)$.

\paragraph{Non-rotational invariant prior} In another example, which we omit here, with the same settings for $\bX, \bW$, we consider the case where $\bY$ is a sparse matrix with entries distributed according to Bernoulli-Rademacher prior. The RIE is not optimal in this setting (since the prior is not bi-rotational invariant), however applying a simple thresholding function on the matrix constructed by RIE yields an estimate with lower MSE. This observation suggests that for the case of general priors, the RIEs can provide a spectral initialization for more efficient estimators. For more details and examples, see section \ref{Y-examples}.

\subsection{Comparing RIEs of matrix factorization and denoising}
The proposed RIEs, namely \eqref{X-optimal-ev} and \eqref{Y-optimal-sv}, simplify greatly when the matrices $\bW, \bY$ are Gaussian, and $\bX$ is a shifted Wigner matrix. We perform experiments with these priors, where for a given observation matrix $\bS$, we look at the RIEs of $\bX$, $\bY$ for the {\it MF problem}, and simultaneously at the RIE of the product $\bX\bY$ as a whole for the {\it denoising problem} with formulas introduced in \cite{pourkamali2023rectangular} (which can also be obtained by taking  $\bX$ to be the identity matrix, see section \ref{rect-RIE}). Figure \ref{MF-c=1-main} illustrates these experiments. In particular the MSE of the denoising-RIE matches well the one of the associated Oracle estimator, and as expected is lower than the MSE of the product of MF-RIEs. 
\begin{figure}
\begin{subfigure}[t]{.32\textwidth}
    \centering
\begin{tikzpicture}[scale = 0.5]

\definecolor{darkgray176}{RGB}{176,176,176}
\definecolor{lightcoral}{RGB}{240,128,128}
\definecolor{lightslategray}{RGB}{119,136,153}
\definecolor{royalblue}{RGB}{65,105,225}

\begin{axis}[
legend cell align={left},
legend style={fill opacity=0.8, draw opacity=1, text opacity=1, draw=white!80!black},
tick align=outside,
tick pos=left,
x grid style={darkgray176},
xlabel={$\kappa$},
xmin=-0.145, xmax=5.245,
xtick style={color=black},
y grid style={darkgray176},
ylabel={${\rm MSE}$},
scaled y ticks=false,
yticklabel style={
  /pgf/number format/precision=3,
  /pgf/number format/fixed},
ymin=0.158989648894357, ymax=0.489776450653492,
ytick style={color=black}
]
\path [draw=lightcoral, semithick]
(axis cs:0.1,0.472679862890143)
--(axis cs:0.1,0.47369940142939);

\path [draw=lightcoral, semithick]
(axis cs:0.3,0.394383432704544)
--(axis cs:0.3,0.395189230169388);

\path [draw=lightcoral, semithick]
(axis cs:0.6,0.322359971106662)
--(axis cs:0.6,0.323545953558721);

\path [draw=lightcoral, semithick]
(axis cs:1,0.271617606398936)
--(axis cs:1,0.272463997304963);

\path [draw=lightcoral, semithick]
(axis cs:2,0.217437387501637)
--(axis cs:2,0.21811034000889);

\path [draw=lightcoral, semithick]
(axis cs:3,0.194396584030318)
--(axis cs:3,0.195341349525913);

\path [draw=lightcoral, semithick]
(axis cs:4,0.182018974565411)
--(axis cs:4,0.18282015642648);

\path [draw=lightcoral, semithick]
(axis cs:5,0.174025412610681)
--(axis cs:5,0.174762125551437);

\path [draw=lightslategray, semithick]
(axis cs:0.1,0.47377559294999)
--(axis cs:0.1,0.474740686937167);

\path [draw=lightslategray, semithick]
(axis cs:0.3,0.395400823592722)
--(axis cs:0.3,0.396203440825488);

\path [draw=lightslategray, semithick]
(axis cs:0.6,0.32478470995362)
--(axis cs:0.6,0.32586208415824);

\path [draw=lightslategray, semithick]
(axis cs:1,0.273584125988031)
--(axis cs:1,0.274333828104676);

\path [draw=lightslategray, semithick]
(axis cs:2,0.218852589334871)
--(axis cs:2,0.219525971584205);

\path [draw=lightslategray, semithick]
(axis cs:3,0.196200360549737)
--(axis cs:3,0.197066901589207);

\path [draw=lightslategray, semithick]
(axis cs:4,0.183443623804713)
--(axis cs:4,0.184262054496322);

\path [draw=lightslategray, semithick]
(axis cs:5,0.175366459606412)
--(axis cs:5,0.176170039823591);

\addplot [semithick, red, mark=triangle*, mark size=3, mark options={solid,rotate=180}, only marks]
table {%
0.1 0.473189632159767
0.3 0.394786331436966
0.6 0.322952962332692
1 0.272040801851949
2 0.217773863755263
3 0.194868966778116
4 0.182419565495946
5 0.174393769081059
};
\addlegendentry{Oracle estimator, ${\bm{\Xi}_X^*}(\bS)$}
\addplot [semithick, royalblue, mark=triangle*, mark size=3, mark options={solid}, only marks]
table {%
0.1 0.474258139943579
0.3 0.395802132209105
0.6 0.32532339705593
1 0.273958977046354
2 0.219189280459538
3 0.196633631069472
4 0.183852839150517
5 0.175768249715002
};
\addlegendentry{RIE, $\widehat{{\bm{\Xi}_X^*}}(\bS)$}
\end{axis}

\end{tikzpicture}
    \caption{\small Estimating $\bX$}
\end{subfigure}
\hfill
\begin{subfigure}[t]{.32\textwidth}
    \centering
\begin{tikzpicture}[scale = 0.5]

\definecolor{darkgray176}{RGB}{176,176,176}
\definecolor{lightcoral}{RGB}{240,128,128}
\definecolor{lightslategray}{RGB}{119,136,153}
\definecolor{royalblue}{RGB}{65,105,225}

\begin{axis}[
legend cell align={left},
legend style={fill opacity=0.8, draw opacity=1, text opacity=1, draw=white!80!black},
tick align=outside,
tick pos=left,
x grid style={darkgray176},
xlabel={$\kappa$},
xmin=-0.145, xmax=5.245,
xtick style={color=black},
y grid style={darkgray176},
ylabel={${\rm MSE}$},
scaled y ticks=false,
yticklabel style={
  /pgf/number format/precision=3,
  /pgf/number format/fixed},
ymin=0.511630570465254, ymax=0.934756886300313,
ytick style={color=black}
]
\path [draw=lightcoral, semithick]
(axis cs:0.1,0.913534806416141)
--(axis cs:0.1,0.914054528076912);

\path [draw=lightcoral, semithick]
(axis cs:0.3,0.801744489184537)
--(axis cs:0.3,0.802333687334179);

\path [draw=lightcoral, semithick]
(axis cs:0.6,0.714432593344145)
--(axis cs:0.6,0.71524679242528);

\path [draw=lightcoral, semithick]
(axis cs:1,0.653954871657974)
--(axis cs:1,0.654674672043728);

\path [draw=lightcoral, semithick]
(axis cs:2,0.587440039833787)
--(axis cs:2,0.588400318916892);

\path [draw=lightcoral, semithick]
(axis cs:3,0.557954986361824)
--(axis cs:3,0.559011283611603);

\path [draw=lightcoral, semithick]
(axis cs:4,0.541548775105845)
--(axis cs:4,0.542902957038355);

\path [draw=lightcoral, semithick]
(axis cs:5,0.530863584821393)
--(axis cs:5,0.531884750800902);

\path [draw=lightslategray, semithick]
(axis cs:0.1,0.915126009006564)
--(axis cs:0.1,0.915523871944174);

\path [draw=lightslategray, semithick]
(axis cs:0.3,0.802308699980285)
--(axis cs:0.3,0.802887691176402);

\path [draw=lightslategray, semithick]
(axis cs:0.6,0.715408417207449)
--(axis cs:0.6,0.716211006714881);

\path [draw=lightslategray, semithick]
(axis cs:1,0.654855088789837)
--(axis cs:1,0.655562701787716);

\path [draw=lightslategray, semithick]
(axis cs:2,0.588214340720012)
--(axis cs:2,0.58915826130104);

\path [draw=lightslategray, semithick]
(axis cs:3,0.559067528213006)
--(axis cs:3,0.560077743866405);

\path [draw=lightslategray, semithick]
(axis cs:4,0.542554740628416)
--(axis cs:4,0.543893438153047);

\path [draw=lightslategray, semithick]
(axis cs:5,0.531831453314418)
--(axis cs:5,0.53284388897055);

\addplot [semithick, red, mark=triangle*, mark size=3, mark options={solid,rotate=180}, only marks]
table {%
0.1 0.913794667246526
0.3 0.802039088259358
0.6 0.714839692884712
1 0.654314771850851
2 0.587920179375339
3 0.558483134986713
4 0.5422258660721
5 0.531374167811148
};
\addlegendentry{Oracle estimator, ${\bm{\Xi}_Y^*}(\bS)$}
\addplot [semithick, royalblue, mark=triangle*, mark size=3, mark options={solid}, only marks]
table {%
0.1 0.915324940475369
0.3 0.802598195578344
0.6 0.715809711961165
1 0.655208895288777
2 0.588686301010526
3 0.559572636039705
4 0.543224089390731
5 0.532337671142484
};
\addlegendentry{RIE, $\widehat{{\bm{\Xi}_Y^*}}(\bS)$}
\end{axis}

\end{tikzpicture}
    \caption{\small Estimating $\bY$}
\end{subfigure}
\hfill
\begin{subfigure}[t]{.32\textwidth}
    \centering
\begin{tikzpicture}[scale = 0.5]

\definecolor{darkgray176}{RGB}{176,176,176}
\definecolor{lightcoral}{RGB}{240,128,128}
\definecolor{lightgreen}{RGB}{144,238,144}
\definecolor{lightslategray}{RGB}{119,136,153}
\definecolor{royalblue}{RGB}{65,105,225}
\definecolor{seagreen}{RGB}{46,139,87}

\begin{axis}[
legend cell align={left},
legend style={fill opacity=0.8, draw opacity=1, text opacity=1, draw=white!80!black},
tick align=outside,
tick pos=left,
x grid style={darkgray176},
xlabel={$\kappa$},
xmin=-0.145, xmax=5.245,
xtick style={color=black},
y grid style={darkgray176},
ylabel={${\rm MSE}$},
scaled y ticks=false,
yticklabel style={
  /pgf/number format/precision=3,
  /pgf/number format/fixed},
ymin=0.0413329418062274, ymax=0.887241708163143,
ytick style={color=black}
]
\path [draw=lightcoral, semithick]
(axis cs:0.1,0.823478241414999)
--(axis cs:0.1,0.823971553047559);

\path [draw=lightcoral, semithick]
(axis cs:0.3,0.584527545396763)
--(axis cs:0.3,0.585374439580424);

\path [draw=lightcoral, semithick]
(axis cs:0.6,0.404711933839447)
--(axis cs:0.6,0.405368118382134);

\path [draw=lightcoral, semithick]
(axis cs:1,0.289423188364281)
--(axis cs:1,0.29004931760945);

\path [draw=lightcoral, semithick]
(axis cs:2,0.171959678819973)
--(axis cs:2,0.172393703279081);

\path [draw=lightcoral, semithick]
(axis cs:3,0.123565211654094)
--(axis cs:3,0.123784104869595);

\path [draw=lightcoral, semithick]
(axis cs:4,0.0968680454232269)
--(axis cs:4,0.0970951684035276);

\path [draw=lightcoral, semithick]
(axis cs:5,0.0797833402769963)
--(axis cs:5,0.0799880080148496);

\path [draw=lightslategray, semithick]
(axis cs:0.1,0.832197639498174)
--(axis cs:0.1,0.832809204799342);

\path [draw=lightslategray, semithick]
(axis cs:0.3,0.586633426361618)
--(axis cs:0.3,0.58747005186858);

\path [draw=lightslategray, semithick]
(axis cs:0.6,0.409743854988)
--(axis cs:0.6,0.410495071384984);

\path [draw=lightslategray, semithick]
(axis cs:1,0.291763049203933)
--(axis cs:1,0.292429209784479);

\path [draw=lightslategray, semithick]
(axis cs:2,0.172775315694239)
--(axis cs:2,0.173210051621457);

\path [draw=lightslategray, semithick]
(axis cs:3,0.124210994710361)
--(axis cs:3,0.124434455425909);

\path [draw=lightslategray, semithick]
(axis cs:4,0.0972604797836797)
--(axis cs:4,0.0974891503771267);

\path [draw=lightslategray, semithick]
(axis cs:5,0.0800523030002821)
--(axis cs:5,0.080257916903967);

\path [draw=lightgreen, semithick]
(axis cs:0.1,0.84834581452111)
--(axis cs:0.1,0.848791309692374);

\path [draw=lightgreen, semithick]
(axis cs:0.3,0.612406464049448)
--(axis cs:0.3,0.613114342032113);

\path [draw=lightgreen, semithick]
(axis cs:0.6,0.439774046638864)
--(axis cs:0.6,0.440276251534087);

\path [draw=lightgreen, semithick]
(axis cs:1,0.318881692133742)
--(axis cs:1,0.319320058520465);

\path [draw=lightgreen, semithick]
(axis cs:2,0.196918530756298)
--(axis cs:2,0.197286537645992);

\path [draw=lightgreen, semithick]
(axis cs:3,0.150676219889456)
--(axis cs:3,0.151002017652665);

\path [draw=lightgreen, semithick]
(axis cs:4,0.123368470669134)
--(axis cs:4,0.123651472373388);

\path [draw=lightgreen, semithick]
(axis cs:5,0.106303616789717)
--(axis cs:5,0.106649453332115);

\addplot [semithick, red, mark=triangle*, mark size=3, mark options={solid,rotate=180}, only marks]
table {%
0.1 0.823724897231279
0.3 0.584950992488593
0.6 0.405040026110791
1 0.289736252986866
2 0.172176691049527
3 0.123674658261845
4 0.0969816069133772
5 0.079885674145923
};
\addlegendentry{Oracle estimator, ${\bm{\Xi}_{XY}^*}(\bS)$}
\addplot [semithick, royalblue, mark=triangle*, mark size=3, mark options={solid}, only marks]
table {%
0.1 0.832503422148758
0.3 0.587051739115099
0.6 0.410119463186492
1 0.292096129494206
2 0.172992683657848
3 0.124322725068135
4 0.0973748150804032
5 0.0801551099521246
};
\addlegendentry{RIE, $\widehat{{\bm{\Xi}_{XY}^*}}(\bS)$}
\addplot [semithick, seagreen, mark=x, mark size=3, mark options={solid}, only marks]
table {%
0.1 0.848568562106742
0.3 0.61276040304078
0.6 0.440025149086475
1 0.319100875327104
2 0.197102534201145
3 0.15083911877106
4 0.123509971521261
5 0.106476535060916
};
\addlegendentry{$\widehat{{\bm{\Xi}_X^*}}(\bS) \widehat{{\bm{\Xi}_Y^*}}(\bS)$}
\end{axis}

\end{tikzpicture}
    \caption{\small Estimating $\bX \bY$}
\end{subfigure}
\caption{\small MSE of factorization problem. MSE is normalized by the norm of the signal. $\bX$ is a shifted Wigner matrix with $c=1$, and both $\bY$ and $\bW$ are $N \times M$ matrices with i.i.d. Gaussian entries of variance $\nicefrac{1}{N}$. RIE is applied to $N=2000, M =4000$. In each run, the observation matrix $\bS$ is generated according to \eqref{MF-model}, and the factors $\bX$, $\bY$ are estimated simultaneously from $\bS$. Results are averaged over 10 runs (error bars are invisible).}
\label{MF-c=1-main}
\end{figure}

\section{Derivation of the explicit RIEs}\label{Derivation}
In this section, we sketch the derivation of our explicit RIE for $\bX$. The RIE for $\bY$ is derived similarly, but requires more involved analysis and is presented in section \ref{Y-RIE-deriv}. For simplicity, we take the SNR parameter in \eqref{MF-model} to be 1, so the model is $\bS = \bX \bY + \bW$. The optimal eigenvalues are constructed as ${\xi_x^*}_i = \sum_{j=1}^N \lambda_j \big(  \bu_i^\intercal \bx_j \big)^2$. We assume that in the large $N$ limit, ${\xi_x^*}_i$ can be approximated by its expectation and we introduce
\begin{equation}\label{eig-over-rel}
    \widehat{{\xi_x^*}}_i = \sum_{j=1}^N \lambda_j \,  \bE \Big[ \big(  \bu_i^\intercal \bx_j \big)^2 \Big]
\end{equation}
where the expectation is over the (left) singular vectors of the observation matrix $\bS$. Therefore, to compute these eigenvalues, we need to find the mean squared overlap $\bE \Big[ \big(  \bu_i^\intercal \bx_j \big)^2 \Big]$ between eigenvectors of $\bX$ and singular vectors of $\bS$. In what follows, we will see that (a rescaling of) this quantity can be expressed in terms of $i$-th singular value of $\bS$ and $j$-th eigenvector of $\bX$ (and the limiting measures, indeed). Thus, we will use the notation $O_X(\gamma_i, \lambda_j) := N \bE \Big[ \big(  \bu_i^\intercal \bx_j \big)^2 \Big]$ in the following. In the next section, we discuss how the overlap can be computed from the resolvent of the "Hermitized" version of $\bS$.

\subsection{Relation between overlap and resolvent}
Construct the matrix $\cS \in \bR^{(N+M) \times (N+M)}$ from the observation matrix:
\begin{equation*}
    \cS = \left[
\begin{array}{cc}
\mathbf{0}_{N\times N} & \bS \\
\bS^\intercal & \mathbf{0}_{M\times M}
\end{array}
\right] 
\end{equation*}
By Theorem 7.3.3 in \cite{horn2012matrix}, $\cS$ has the following eigen-decomposition:
\begin{equation}
  \cS  = \left[
\begin{array}{ccc}
\hat{\bU}_S & \hat{\bU}_S &  \mathbf{0} \\
\hat{\bV}_S^{(1)} & -\hat{\bV}_S^{(1)} &  \bV_S^{(2)}
\end{array}
\right] \left[
\begin{array}{ccc}
\bGam_N & \mathbf{0} & \mathbf{0}\\
\mathbf{0} & -\bGam_N &  \mathbf{0}\\
\mathbf{0} & \mathbf{0} & \mathbf{0} 
\end{array}
\right]  \left[
\begin{array}{ccc}
\hat{\bU}_S & \hat{\bU}_S &  \mathbf{0} \\
\hat{\bV}_S^{(1)} & -\hat{\bV}_S^{(1)} &  \bV_S^{(2)}
\end{array}
\right]^\intercal
\label{EVD of S}
\end{equation}
with $\bV_S = \left[
\begin{array}{cc}
\bV_S^{(1)} & \bV_S^{(2)}
\end{array}
\right]$ in which $\bV_S^{(1)} \in \bR^{M \times N}$. And, $\hat{\bV}_S^{(1)} = \frac{1}{\sqrt{2}} \bV_S^{(1)}$, $\hat{\bU}_S= \frac{1}{\sqrt{2}} \bU_S$. Eigenvalues of $\cS$ are signed singular values of $\bS$, therefore the limiting eigenvalue distribution of $\cS$ (ignoring zero eigenvalues) is the same as the limiting symmetrized singular value distribution of $\bS$.
Define the resolvent of $\cS$, 
\begin{equation*}
    \bG_{\mathcal{S}}(z) = (z \bI - \cS)^{-1}
\end{equation*}
We assume that as $N \to \infty$ and $z$ is not too close to the real axis, the matrix $\bG_{\mathcal{S}}(z)$ concentrates around its mean. Consequently, the value of $\bG_{\mathcal{S}}(z)$ becomes uncorrelated with the particular realization of $\bS$. Specifically, as $N \to \infty$ , $\bG_{\mathcal{S}}(z)$ converges to a deterministic matrix for any fixed value of $z \in \mathbb{C}\backslash\bR$ (independent of N). Denote the eigenvectors of $\cS$ by $\bs_i \in \bR^{M+N}$, $i = 1, \dots,M+N$. For $z = x - \ci \epsilon$ with $x \in \bR$ and small $\epsilon$, we have:
\begin{equation*}
    \bG_{\mathcal{S}}(x- \ci \epsilon) = \sum_{k=1}^{2N} \frac{x + \ci \epsilon }{(x - \tilde{\gamma}_k)^2+\epsilon^2} \bs_k \bs_k^\intercal + \frac{x + \ci \epsilon }{x^2+\epsilon^2} \sum_{k=2N + 1}^{N +M} \bs_k \bs_k^\intercal
\end{equation*}
where $\tilde{\gamma}_k$ are the eigenvalues of $\cS$, which are in fact the (signed) singular values of $\bS$, $\tilde{\gamma}_1 = \gamma_1, \hdots, \tilde{\gamma}_N=\gamma_N, \tilde{\gamma}_{N+1}= - \gamma_1, \hdots, \tilde{\gamma}_{2N}=-\gamma_N$. 

Define the vectors $\tilde{\bx}_i = [ \bx_i^\intercal ,  \mathbf{0}_{M} ]^\intercal$ for $\bx_i$ eigenvectors of $\bX$. We have
\begin{equation}
\begin{split}
    \tilde{\bx}_i^\intercal \big( {\rm Im}\, \bG_{\mathcal{S}}(x - \ci \epsilon) \big) \tilde{\bx}_i = \sum_{k=1}^{2 N} \frac{\epsilon }{(x - \tilde{\gamma}_k)^2+\epsilon^2} \big( \tilde{\bx}_i^\intercal \bs_k \big)^2 + \frac{\epsilon }{x^2+\epsilon^2} \sum_{k=2N + 1}^{N +M} \big( \tilde{\bx}_i^\intercal \bs_k \big)^2
\end{split}
\label{Eq 16}
\end{equation}
Given the structure of $\bs_k$'s in \eqref{EVD of S},  $\big( \tilde{\bx}_i^\intercal \bs_j \big)^2 = \frac{1}{2} \big( \bx_i^\intercal \bu_j \big)^2 = \big( \tilde{\bx}_i^\intercal \bs_{j+N} \big)^2$ for $1 \leq j \leq N$, and the second sum in \eqref{Eq 16} is zero. We assume that in the limit of large N this quantity concentrates on $O_X(\gamma_j, \lambda_i)$ and depends only on the singular values and eigenvalue pairs $(\gamma_j, \lambda_i)$. We thus have:
\begin{equation}
    \tilde{\bx}_i^\intercal \big( {\rm Im}\, \bG_{\mathcal{S}}(x - \ci \epsilon) \big) \tilde{\bx}_i \xrightarrow[]{N \to \infty} \int_\bR \frac{\epsilon}{(x - t)^2+ \epsilon^2} O_X(t,\lambda_i) \bar{\mu}_{S}(t) \, dt
\end{equation}
where the overlap function $O_X(t,\lambda_i) $ is extended (continuously) to arbitrary values within the support of $\bar{\mu}_{S}$ (the symmetrized limiting singular value distribution of $\bS$) with the property that $O_X(t,\lambda_i) = O_X(-t,\lambda_i)$ for $ t \in {\rm supp} (\mu_S)$ . Sending $\epsilon \to 0$, we find 
\begin{equation}
    \tilde{\bx}_i^\intercal \big( {\rm Im}\, \bG_{\mathcal{S}}(x - \ci \epsilon) \big)\tilde{\bx}_i \to \pi \bar{\mu}_{S}(x) O_X(x, \lambda_i)
    \label{X-resolvent-overlap}
\end{equation}
This is a crucial relation as it allows us to study the overlap by means of the resolvent of $\cS$. In the next section, we establish a connection between this resolvent and the signal $\bX$, which enables us to determine the optimal eigenvalues values $\widehat{{\xi_x^*}}_i$ in terms of the singular values of  $\bS$.

\subsection{Resolvent relation}
To derive the resolvent relation between $\bS$ and $\bX$, we fix the matrix $\bX$ and consider the model 
\begin{equation*}
    \bS = \bX \bU_1 \bY \bV_1^\intercal + \bU_2 \bW \bV_2^\intercal
\end{equation*}
with $\bY, \bW \in \bR^{N \times M}$ fixed matrices with limiting singular value distribution $\mu_Y, \mu_W$, and $\bU_1, \bU_2 \in \bR^{N \times N}, \bV_1, \bV_2 \in \bR^{M \times M}$ independent random Haar matrices. Indeed, if we  substitute the SVD of the matrices $\bY, \bW$ in model \eqref{MF-model} we find the latter model. Now, the average over the singular vectors of $\bS$ (with fixed $\bX$) is equivalent to the average over the matrices $\bU_1, \bU_2 , \bV_1, \bV_2$. In section \ref{X-resolvent-rel}, using the Replica trick, we derive the following relation in the limit $N \to \infty$:
\begin{equation}
    \big \langle \bG_{\mathcal{S}} (z) \big \rangle = \left[
\begin{array}{cc}
\zeta_3^{-1} \bG_{X^2} \big( \frac{z - \zeta_1 }{\zeta_3} \big)&  \mathbf{0}  \\
\mathbf{0} & (z-\zeta_2)^{-1} \bI_M
\end{array}
\right]
\label{X-resolvent relation}
\end{equation}
with $\zeta_1, \zeta_2, \zeta_3$ satisfying set of equations \eqref{X-sol-app}. $\langle . \rangle$ is the expectation w.r.t. the singular vectors of $\bS$ (or equivalently over $\bU_1, \bU_2 , \bV_1, \bV_2$), and $\bG_{X^2}$ is the resolvent of $\bX^2$. As stated earlier, we assume that the resolvent $\bG_{\mathcal{S}} (z) $ concentrates in the limit $ N \to \infty$, therefore we drop the brackets in the following computation.

\subsection{Overlaps and optimal eigenvalues}
From \eqref{X-resolvent-overlap}, \eqref{X-resolvent relation}, we find:
\begin{equation}
\begin{split}
    O_X(\gamma, \lambda_i) &\approx \frac{1}{\pi \bar{\mu}_{S}(\gamma)} \, {\rm Im} \, \lim_{z \to \gamma - \ci 0^+} \, \bx_i^\intercal \, \zeta_3^{-1} \bG_{X^2} \big( \frac{z - \zeta_1 }{\zeta_3} \big) \, \bx_i \\
    &= \frac{1}{\pi \bar{\mu}_{S}(\gamma)} \, {\rm Im} \, \lim_{z \to \gamma - \ci 0^+} \,  \frac{1}{ z - \zeta_1  - \zeta_3 \lambda_i^2}
\end{split}
\label{X-overlap-eq}
\end{equation}
In Fig. \ref{fig:X-Overlap-Wigner X} we illustrate that the theoretical predictions \eqref{X-overlap-eq} are in good agreement with numerical simulations for a particular case of $\bX$ a Wigner matrix, and $\bY, \bW$ with i.i.d. Gaussian entries.
\begin{SCfigure}[4]
    \input{Figures/X-est/Overlap_WignerX_YZGauss.tex}
   \caption{\small Comparison of the theoretical prediction \eqref{X-overlap-eq} of the rescaled overlap with the numerical simulation.  The rescaled overlap between $200$-th and $800$-th left singular vector of $\bS$ and the eigenvectors of $\bX$ is illustrated. $\bX = \bX^{\intercal} \in \bR^{N \times N}$ has i.i.d. Gaussian entries with variance $\nicefrac{1}{\sqrt{N}}$ and is fixed. Both $\bY$ and $\bZ$ are $N \times M$ matrices with i.i.d. Gaussian entries of variance $\nicefrac{1}{N}$. The simulation results are average of 1000 experiments with fixed $\bX$, and $N = 1000, M =2000$. Some of the simulation points are dropped for clarity.\\
   One can see that the overlap is an even function of eigenvalues $\lambda_i$, so the optimal eigenvalues ${\xi_x^*}_i = \sum_{j=1}^N \lambda_j \, \big(  \bu_i^\intercal \bx_j \big)^2$ are all zero, as discussed in remark \ref{symmetrix-rho}.}
   \label{fig:X-Overlap-Wigner X}
\end{SCfigure}

Once we have the overlap, we can compute the optimal eigenvalues to be
\begin{equation}
\begin{split}
    {\widehat{\xi_x^*}}_i &\approx \frac{1}{N}  \sum_{j=1}^N \lambda_j O_X(\gamma_i, \lambda_j) \approx \frac{1}{\pi \bar{\mu}_{S}(\gamma_i) } \, {\rm Im} \, \lim_{z \to \gamma_i - \ci 0^+}   \frac{1}{N}  \sum_{j=1}^N \frac{\lambda_j}{ z - \zeta_1  - \zeta_3 \lambda_j^2}
\end{split}
\label{X-optimal-ev-estimator}
\end{equation}
With a bit of algebra, we find the estimator in \eqref{X-optimal-ev} in the limit $N \to \infty$, see section \ref{X-overlap-ev}.
\section*{Acknowledgments} 
We are thankful to Jean Barbier for interesting discussions on related problems. The work of F. P has been supported by Swiss National Science Foundation grant no 200021-204119. 
\bibliographystyle{unsrt}
\bibliography{References}

\newpage 
\appendix

\section{Posterior mean estimator is in the RIE class}
In this section, we show that for rotational invariant priors, the posterior mean estimator is inside the RIE class. For each of the estimators of $\bX, \bY$, we present an equivalent definition of the RIE, then we show that posterior mean estimator satisfies this definition.

\subsection{X Estimator} \label{X-posterior-RIE}
\begin{lemma}\label{X-RIE-lemma}
Given the observation matrix $\bS$, let $\hat{\bX}(\bS)$ be an estimator of $\bX$.  Then $\hat{\bX}(\bS)$ is a RIE if and only if for any  orthogonal matrices $\bU \in \bR^{N \times N}, \bV \in \bR^{M \times M}$:
    \begin{equation}
        \hat{\bX}( \bU \bS \bV^\intercal) = \bU \hat{\bX}( \bS ) \bU^\intercal
        \label{RIE-property-X}
    \end{equation}
\end{lemma}
\begin{proof}
If $\hat{\bX}(\bS)$ is a RIE, then the property \eqref{RIE-property-X} clearly follows from the definition \eqref{X-RIE-class}. Now we turn to the converse.

Suppose that an estimator $\hat{\bX}(\bS)$ satisfies \eqref{RIE-property-X}. First, we show that if the observation matrix is diagonal, then the estimator is also diagonal. Consider the observation matrix to be $\bS^{\rm diag} = \left[
\begin{array}{c|c}
\rm{diag}(s_1, \dots, s_N) & \mathbf{0}_{N \times (M-N)} 
\end{array}
\right]$. Let $\bI_k^- \in \bR^{N \times N}, \bJ_k^- \in \bR^{M \times M}$ be diagonal matrices with diagonal entries all one except the $k$-th entry which is $-1$. Note that for $1 \leq k \leq N$, we have $\bS^{\rm diag} = \bI_k^- \bS^{\rm diag} \bJ_k^-$. Moreover, matrices $\bI_k^-, \bJ_k^-$ are indeed orthogonal. For any $1 \leq k \leq N$, from the property we have:
\begin{equation}
    \hat{\bX}(\bS^{\rm diag}) =  \hat{\bX}(\bI_k^- \bS^{\rm diag} \bJ_k^-) = \bI_k^-  \hat{\bX}(\bS^{\rm diag}) \bI_k^-
\end{equation}
This implies that all entries on the $k$-th row and $k$-th column of $\hat{\bX}(\bS^{\rm diag})$ are zero except the $k$-th entry on the diagonal. Since this holds for any $k$, we conclude that $\hat{\bX}(\bS^{\rm diag})$ is diagonal.

Now, for a given general observation matrix with SVD $\bS = \bU_S \bGam \bV_S^\intercal$, put $\bU = \bU_S^\intercal, \bV = \bV_S^\intercal$ in the property \eqref{RIE-property-X}. We have:
\begin{equation*}
    \hat{\bX}( \bGam ) = \bU_S^\intercal \hat{\bX}( \bS ) \bU_S
\end{equation*}
From the argument above, the matrix on the lhs is diagonal. Consequently, the matrix $\bU_S^\intercal \hat{\bX}( \bS ) \bU_S$ is diagonal which implies that the columns of $\bU_S$ are eigenvectors of $\hat{\bX}( \bS )$. Therefore, $\hat{\bX}( \bS ) $ is a RIE.
\end{proof}

Now, we prove that the posterior mean estimator $\hat{\bX}^*(\bS) = \bE[ \bX | \bS ]$ has the property \eqref{RIE-property-X}, and therefore belongs to the RIE class. For simplicity, we drop the SNR factor $\sqrt{\kappa}$. For any orthogonal matrices $\bU \in \bR^{N \times N}, \bV \in \bR^{M \times M}$, we have:
\begin{equation*}
\begin{split}
     \bE[ \bX |  \bU \bS \bV^\intercal] &= \frac{\int d \bY \, d \tilde{\bX} \, \tilde{\bX} \, P_X(\tilde{\bX}) P_Y(\bY) P_W(  \bU \bS \bV^\intercal - \tilde{\bX} \bY ) }  {\int d \bY \, d \tilde{\bX} \,  P_X(\tilde{\bX}) P_Y(\bY) P_W(  \bU \bS \bV^\intercal - \tilde{\bX} \bY )} \\
    &\stackrel{\text{(a)}}{=} \frac{\int d \bY \, d \tilde{\bX} \, \bU \tilde{\bX} \bU^\intercal \, P_X(\tilde{\bX}) P_Y(\bY) P_W( \bU \bS \bV^\intercal - \bU \tilde{\bX} \bU^\intercal \bY ) }  {\int d \bY \, d \tilde{\bX} \,  P_X(\tilde{\bX}) P_Y(\bY) P_W( \bU \bS \bV^\intercal - \bU \tilde{\bX} \bU^\intercal \bY )}  \\
    &\stackrel{\text{(b)}}{=} \frac{\int d \bY \, d \tilde{\bX} \, \bU \tilde{\bX} \bU^\intercal \, P_X(\tilde{\bX}) P_Y(\bY) P_W( \bU \bS \bV^\intercal - \bU \tilde{\bX} \bU^\intercal \bU \bY \bV^\intercal ) }  {\int d \bY \, d \tilde{\bX} \,  P_X(\tilde{\bX}) P_Y(\bY) P_W( \bU \bS \bV^\intercal - \bU \tilde{\bX} \bU^\intercal \bU \bY \bV^\intercal )}  \\
    &\stackrel{\text{(c)}}{=} \bU  \Big\{ \frac{\int d \bY \, d \tilde{\bX} \, \tilde{\bX}  \, P_X(\tilde{\bX}) P_Y(\bY) P_W(  \bS -  \tilde{\bX} \bY ) }  {\int d \bY \, d \tilde{\bX} \,  P_X(\tilde{\bX}) P_Y(\bY) P_W(  \bS - \tilde{\bX} \bY )} \Big\} \bU^\intercal \\
    &=  \bU \bE[ \bX |\bS]  \bU^\intercal
\end{split}
\end{equation*}
where in (a), we changed variables $\tilde{\bX} \to \bU \tilde{\bX} \bU^\intercal$, used $|\det \bU | =1$, and rotational invariance of $P_X$, $P_X(\tilde{\bX}) = P_X(  \bU \tilde{\bX} \bU^\intercal )$. In (b), we changed variables $\bY \to \bU \bY \bV^\intercal$, used  $|\det \bU | = |\det \bV |  = 1$, and bi-rotational invariance of $P_Y$, $P_Y(\bY) = P_Y(  \bU \bY \bV^\intercal )$. In (c), we used the bi-rotational invariance property of $P_W$, namely $P_W( \bU \bS \bV^\intercal - \bU \tilde{\bX} \bY \bV^\intercal ) = P_W( \bS -\tilde{\bX} \bY ) $.

\subsection{Y Estimator}\label{Y-posterior-RIE}
\begin{lemma}\label{Y-RIE-lemma}
Given the observation matrix $\bS$, let $\hat{\bY}(\bS)$ be an estimator for $\bY$. Then $\hat{\bY}(\bS)$ is a RIE if and only if for any  orthogonal matrices $\bU \in \bR^{N \times N}, \bV \in \bR^{M \times M}$:
    \begin{equation}
        \hat{\bY}( \bU \bS \bV^\intercal) = \bU \hat{\bY}( \bS ) \bV^\intercal
        \label{RIE-property-Y}
    \end{equation}
\end{lemma}
\begin{proof}
If $\hat{\bY}(\bS)$ is a RIE, then this property clearly follows from the definition \eqref{Y-RIE-class}. Let us now show the converse.

Suppose that an estimator $\hat{\bY}(\bS)$ satisfies \eqref{RIE-property-Y}. First, we show that if the observation matrix is diagonal, then the estimator is also diagonal. Consider the observation matrix to be $\bS^{\rm diag} = \left[
\begin{array}{c|c}
\rm{diag}(s_1, \dots, s_N) & \mathbf{0}_{N \times (M-N)} 
\end{array}
\right]$. Let $\bI_k^- \in \bR^{N \times N}, \bJ_k^- \in \bR^{M \times M}$ be diagonal matrices with diagonal entries all one except the $k$-th entry which is $-1$. Note that for $1 \leq k \leq N$, we have $\bS^{\rm diag} = \bI_k^- \bS^{\rm diag} \bJ_k^-$. Moreover, matrices $\bI_k^-, \bJ_k^-$ are indeed orthogonal. For any $1 \leq k \leq N$, from the property we have:
\begin{equation}
    \hat{\bY}(\bS^{\rm diag}) = \hat{\bY}( \bI_k^- \bS^{\rm diag} \bJ_k^- )= \bI_k^-  \hat{\bY}(\bS^{\rm diag}) \bJ_k^-
\end{equation}
This implies that all entries on the $k$-th row and $k$-th column of $\hat{\bY}(\bS^{\rm diag})$ is zero except the $k$-th entry on the diagonal. Since this holds for any $k$, we conclude that $\hat{\bY}(\bS^{\rm diag})$ is diagonal.

Now, for a given general observation matrix $\bS = \bU_S \bGam \bV_S^\intercal$, put $\bU = \bU_S^\intercal, \bV = \bV_S^\intercal$ in the property \eqref{RIE-property-Y}. We have:
\begin{equation*}
    \hat{\bY}( \bGam ) = \bU_S^\intercal \hat{\bY}( \bS ) \bV_S
\end{equation*}
From the argument above, the matrix on the lhs is diagonal. Consequently, the matrix $\bU_S^\intercal \hat{\bY}( \bS ) \bV_S$ is diagonal which implies that the columns of $\bU_S, \bV_S$ are the left and right singular vectors of $\hat{\bY}( \bS )$. Therefore, $\hat{\bY}( \bS ) $ is a RIE.
\end{proof}

Now, we prove that the posterior mean estimator $\hat{\bY}^*(\bS) = \bE[ \bY | \bS ]$ has the property \eqref{RIE-property-Y}, and it is inside the RIE class. For simplicity, we drop the SNR factor $\sqrt{\kappa}$. For any orthogonal matrices $\bU \in \bR^{N \times N}, \bV \in \bR^{M \times M}$, we have:
\begin{equation*}
\begin{split}
     \bE[ \bY |  \bU \bS \bV^\intercal] &= \frac{\int d \bX \, d \tilde{\bY} \, \tilde{\bY} \, P_X(\bX) P_Y(\tilde{\bY}) P_W(  \bU \bS \bV^\intercal - \bX \tilde{\bY} ) }  {\int d \bX \, d \tilde{\bY} \,  P_X(\bX) P_Y(\tilde{\bY}) P_W(  \bU \bS \bV^\intercal - \bX \tilde{\bY} )} \\
    &\stackrel{\text{(a)}}{=} \frac{\int d \bX \, d \tilde{\bY} \, \bU \tilde{\bY} \bV^\intercal \, P_X(\bX) P_Y(\tilde{\bY}) P_W( \bU \bS \bV^\intercal - \bX \bU \tilde{\bY} \bV^\intercal ) }  {\int d \bX \, d \tilde{\bY} \,  P_X(\bX) P_Y(\tilde{\bY}) P_W( \bU \bS \bV^\intercal - \bX \bU \tilde{\bY} \bV^\intercal  )}  \\
    &\stackrel{\text{(b)}}{=} \frac{\int d \bX \, d \tilde{\bY} \, \bU \tilde{\bY} \bV^\intercal \, P_X(\bX) P_Y(\tilde{\bY}) P_W( \bU \bS \bV^\intercal - \bU \bX \bU^\intercal \bU \tilde{\bY} \bV^\intercal ) }  {\int d \bX \, d \tilde{\bY} \,  P_X(\bX) P_Y(\tilde{\bY}) P_W( \bU \bS \bV^\intercal - \bU \bX \bU^\intercal \bU \tilde{\bY} \bV^\intercal )}  \\
    &\stackrel{\text{(c)}}{=} \bU  \Big\{ \frac{\int d \bX \, d \tilde{\bY} \, \tilde{\bY}  \, P_X(\bX) P_Y(\tilde{\bY}) P_W(  \bS -  \bX \tilde{\bY} ) }  {\int d \bX \, d \tilde{\bY} \,  P_X(\bX) P_Y(\tilde{\bY}) P_W(  \bS - \bX \tilde{\bY} )} \Big\} \bV^\intercal \\
    &=  \bU \bE[ \bY |\bS]  \bV^\intercal
\end{split}
\end{equation*}
where in (a), we changed variables $\tilde{\bY} \to \bU \tilde{\bY} \bV^\intercal$, used $|\det \bU | = |\det \bV | = 1$, and bi-rotational invariance of $P_Y$, $P_Y(\tilde{\bY}) = P_Y(  \bU \tilde{\bY} \bV^\intercal )$. In (b),  we changed variables $\bX \to \bU \bX \bU^\intercal$, used $|\det \bU | =1$, and rotational invariance of $P_X$, $P_X(\bX) = P_X(  \bU \bX \bU^\intercal )$. In (c), we used the bi-rotational invariance property of $P_W$, namely $P_W( \bU \bS \bV^\intercal - \bU \bX \tilde{\bY} \bV^\intercal ) = P_W( \bS - \bX \tilde{\bY} ) $.

\clearpage
\section{The replica method for deriving the resolvent relation}\label{replica}
In this section we present the replica method used to obtain the resolvent relation. For simplicity of notation we use $\bG(z) \equiv \bG_{\mathcal{S}}(z)$ for the resolvent of a random matrix $\bm{\mathcal{S}}$.

First, we express the entries of the resolvent $\bG(z)$ using the Gaussian integral representation of an inverse matrix \cite{zinn2021quantum}:
\begin{equation}
\begin{split}
    G_{ij}(z) &= \sqrt{\frac{1}{( 2 \pi )^{N+M} \det \,  (z \bI - \cS) } }  \int \Big( \prod_{k=1}^{M+N}  d \eta_k  \Big) \, \eta_i \eta_j \,  \exp \Big\{ -\frac{1}{2}  \bbeta^\intercal \big( z \bI - \cS \big) \bbeta \Big\} \\
    &= \ddfrac{\int \Big( \prod_{k=1}^{M+N}  d \eta_k  \Big) \, \eta_i \eta_j \,   \exp \Big\{ -\frac{1}{2} \bbeta^\intercal \big( z \bI - \cS \big) \bbeta \Big\}}{\int \Big( \prod_{k=1}^{M+N}  d \eta_k  \Big) \,    \exp \Big\{ -\frac{1}{2} \bbeta^\intercal \big( z \bI - \cS \big) \bbeta \Big\}}
\end{split}
\label{Inverse-G-integral}
\end{equation}
For $z$ not close to the real axis, the resolvent is expected to exhibit self-averaging behavior in the limit of large N, meaning that it will not depend on the particular matrix realization. Thus, we can examine the resolvent  $\bG_{\mathcal{S}}(z)$ by analyzing its ensemble average, denoted by $\langle . \rangle$ in the following.
\begin{equation}
    \big\langle G_{ij}(z) \big\rangle = \bigg\langle \frac{1}{\mathcal{Z}} \, \int \Big( \prod_{k=1}^{M+N}  d \eta_k  \Big) \, \eta_i \eta_j \,   \exp \Big\{ -\frac{1}{2} \bbeta^\intercal \big( z \bI - \cS \big) \bbeta \Big\} \bigg\rangle
    \label{Inverse-G-integral-average}
\end{equation}
where $\mathcal{Z}$ is the denominator in \eqref{Inverse-G-integral}. Computing the average is, in general, non-trivial. However, the replica method provides us with a technique to overcome this issue by employing the following identity:
\begin{equation}
    \begin{split}
        \big\langle G_{ij}(z) \big\rangle &=  \lim_{n \to 0} \bigg\langle \mathcal{Z}^{n-1} \, \int \Big( \prod_{k=1}^{M+N}  d \eta_k  \Big) \, \eta_i \eta_j \,   \exp \Big\{ -\frac{1}{2} \bbeta^\intercal \big( z \bI - \cS \big) \bbeta \Big\} \bigg\rangle \\
        &=  \lim_{n \to 0} \bigg\langle  \, \int \Big( \prod_{k=1}^{M+N} \prod_{\tau=1}^{n}  d \eta^{(\tau)}_k  \Big) \, \eta^{(1)}_i \eta^{(1)}_j \,   \exp \Big\{ -\frac{1}{2} \sum_{\tau =1}^n {\bbeta^{(\tau)}}^\intercal \big( z \bI - \cS \big) \bbeta^{(\tau)} \Big\} \bigg\rangle
    \end{split}
    \label{resolvent-replica}
\end{equation}
So, the problem now is reduced to computation of an average over $n$ copies (or replicas) of the initial system \eqref{Inverse-G-integral}. After computing the average value (the bracket) in \eqref{resolvent-replica}, we can perform an analytical continuation of the result to real values of $n$ and then take the limit $n \to 0$. Throughout, we assume as is common in the replica method, that the analytical continuation can be done with only $n$ different sets of points. Of course, this is a totally uncontrolled step that comes with no guarantees. 
\clearpage
\section{Derivation of the RIE for $\bX$}\label{X-derivation}
In this section, we consider estimating $\bX$, and treat both $\bY$ and $\bW$ as noise. We consider $\bX$ to be fixed, and the observation model:
\begin{equation}
    \bS = \bX \bU_1 \bY \bV_1^\intercal + \bU_2 \bW \bV_2^\intercal
    \label{X-model}
\end{equation}
where $\bY, \bW \in \bR^{N \times M}$ are fixed matrices with limiting singular value distribution $\mu_Y, \mu_W$, and $\bU_1, \bU_2 \in \bR^{N \times N}, \bV_1, \bV_2 \in \bR^{M \times M}$ are independent random Haar matrices. 

Construct the hermitization $\cS \in \bR^{(N+M) \times (N+M)}$ from $\bS$ as
\begin{equation*}
    \cS = \left[
\begin{array}{cc}
\mathbf{0}_{N\times N} & \bS \\
\bS^\intercal & \mathbf{0}_{M\times M}
\end{array}
\right]
\end{equation*}
For simplicity of notation, we use $\bT \equiv \bX \bU_1 \bY \bV_1^\intercal$,  $\cT \in \bR^{(N+M) \times (N+M)}$ the hermitization of $\bT$, and $\widetilde{\cW}$ denotes the hermitization of the matrix $\bU_2 \bW \bV_2^\intercal$.

\subsection{Resolvent relation}\label{X-resolvent-rel}
We want to find a relation between $\bG(z) \equiv \bG_{\mathcal{S}}(z)$, and the signal matrix $\bX$. From \eqref{resolvent-replica}, we have
\begin{equation}
    \begin{split}
         \langle G_{i j}(z) \rangle &= \lim_{n \to \infty} \int \Big( \prod_{k=1}^{N+M} \prod_{\tau =1}^n d \eta_k^{(\tau)} \Big)  \eta_i^{(1)} \eta_j^{(1)} \, \Big\langle \exp \big\{ -\frac{1}{2} \sum_{\tau =1}^n  {\bbeta^{(\tau)}}^\intercal ( z \bI - \cS) \bbeta^{(\tau)} \big\} \Big\rangle_{\bU_1,\bU_2,\bV_1, \bV_2} \\
         &= \lim_{n \to \infty} \int \big( \prod_{k=1}^{N+M} \prod_{\tau =1}^n d \eta_k^{(\tau)} \big) \, \eta_i^{(1)} \eta_j^{(1)} \exp \big\{-\frac{z}{2} \sum_{\tau =1}^n {\bbeta^{(\tau)}}^\intercal \bbeta^{(\tau)} \big\} \\ 
         &\hspace{2cm} \times \Big\langle \exp \big\{ \frac{1}{2} \sum_{\tau =1}^n  {\bbeta^{(\tau)}}^\intercal  \cT \bbeta^{(\tau)} \big\} \Big\rangle_{\bU_1, \bV_1} \Big\langle \exp \big\{ \frac{1}{2} \sum_{\tau =1}^n  {\bbeta^{(\tau)}}^\intercal  \widetilde{\cW} \bbeta^{(\tau)} \big\} \Big\rangle_{\bU_2, \bV_2}
    \end{split}
    \label{X-Gaussian Integral}
\end{equation}
Split each replica $\bbeta^{(\tau)}$ into two vectors $\ba^{(\tau)} \in \bR^N, \bb^{(\tau)} \in \bR^M$, $\bbeta^{(\tau)} = \left[
\begin{array}{c}
\ba^{(\tau)} \\
\bb^{(\tau)}
\end{array}
\right]$. The exponent in the first bracket in \eqref{X-Gaussian Integral} can be written as:
\begin{equation}
    \begin{split}
        {\bbeta^{(\tau)}}^\intercal \cT \bbeta^{(\tau)} &= {\ba^{(\tau)}}^\intercal \bX \bU_1 \bY \bV_1^\intercal \bb^{(\tau)}  + {\bb^{(\tau)}}^\intercal \bV_1 \bY^\intercal \bU_1^\intercal \bX \ba^{(\tau)} \\
        &= 2 {\ba^{(\tau)}}^\intercal \bX \bU_1 \bY \bV_1^\intercal \bb^{(\tau)} \\
        &= 2 \Tr \bb^{(\tau)} {\ba^{(\tau)}}^\intercal \bX \bU_1 \bY \bV_1^\intercal  
    \end{split}
    \label{X-first bracket}
\end{equation}
Using the formula for the rectangular spherical integral \cite{benaych2011rectangular} (see Theorem \ref{rect-sphericla-integral} in \ref{spherical integral app}), we find:
\begin{equation}
    \Big\langle \exp \big\{  \sum_{\tau =1}^n  \Tr \bb^{(\tau)} {\ba^{(\tau)}}^\intercal \bX \bU_1 \bY \bV_1^\intercal   \big\} \Big\rangle_{\bU_1, \bV_1}  \approx \exp \Big\{ \frac{N}{2} \sum_{\tau=1}^n \mathcal{Q}_{\mu_Y}^{(\alpha)} \big(\frac{1}{N M} \| \bX \ba^{(\tau)} \|^2 \| \bb^{(\tau)} \|^2 \big) \Big\}
    \label{X-first bracket spherical}
\end{equation}
with $\mathcal{Q}_{\mu_Y}^{(\alpha)}(x) = \int_0^x \frac{\mathcal{C}_{\mu_Y}^{(\alpha)}(t)}{t} \, dt$. In \eqref{X-first bracket spherical}, we used that $\bb^{(\tau)} {\ba^{(\tau)}}^\intercal \bX$ is a rank-one matrix with non-zero singular value $\| \bb^{(\tau)} \| \| \bX {\ba^{(\tau)}} \|$.

Similarly, for the second bracket in \eqref{X-Gaussian Integral} we can write:
\begin{equation}
    \begin{split}
        {\bbeta^{(\tau)}}^\intercal \widetilde{\cW} \bbeta^{(\tau)} &= {\ba^{(\tau)}}^\intercal  \bU_2 \bW \bV_2^\intercal \bb^{(\tau)}  + {\bb^{(\tau)}}^\intercal \bV_2 \bW^\intercal \bU_2^\intercal \ba^{(\tau)} \\
        &= 2 {\ba^{(\tau)}}^\intercal  \bU_2 \bW \bV_2^\intercal \bb^{(\tau)} \\
        &= 2 \Tr \bb^{(\tau)} {\ba^{(\tau)}}^\intercal  \bU_2 \bW \bV_2^\intercal  
    \end{split}
    \label{X-second bracket}
\end{equation}
which using the formula of rectangular spherical integrals, implies
\begin{equation}
    \Big\langle \exp \big\{  \sum_{\tau =1}^n  \Tr \bb^{(\tau)} {\ba^{(\tau)}}^\intercal  \bU_2 \bW \bV_2^\intercal   \big\} \Big\rangle_{\bU_2, \bV_2}  \approx \exp \Big\{ \frac{N}{2} \sum_{\tau=1}^n  \mathcal{Q}_{\mu_W}^{(\alpha)} \big(\frac{1}{N M} \| \ba^{(\tau)} \|^2 \| \bb^{(\tau)} \|^2 \big) \Big\}
    \label{X-second bracket spherical}
\end{equation}

From \eqref{X-Gaussian Integral}, \eqref{X-first bracket spherical}, \eqref{X-second bracket spherical}, we find:
\begin{equation}
    \begin{split}
         \langle & G_{i j}(z) \rangle =  \lim_{n \to \infty} \int  \big( \prod_{k=1}^{N+M} \prod_{\tau =1}^n d \eta_k^{(\tau)} \big) \, \eta_i^{(1)} \eta_j^{(1)} \\
         &\times \exp \bigg\{-\frac{1}{2} \sum_{\tau =1}^n z \|\bbeta^{(\tau)}\|^2 - N  \mathcal{Q}_{\mu_Y}^{(\alpha)} \Big(\frac{ \| \bX \ba^{(\tau)} \|^2 \| \bb^{(\tau)} \|^2}{N M} \Big) - N \mathcal{Q}_{\mu_W}^{(\alpha)} \Big(\frac{\| \ba^{(\tau)} \|^2 \| \bb^{(\tau)} \|^2}{N M}  \Big) \bigg\}
    \end{split}
    \label{X-Gaussian Integral 2}
\end{equation}
Now, we introduce delta functions  $\delta \big( p_1^{(\tau)} - \frac{\| \ba^{(\tau)} \|^2}{N}  \big)$, $\delta \big( p_2^{(\tau)} - \frac{\| \bb^{(\tau)} \|^2}{M}  \big)$, and $\delta \big( p_3^{(\tau)} - \frac{\| \bX \ba^{(\tau)} \|^2}{N}  \big)$, and using them, the integral in \eqref{X-Gaussian Integral 2} can be written as (for brevity we drop the limit term):
\begin{equation}
    \begin{split}
         \langle G_{i j}(z) \rangle = \int \big( \prod_{k=1}^{N+M} &\prod_{\tau =1}^n d \eta_k^{(\tau)} \big) \big( \prod_{\tau =1}^n d p^{(\tau)}_1 \, d p^{(\tau)}_2 \, d p^{(\tau)}_3 \big)  \, \eta_i^{(1)} \eta_j^{(1)}\\
         &\times \prod_{\tau =1}^n \delta \Big( p_1^{(\tau)} - \frac{\| \ba^{(\tau)} \|^2}{N}  \Big) \, \delta \Big( p_2^{(\tau)} - \frac{ \| \bb^{(\tau)} \|^2}{M} \Big) \, \delta \Big( p_3^{(\tau)} - \frac{\| \bX \ba^{(\tau)} \|^2}{N}  \Big) \\
         &\times  \exp \Big\{-\frac{1}{2} \sum_{\tau =1}^n z \|\bbeta^{(\tau)}\|^2 - N  \mathcal{Q}_{\mu_Y}^{(\alpha)} (p^{(\tau)}_2 p^{(\tau)}_3) - N \mathcal{Q}_{\mu_W}^{(\alpha)} (p^{(\tau)}_1 p^{(\tau)}_2) \Big\}
    \end{split}
    \label{X-Gaussian Integral delta}
\end{equation}
In the next step, we replace each delta with its Fourier transform,
$\delta \big(p_1^{\tau} - \frac{1}{N}\|\ba^{\tau}\|^2\big) \propto \int \, d \zeta_1^{\tau}  \exp  \Big\{ -\frac{N}{2} \zeta_1^{\tau} \big( p_1^{\tau} - \frac{1}{N}\|\ba^{\tau}\|^2 \big) \Big\}$. After rearranging, we find:
\begin{equation}
    \begin{split}
         \langle G_{i j}(z) \rangle \propto \int&  \big( \prod_{\tau =1}^n d p^{(\tau)}_1 \, d p^{(\tau)}_2 \, d p^{(\tau)}_3 \, d \zeta^{(\tau)}_1 \, d \zeta^{(\tau)}_2 \, d \zeta^{(\tau)}_3 \big) \\
         &\hspace{-13pt}\times \exp \Big\{ \frac{N}{2} \sum_{\tau =1}^n   \mathcal{Q}_{\mu_Y}^{(\alpha)} (p^{(\tau)}_2 p^{(\tau)}_3) +  \mathcal{Q}_{\mu_W}^{(\alpha)} (p^{(\tau)}_1 p^{(\tau)}_2) - \zeta_1^{(\tau)}p_1^{(\tau)} - \frac{1}{\alpha} \zeta_2^{(\tau)}p_2^{(\tau)} - \zeta_3^{(\tau)}p_3^{(\tau)} \Big\} \\
         &\hspace{-20pt}\times \int \big( \prod_{k=1}^{N+M} \prod_{\tau =1}^n d \eta_k^{(\tau)} \big) \, \eta_i^{(1)} \eta_j^{(1)} \\
         &\times \exp \Big\{-\frac{1}{2} \sum_{\tau =1}^n z \|\bbeta^{(\tau)}\|^2 - \zeta_1^{(\tau)} \| \ba^{(\tau)} \|^2 - \zeta_2^{(\tau)} \| \bb^{(\tau)} \|^2 - \zeta_3^{(\tau)} \| \bX \ba^{(\tau)} \|^2 \Big\}
    \end{split}
    \label{X-Gaussian Integral delta Fourier}
\end{equation}
The inner integral in \eqref{X-Gaussian Integral delta Fourier} is a Gaussian integral, and can be written as:
\begin{equation}
\begin{split}
    \int \big( \prod_{k=1}^{N+M} \prod_{\tau =1}^n d \eta_k^{(\tau)} &\big) \,  \eta_i^{(1)} \eta_j^{(1)} \\
    &\times \exp \Bigg\{ \sum_{\tau =1}^n -\frac{1}{2} {\bbeta^{(\tau)}}^\intercal \left[
\begin{array}{cc}
(z - \zeta_1^{(\tau)}) \bI_N - \zeta_3^{(\tau)} \bX^2 & \mathbf{0}  \\
\mathbf{0} & (z-\zeta_2^{(\tau)}) \bI_M 
\end{array}
\right] \bbeta^{(\tau)} \Bigg\}
\end{split}
\label{X-Gaussian}
\end{equation}
Denote the matrix in the exponent by $\bC_X^{(\tau)}$. Its determinant reads:
\begin{equation*}
    \det \bC_X^{(\tau)} = (z-\zeta_2^{(\tau)})^M \prod_{k=1}^N (z - \zeta_1^{(\tau)} - \zeta_3^{(\tau)} \lambda_k^2 )
\end{equation*}
where $\lambda_k$'s are eigenvalues of $\bX$. So replacing the formula for the Gaussian integrals, \eqref{X-Gaussian Integral delta Fourier} can be written as:
\begin{equation}
    \begin{split}
         \langle G_{i j}(z) \rangle \propto \int \big( \prod_{\tau =1}^n d p^{(\tau)}_1 \, d p^{(\tau)}_2 \, d p^{(\tau)}_3 \, d \zeta^{(\tau)}_1 \, d \zeta^{(\tau)}_2 \, d \zeta^{(\tau)}_3 &\big) \Big( {\bC_X^{(1)}}^{-1} \Big)_{i j} \\
         &\times \exp \big\{ -\frac{N n}{2} F^X_0 ( \bm{p}_1, \bm{p}_2, \bm{p}_3,  \bm{\zeta}_1, \bm{\zeta}_2, \bm{\zeta}_3 ) \big\}
    \end{split}
    \label{X-Saddle Point Integral}
\end{equation}
with
\begin{equation}
\begin{split}
    F^X_0 ( \bm{p}_1, \bm{p}_2, \bm{p}_3,  \bm{\zeta}_1, \bm{\zeta}_2, \bm{\zeta}_3 ) &= \frac{1}{n} \sum_{\tau =1}^n \bigg[ \frac{1}{N} \sum_{k=1}^{N} \ln (z - \zeta_1^{(\tau)} - \zeta_3^{(\tau)} \lambda_k^2 ) + \frac{1}{\alpha} \ln (z - \zeta_2^{(\tau)}) \\
    &\hspace{-4 pt} - \mathcal{Q}_{\mu_Y}^{(\alpha)} (p^{(\tau)}_2 p^{(\tau)}_3) - \mathcal{Q}_{\mu_W}^{(\alpha)}(p^{(\tau)}_1 p^{(\tau)}_2) +\zeta_1^{(\tau)}p_1^{(\tau)} + \frac{1}{\alpha} \zeta_2^{(\tau)}p_2^{(\tau)} + \zeta_3^{(\tau)}p_3^{(\tau)} \bigg]
    \end{split}
    \label{Free energy X}
\end{equation}
In the large $N$ limit, the integral in \eqref{X-Saddle Point Integral} can be computed using the saddle-points of the function $F^X_0$. In the evaluation of this integral, we use the \textit{replica symmetric} ansatz that assumes a saddle-point of the form:
\begin{equation*}
    \forall \tau \in \{1, \cdots, n\}: \quad \begin{cases}
        p_1^{\tau} = p_1, \quad p_2^{\tau} = p_2, \quad p_3^{\tau} = p_3 \\
        \zeta_1^{\tau} = \zeta_1, \quad \zeta_2^{\tau} = \zeta_2, \quad \zeta_3^{\tau} = \zeta_3
    \end{cases}
\end{equation*}
The saddle point is a solution of the set of equations:
\begin{equation}
    \begin{cases}
    \zeta_1^* = \frac{\mathcal{C}_{\mu_W}^{(\alpha)}(p_1^* p_2^*)}{p_1^*}, \quad \zeta_2^* = \frac{\alpha}{p_2^*} \big( \mathcal{C}_{\mu_W}^{(\alpha)}(p_1^* p_2^*) + \mathcal{C}_{\mu_Y}^{(\alpha)}(p_2^* p_3^*) \big), \quad \zeta_3^* = \frac{\mathcal{C}_{\mu_Y}^{(\alpha)}(p_2^* p_3^*)}{p_3^*} \\
    \\
    p_1^* = \frac{1}{\zeta_3^*} \mathcal{G}_{\rho_{X^2}} \big( \frac{z - \zeta_1^* }{\zeta_3^*} \big), \quad p_2^* = \frac{1}{z - \zeta_2^*}, \quad p_3^* = \frac{z - \zeta_1^*}{{\zeta_3^*}^2} \mathcal{G}_{\rho_{X^2}} \big( \frac{z - \zeta_1^* }{\zeta_3^*} \big) - \frac{1}{\zeta_3^*}
    \end{cases}
    \label{X-sol-app}
\end{equation}

Now, since the relation \eqref{X-Saddle Point Integral} and the solutions \eqref{X-sol-app} hold for arbitrary indices $i,j$, we can state the relation in matrix form. The inverse of ${\bC_X^*}^{-1}$, and the block structure of $\bG_{\mathcal{S}}(z)$ are computed in sections \ref{Matrix-tool}. From \eqref{resolvent_S}, \eqref{CX-inv} we have (for sufficiently large $N$):
\begin{equation}
\begin{split}
    \big\langle \bG_{\mathcal{S}}(z) \big\rangle_{\bU_1, \bU_2, \bV_1, \bV_2} &= \Bigg\langle \left[
\begin{array}{cc}
\frac{1}{z} \bI_N + \frac{1}{z} \bS \bG_{S^\intercal S}(z^2) \bS^\intercal & \bS \bG_{S^\intercal S}(z^2) \\
\bG_{S^\intercal S}(z^2) \bS^\intercal & z \bG_{S^\intercal S}(z^2)
\end{array}
\right] \Bigg\rangle \\
&= \left[
\begin{array}{cc}
 \frac{1}{\zeta_3^*}\bG_{X^2} \big( \frac{z - \zeta_1^*}{\zeta_3^*} \big) & \mathbf{0}  \\
\mathbf{0} & \frac{1}{z-\zeta_2^*} \bI_M 
\end{array}
\right]
\end{split}
\label{X-resolvent-relation-app}
\end{equation}

With this relation, we proceed to   simplify the equations \eqref{X-sol-app}. 

The normalized trace of the upper-left blocks of $\big\langle \bG_{\mathcal{S}}(z) \big\rangle_{\bU_1, \bU_2, \bV_1, \bV_2}$ is:
\begin{equation}
\begin{split}
    \frac{1}{N} \sum_{k=1}^N \big[ \frac{1}{z} + \frac{1}{z} \frac{\gamma_k^2}{z^2 - \gamma_k^2} \big] &= \frac{1}{z} \frac{1}{N} \sum_{k=1}^N \big[ 1 + \frac{\gamma_k^2}{z^2 - \gamma_k^2} \big] \\
    &= z \frac{1}{N} \sum_{k=1}^N \frac{1}{z^2 - \gamma_k^2} \\
    &= \frac{1}{2 N} \sum_{k=1}^N \big[  \frac{1}{z - \gamma_k} + \frac{1}{z + \gamma_k} \big] = \mathcal{G}_{\bar{\mu}_S} (z)
\end{split}
\label{trace-GS-X-first}
\end{equation}
and the normalized trace of the upper-left block in ${\bC_X^*}^{-1}$ is $\frac{1}{\zeta_3^*} \mathcal{G}_{\rho_{X^2}} \big( \frac{z - \zeta_1^*}{\zeta_3^*} \big) = p_1^*$. Therefore, we have $p_1^* = \mathcal{G}_{\bar{\mu}_S} (z)$.

The normalized trace of lower-right block of $\big\langle \bG_{\mathcal{S}}(z) \big\rangle_{\bU_1, \bU_2, \bV_1, \bV_2}$ reads:
\begin{equation}
\begin{split}
    \frac{1}{M} z \Big[ \sum_{k = 1}^N \frac{1}{z^2 - \gamma_k^2} + (M-N) \frac{1}{z^2} \Big] &= \frac{N}{M} \mathcal{G}_{\bar{\mu}_S} (z) + \frac{M-N}{M} \frac{1}{z} = \alpha \mathcal{G}_{\bar{\mu}_S} (z) + (1-\alpha) \frac{1}{z}
\end{split}
\label{trace-GS-X-last}
\end{equation}
and the normalized trace of the lower-right  block in ${\bC_X^*}^{-1}$ is $\frac{1}{z-\zeta_2^*} = p_2^*$. Therefore, we have $p_2^* = \alpha \mathcal{G}_{\bar{\mu}_S} (z) + (1-\alpha) \frac{1}{z}$. Moreover, we also have that $\zeta_2^* = \alpha z \frac{z \mathcal{G}_{\bar{\mu}_S} (z) - 1 }{\alpha z \mathcal{G}_{\bar{\mu}_S} (z) + 1- \alpha}$.

Therefore, the saddle point equations \eqref{X-sol-app} can be rewritten in a simplified form, which does not involve $\rho_{X^2}$,  as:
\begin{equation}
    \begin{cases}
    \zeta_1^* = \frac{\mathcal{C}_{\mu_W}^{(\alpha)}(p_1^* p_2^*)}{p_1^*}, \quad \zeta_2^* = \alpha z \frac{z \mathcal{G}_{\bar{\mu}_S} (z) - 1 }{\alpha z \mathcal{G}_{\bar{\mu}_S} (z) + 1- \alpha}, \quad \zeta_3^* = \frac{\mathcal{C}_{\mu_Y}^{(\alpha)}(p_2^* p_3^*)}{p_3^*} \\
    \\
    p_1^* = \mathcal{G}_{\bar{\mu}_S} (z), \quad p_2^* = \alpha \mathcal{G}_{\bar{\mu}_S} (z) + (1-\alpha) \frac{1}{z}, \quad p_3^* = \frac{z - \zeta_1^*}{\zeta_3^*} \mathcal{G}_{\bar{\mu}_S} (z) - \frac{1}{\zeta_3^*} 
    \end{cases}
    \label{X-sol-app-sec}
\end{equation}
Note that $\zeta_1^*, \zeta_2^*$ can be computed from the observation matrix, and we only need to find $\zeta_3^*$ satisfying the following equation:
\begin{equation}
    (z - \zeta_1^*) \mathcal{G}_{\bar{\mu}_S} (z) - 1 = \mathcal{C}_{\mu_Y}^{(\alpha)}\Big( \frac{1}{\zeta_3^*} \big[\alpha \mathcal{G}_{\bar{\mu}_S} (z) + \frac{1-\alpha}{z} \big] \big[(z - \zeta_1^*) \mathcal{G}_{\bar{\mu}_S} (z) - 1 \big] \Big)
    \label{zeta3-X-est-app}
\end{equation}

\subsection{Overlaps and optimal eigenvalues}\label{X-overlap-ev}
We restate the relation between the resolvent and the overlaps from the main text \eqref{X-resolvent-overlap}. For $\tilde{\bx}_i = [ \bx_i^\intercal ,  \mathbf{0}_{M} ]^\intercal$ with $\bx_i$ eigenvectors of $\bX$, we have:
\begin{equation}
    \tilde{\bx}_i^\intercal \big( {\rm Im}\, \bG_{\mathcal{S}}(x - \ci \epsilon) \big)\tilde{\bx}_i \approx \pi \bar{\mu}_{S}(x) O_X(x, \lambda_i)
    \label{X-resolvent-overlap-app}
\end{equation}
From \eqref{X-resolvent-overlap-app}, \eqref{X-resolvent-relation-app}, we find:
\begin{equation}
\begin{split}
    O_X(\gamma, \lambda_i) &\approx \frac{1}{\pi \bar{\mu}_{S}(\gamma)} \, {\rm Im} \, \lim_{z \to \gamma - \ci 0^+} \, \bx_i^\intercal \, {\zeta_3^*}^{-1} \bG_{X^2} \big( \frac{z - \zeta^*_1 }{\zeta^*_3} \big) \, \bx_i \\
    &= \frac{1}{\pi \bar{\mu}_{S}(\gamma)} \, {\rm Im} \, \lim_{z \to \gamma - \ci 0^+} \,  \frac{1}{ z - \zeta^*_1  - \zeta^*_3 \lambda_i^2}
\end{split}
\label{X-overlap-eq-app}
\end{equation}

Once we have the overlap, we can compute the optimal eigenvalues from \eqref{eig-over-rel} in section \ref{Derivation}. Note that, until now we had absorbed $\sqrt{\kappa}$ into $\bX$. Therefore, we should use \eqref{X-overlap-eq-app} with $O_X(\gamma, \sqrt\kappa\lambda_i)$. This leads to: 
\begin{equation}
\begin{split}
    \widehat{\xi_x^*}_i &\approx \frac{1}{N}  \sum_{j=1}^N \lambda_j O_X(\gamma_i, \sqrt\kappa\lambda_j) \\
    &\approx \frac{1}{\pi \bar{\mu}_{S}(\gamma_i) } \, {\rm Im} \, \lim_{z \to \gamma_i - \ci 0^+}  \, \frac{1}{N}  \sum_{j=1}^N \frac{\lambda_j}{ z - \zeta_1^*  - \zeta_3^* \kappa \lambda_j^2} \\
    &= \frac{1}{\pi \bar{\mu}_{S}(\gamma_i) } \, {\rm Im} \, \lim_{z \to \gamma_i - \ci 0^+}  \, \frac{1}{\kappa \zeta_3^*} \frac{1}{N}  \sum_{j=1}^N \frac{\lambda_j}{ \frac{z-\zeta_1^*}{\kappa \zeta_3^*} -  \lambda_j^2} \\
    &= \frac{1}{\kappa \pi \bar{\mu}_{S}(\gamma_i) } \, {\rm Im} \,  \lim_{z \to \gamma_i - \ci 0^+}   \, \frac{1}{\zeta_3^*} \bigg( \frac{1}{2} \frac{1}{N}  \sum_{j=1}^N \frac{1}{ \sqrt{\frac{z-\zeta_1^*}{\kappa \zeta_3^*}} -  \lambda_j} - \frac{1}{2} \frac{1}{N}  \sum_{j=1}^N \frac{1}{ \sqrt{\frac{z-\zeta_1^*}{\kappa \zeta_3^*}} +  \lambda_j} \bigg) \\
    &\approx \frac{1}{\kappa \pi \bar{\mu}_{S}(\gamma_i) } \, {\rm Im} \, \lim_{z \to \gamma_i - \ci 0^+}   \bigg\{ \frac{1}{2} \frac{1}{\zeta_3^*} \mathcal{G}_{\rho_X}\Big(\sqrt{\frac{z-\zeta_1^*}{\kappa \zeta_3^*}}\Big) - \frac{1}{2} \frac{1}{\zeta_3^*} \mathcal{G}_{\rho_{-X}}\Big(\sqrt{\frac{z-\zeta_1^*}{\kappa \zeta_3^*}}\Big) \bigg\} \\
    &= \frac{1}{2 \kappa \pi \bar{\mu}_{S}(\gamma_i) } \, {\rm Im}  \, \lim_{z \to \gamma_i - \ci 0^+}  \,  \bigg\{  \frac{1}{\zeta_3^*}  \Big[\mathcal{G}_{\rho_X}\Big(\sqrt{\frac{z-\zeta_1^*}{\kappa \zeta_3^*}}\Big) +  \mathcal{G}_{\rho_{X}}\Big(-\sqrt{\frac{z-\zeta_1^*}{\kappa \zeta_3^*}}\Big) \Big] \bigg\}
\end{split}
\label{X-optimal-ev-app}
\end{equation}

\subsubsection{Estimating $\bX^2$}
The resolvent relation we have found in \eqref{X-resolvent-relation-app} is in terms of $\bG_{X^2}$. Therefore, like other RIEs in other problems \cite{bun2016rotational, pourkamali2023rectangular}, we can express the estimator for $\bX^2$ without any knowledge about $\rho_X$ or $\rho_{X^2}$. One can see that, the optimal RIE for $\bX^2$ is constructed in the same way as for $\bX$ with eigenvalues denoted by $\widehat{\xi_{x^2}^*}_i$. To compute the optimal eigenvalues, we absorb $\sqrt{\kappa}$ into $\bX$ and we use the exact expression in \eqref{X-overlap-eq-app}. In the end, we only need to divide by $\kappa$ to find an estimator for the true $\bX^2$.
\begin{equation}
\begin{split}
    \widehat{\xi_{x^2}^*}_i &\approx \frac{1}{N}  \sum_{j=1}^N \lambda_j^2 O_X(\gamma_i, \lambda_j) \\
    &\approx \frac{1}{\pi \bar{\mu}_{S}(\gamma_i) } \, {\rm Im} \, \lim_{z \to \gamma_i - \ci 0^+}  \, \frac{1}{N}  \sum_{j=1}^N \frac{\lambda_j^2}{ z - \zeta_1^*  - \zeta_3^* \lambda_j^2} \\
    &= \frac{1}{\pi \bar{\mu}_{S}(\gamma_i) } \, {\rm Im} \, \lim_{z \to \gamma_i - \ci 0^+}  \, \frac{1}{\zeta_3^*} \frac{1}{N}  \sum_{j=1}^N \frac{\lambda_j^2}{ \frac{z-\zeta_1^*}{\zeta_3^*} -  \lambda_j^2} \\
    &= \frac{1}{\pi \bar{\mu}_{S}(\gamma_i) } \, {\rm Im} \, \lim_{z \to \gamma_i - \ci 0^+}  \, - \frac{1}{\zeta_3^*} \frac{1}{N}  \sum_{j=1}^N \frac{\frac{z-\zeta_1^*}{\zeta_3^*} - \lambda_j^2 - \frac{z-\zeta_1^*}{\zeta_3^*}}{ \frac{z-\zeta_1^*}{\zeta_3^*} -  \lambda_j^2} \\
    &= \frac{1}{\pi \bar{\mu}_{S}(\gamma_i) } \, {\rm Im} \, \lim_{z \to \gamma_i - \ci 0^+}  \, - \frac{1}{\zeta_3^*} \frac{1}{N}  \sum_{j=1}^N \Big[ 1 - \frac{z-\zeta_1^*}{\zeta_3^*} \frac{1}{ \frac{z-\zeta_1^*}{\zeta_3^*} -  \lambda_j^2} \Big] \\
    &\approx \frac{1}{\pi \bar{\mu}_{S}(\gamma_i) } \, {\rm Im} \, \lim_{z \to \gamma_i - \ci 0^+}  \, - \frac{1}{\zeta_3^*} +  \frac{z-\zeta_1^*}{{\zeta_3^*}^2} \mathcal{G}_{\rho_{X^2}} \big( \frac{z-\zeta_1^*}{\zeta_3^*}\big) \\
    &\stackrel{\text{(a)}}{=} \frac{1}{\pi \bar{\mu}_{S}(\gamma_i) } \, {\rm Im} \, \lim_{z \to \gamma_i - \ci 0^+}  \, p_3^* \\
    &\stackrel{\text{(b)}}{=} \frac{1}{\pi \bar{\mu}_{S}(\gamma_i) } \, {\rm Im} \, \lim_{z \to \gamma_i - \ci 0^+}  \, \frac{z - \zeta_1^*}{\zeta_3^*} \mathcal{G}_{\bar{\mu}_S} (z) - \frac{1}{\zeta_3^*} 
\end{split}
\label{X2-optimal-ev-app-comp}
\end{equation}
where in (a) we used \eqref{X-sol-app}, and for (b) we used \eqref{X-sol-app-sec}. Thus, the optimal eigenvalues for $\bX^2$ read:
\begin{equation}
    \widehat{\xi_{x^2}^*}_i = \frac{1}{\kappa} \frac{1}{\pi \bar{\mu}_{S}(\gamma_i) } \, {\rm Im} \, \lim_{z \to \gamma_i - \ci 0^+}  \, \frac{z - \zeta_1^*}{\zeta_3^*} \mathcal{G}_{\bar{\mu}_S} (z) - \frac{1}{\zeta_3^*} 
    \label{X2-optimal-ev-app}
\end{equation}
Note that the parameters $\zeta_1^*, \zeta_3^*$ can be computed from \eqref{X-sol-app-sec}, \eqref{zeta3-X-est-app}, without the knowledge of $\rho_X$ or $\rho_{X^2}$.

\begin{remark}
The main barrier to find an estimator for $\bX$ is that the resolvent relation \eqref{X-resolvent-relation-app} is in terms of $\mathcal{G}_{\rho_{X^2}}$. Moreover, in the estimator for $\bX$, second equality in \eqref{X-optimal-ev-app}, we have the sum $\sum_{j=1}^N \frac{\lambda_j}{ z - \zeta_1^*  - \kappa \zeta_3^* \lambda_j^2}$ which cannot be written in terms of $\mathcal{G}_{\rho_{X^2}}$.
\end{remark}

\begin{remark}\label{est-sqXX}
    If we add the assumption that the matrix $\bX$ is positive semi-definite, without any further knowledge on the prior, we can use $\sqrt{\widehat{\xi_{x^2}^*}_i}$ for the eigenvalues of $\bm{\Xi}_X(\bS)$. However, note that, this estimator is sub-optimal for $\bX$ as $\sqrt{\sum_{j=1}^N \lambda_j^2 \, \big(  \bu_i^\intercal \bx_j \big)^2} \neq \sum_{j=1}^N \lambda_j \, \big(  \bu_i^\intercal \bx_j \big)^2$. 
\end{remark}

\subsection{Numerical Examples}\label{X-examples}
In this section, we will illustrate the derived formulas \eqref{X-resolvent-relation-app}, \eqref{X-overlap-eq-app}, and \eqref{X-optimal-ev-app} with numerical experiments.

We consider matrices $\bY, \bW \in \bR^{N \times M}$ to have i.i.d. Gaussian entries, so $\mathcal{C}^{(\alpha)}_{\mu_Y}(z) = \mathcal{C}^{(\alpha)}_{\mu_W}(z) = \frac{1}{\alpha} z$ which leads to a simplification of saddle point equations \eqref{X-sol-app-sec}: 
\begin{equation}
    \begin{cases}
    \zeta_1^* = \frac{1}{\alpha} p_2^*, \quad \zeta_2^* = \alpha z \frac{z \mathcal{G}_{\bar{\mu}_S} (z) - 1 }{\alpha z \mathcal{G}_{\bar{\mu}_S} (z) + 1- \alpha}, \quad \zeta_3^* = \frac{1}{\alpha} p_2^* \\
    p_1^* = \mathcal{G}_{\bar{\mu}_S} (z), \quad p_2^* = \alpha \mathcal{G}_{\bar{\mu}_S} (z) + (1-\alpha) \frac{1}{z}, \quad p_3^* = \frac{z - \zeta_1^*}{\zeta_3^*} \mathcal{G}_{\bar{\mu}_S} (z) - \frac{1}{\zeta_3^*} 
    \end{cases}
    \label{X-Y,W-Gauss}
\end{equation}

\subsubsection{Resolvent relation}
We take $\kappa =1$. In model \eqref{X-model}, without loss of generality we can consider $\bX$ to be diagonal. In figures \ref{fig: X - Resolvent-Relation - first example} and \ref{fig: X - Resolvent-Relation - second example} respectively, we consider the $\bX$ to be a diagonal matrix obtained by taking the eigenvalues of a Wigner matrix and a Wishart matrix respectively.


Note that $\mu_S$ and $\mathcal{G}_{\bar{\mu}_S} (z)$ can be computed analytically using tools from random matrix theory, but the computation is highly involved. In our experiments, we use instead a numerical estimation of $\mathcal{G}_{\bar{\mu}_S} (z)$ obtained from the observation matrix with the help of a Cauchy kernel  to compute the parameters  $\zeta^*_1, \zeta^*_3$ (see \cite{potters2020first} for details on the Cauchy kernel method).

Unlike the simpler models \cite{bun2017cleaning} for which the fluctuations are derived to be of the order $\nicefrac{1}{\sqrt{N}}$, based on our derivation we cannot assess the order of fluctuations. However, from our numerics we observe that the fluctuations are of the order $o(N)$. Moreover, fluctuations near the edge points of density are larger (in particular for the last row in both figures \ref{fig: X - Resolvent-Relation - first example}, \ref{fig: X - Resolvent-Relation - second example}), which is due to the fact that the limiting measures have higher fluctuations on their edge-points. 

Another observation, from comparison of figures \ref{fig: X - Resolvent-Relation - first example}, \ref{fig: X - Resolvent-Relation - second example}, is that the fluctuations for the first example are relatively larger than the second one. One possible guess could be that this is due to the symmetry of $\rho_X$ in the first example. However based on more extensive numerical observations (which we omit here) we speculate that this issue is in fact related to the existence of small eigenvalues of $\bX$. In other words, if $\bX$ has eigenvalue $0$ or small eigenvalues, we have higher fluctuations in the relation \eqref{X-resolvent-overlap-app}. 
\begin{figure}
\begin{subfigure}[t]{\textwidth}
  \begin{subfigure}[t]{.4\textwidth}
    \centering
    \input{Figures/App/X-est/Resolvent_Relation/X_ResolventRelatation_real_entry1}
  \end{subfigure}
  \hfil
  \begin{subfigure}[t]{.4\textwidth}
  \centering
      \input{Figures/App/X-est/Resolvent_Relation/X_ResolventRelatation_imag_entry1}
  \end{subfigure}
  \caption{{\small Entry $ i = j = 1$, first diagonal entry in the upper-left block}}
  \end{subfigure}
  \\
  \begin{subfigure}[t]{\textwidth}
  \begin{subfigure}[t]{.4\textwidth}
    \centering
    \input{Figures/App/X-est/Resolvent_Relation/X_ResolventRelatation_real_entry2}
  \end{subfigure}
  \hfil
  \begin{subfigure}[t]{.4\textwidth}
  \centering
      \input{Figures/App/X-est/Resolvent_Relation/X_ResolventRelatation_imag_entry2}
  \end{subfigure}
  \caption{{\small Entry $ i = j = 5000$, last diagonal entry in the lower-right block}}
  \end{subfigure}
  \\
  \begin{subfigure}[t]{\textwidth}
  \begin{subfigure}[t]{.4\textwidth}
    \centering
    \input{Figures/App/X-est/Resolvent_Relation/X_ResolventRelatation_real_entry3}
  \end{subfigure}
  \hfil
  \begin{subfigure}[t]{.4\textwidth}
  \centering
      \input{Figures/App/X-est/Resolvent_Relation/X_ResolventRelatation_imag_entry3}
  \end{subfigure}
  \caption{{\small Entry $ i = 4999,  j = 5000$, a non-diagonal entry in the lower-right block}}
  \end{subfigure}
  \\
  \begin{subfigure}[t]{\textwidth}
  \begin{subfigure}[t]{.4\textwidth}
    \centering
    \input{Figures/App/X-est/Resolvent_Relation/X_ResolventRelatation_real_entry4}
  \end{subfigure}
  \hfil
  \begin{subfigure}[t]{.4\textwidth}
  \centering
      \input{Figures/App/X-est/Resolvent_Relation/X_ResolventRelatation_imag_entry4}
  \end{subfigure}
  \caption{{\small Entry $ i = 1, j = 5000$, an entry in the upper-right block}}
  \end{subfigure}
  \\
  \begin{subfigure}[t]{\textwidth}
  \begin{subfigure}[t]{.4\textwidth}
    \centering
    \input{Figures/App/X-est/Resolvent_Relation/X_ResolventRelatation_real_entry5}
  \end{subfigure}
  \hfil
  \begin{subfigure}[t]{.4\textwidth}
  \centering
      \input{Figures/App/X-est/Resolvent_Relation/X_ResolventRelatation_imag_entry5}
  \end{subfigure}
  \caption{{\small Entry $ i = 1, j = 2$, a non-diagonal entry in the upper-left block}}
  \end{subfigure}
    \caption{\small Illustration of \eqref{X-resolvent-relation-app}. $\bX$ is diagonal matrix from the eigenvalues of a Wigner matrix and $\bY, \bZ$ are Gaussian matrices with $ N=2000, M = 3000$. The empirical estimate of $\bG_{\mathcal{S}}(z)$ (dashed blue line) is computed for $ z = \gamma_i - \ci \sqrt{\frac{1}{2N}}$ for $1 \leq i \leq N$. Theoretical estimate (solid orange line) computed from the rhs of \eqref{X-resolvent-relation-app} with parameters obtained from the generated matrix. Note that, the theoretical estimate has also fluctuations because the parameters $\zeta^*_1, \zeta^*_3$ are given by the numerical estimate of $\mathcal{G}_{\bar{\mu}_S}(z)$.
    }\label{fig: X - Resolvent-Relation - first example}
\end{figure}

\begin{figure}
\begin{subfigure}[t]{\textwidth}
  \begin{subfigure}[t]{.4\textwidth}
    \centering
    \input{Figures/App/X-est/Resolvent_Relation/X_2nd_ResolventRelatation_real_entry1}
  \end{subfigure}
  \hfil
  \begin{subfigure}[t]{.4\textwidth}
  \centering
      \input{Figures/App/X-est/Resolvent_Relation/X_2nd_ResolventRelatation_imag_entry1}
  \end{subfigure}
  \caption{{\small Entry $ i = j = 1$, first diagonal entry in the upper-left block}}
  \end{subfigure}
  \\
  \begin{subfigure}[t]{\textwidth}
  \begin{subfigure}[t]{.4\textwidth}
    \centering
    \input{Figures/App/X-est/Resolvent_Relation/X_2nd_ResolventRelatation_real_entry2}
  \end{subfigure}
  \hfil
  \begin{subfigure}[t]{.4\textwidth}
  \centering
      \input{Figures/App/X-est/Resolvent_Relation/X_2nd_ResolventRelatation_imag_entry2}
  \end{subfigure}
  \caption{{\small Entry $ i = j = 5000$, last diagonal entry in the lower-right block}}
  \end{subfigure}
  \\
  \begin{subfigure}[t]{\textwidth}
  \begin{subfigure}[t]{.4\textwidth}
    \centering
    \input{Figures/App/X-est/Resolvent_Relation/X_2nd_ResolventRelatation_real_entry3}
  \end{subfigure}
  \hfil
  \begin{subfigure}[t]{.4\textwidth}
  \centering
      \input{Figures/App/X-est/Resolvent_Relation/X_2nd_ResolventRelatation_imag_entry3}
  \end{subfigure}
  \caption{{\small Entry $ i = 4999,  j = 5000$, a non-diagonal entry in the lower-right block}}
  \end{subfigure}
  \\
  \begin{subfigure}[t]{\textwidth}
  \begin{subfigure}[t]{.4\textwidth}
    \centering
    \input{Figures/App/X-est/Resolvent_Relation/X_2nd_ResolventRelatation_real_entry4}
  \end{subfigure}
  \hfil
  \begin{subfigure}[t]{.4\textwidth}
  \centering
      \input{Figures/App/X-est/Resolvent_Relation/X_2nd_ResolventRelatation_imag_entry4}
  \end{subfigure}
  \caption{{\small Entry $ i = 1, j = 5000$, an entry in the upper-right block}}
  \end{subfigure}
  \\
  \begin{subfigure}[t]{\textwidth}
  \begin{subfigure}[t]{.4\textwidth}
    \centering
    \input{Figures/App/X-est/Resolvent_Relation/X_2nd_ResolventRelatation_real_entry5}
  \end{subfigure}
  \hfil
  \begin{subfigure}[t]{.4\textwidth}
  \centering
      \input{Figures/App/X-est/Resolvent_Relation/X_2nd_ResolventRelatation_imag_entry5}
  \end{subfigure}
  \caption{{\small Entry $ i = 1, j = 2$, a non-diagonal entry in the upper-left block}}
  \end{subfigure}
    \caption{\small Illustration of \eqref{X-resolvent-relation-app}.$\bX$ is diagonal matrix from the eigenvalues of a Wishart matrix with aspect ratio $\nicefrac{1}{2}$ and $\bY, \bZ$ are Gaussian matrices with $ N=2000, M = 3000$. The empirical estimate of $\bG_{\mathcal{S}}(z)$ (dashed blue line) is computed for $ z = \gamma_i - \ci \sqrt{\frac{1}{2N}}$ for $1 \leq i \leq N$. The Theoretical estimate (solid orange line) is computed from the rhs of \eqref{X-resolvent-relation-app} with parameters obtained from the generated matrix. Note that, the theoretical estimate has also fluctuations because the parameters $\zeta^*_1, \zeta^*_3$ are given by the numerical estimate of $\mathcal{G}_{\bar{\mu}_S}(z)$.
    }\label{fig: X - Resolvent-Relation - second example}
\end{figure}

\subsubsection{Overlaps}
To illustrate the formula for the overlap \eqref{X-overlap-eq-app}, we fix the matrix $\bX$ and run experiments over various realization of the model \eqref{X-model}. For each experiment, we record the overlap of $k$-th left singular vector of $\bS$ and the eigenvectors of $\bX$. To compute the theoretical prediction, we find $\zeta^*_1 = \zeta^*_3$ for $z = \bar{\gamma}_k - \ci 0^+$ where $\bar{\gamma}_k$ is the average of $k$-th singular value of $\bS$ in the experiments.

To find $\zeta^*_1 = \zeta^*_3$, we use the set of equations \eqref{X-sol-app} which for $\bY, \bW$ Gaussian can be written as:
\begin{equation}
    \begin{cases}
    \zeta_1^* = \frac{1}{\alpha} p_2^*, \quad \zeta_2^* =  p_1^* + p_3^*, \quad \zeta_3^* = \frac{1}{\alpha} p_2^*  \\
    \\
    p_1^* = \frac{1}{\zeta_1^*} \mathcal{G}_{\rho_{X^2}} \big( \frac{z}{\zeta_1^*} - 1 \big), \quad p_2^* = \frac{1}{z - \zeta_2^*}, \quad p_3^* = \frac{z - \zeta_1^*}{{\zeta_1^*}^2} \mathcal{G}_{\rho_{X^2}} \big( \frac{z}{\zeta_1^*} - 1 \big) - \frac{1}{\zeta_1^*}
    \end{cases}
    \label{X-sol-Y,Z G}
\end{equation}
Now we proceed to simplify the solution above:
\begin{equation*}
    \zeta_2^* =  p_1^* + p_3^* = \frac{z}{{\zeta_1^*}^2} \mathcal{G}_{\rho_{X^2}} \big( \frac{z}{\zeta_1^*} - 1 \big) - \frac{1}{\zeta_1^*}
\end{equation*}
\begin{equation*}
    p_2^* =  \frac{1}{z - \zeta_2^*} = \frac{\zeta_1^*}{\zeta_1^* z -  \frac{z}{\zeta_1^*} \mathcal{G}_{\rho_{X^2}} \big( \frac{z}{\zeta_1^*} - 1 \big) + 1}
\end{equation*}
\begin{equation}
\begin{split}
    \zeta_1^* = \frac{1}{\alpha} p_2^* &\Longrightarrow  \zeta_1^* z -  \frac{z}{\zeta_1^*} \mathcal{G}_{\rho_{X^2}} \big( \frac{z}{\zeta_1^*} - 1 \big) + 1 = \frac{1}{\alpha} \\
    &\Rightarrow \mathcal{G}_{\rho_{X^2}} \big( \frac{z}{\zeta_1^*} - 1 \big) = {\zeta_1^*}^2 + \big(1-\frac{1}{\alpha} \big) \frac{\zeta_1^*}{z} \\
    &\Rightarrow  \frac{z}{\zeta_1^*} - 1 = \mathcal{G}_{\rho_{X^2}}^{-1} \Big( {\zeta_1^*}^2 + \big(1-\frac{1}{\alpha} \big) \frac{\zeta_1^*}{z} \Big)\\
    &\Rightarrow  \frac{z}{\zeta_1^*} - 1 - \frac{1}{ {\zeta_1^*}^2 + \big(1-\frac{1}{\alpha} \big) \frac{\zeta_1^*}{z}} = \mathcal{R}_{\rho_{X^2}} \Big( {\zeta_1^*}^2 + \big(1-\frac{1}{\alpha} \big) \frac{\zeta_1^*}{z} \Big)
\end{split}
\label{X-self-consistent zeta}
\end{equation}
Thus, $\zeta^*_1$ is the solution to \eqref{X-self-consistent zeta}. For each example, we solve this equation and compare the obtained theoretical overlap against the average over the experiments.

\begin{figure}
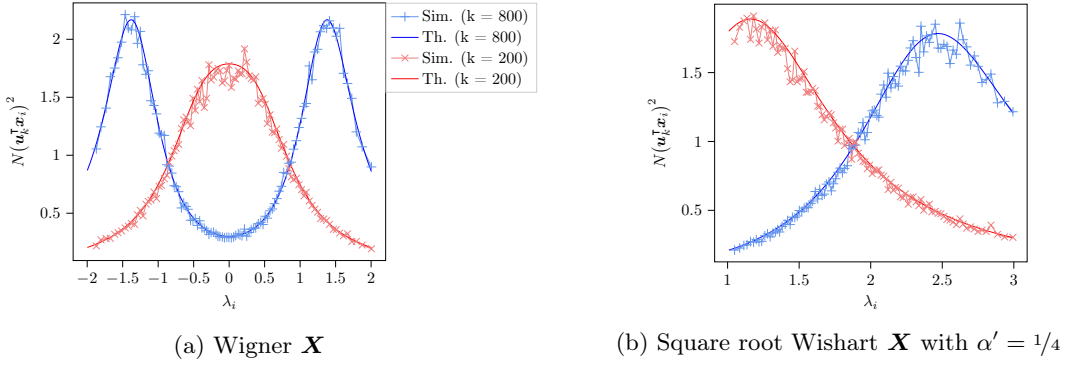

    \centering
\begin{subfigure}[t]{.4\textwidth}
    \centering
    \input{Figures/App/X-est/Overlap/Overlap_WignerX_YZGauss}
    \vspace{-11pt}
    \caption{\small Wigner $\bX$}
    \label{fig:X-Overlap-Wigner X-app}
\end{subfigure}
\hfil
\begin{subfigure}[t]{.4\textwidth}
    \centering
    \input{Figures/App/X-est/Overlap/Overlap_sqrtWishartX_YZGauss}
    \caption{\small Square root Wishart $\bX$ with $\alpha'=\nicefrac{1}{4}$}
    \label{fig:X-Overlap-sqrtWishart X}
\end{subfigure}
\caption{\small Computation of the rescaled overlap. Both $\bY$ and $\bW$ are $N \times M$ matrices with i.i.d. Gaussian entries of variance $1/N$, and aspect ratio $N/M = 1/2$. The simulation results are averaged over 1000 experiments with fixed $\bX$, and $N = 1000, M =2000$. Some of the simulation points are dropped for clarity.}
\end{figure}

\paragraph{Wigner $\bX$.} Let $\bX \in \bR^{N \times N}$ be a Wigner matrix, then $\mathcal{R}_{\rho_{X^2}}(z) = \frac{1}{1-z}$. Solving \eqref{X-self-consistent zeta}, we can compute the overlap using \eqref{X-overlap-eq-app}. In Fig. \ref{fig:X-Overlap-Wigner X-app}, we compare the theoretical computation with simulations for $N =1000, M = 2000$. As in previous cases $\bar{\mu}_{S}(\gamma)$ is approximated using a Cauchy kernel \cite{potters2020first}. 

\paragraph{Square root Wishart $\bX$.} Let $\bX \in \bR^{N \times N}$ be the square root of a Wishart matrix $\bX = \sqrt{\frac{1}{N} \bH \bH^\intercal}$ with $\bH \in \bR^{N \times N'}$ having i.i.d. Gaussian entries. Then $\mathcal{R}_{\rho_{X^2}}(z) = \frac{1}{\alpha'}\frac{1}{1-z}$, $\alpha' = \nicefrac{N}{N'}$. Solving \eqref{X-self-consistent zeta}, we can compute the overlap using \eqref{X-overlap-eq-app}. In Fig. \ref{fig:X-Overlap-sqrtWishart X}, we compare the theoretical computation with simulations for $N =1000, N'=4000, M = 2000$.

\subsubsection{RIE performance}\label{App-X-RIE}
In this section, we investigate the performance of our proposed estimators for $\bX$. We compare performances of the optimal RIE \eqref{X-optimal-ev-app} with the one of Oracle estimator \eqref{X-oracle-estimator}. Moreover, we illustrate the performance of the estimator for $\bX^2$ \eqref{X2-optimal-ev-app-comp}, and the sub-optimal estimator of $\bX$ derived from it, see remark \ref{est-sqXX}. 

For $\bY, \bW$ with Gaussian i.i.d. entries, \eqref{X2-optimal-ev-app} simplifies to:
\begin{equation}
\begin{split}
    \widehat{\xi_{x^2}^*}_i &= \frac{1}{\kappa} \frac{1}{\pi \bar{\mu}_{S}(\gamma_i) } \, {\rm Im} \, \lim_{z \to \gamma_i - \ci 0^+}  \, \frac{z - \zeta_1^*}{\zeta_3^*} \mathcal{G}_{\bar{\mu}_S} (z) - \frac{1}{\zeta_3^*} \\
    &= \frac{1}{\kappa} \frac{1}{\pi \bar{\mu}_{S}(\gamma_i) } \, {\rm Im} \, \lim_{z \to \gamma_i - \ci 0^+}  \, \frac{z}{\zeta_1^*} \mathcal{G}_{\bar{\mu}_S} (z) - \mathcal{G}_{\bar{\mu}_S} - \frac{1}{\zeta_1^*} \\
    &= \frac{1}{\kappa} \frac{1}{\pi \bar{\mu}_{S}(\gamma_i) } \, {\rm Im} \, \lim_{z \to \gamma_i - \ci 0^+}  \, \frac{z}{\mathcal{G}_{\bar{\mu}_S} (z) + \frac{1-\alpha}{\alpha} \frac{1}{z}} \mathcal{G}_{\bar{\mu}_S} (z) - \mathcal{G}_{\bar{\mu}_S}(z) - \frac{1}{\mathcal{G}_{\bar{\mu}_S} (z) + \frac{1-\alpha}{\alpha} \frac{1}{z}} \\
    &= \frac{1}{\kappa} \frac{1}{\pi \bar{\mu}_{S}(\gamma_i) } \, {\rm Im} \, \bigg\{ \frac{\gamma_i}{\pi \sH [\bar{\mu}_S](\gamma_i) + \pi \ci \bar{\mu}_S(\gamma_i) + \frac{1-\alpha}{\alpha} \frac{1}{\gamma_i}} \big(\pi \sH [\bar{\mu}_S](\gamma_i) + \pi \ci \bar{\mu}_S(\gamma_i) \big) \\
    & \hspace{3cm} - \big(\pi \sH [\bar{\mu}_S](\gamma_i) + \pi \ci \bar{\mu}_S(\gamma_i) \big) - \frac{1}{\pi \sH [\bar{\mu}_S](\gamma_i) + \pi \ci \bar{\mu}_S(\gamma_i)  + \frac{1-\alpha}{\alpha} \frac{1}{\gamma_i }} \bigg\} \\
    &= \frac{1}{\kappa} \frac{1}{\pi \bar{\mu}_{S}(\gamma_i) } \, \pi \bar{\mu}_{S}(\gamma_i) \bigg( -1 + \frac{1}{\alpha \Big( \pi^2 \bar{\mu}_{S}(\gamma_i)^2 + \big(\pi \sH [\bar{\mu}_S](\gamma_i) + \frac{-1+\frac{1}{\alpha}}{\gamma_i} \big)^2 \Big)   } \bigg) \\
    &= \frac{1}{\kappa} \, \Bigg[ -1 + \frac{1}{\alpha \Big( \pi^2 \bar{\mu}_{S}(\gamma_i)^2 + \big(\pi \sH [\bar{\mu}_S](\gamma_i) + \frac{-1+\frac{1}{\alpha}}{\gamma_i} \big)^2 \Big)   } \Bigg]
\end{split}
    \label{X2-optimal-ev-Y,W-G}
\end{equation}

For our first example, we consider two priors for $\bX$:

\paragraph{Shifted Wigner $\bX$.} We consider $\bX = \bF + c \bI$ where $\bF = \bF^\intercal \in \bR^{N \times N}$ has i.i.d. entries with variance $\nicefrac{1}{N}$, and $c \neq 0$ is a real number. Then, the spectrum of $\bX$ is a shifted version of the Wigner law
\begin{equation*}
    \rho_X(\lambda) = \frac{\sqrt{4 - (\lambda - c)^2}}{2 \pi}, \quad \text{for } c - 2 < \lambda < c+2,
\end{equation*}
and the Stieltjes transform reads:
\begin{equation*}
    \mathcal{G}_{\rho_X}(z) = \frac{z - c - \sqrt{(z-2-c)(z+2-c)}}{2}
\end{equation*}

\paragraph{Wishart $\bX$.} Take  $\bX = \frac{1}{N} \bH \bH^\intercal$ with $\bH \in \bR^{N \times N'}$ having i.i.d. Gaussian entries, with $\nicefrac{N}{N'} = \alpha' \leq 1$. Then, the spectrum of $\bX$ is the renowned \textit{Marchenko-Pastur} distribution:
\begin{equation*}
    \rho_X(\lambda) = \frac{\sqrt{\Big[ \lambda - \big( \frac{1}{\sqrt{\alpha'}} - 1 \big)^2 \Big] \Big[ \big( \frac{1}{\sqrt{\alpha'}} + 1 \big)^2 - \lambda \Big] }}{2 \pi \lambda}, \quad \text{for } \big( \frac{1}{\sqrt{\alpha'}} - 1 \big)^2 < \lambda < \big( \frac{1}{\sqrt{\alpha'}} + 1 \big)^2,
\end{equation*}
and the Stieltjes transform reads:
\begin{equation*}
    \mathcal{G}_{\rho_X}(z) = \frac{z - \big(\frac{1}{\alpha'} - 1 \big) - \sqrt{\Big[ z - \big( \frac{1}{\sqrt{\alpha'}} - 1 \big)^2 \Big] \Big[ z - \big( \frac{1}{\sqrt{\alpha'}} + 1 \big)^2  \Big] }}{2 z}
\end{equation*}

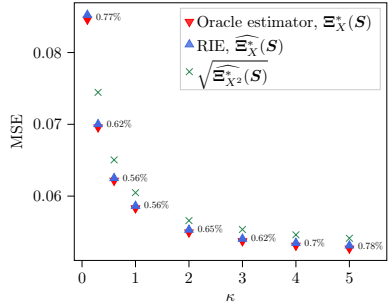
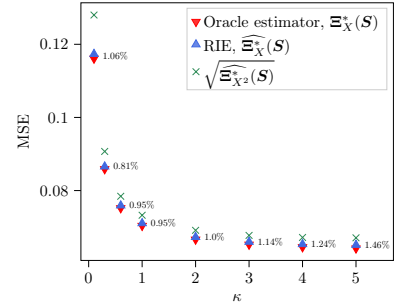
\begin{figure}
    \centering
\begin{subfigure}[t]{.4\textwidth}
    \centering
\begin{tikzpicture}[scale = 0.6]

\definecolor{darkgray176}{RGB}{176,176,176}
\definecolor{lightcoral}{RGB}{240,128,128}
\definecolor{lightgreen}{RGB}{144,238,144}
\definecolor{lightslategray}{RGB}{119,136,153}
\definecolor{royalblue}{RGB}{65,105,225}
\definecolor{seagreen}{RGB}{46,139,87}

\begin{axis}[
legend cell align={left},
legend style={fill opacity=0.8, draw opacity=1, text opacity=1, draw=white!80!black},
tick align=outside,
tick pos=left,
unbounded coords=jump,
x grid style={darkgray176},
xlabel={$\kappa$},
xmin=-0.145, xmax=5.7,
xtick style={color=black},
y grid style={darkgray176},
ylabel={${\rm MSE}$},
scaled y ticks=false,
yticklabel style={
  /pgf/number format/precision=3,
  /pgf/number format/fixed},
ymin=0.051040315258303, ymax=0.0869307379443293,
ytick style={color=black}
]
\path [draw=lightcoral, semithick]
(axis cs:0.1,0.0845193537313143)
--(axis cs:0.1,0.0846644640746536);

\path [draw=lightcoral, semithick]
(axis cs:0.3,0.0695232661561142)
--(axis cs:0.3,0.0696404298152873);

\path [draw=lightcoral, semithick]
(axis cs:0.6,0.0620853120295677)
--(axis cs:0.6,0.0622220281087357);

\path [draw=lightcoral, semithick]
(axis cs:1,0.0582493171426284)
--(axis cs:1,0.0583653128152994);

\path [draw=lightcoral, semithick]
(axis cs:2,0.0548931999231077)
--(axis cs:2,0.0550063880614862);

\path [draw=lightcoral, semithick]
(axis cs:3,0.0536561439020168)
--(axis cs:3,0.0537878803941807);

\path [draw=lightcoral, semithick]
(axis cs:4,0.0530671972952112)
--(axis cs:4,0.0531484793521173);

\path [draw=lightcoral, semithick]
(axis cs:5,0.0526716981076679)
--(axis cs:5,0.0527293543702968);

\path [draw=lightslategray, semithick]
(axis cs:0.1,0.0851817347604527)
--(axis cs:0.1,0.0852993550949645);

\path [draw=lightslategray, semithick]
(axis cs:0.3,0.0699414435397982)
--(axis cs:0.3,0.0700836759767609);

\path [draw=lightslategray, semithick]
(axis cs:0.6,0.0624288224880464)
--(axis cs:0.6,0.0625695507829693);

\path [draw=lightslategray, semithick]
(axis cs:1,0.0585657670402489)
--(axis cs:1,0.058696639350407);

\path [draw=lightslategray, semithick]
(axis cs:2,0.055237619640973)
--(axis cs:2,0.0553816296920481);

\path [draw=lightslategray, semithick]
(axis cs:3,0.0539805797357261)
--(axis cs:3,0.0541243469297613);

\path [draw=lightslategray, semithick]
(axis cs:4,0.0534134150952669)
--(axis cs:4,0.0535495079417727);

\path [draw=lightslategray, semithick]
(axis cs:5,0.053042543430421)
--(axis cs:5,0.0531852567982512);

\path [draw=lightgreen, semithick]
;

\path [draw=lightgreen, semithick]
(axis cs:0.3,0.0743699542750373)
--(axis cs:0.3,0.0745330186477573);

\path [draw=lightgreen, semithick]
(axis cs:0.6,0.0649735577649277)
--(axis cs:0.6,0.0651099368335978);

\path [draw=lightgreen, semithick]
(axis cs:1,0.0604351863369086)
--(axis cs:1,0.0605325261852758);

\path [draw=lightgreen, semithick]
(axis cs:2,0.0565337603591625)
--(axis cs:2,0.0566446141578634);

\path [draw=lightgreen, semithick]
(axis cs:3,0.0552918955288531)
--(axis cs:3,0.0554075290942125);

\path [draw=lightgreen, semithick]
(axis cs:4,0.0545694907933162)
--(axis cs:4,0.0546560920905964);

\path [draw=lightgreen, semithick]
(axis cs:5,0.054108529925749)
--(axis cs:5,0.0541487344274003);

\addplot [semithick, red, mark=triangle*, mark size=3, mark options={solid,rotate=180}, only marks]
table {%
0.1 0.0845919089029839
0.3 0.0695818479857008
0.6 0.0621536700691517
1 0.0583073149789639
2 0.054949793992297
3 0.0537220121480988
4 0.0531078383236643
5 0.0527005262389823
};
\addlegendentry{Oracle estimator, ${\bm{\Xi}_X^*}(\bS)$}
\addplot [semithick, royalblue, mark=triangle*, mark size=3, mark options={solid}, only marks]
table {%
0.1 0.0852405449277086
0.3 0.0700125597582796
0.6 0.0624991866355078
1 0.0586312031953279
2 0.0553096246665105
3 0.0540524633327437
4 0.0534814615185198
5 0.0531139001143361
};
\addlegendentry{RIE, $\widehat{{\bm{\Xi}_X^*}}(\bS)$}
\addplot [semithick, seagreen, mark=x, mark size=3, mark options={solid}, only marks]
table {%
0.1 nan
0.3 0.0744514864613973
0.6 0.0650417472992628
1 0.0604838562610922
2 0.056589187258513
3 0.0553497123115328
4 0.0546127914419563
5 0.0541286321765747
};
\addlegendentry{ $\sqrt{\widehat{{\bm{\Xi}_{X^2}^*}}(\bS)}$}
\draw (axis cs:0.2,0.0845919089029839) node[
  scale=0.6,
  anchor=base west,
  text=black,
  rotate=0.0
]{0.77\%};
\draw (axis cs:0.4,0.0695818479857008) node[
  scale=0.6,
  anchor=base west,
  text=black,
  rotate=0.0
]{0.62\%};
\draw (axis cs:0.7,0.0621536700691517) node[
  scale=0.6,
  anchor=base west,
  text=black,
  rotate=0.0
]{0.56\%};
\draw (axis cs:1.1,0.0583073149789639) node[
  scale=0.6,
  anchor=base west,
  text=black,
  rotate=0.0
]{0.56\%};
\draw (axis cs:2.1,0.054949793992297) node[
  scale=0.6,
  anchor=base west,
  text=black,
  rotate=0.0
]{0.65\%};
\draw (axis cs:3.1,0.0537220121480988) node[
  scale=0.6,
  anchor=base west,
  text=black,
  rotate=0.0
]{0.62\%};
\draw (axis cs:4.1,0.0531078383236643) node[
  scale=0.6,
  anchor=base west,
  text=black,
  rotate=0.0
]{0.7\%};
\draw (axis cs:5.1,0.0527005262389823) node[
  scale=0.6,
  anchor=base west,
  text=black,
  rotate=0.0
]{0.78\%};

\end{axis}

\end{tikzpicture}
    \caption{Shifted Wigner, $c=3$}
\end{subfigure}
\hfil
\begin{subfigure}[t]{.4\textwidth}
    \centering
\begin{tikzpicture}[scale = 0.6]

\definecolor{darkgray176}{RGB}{176,176,176}
\definecolor{lightcoral}{RGB}{240,128,128}
\definecolor{lightgreen}{RGB}{144,238,144}
\definecolor{lightslategray}{RGB}{119,136,153}
\definecolor{royalblue}{RGB}{65,105,225}
\definecolor{seagreen}{RGB}{46,139,87}

\begin{axis}[
legend cell align={left},
legend style={fill opacity=0.8, draw opacity=1, text opacity=1, draw=white!80!black},
tick align=outside,
tick pos=left,
x grid style={darkgray176},
xlabel={$ \kappa $},
xmin=-0.145, xmax=5.7,
xtick style={color=black},
y grid style={darkgray176},
ylabel={MSE},
yticklabel style={
  /pgf/number format/precision=3,
  /pgf/number format/fixed},
ymin=0.0608968192932828, ymax=0.131350065473814,
ytick style={color=black}
]
\path [draw=lightcoral, semithick]
(axis cs:0.1,0.11597352696668)
--(axis cs:0.1,0.116226286174364);

\path [draw=lightcoral, semithick]
(axis cs:0.3,0.0858004833813655)
--(axis cs:0.3,0.085934401854847);

\path [draw=lightcoral, semithick]
(axis cs:0.6,0.0751090320297627)
--(axis cs:0.6,0.0752794051870071);

\path [draw=lightcoral, semithick]
(axis cs:1,0.0702933183594271)
--(axis cs:1,0.0704626460099733);

\path [draw=lightcoral, semithick]
(axis cs:2,0.0665407707982209)
--(axis cs:2,0.0666836105777739);

\path [draw=lightcoral, semithick]
(axis cs:3,0.0651400173923888)
--(axis cs:3,0.0653269798137931);

\path [draw=lightcoral, semithick]
(axis cs:4,0.0644517949655576)
--(axis cs:4,0.0645970281938319);

\path [draw=lightcoral, semithick]
(axis cs:5,0.0640992395742161)
--(axis cs:5,0.064272661692868);

\path [draw=lightslategray, semithick]
(axis cs:0.1,0.117182008302904)
--(axis cs:0.1,0.117473960098773);

\path [draw=lightslategray, semithick]
(axis cs:0.3,0.0864583860418636)
--(axis cs:0.3,0.0866594082830361);

\path [draw=lightslategray, semithick]
(axis cs:0.6,0.0758041347484414)
--(axis cs:0.6,0.0760078225706573);

\path [draw=lightslategray, semithick]
(axis cs:1,0.0709352696450034)
--(axis cs:1,0.0711644066024661);

\path [draw=lightslategray, semithick]
(axis cs:2,0.0671669121333483)
--(axis cs:2,0.0673852905425038);

\path [draw=lightslategray, semithick]
(axis cs:3,0.0658746868555549)
--(axis cs:3,0.0660754618122143);

\path [draw=lightslategray, semithick]
(axis cs:4,0.0652096219585044)
--(axis cs:4,0.0654422904068857);

\path [draw=lightslategray, semithick]
(axis cs:5,0.0650395395714377)
--(axis cs:5,0.0652097566820946);

\path [draw=lightgreen, semithick]
(axis cs:0.1,0.127857616228765)
--(axis cs:0.1,0.128147645192881);

\path [draw=lightgreen, semithick]
(axis cs:0.3,0.0906377780377246)
--(axis cs:0.3,0.0907596986850807);

\path [draw=lightgreen, semithick]
(axis cs:0.6,0.078360828879579)
--(axis cs:0.6,0.078517069849001);

\path [draw=lightgreen, semithick]
(axis cs:1,0.0731690751296091)
--(axis cs:1,0.0733393839785486);

\path [draw=lightgreen, semithick]
(axis cs:2,0.0690622773897066)
--(axis cs:2,0.0691960380512968);

\path [draw=lightgreen, semithick]
(axis cs:3,0.0676585519485961)
--(axis cs:3,0.0678269439007362);

\path [draw=lightgreen, semithick]
(axis cs:4,0.0671052684798546)
--(axis cs:4,0.0672516785190221);

\path [draw=lightgreen, semithick]
(axis cs:5,0.067000845616663)
--(axis cs:5,0.0671602188686105);

\addplot [semithick, red, mark=triangle*, mark size=3, mark options={solid,rotate=180}, only marks]
table {%
0.1 0.116099906570522
0.3 0.0858674426181063
0.6 0.0751942186083849
1 0.0703779821847002
2 0.0666121906879974
3 0.0652334986030909
4 0.0645244115796947
5 0.064185950633542
};
\addlegendentry{Oracle estimator, ${\bm{\Xi}_X^*}(\bS)$}
\addplot [semithick, royalblue, mark=triangle*, mark size=3, mark options={solid}, only marks]
table {%
0.1 0.117327984200839
0.3 0.0865588971624499
0.6 0.0759059786595493
1 0.0710498381237347
2 0.067276101337926
3 0.0659750743338846
4 0.0653259561826951
5 0.0651246481267661
};
\addlegendentry{RIE, $\widehat{{\bm{\Xi}_X^*}}(\bS)$}
\addplot [semithick, seagreen, mark=x, mark size=3, mark options={solid}, only marks]
table {%
0.1 0.128002630710823
0.3 0.0906987383614027
0.6 0.07843894936429
1 0.0732542295540789
2 0.0691291577205017
3 0.0677427479246661
4 0.0671784734994383
5 0.0670805322426368
};
\addlegendentry{$\sqrt{\widehat{{\bm{\Xi}_{X^2}^*}}(\bS)}$}
\draw (axis cs:0.2,0.116099906570522) node[
  scale=0.6,
  anchor=base west,
  text=black,
  rotate=0.0
]{1.06\%};
\draw (axis cs:0.4,0.0858674426181063) node[
  scale=0.6,
  anchor=base west,
  text=black,
  rotate=0.0
]{0.81\%};
\draw (axis cs:0.7,0.0751942186083849) node[
  scale=0.6,
  anchor=base west,
  text=black,
  rotate=0.0
]{0.95\%};
\draw (axis cs:1.1,0.0703779821847002) node[
  scale=0.6,
  anchor=base west,
  text=black,
  rotate=0.0
]{0.95\%};
\draw (axis cs:2.1,0.0666121906879974) node[
  scale=0.6,
  anchor=base west,
  text=black,
  rotate=0.0
]{1.0\%};
\draw (axis cs:3.1,0.0652334986030909) node[
  scale=0.6,
  anchor=base west,
  text=black,
  rotate=0.0
]{1.14\%};
\draw (axis cs:4.1,0.0645244115796947) node[
  scale=0.6,
  anchor=base west,
  text=black,
  rotate=0.0
]{1.24\%};
\draw (axis cs:5.1,0.064185950633542) node[
  scale=0.6,
  anchor=base west,
  text=black,
  rotate=0.0
]{1.46\%};
\end{axis}

\end{tikzpicture}
    \caption{Wishart, $\alpha' = \nicefrac{1}{4}$}
\end{subfigure}
\caption{\small Estimating $\bX$. The MSE is normalized by the norm of the signal, $\| \bX \|_{\rm F}^2$. Both $\bY$ and $\bW$ are $N \times M$ matrices with i.i.d. Gaussian entries of variance $1/N$, and aspect ratio $N/M = 1/2$. The RIE is applied to $N=2000, M =4000$, and the results are averaged over 10 runs (error bars are invisible). Average relative error between RIE $\widehat{{\bm{\Xi}_X^*}}(\bS)$ and Oracle estimator is also reported.}
\label{Wigner-Wishart-X}
\end{figure}

In Figure \ref{Wigner-Wishart-X}, the MSE of Oracle estimator, RIE \eqref{X-optimal-ev-app}, and $\sqrt{\bX^2}$-RIE is illustrated for shifted Wigner $\bX$ with $c=3$, and Wishart with aspect-ratio $\alpha' = \nicefrac{1}{4}$. We see that the performance of RIE is close to the one of Oracle estimator, which implies the optimality of the proposed estimator \eqref{X-optimal-ev-app}. Moreover, we observe the sub-optimality of estimating $\bX$ using $\sqrt{\widehat{{\bm{\Xi}_{X^2}^*}}(\bS)}$. Note that, in the low-SNR regime, the estimated eigenvalues $\widehat{\xi_{x^2}^*}_i$ might be negative which makes the estimator $\sqrt{\widehat{{\bm{\Xi}_{X^2}^*}}(\bS)}$ undefined, so the MSE is not depicted in this case.

\begin{figure}
    \centering
\begin{subfigure}[t]{.4\textwidth}
    \centering
\begin{tikzpicture}[scale = 0.6]

\definecolor{darkgray176}{RGB}{176,176,176}
\definecolor{lightcoral}{RGB}{240,128,128}
\definecolor{lightslategray}{RGB}{119,136,153}
\definecolor{royalblue}{RGB}{65,105,225}

\begin{axis}[
legend cell align={left},
legend style={fill opacity=0.8, draw opacity=1, text opacity=1, draw=white!80!black},
tick align=outside,
tick pos=left,
x grid style={darkgray176},
xlabel={$\kappa$},
xmin=-0.145, xmax=5.7,
xtick style={color=black},
y grid style={darkgray176},
ylabel={MSE},
scaled y ticks=false,
yticklabel style={
  /pgf/number format/precision=3,
  /pgf/number format/fixed},
ymin=0.143326472357344, ymax=0.256687250181864,
ytick style={color=black}
]
\path [draw=lightcoral, semithick]
(axis cs:0.1,0.227520905292615)
--(axis cs:0.1,0.227786849060696);

\path [draw=lightcoral, semithick]
(axis cs:0.3,0.188331502008258)
--(axis cs:0.3,0.188640749274391);

\path [draw=lightcoral, semithick]
(axis cs:0.6,0.170141119563316)
--(axis cs:0.6,0.170446380516286);

\path [draw=lightcoral, semithick]
(axis cs:1,0.161119644893602)
--(axis cs:1,0.161325570851976);

\path [draw=lightcoral, semithick]
(axis cs:2,0.153429181835853)
--(axis cs:2,0.153662109506614);

\path [draw=lightcoral, semithick]
(axis cs:3,0.150634449731166)
--(axis cs:3,0.150977921533952);

\path [draw=lightcoral, semithick]
(axis cs:4,0.149371986485071)
--(axis cs:4,0.149556922533036);

\path [draw=lightcoral, semithick]
(axis cs:5,0.148479234985732)
--(axis cs:5,0.148640216526243);

\path [draw=lightslategray, semithick]
(axis cs:0.1,0.251001264744588)
--(axis cs:0.1,0.251534487553477);

\path [draw=lightslategray, semithick]
(axis cs:0.3,0.19294555221408)
--(axis cs:0.3,0.193306024579295);

\path [draw=lightslategray, semithick]
(axis cs:0.6,0.172106529819885)
--(axis cs:0.6,0.172411212965308);

\path [draw=lightslategray, semithick]
(axis cs:1,0.162252815714697)
--(axis cs:1,0.162494970991107);

\path [draw=lightslategray, semithick]
(axis cs:2,0.154077775580094)
--(axis cs:2,0.154296277476659);

\path [draw=lightslategray, semithick]
(axis cs:3,0.151338003376845)
--(axis cs:3,0.151655350024016);

\path [draw=lightslategray, semithick]
(axis cs:4,0.149935111436974)
--(axis cs:4,0.150161914254772);

\path [draw=lightslategray, semithick]
(axis cs:5,0.149006513964046)
--(axis cs:5,0.149167839407386);

\addplot [semithick, red, mark=triangle*, mark size=3, mark options={solid,rotate=180}, only marks]
table {%
0.1 0.227653877176655
0.3 0.188486125641325
0.6 0.170293750039801
1 0.161222607872789
2 0.153545645671233
3 0.150806185632559
4 0.149464454509053
5 0.148559725755987
};
\addlegendentry{Oracle estimator, ${\bm{\Xi}_{X^2}^*}(\bS)$}
\addplot [semithick, royalblue, mark=triangle*, mark size=3, mark options={solid}, only marks]
table {%
0.1 0.251267876149033
0.3 0.193125788396687
0.6 0.172258871392596
1 0.162373893352902
2 0.154187026528376
3 0.15149667670043
4 0.150048512845873
5 0.149087176685716
};
\addlegendentry{RIE, $\widehat{{\bm{\Xi}_{X^2}^*}}(\bS)$}
\draw (axis cs:0.2,0.227653877176655) node[
  scale=0.6,
  anchor=base west,
  text=black,
  rotate=0.0
]{10.37\%};
\draw (axis cs:0.4,0.188486125641325) node[
  scale=0.6,
  anchor=base west,
  text=black,
  rotate=0.0
]{2.46\%};
\draw (axis cs:0.7,0.170293750039801) node[
  scale=0.6,
  anchor=base west,
  text=black,
  rotate=0.0
]{1.15\%};
\draw (axis cs:1.1,0.161222607872789) node[
  scale=0.6,
  anchor=base west,
  text=black,
  rotate=0.0
]{0.71\%};
\draw (axis cs:2.1,0.153545645671233) node[
  scale=0.6,
  anchor=base west,
  text=black,
  rotate=0.0
]{0.42\%};
\draw (axis cs:3.1,0.150806185632559) node[
  scale=0.6,
  anchor=base west,
  text=black,
  rotate=0.0
]{0.46\%};
\draw (axis cs:4.1,0.149464454509053) node[
  scale=0.6,
  anchor=base west,
  text=black,
  rotate=0.0
]{0.39\%};
\draw (axis cs:5.1,0.148559725755987) node[
  scale=0.6,
  anchor=base west,
  text=black,
  rotate=0.0
]{0.36\%};
\end{axis}

\end{tikzpicture}
    \caption{Shifted Wigner, $c=3$}
\end{subfigure}
\hfil
\begin{subfigure}[t]{.4\textwidth}
    \centering
\begin{tikzpicture}[scale = 0.6]

\definecolor{darkgray176}{RGB}{176,176,176}
\definecolor{lightcoral}{RGB}{240,128,128}
\definecolor{lightslategray}{RGB}{119,136,153}
\definecolor{royalblue}{RGB}{65,105,225}

\begin{axis}[
legend cell align={left},
legend style={fill opacity=0.8, draw opacity=1, text opacity=1, draw=white!80!black},
tick align=outside,
tick pos=left,
x grid style={darkgray176},
xlabel={$\kappa$},
xmin=-0.145, xmax=5.7,
xtick style={color=black},
y grid style={darkgray176},
ylabel={MSE},
scaled y ticks=false,
yticklabel style={
  /pgf/number format/precision=3,
  /pgf/number format/fixed},
ymin=0.156649019671993, ymax=0.272180296522729,
ytick style={color=black}
]
\path [draw=lightcoral, semithick]
(axis cs:0.1,0.26100768008829)
--(axis cs:0.1,0.261712068939429);

\path [draw=lightcoral, semithick]
(axis cs:0.3,0.200057679095148)
--(axis cs:0.3,0.200493806410718);

\path [draw=lightcoral, semithick]
(axis cs:0.6,0.180688279291477)
--(axis cs:0.6,0.181110001867933);

\path [draw=lightcoral, semithick]
(axis cs:1,0.172454440548702)
--(axis cs:1,0.172846289244156);

\path [draw=lightcoral, semithick]
(axis cs:2,0.165818792335782)
--(axis cs:2,0.166298537078208);

\path [draw=lightcoral, semithick]
(axis cs:3,0.163711757446475)
--(axis cs:3,0.164084521211202);

\path [draw=lightcoral, semithick]
(axis cs:4,0.162545023659796)
--(axis cs:4,0.1630765738001);

\path [draw=lightcoral, semithick]
(axis cs:5,0.161900441347026)
--(axis cs:5,0.162435229207122);

\path [draw=lightslategray, semithick]
(axis cs:0.1,0.266263719550169)
--(axis cs:0.1,0.266928874847695);

\path [draw=lightslategray, semithick]
(axis cs:0.3,0.201257007293457)
--(axis cs:0.3,0.201696888475037);

\path [draw=lightslategray, semithick]
(axis cs:0.6,0.18136196719822)
--(axis cs:0.6,0.181762517868795);

\path [draw=lightslategray, semithick]
(axis cs:1,0.172979245615581)
--(axis cs:1,0.17333759443499);

\path [draw=lightslategray, semithick]
(axis cs:2,0.166324375052594)
--(axis cs:2,0.166799585116673);

\path [draw=lightslategray, semithick]
(axis cs:3,0.164221732687424)
--(axis cs:3,0.16458334583123);

\path [draw=lightslategray, semithick]
(axis cs:4,0.163081210400834)
--(axis cs:4,0.163542835490589);

\path [draw=lightslategray, semithick]
(axis cs:5,0.162441637772498)
--(axis cs:5,0.162974232454269);

\addplot [semithick, red, mark=triangle*, mark size=3, mark options={solid,rotate=180}, only marks]
table {%
0.1 0.261359874513859
0.3 0.200275742752933
0.6 0.180899140579705
1 0.172650364896429
2 0.166058664706995
3 0.163898139328839
4 0.162810798729948
5 0.162167835277074
};
\addlegendentry{Oracle estimator, ${\bm{\Xi}_{X^2}^*}(\bS)$}
\addplot [semithick, royalblue, mark=triangle*, mark size=3, mark options={solid}, only marks]
table {%
0.1 0.266596297198932
0.3 0.201476947884247
0.6 0.181562242533508
1 0.173158420025286
2 0.166561980084633
3 0.164402539259327
4 0.163312022945712
5 0.162707935113383
};
\addlegendentry{RIE, $\widehat{{\bm{\Xi}_{X^2}^*}}(\bS)$}
\draw (axis cs:0.2,0.261359874513859) node[
  scale=0.6,
  anchor=base west,
  text=black,
  rotate=0.0
]{2.0\%};
\draw (axis cs:0.4,0.200275742752933) node[
  scale=0.6,
  anchor=base west,
  text=black,
  rotate=0.0
]{0.6\%};
\draw (axis cs:0.7,0.180899140579705) node[
  scale=0.6,
  anchor=base west,
  text=black,
  rotate=0.0
]{0.37\%};
\draw (axis cs:1.1,0.172650364896429) node[
  scale=0.6,
  anchor=base west,
  text=black,
  rotate=0.0
]{0.29\%};
\draw (axis cs:2.1,0.166058664706995) node[
  scale=0.6,
  anchor=base west,
  text=black,
  rotate=0.0
]{0.3\%};
\draw (axis cs:3.1,0.163898139328839) node[
  scale=0.6,
  anchor=base west,
  text=black,
  rotate=0.0
]{0.31\%};
\draw (axis cs:4.1,0.162810798729948) node[
  scale=0.6,
  anchor=base west,
  text=black,
  rotate=0.0
]{0.31\%};
\draw (axis cs:5.1,0.162167835277074) node[
  scale=0.6,
  anchor=base west,
  text=black,
  rotate=0.0
]{0.33\%};
\end{axis}

\end{tikzpicture}
    \caption{Wishart, $\alpha' = \nicefrac{1}{4}$}
\end{subfigure}
\caption{\small Estimating $\bX^2$. The MSE is normalized by the norm of the signal, $\| \bX^2 \|_{\rm F}^2$. Both $\bY$ and $\bW$ are $N \times M$ matrices with i.i.d. Gaussian entries of variance $1/N$, and aspect ratio $N/M = 1/2$. The RIE is applied to $N=2000, M =4000$, and the results are averaged over 10 runs (error bars are invisible). Average relative error between RIE $\widehat{{\bm{\Xi}_X^*}}(\bS)$ and Oracle estimator is also reported.}
\label{Wigner-Wishart-XX}
\end{figure}
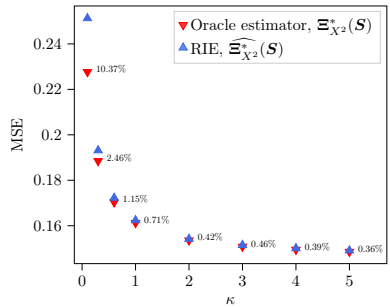
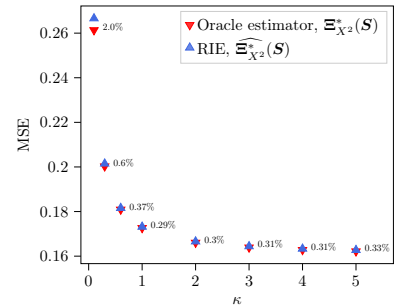

In Figure \ref{Wigner-Wishart-XX}, the MSE of estimating $\bX^2$ is shown. We see that in the high-SNR regimes the RIE \eqref{X2-optimal-ev-Y,W-G} has the same performance as the Oracle estimator.

\begin{figure}
    \centering
\begin{subfigure}[t]{.4\textwidth}
    \centering
\begin{tikzpicture}[scale = 0.6]

\definecolor{darkgray176}{RGB}{176,176,176}
\definecolor{darkorange}{RGB}{255,140,0}
\definecolor{lightcoral}{RGB}{240,128,128}
\definecolor{lightgreen}{RGB}{144,238,144}
\definecolor{lightslategray}{RGB}{119,136,153}
\definecolor{orange}{RGB}{255,165,0}
\definecolor{royalblue}{RGB}{65,105,225}
\definecolor{seagreen}{RGB}{46,139,87}

\begin{axis}[
legend cell align={left},
legend style={fill opacity=0.8, draw opacity=1, text opacity=1, draw=white!80!black},
tick align=outside,
tick pos=left,
x grid style={darkgray176},
xlabel={$\kappa$},
xmin=-0.145, xmax=5.3,
xtick style={color=black},
y grid style={darkgray176},
ylabel={${\rm MSE}$},
ymin=0.09, ymax=0.947818668389744,
ytick style={color=black}
]
\path [draw=lightcoral, semithick]
(axis cs:0.1,0.485551153402528)
--(axis cs:0.1,0.511583479790304);

\path [draw=lightcoral, semithick]
(axis cs:0.3,0.472255460846842)
--(axis cs:0.3,0.489487911823584);

\path [draw=lightcoral, semithick]
(axis cs:0.6,0.444136674492931)
--(axis cs:0.6,0.450444359117777);

\path [draw=lightcoral, semithick]
(axis cs:1,0.398036844632548)
--(axis cs:1,0.41674225655726);

\path [draw=lightcoral, semithick]
(axis cs:2,0.292196181198336)
--(axis cs:2,0.305368224045935);

\path [draw=lightcoral, semithick]
(axis cs:3,0.224398018404295)
--(axis cs:3,0.230957604693773);

\path [draw=lightcoral, semithick]
(axis cs:4,0.173084418913466)
--(axis cs:4,0.177895993281815);

\path [draw=lightcoral, semithick]
(axis cs:5,0.138285281175495)
--(axis cs:5,0.141906947876347);

\path [draw=lightslategray, semithick]
(axis cs:0.1,0.486533254585728)
--(axis cs:0.1,0.513242223058339);

\path [draw=lightslategray, semithick]
(axis cs:0.3,0.473204209066873)
--(axis cs:0.3,0.490644493486444);

\path [draw=lightslategray, semithick]
(axis cs:0.6,0.445022941117011)
--(axis cs:0.6,0.451268494442006);

\path [draw=lightslategray, semithick]
(axis cs:1,0.398917722190439)
--(axis cs:1,0.418300783117397);

\path [draw=lightslategray, semithick]
(axis cs:2,0.294201798550047)
--(axis cs:2,0.305758697575281);

\path [draw=lightslategray, semithick]
(axis cs:3,0.225077986306194)
--(axis cs:3,0.232544685989537);

\path [draw=lightslategray, semithick]
(axis cs:4,0.174451745032725)
--(axis cs:4,0.17929598711759);

\path [draw=lightslategray, semithick]
(axis cs:5,0.140132669224491)
--(axis cs:5,0.142949166849548);

\path [draw=orange, semithick]
(axis cs:0.1,0.893570053797865)
--(axis cs:0.1,0.906796013423193);

\path [draw=orange, semithick]
(axis cs:0.3,0.8769564094023)
--(axis cs:0.3,0.88703022438502);

\path [draw=orange, semithick]
(axis cs:0.6,0.830965222231841)
--(axis cs:0.6,0.848056052843522);

\path [draw=orange, semithick]
(axis cs:1,0.731488240581534)
--(axis cs:1,0.743610590039239);

\path [draw=orange, semithick]
(axis cs:2,0.475124660994726)
--(axis cs:2,0.479597073653208);

\path [draw=orange, semithick]
(axis cs:3,0.323451718596226)
--(axis cs:3,0.327192809867025);

\path [draw=orange, semithick]
(axis cs:4,0.244530555424804)
--(axis cs:4,0.245852280354402);

\path [draw=orange, semithick]
(axis cs:5,0.195593351067318)
--(axis cs:5,0.197191056686807);

\path [draw=lightgreen, semithick]
(axis cs:0.1,0.894367717249015)
--(axis cs:0.1,0.90926945947478);

\path [draw=lightgreen, semithick]
(axis cs:0.3,0.878903347577764)
--(axis cs:0.3,0.888159694108267);

\path [draw=lightgreen, semithick]
(axis cs:0.6,0.834024766051021)
--(axis cs:0.6,0.85021379338804);

\path [draw=lightgreen, semithick]
(axis cs:1,0.735301780254132)
--(axis cs:1,0.746013353017704);

\path [draw=lightgreen, semithick]
(axis cs:2,0.479217448362233)
--(axis cs:2,0.485206369778975);

\path [draw=lightgreen, semithick]
(axis cs:3,0.327248733527804)
--(axis cs:3,0.342165383307345);

\path [draw=lightgreen, semithick]
(axis cs:4,0.248536209394181)
--(axis cs:4,0.251303691237377);

\path [draw=lightgreen, semithick]
(axis cs:5,0.199314234480497)
--(axis cs:5,0.206082366101679);

\addplot [semithick, red, mark=triangle*, mark size=3, mark options={solid,rotate=180}, only marks]
table {%
0.1 0.498567316596416
0.3 0.480871686335213
0.6 0.447290516805354
1 0.407389550594904
2 0.298782202622136
3 0.227677811549034
4 0.17549020609764
5 0.140096114525921
};
\addlegendentry{$p=0.5$, Oracle  ${\bm{\Xi}_X^*}(\bS)$}
\addplot [semithick, royalblue, mark=triangle*, mark size=3, mark options={solid}, only marks]
table {%
0.1 0.499887738822034
0.3 0.481924351276658
0.6 0.448145717779508
1 0.408609252653918
2 0.299980248062664
3 0.228811336147866
4 0.176873866075158
5 0.141540918037019
};
\addlegendentry{$p=0.5$, RIE $\widehat{{\bm{\Xi}_X^*}}(\bS)$}
\addplot [semithick, darkorange, mark=triangle*, mark size=3, mark options={solid,rotate=180}, only marks]
table {%
0.1 0.900183033610529
0.3 0.88199331689366
0.6 0.839510637537681
1 0.737549415310387
2 0.477360867323967
3 0.325322264231625
4 0.245191417889603
5 0.196392203877062
};
\addlegendentry{$p=0.9$, Oracle ${\bm{\Xi}_X^*}(\bS)$}
\addplot [semithick, seagreen, mark=triangle*, mark size=3, mark options={solid}, only marks]
table {%
0.1 0.901818588361897
0.3 0.883531520843016
0.6 0.842119279719531
1 0.740657566635918
2 0.482211909070604
3 0.334707058417574
4 0.249919950315779
5 0.202698300291088
};
\addlegendentry{$p=0.9$, RIE $\widehat{{\bm{\Xi}_X^*}}(\bS)$}
\draw (axis cs:-0.1,0.518567316596416) node[
  scale=0.6,
  anchor=base west,
  text=black,
  rotate=0.0
]{0.26\%};
\draw (axis cs:0.1,0.440871686335213) node[
  scale=0.6,
  anchor=base west,
  text=black,
  rotate=0.0
]{0.22\%};
\draw (axis cs:0.4,0.467290516805354) node[
  scale=0.6,
  anchor=base west,
  text=black,
  rotate=0.0
]{0.19\%};
\draw (axis cs:0.8,0.367389550594904) node[
  scale=0.6,
  anchor=base west,
  text=black,
  rotate=0.0
]{0.3\%};
\draw (axis cs:1.8,0.258782202622136) node[
  scale=0.6,
  anchor=base west,
  text=black,
  rotate=0.0
]{0.41\%};
\draw (axis cs:2.8,0.187677811549034) node[
  scale=0.6,
  anchor=base west,
  text=black,
  rotate=0.0
]{0.5\%};
\draw (axis cs:3.8,0.13549020609764) node[
  scale=0.6,
  anchor=base west,
  text=black,
  rotate=0.0
]{0.79\%};
\draw (axis cs:4.8,0.100096114525921) node[
  scale=0.6,
  anchor=base west,
  text=black,
  rotate=0.0
]{1.04\%};
\draw (axis cs:-0.1,0.860183033610529) node[
  scale=0.6,
  anchor=base west,
  text=black,
  rotate=0.0
]{0.18\%};
\draw (axis cs:0.1,0.90199331689366) node[
  scale=0.6,
  anchor=base west,
  text=black,
  rotate=0.0
]{0.17\%};
\draw (axis cs:0.4,0.799510637537681) node[
  scale=0.6,
  anchor=base west,
  text=black,
  rotate=0.0
]{0.31\%};
\draw (axis cs:0.8,0.757549415310387) node[
  scale=0.6,
  anchor=base west,
  text=black,
  rotate=0.0
]{0.42\%};
\draw (axis cs:1.8,0.507360867323967) node[
  scale=0.6,
  anchor=base west,
  text=black,
  rotate=0.0
]{1.02\%};
\draw (axis cs:2.8,0.355322264231625) node[
  scale=0.6,
  anchor=base west,
  text=black,
  rotate=0.0
]{2.9\%};
\draw (axis cs:3.8,0.275191417889603) node[
  scale=0.6,
  anchor=base west,
  text=black,
  rotate=0.0
]{1.93\%};
\draw (axis cs:4.8,0.226392203877062) node[
  scale=0.6,
  anchor=base west,
  text=black,
  rotate=0.0
]{3.21\%};

\end{axis}

\end{tikzpicture}
    \caption{Estimating $\bX$}
\end{subfigure}
\hfil
\begin{subfigure}[t]{.4\textwidth}
    \centering
\begin{tikzpicture}[scale = 0.6]

\definecolor{darkgray176}{RGB}{176,176,176}
\definecolor{darkorange}{RGB}{255,140,0}
\definecolor{lightcoral}{RGB}{240,128,128}
\definecolor{lightgreen}{RGB}{144,238,144}
\definecolor{lightslategray}{RGB}{119,136,153}
\definecolor{orange}{RGB}{255,165,0}
\definecolor{royalblue}{RGB}{65,105,225}
\definecolor{seagreen}{RGB}{46,139,87}

\begin{axis}[
legend cell align={left},
legend style={fill opacity=0.8, draw opacity=1, text opacity=1, draw=white!80!black},
tick align=outside,
tick pos=left,
x grid style={darkgray176},
xlabel={$\kappa$},
xmin=0.38, xmax=5.25,
xtick style={color=black},
y grid style={darkgray176},
ylabel={${\rm MSE}$},
ymin=0.09, ymax=0.952391367659243,
ytick style={color=black}
]
\path [draw=lightcoral, semithick]
(axis cs:0.6,0.444136674492931)
--(axis cs:0.6,0.450444359117777);

\path [draw=lightcoral, semithick]
(axis cs:1,0.398036844632548)
--(axis cs:1,0.41674225655726);

\path [draw=lightcoral, semithick]
(axis cs:2,0.292196181198337)
--(axis cs:2,0.305368224045935);

\path [draw=lightcoral, semithick]
(axis cs:3,0.224398018404295)
--(axis cs:3,0.230957604693773);

\path [draw=lightcoral, semithick]
(axis cs:4,0.173084418913466)
--(axis cs:4,0.177895993281815);

\path [draw=lightcoral, semithick]
(axis cs:5,0.138285281175495)
--(axis cs:5,0.141906947876347);

\path [draw=lightslategray, semithick]
(axis cs:0.6,0.458371986585227)
--(axis cs:0.6,0.464784262764528);

\path [draw=lightslategray, semithick]
(axis cs:1,0.403844346661683)
--(axis cs:1,0.422834459721427);

\path [draw=lightslategray, semithick]
(axis cs:2,0.294186556275512)
--(axis cs:2,0.307356178265996);

\path [draw=lightslategray, semithick]
(axis cs:3,0.225682114141033)
--(axis cs:3,0.232279616121979);

\path [draw=lightslategray, semithick]
(axis cs:4,0.174219980222534)
--(axis cs:4,0.1790679120443);

\path [draw=lightslategray, semithick]
(axis cs:5,0.139225335356731)
--(axis cs:5,0.142782443123703);

\path [draw=orange, semithick]
(axis cs:0.6,0.830965222231841)
--(axis cs:0.6,0.848056052843522);

\path [draw=orange, semithick]
(axis cs:1,0.731488240581533)
--(axis cs:1,0.74361059003924);

\path [draw=orange, semithick]
(axis cs:2,0.475124660994726)
--(axis cs:2,0.479597073653208);

\path [draw=orange, semithick]
(axis cs:3,0.323451718596226)
--(axis cs:3,0.327192809867025);

\path [draw=orange, semithick]
(axis cs:4,0.244530555424804)
--(axis cs:4,0.245852280354402);

\path [draw=orange, semithick]
(axis cs:5,0.195593351067318)
--(axis cs:5,0.197191056686807);

\path [draw=lightgreen, semithick]
(axis cs:0.6,0.887589354272516)
--(axis cs:0.6,0.913624411160017);

\path [draw=lightgreen, semithick]
(axis cs:1,0.752934944456181)
--(axis cs:1,0.766903197974207);

\path [draw=lightgreen, semithick]
(axis cs:2,0.482094009737364)
--(axis cs:2,0.487205133393487);

\path [draw=lightgreen, semithick]
(axis cs:3,0.32666362898756)
--(axis cs:3,0.330797078100414);

\path [draw=lightgreen, semithick]
(axis cs:4,0.246528959721626)
--(axis cs:4,0.24792720815816);

\path [draw=lightgreen, semithick]
(axis cs:5,0.197109600755716)
--(axis cs:5,0.198820294692628);

\addplot [semithick, red, mark=triangle*, mark size=3, mark options={solid,rotate=180}, only marks]
table {%
0.6 0.447290516805354
1 0.407389550594904
2 0.298782202622136
3 0.227677811549034
4 0.17549020609764
5 0.140096114525921
};
\addlegendentry{$p=0.5$, Oracle  ${\bm{\Xi}_{X^2}^*}(\bS)$ }
\addplot [semithick, royalblue, mark=triangle*, mark size=3, mark options={solid}, only marks]
table {%
0.6 0.461578124674877
1 0.413339403191555
2 0.300771367270754
3 0.228980865131506
4 0.176643946133417
5 0.141003889240217
};
\addlegendentry{$p=0.5$, RIE $\widehat{{\bm{\Xi}_{X^2}^*}}(\bS)$}
\addplot [semithick, darkorange, mark=triangle*, mark size=3, mark options={solid,rotate=180}, only marks]
table {%
0.6 0.839510637537681
1 0.737549415310387
2 0.477360867323967
3 0.325322264231625
4 0.245191417889603
5 0.196392203877062
};
\addlegendentry{$p=0.9$, Oracle  ${\bm{\Xi}_{X^2}^*}(\bS)$ }
\addplot [semithick, seagreen, mark=triangle*, mark size=3, mark options={solid}, only marks]
table {%
0.6 0.900606882716266
1 0.759919071215194
2 0.484649571565426
3 0.328730353543987
4 0.247228083939893
5 0.197964947724172
};
\addlegendentry{$p=0.9$, RIE $\widehat{{\bm{\Xi}_{X^2}^*}}(\bS)$}
\draw (axis cs:0.4,0.477290516805354) node[
  scale=0.6,
  anchor=base west,
  text=black,
  rotate=0.0
]{3.19\%};
\draw (axis cs:0.8,0.367389550594904) node[
  scale=0.6,
  anchor=base west,
  text=black,
  rotate=0.0
]{1.46\%};
\draw (axis cs:1.8,0.258782202622136) node[
  scale=0.6,
  anchor=base west,
  text=black,
  rotate=0.0
]{0.67\%};
\draw (axis cs:2.8,0.187677811549034) node[
  scale=0.6,
  anchor=base west,
  text=black,
  rotate=0.0
]{0.57\%};
\draw (axis cs:3.8,0.13549020609764) node[
  scale=0.6,
  anchor=base west,
  text=black,
  rotate=0.0
]{0.66\%};
\draw (axis cs:4.8,0.100096114525921) node[
  scale=0.6,
  anchor=base west,
  text=black,
  rotate=0.0
]{0.65\%};
\draw (axis cs:0.4,0.799510637537681) node[
  scale=0.6,
  anchor=base west,
  text=black,
  rotate=0.0
]{7.27\%};
\draw (axis cs:0.8,0.687549415310387) node[
  scale=0.6,
  anchor=base west,
  text=black,
  rotate=0.0
]{3.03\%};
\draw (axis cs:1.8,0.507360867323967) node[
  scale=0.6,
  anchor=base west,
  text=black,
  rotate=0.0
]{1.53\%};
\draw (axis cs:2.8,0.355322264231625) node[
  scale=0.6,
  anchor=base west,
  text=black,
  rotate=0.0
]{1.05\%};
\draw (axis cs:3.8,0.275191417889603) node[
  scale=0.6,
  anchor=base west,
  text=black,
  rotate=0.0
]{0.83\%};
\draw (axis cs:4.8,0.226392203877062) node[
  scale=0.6,
  anchor=base west,
  text=black,
  rotate=0.0
]{0.8\%};
\end{axis}

\end{tikzpicture}
    \caption{Estimating $\bX^2$}
    \label{Bernoulli-XX}
\end{subfigure}
\caption{\small Estimating $\bX$ and $\bX^2$ with Bernoulli spectral prior distribution. The MSE is normalized by the norm of the signal, $\| \bX \|_{\rm F}^2 = \| \bX^2 \|_{\rm F}^2$. Both $\bY$ and $\bW$ are $N \times M$ matrices with i.i.d. Gaussian entries of variance $1/N$, and aspect ratio $N/M = 1/2$. The RIE is applied to $N=2000, M =4000$, and the results are averaged over 10 runs (error bars are invisible). Average relative error between RIE $\widehat{{\bm{\Xi}_X^*}}(\bS)$ and Oracle estimator is also reported.}
\label{Bernoulli-X,XX}
\end{figure}
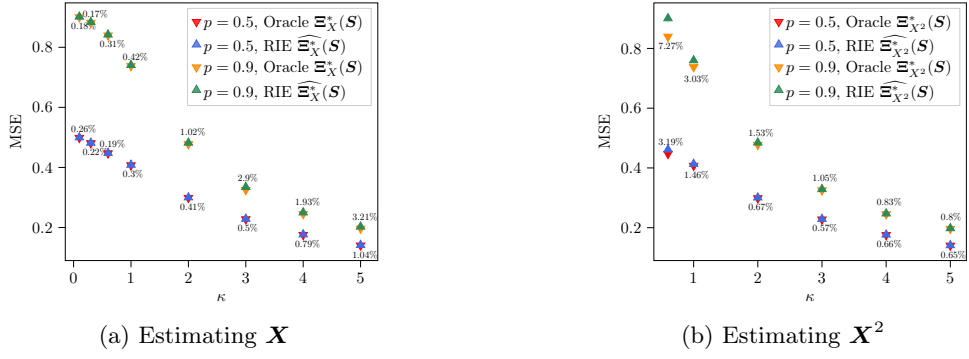

\paragraph{Bernoulli spectral distribution.} In this case, the matrix $\bX$ is constructed as $\bX = \bU_X \bm{\Lambda} \bU_X^\intercal$ with $\bU_X$ a $N \times N$ orthogonal matrix distributed according to Haar measure on orthogonal matrices, and $\bm{\Lambda} = {\rm diag}( \blam )$ where $\blam$ has i.i.d. Bernoulli elements. Thus, $\rho_X = p \delta_0 + (1-p) \delta_{+1}$ for $p \in (0,1)$, and the Stieltjes transform is:
\begin{equation*}
     \mathcal{G}_{\rho_X}(z) = p \frac{1}{z} + (1-p) \frac{1}{z - 1}
\end{equation*}
For this prior, we have that $\bX = \bX^2$, so both estimators $\widehat{\bm{\Xi}^*_X}(\bS)$ and $\widehat{\bm{\Xi}^*_{X^2}}(\bS)$ should have the same performance. However, note that $\widehat{\bm{\Xi}^*_{X^2}}(\bS)$ does not use any knowledge of $\rho_X$. In Figure \ref{Bernoulli-X,XX}, the MSE is illustrated for these two estimators for two sparsity parameter, $p = 0.5$ and  $0.9$. We observe that, except for for the low-SNR regimes, both estimators have the same MSE. The poor performance of $\widehat{\bm{\Xi}^*_{X^2}}(\bS)$ in the low-SNR regimes might be due to the fact that, some of the estimated eigenvalues $\widehat{\xi^*_{x^2}}_i$ are negative although the true eigenvalue is $0$. This makes the estimation more difficult for the sparser prior, see Figure \ref{Bernoulli-XX}. However, this problem is resolved in  $\widehat{\bm{\Xi}^*_X}(\bS)$ by taking the knowledge of $\mathcal{G}_{\rho_X}(z)$ into account. 

\begin{figure}
    \centering
\begin{tikzpicture}[scale = 0.7]

\definecolor{darkgray176}{RGB}{176,176,176}
\definecolor{lightcoral}{RGB}{240,128,128}
\definecolor{lightslategray}{RGB}{119,136,153}
\definecolor{royalblue}{RGB}{65,105,225}

\begin{axis}[
legend cell align={left},
legend style={fill opacity=0.8, draw opacity=1, text opacity=1, draw=white!80!black},
tick align=outside,
tick pos=left,
x dir=reverse,
x grid style={darkgray176},
xlabel={$\alpha$},
xmin=0.2125, xmax=1.0375,
xtick style={color=black},
y grid style={darkgray176},
ylabel={MSE},
scaled y ticks=false,
yticklabel style={
  /pgf/number format/precision=3,
  /pgf/number format/fixed},
ymin=0.032838448086844, ymax=0.073239843754072,
ytick style={color=black}
]
\path [draw=lightcoral, semithick]
(axis cs:1,0.0709636407947644)
--(axis cs:1,0.0711022647103567);

\path [draw=lightcoral, semithick]
(axis cs:0.666666666666667,0.0607107322466864)
--(axis cs:0.666666666666667,0.0608025589119827);

\path [draw=lightcoral, semithick]
(axis cs:0.5,0.0526807664231779)
--(axis cs:0.5,0.0527811361257995);

\path [draw=lightcoral, semithick]
(axis cs:0.4,0.0465273183179775)
--(axis cs:0.4,0.0466255080994465);

\path [draw=lightcoral, semithick]
(axis cs:0.333333333333333,0.0417529229306508)
--(axis cs:0.333333333333333,0.0418176890418488);

\path [draw=lightcoral, semithick]
(axis cs:0.285714285714286,0.0378472413541507)
--(axis cs:0.285714285714286,0.0379161939463366);

\path [draw=lightcoral, semithick]
(axis cs:0.25,0.0346748751626271)
--(axis cs:0.25,0.0347527023045141);

\path [draw=lightslategray, semithick]
(axis cs:1,0.0712674128701964)
--(axis cs:1,0.0714034166782889);

\path [draw=lightslategray, semithick]
(axis cs:0.666666666666667,0.0610238983164144)
--(axis cs:0.666666666666667,0.0611133257202843);

\path [draw=lightslategray, semithick]
(axis cs:0.5,0.053001927433073)
--(axis cs:0.5,0.0531235227755192);

\path [draw=lightslategray, semithick]
(axis cs:0.4,0.0468788369627049)
--(axis cs:0.4,0.0469815957403161);

\path [draw=lightslategray, semithick]
(axis cs:0.333333333333333,0.0420886539025169)
--(axis cs:0.333333333333333,0.0421664418322596);

\path [draw=lightslategray, semithick]
(axis cs:0.285714285714286,0.0381898359523545)
--(axis cs:0.285714285714286,0.0382780567254654);

\path [draw=lightslategray, semithick]
(axis cs:0.25,0.0350309065749222)
--(axis cs:0.25,0.0351235282400878);

\addplot [semithick, red, mark=triangle*, mark size=3, mark options={solid,rotate=180}, only marks]
table {%
1 0.0710329527525605
0.666666666666667 0.0607566455793346
0.5 0.0527309512744887
0.4 0.046576413208712
0.333333333333333 0.0417853059862498
0.285714285714286 0.0378817176502436
0.25 0.0347137887335706
};
\addlegendentry{Oracle estimator, ${\bm{\Xi}_X^*}(\bS)$}
\addplot [semithick, royalblue, mark=triangle*, mark size=3, mark options={solid}, only marks]
table {%
1 0.0713354147742426
0.666666666666667 0.0610686120183493
0.5 0.0530627251042961
0.4 0.0469302163515105
0.333333333333333 0.0421275478673883
0.285714285714286 0.03823394633891
0.25 0.035077217407505
};
\addlegendentry{RIE, $\widehat{{\bm{\Xi}_X^*}}(\bS)$}
\draw (axis cs:0.98,0.0710329527525605) node[
  scale=0.6,
  anchor=base west,
  text=black,
  rotate=0.0
]{0.43\%};
\draw (axis cs:0.766666666666667,0.0607566455793346) node[
  scale=0.6,
  anchor=base west,
  text=black,
  rotate=0.0
]{0.51\%};
\draw (axis cs:0.6,0.0527309512744887) node[
  scale=0.6,
  anchor=base west,
  text=black,
  rotate=0.0
]{0.63\%};
\draw (axis cs:0.5,0.046576413208712) node[
  scale=0.6,
  anchor=base west,
  text=black,
  rotate=0.0
]{0.76\%};
\draw (axis cs:0.433333333333333,0.0417853059862498) node[
  scale=0.6,
  anchor=base west,
  text=black,
  rotate=0.0
]{0.82\%};
\draw (axis cs:0.385714285714286,0.0378817176502436) node[
  scale=0.6,
  anchor=base west,
  text=black,
  rotate=0.0
]{0.93\%};
\draw (axis cs:0.35,0.0347137887335706) node[
  scale=0.6,
  anchor=base west,
  text=black,
  rotate=0.0
]{1.05\%};
\end{axis}

\end{tikzpicture}
    \caption{\small MSE of estimating $\bX$ as a function of aspect-ratio $\alpha$, prior on $\bX$ is shifted Wigner with $c =3$, and $\kappa = 5$. MSE is normalized by the norm of the signal, $\| \bX \|_{\rm F}^2$. Both $\bY$ and $\bW$ are $N \times M$ matrices with i.i.d. Gaussian entries of variance $1/N$. The RIE is applied to $N=2000, M = \nicefrac{1}{\alpha} N$, and the results are averaged over 10 runs (error bars are invisible). Average relative error between RIE $\widehat{{\bm{\Xi}_X^*}}(\bS)$ and Oracle estimator is also reported.}
    \label{X-ar}
\end{figure}
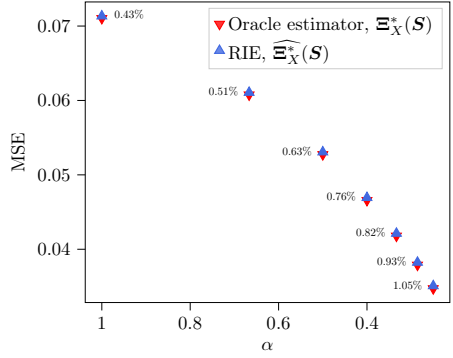

\paragraph{Effect of aspect-ratio $\alpha$.} In Figure \ref{X-ar}, we consider $\bX$ to be shifted Wigner with $c = 3$, and the MSE is depicted for various values of the aspect-ratio $\alpha$. As expected,  as $M$ increases ($\alpha$ decreases) and we have more observation or more data samples,  the estimation error decreases.  

\clearpage

\section{Estimating  $\bY$}\label{Y-RIE-deriv}
In this section, we present the derivation of the optimal RIE for $\bY$. For simplicity, the SNR parameter in \eqref{MF-model} is absorbed into $\bY$, so the model is $\bS = \bX \bY + \bW$. Therefore, the final estimator should be divided by $\nicefrac{1}{\sqrt{\kappa}}$ to give an estimate of the original $\bY$. 

The optimal singular values are constructed as ${\xi_y^*}_i = \sum_{j=1}^N \sigma_j  \, \big(  \bu_i^\intercal \by_j^{(l)}\big) \big(  \bv_i^\intercal \by_j^{(r)} \big)$. We assume that, for large $N$, ${\xi_y^*}_i$ can be approximated by its expectation:
\begin{equation*}
    \widehat{{\xi_y^*}}_i \approx \sum_{j=1}^N \sigma_j \,  \bE \Big[ \big(  \bu_i^\intercal \by_j^{(l)}\big) \big(  \bv_i^\intercal \by_j^{(r)} \big) \Big]
\end{equation*}
where the expectation is over the singular vectors of the observation matrix $\bS$. Therefore, to compute the optimal singular values, we need to find the mean overlap $ \bE \Big[ \big(  \bu_i^\intercal \by_j^{(l)}\big) \big(  \bv_i^\intercal \by_j^{(r)} \big) \Big]$ between singular vectors of $\bY$ and singular vectors of $\bS$. In the following we will see that (a rescaling of) this quantity can be expressed in terms of $i$-th singular value of $\bS$ and $j$-th singular value of $\bY$ (and the limiting measures, indeed). Thus, we will use the notation $O_Y(\gamma_i, \sigma_j) :=  N \bE \Big[ \big(  \bu_i^\intercal \by_j^{(l)}\big) \big(  \bv_i^\intercal \by_j^{(r)} \big) \Big]$ in what follows. In the nest section, we discuss how the overlap can be computed from the resolvent of the Hermitized matrix of $\bS$.

\subsection{Relation between overlap and the resolvent}
Construct the matrix $\cS \in \bR^{(N+M) \times (N+M)}$ from the observation matrix:
\begin{equation*}
    \cS = \left[
\begin{array}{cc}
\mathbf{0}_{N\times N} & \bS \\
\bS^\intercal & \mathbf{0}_{M\times M}
\end{array}
\right] 
\end{equation*}
By Theorem 7.3.3 in \cite{horn2012matrix}, $\cS$ has the following eigen-decomposition:
\begin{equation}
  \cS  = \left[
\begin{array}{ccc}
\hat{\bU}_S & \hat{\bU}_S &  \mathbf{0} \\
\hat{\bV}_S^{(1)} & -\hat{\bV}_S^{(1)} &  \bV_S^{(2)}
\end{array}
\right] \left[
\begin{array}{ccc}
\bGam_N & \mathbf{0} & \mathbf{0}\\
\mathbf{0} & -\bGam_N &  \mathbf{0}\\
\mathbf{0} & \mathbf{0} & \mathbf{0} 
\end{array}
\right]  \left[
\begin{array}{ccc}
\hat{\bU}_S & \hat{\bU}_S &  \mathbf{0} \\
\hat{\bV}_S^{(1)} & -\hat{\bV}_S^{(1)} &  \bV_S^{(2)}
\end{array}
\right]^\intercal
\label{EVD of S - app}
\end{equation}
with $\bV_S = \left[
\begin{array}{cc}
\bV_S^{(1)} & \bV_S^{(2)}
\end{array}
\right]$ in which $\bV_S^{(1)} \in \bR^{M \times N}$. And, $\hat{\bV}_S^{(1)} = \frac{1}{\sqrt{2}} \bV_S^{(1)}$, $\hat{\bU}_S= \frac{1}{\sqrt{2}} \bU_S$. Eigenvalues of $\cS$ are signed singular values of $\bS$, therefore the limiting eigenvalue distribution of $\cS$ (ignoring zero eigenvalues) is the same as the limiting symmetrized singular value distribution of $\bS$.

Define the resolvent of $\cS$ 
\begin{equation*}
    \bG_{\mathcal{S}}(z) = \big(
\begin{array}{c}
z \bI - \cS 
\end{array}
\big)^{-1}
\end{equation*}

Denote the eigenvectors of $\cS$ by $\bs_i \in \bR^{M+N}$, $i = 1, \dots,M+N$. For $z = x - \ci \epsilon$ with $x \in \bR$ and $\epsilon \gg \frac{1}{N}$, we have:
\begin{equation*}
    \bG_{\mathcal{S}}(x- \ci \epsilon) = \sum_{k=1}^{2N} \frac{x + \ci \epsilon }{(x - \tilde{\gamma}_k)^2+\epsilon^2} \bs_k \bs_k^\intercal + \frac{x + \ci \epsilon }{x^2+\epsilon^2} \sum_{k=2N+1}^{N+M}  \bs_k \bs_k^\intercal
\end{equation*}
where $\tilde{\gamma}_k$ are the eigenvalues of $\cS$, which are in fact the (signed) singular values of $\bS$, $\tilde{\gamma}_1 = \gamma_1, \hdots, \tilde{\gamma}_N=\gamma_N, \tilde{\gamma}_{N+1}= - \gamma_1, \hdots, \tilde{\gamma}_{2N}=-\gamma_N$.

Define the vectors  $\br_i = \left[
\begin{array}{c}
\mathbf{0}_N \\
\by_i^{(r)}
\end{array}
\right]$, $\bl_i =\left[
\begin{array}{c}
\by_i^{(l)} \\
\mathbf{0}_M
\end{array}
\right]$ for $\by_i^{(r)}, \by_i^{(l)}$ right/ left singular vectors of $\bY$, we have
\begin{equation}
\begin{split}
    \br_i^\intercal \big( {\rm Im}\, \bG_{\mathcal{S}}(x - \ci \epsilon) \big) \bl_i = \sum_{k=1}^{2 N} \frac{\epsilon }{(x - \tilde{\gamma}_k)^2+\epsilon^2} \big( \br_i^\intercal \bs_k \big) \big( \bl_i^\intercal \bs_k \big) + \frac{x + \ci \epsilon }{x^2+\epsilon^2} \sum_{k=2N+1}^{N+M}  \big( \br_i^\intercal \bs_k \big) \big( \bl_i^\intercal \bs_k \big)
\end{split}
\end{equation}
Given the structure of $\bs_k$'s in \eqref{EVD of S - app},  we have:
\begin{equation*}
  \big( \br_i^\intercal \bs_k \big) \big( \bl_i^\intercal \bs_k \big) = \begin{cases}
      \frac{1}{2}  \big(  \bu_k^\intercal \by_i^{(l)}\big) \big(  \bv_k^\intercal \by_i^{(r)} \big) & {\rm for } \hspace{5pt} 1 \leq k \leq N \\
      - \frac{1}{2}  \big(  \bu_{k-N}^\intercal \by_i^{(l)}\big) \big(  \bv_{k-N}^\intercal \by_i^{(r)} \big)  & {\rm for }  \hspace{5pt} N+1 \leq k \leq 2N \\
      0 & {\rm for }  \hspace{5pt} 2N+1 \leq k \leq N+M
  \end{cases}
\end{equation*}
In the limit of large N, the latter quantity is also self-averaging, due to the fact that as $N \to \infty$, these overlaps exhibit asymptotic independence, enabling the law of large numbers to be applied here. We can thus state that:
\begin{equation}
    \br_i^\intercal \big( {\rm Im}\, \bG_{\mathcal{S}}(x - \ci \epsilon) \big) \bl_i \xrightarrow[]{N \to \infty} \int_\bR \frac{\epsilon}{(x - t)^2+ \epsilon^2} O_Y(t,\sigma_i) \bar{\mu}_{S}(t) \, dt
\end{equation}
where the overlap function $O_Y(t,\lambda_i) $ is extended (continuously) to arbitrary values within the support of $\bar{\mu}_{S}$ with the property that $O_Y(-t,\lambda_i) = - O_Y(t,\lambda_i)$ for $ t \in {\rm supp} (\mu_S)$ . Sending $\epsilon \to 0$, we find 
\begin{equation}
    \br_i^\intercal \big( {\rm Im}\, \bG_{\mathcal{S}}(x - \ci \epsilon) \big) \bl_i \approx \pi \bar{\mu}_{S}(x) O_Y(x, \sigma_i)
    \label{Y-resolvent-overlap}
\end{equation}
In the next section, we establish a connection between the  resolvent $\bG_{\mathcal{S}}(z)$ and the signal $\bY$, which enables us to determine the overlap and consequently the optimal singular values values $\widehat{{\xi_y^*}}_i$ in terms of the singular values of the observation matrix $\bS$.

\subsection{Resolvent relation for  $\bY$}
In this section, we consider estimating $\bY$, and treat both $\bX$ and $\bW$ as noise. We consider the model to be:
\begin{equation}
    \bS = \bO \bX \bO^\intercal \bY + \bU \bW \bV^\intercal
    \label{Y-model}
\end{equation}
where $\bX = \bX^\intercal \in \bR^{N \times N}, \bW \in \bR^{N \times M}$ are fixed matrices with limiting eigenvalue/singular value distribution $\rho_X, \mu_W$, and $\bO, \bU \in \bR^{N \times N}, \bV \in \bR^{M \times M}$ are independent random Haar matrices.
For simplicity of notation, we use $\bT \equiv \bO \bX \bO^\intercal \bY$, and $\cT \in \bR^{(N+M) \times (N+M)}$ the hermitization of $\bT$. And $\widetilde{\cW}$ denotes the hermitization of the matrix $\bU \bW \bV^\intercal$.

As in the case for $\bX$, we express the entries of $\bG(z) \equiv \bG_{\mathcal{S}}(z) $ using Gaussian integral representation, and after applying the replica trick \eqref{resolvent-replica}, we find:
\begin{equation}
    \begin{split}
         \langle G_{i j}(z) \rangle &=  \lim_{n \to \infty} \int \Big( \prod_{k=1}^{N+M} \prod_{\tau =1}^n d \eta_k^{(\tau)} \Big)  \eta_i^{(1)} \eta_j^{(1)} \, \Big\langle \exp \big\{ -\frac{1}{2} \sum_{\tau =1}^n  {\bbeta^{(\tau)}}^\intercal ( z \bI - \cS) \bbeta^{(\tau)} \big\} \Big\rangle_{\bO, \bU, \bV} \\
         &= \lim_{n \to \infty} \int \big( \prod_{k=1}^{N+M} \prod_{\tau =1}^n d \eta_k^{(\tau)} \big) \, \eta_i^{(1)} \eta_j^{(1)} \exp \big\{-\frac{z}{2} \sum_{\tau =1}^n {\bbeta^{(\tau)}}^\intercal \bbeta^{(\tau)} \big\} \\
         &\hspace{2 cm} \times \Big\langle \exp \big\{ \frac{1}{2} \sum_{\tau =1}^n  {\bbeta^{(\tau)}}^\intercal  \cT \bbeta^{(\tau)} \big\} \Big\rangle_{\bO} \Big\langle \exp \big\{ \frac{1}{2} \sum_{\tau =1}^n  {\bbeta^{(\tau)}}^\intercal  \widetilde{\cW} \bbeta^{(\tau)} \big\} \Big\rangle_{\bU, \bV} \\
    \end{split}
    \label{Y-Gaussian Integral}
\end{equation}
Split each replica $\bbeta^{(\tau)}$ into two vectors $\ba^{(\tau)} \in \bR^N, \bb^{(\tau)} \in \bR^M$, $\bbeta^{(\tau)} = \left[
\begin{array}{c}
\ba^{(\tau)} \\
\bb^{(\tau)}
\end{array}
\right]$. The exponent in the first bracket in \eqref{Y-Gaussian Integral} can be written as :
\begin{equation}
    \begin{split}
         {\bbeta^{(\tau)}}^\intercal \cT \bbeta^{(\tau)} &= {\ba^{(\tau)}}^\intercal \bO \bX \bO^\intercal \bY \bb^{(\tau)}  + {\bb^{(\tau)}}^\intercal \bY^\intercal \bO \bX \bO^\intercal \ba^{(\tau)} \\
        &= \Tr \bO \bX \bO^\intercal \big( \underbrace{ \bY \bb^{(\tau)} {\ba^{(\tau)}}^\intercal + \ba^{(\tau)} {\bb^{(\tau)}}^\intercal \bY^\intercal}_\textrm{$\tilde{\bY}^{(\tau)}$} \big)
    \end{split}
    \label{Y- first bracket expansion}
\end{equation}
where $\tilde{\bY}^{(\tau)}$ is a symmetric $N \times N$ matrix with two non-zero eigenvalues  $ {\ba^{(\tau)}}^\intercal \bY \bb^{(\tau)} \pm \|\ba^{(\tau)}\| \|\bY \bb^{(\tau)}\|$ by lemma \ref{eigenvalue of xy^T + y^Tx}.

Using the formula for the spherical integral \cite{guionnet2005fourier} (see Theorem \ref{sphericla-integral} in \ref{spherical integral app}), we find:
\begin{equation}
\begin{split}
    \Big\langle \exp \big\{ \frac{1}{2} \sum_{\tau=1}^n \Tr  \bO \bX \bO^\intercal \tilde{\bY}^{(\tau)}  \big\}  \Big\rangle_{\bO} \approx \exp \bigg\{ \frac{N}{2} &\sum_{\tau=1}^n  \mathcal{P}_{\rho_X} \Big(\frac{1}{N} \big( {\ba^{(\tau)}}^\intercal \bY \bb^{(\tau)} + \|\ba^{(\tau)}\| \|\bY \bb^{(\tau)}\|   \big) \Big) \\
    &+ \mathcal{P}_{\rho_X} \Big(\frac{1}{N} \big( {\ba^{(\tau)}}^\intercal \bY \bb^{(\tau)} - \|\ba^{(\tau)}\| \|\bY \bb^{(\tau)}\|   \big) \Big) \bigg\}
\end{split}
    \label{Y-first bracket spherical}
\end{equation}

By the same computation as previous section, for the second bracket we have:
\begin{equation}
    \Big\langle \exp \big\{  \sum_{\tau =1}^n  \Tr \bb^{(\tau)} {\ba^{(\tau)}}^\intercal  \bU \bW \bV^\intercal   \big\} \Big\rangle_{\bU, \bV}  \approx \exp \Big\{ \frac{N}{2} \sum_{\tau=1}^n \mathcal{Q}_{\mu_W}^{(\alpha)} \big(\frac{1}{N M} \| \ba^{(\tau)} \|^2 \| \bb^{(\tau)} \|^2 \big) \Big\}
    \label{Y-second bracket spherical}
\end{equation}

From \eqref{Y-Gaussian Integral}, \eqref{Y-first bracket spherical}, \eqref{Y-second bracket spherical}, we find:
\begin{equation}
    \begin{split}
         \langle G_{i j}(z) \rangle &= \lim_{n\to \infty} \, \int \big( \prod_{k=1}^{N+M} \prod_{\tau =1}^n d \eta_k^{(\tau)} \big) \, \eta_i^{(1)} \eta_j^{(1)} \\
         & \times \exp \Bigg\{  -\frac{1}{2} \sum_{\tau =1}^n \bigg[ z \|{\bbeta^{(\tau)}}\|^2 - N \mathcal{Q}_{\mu_W}^{(\alpha)} \big(\frac{1}{N M} \| \ba^{(\tau)} \|^2 \| \bb^{(\tau)} \|^2 \big) \\
         & \hspace{2cm}  -N \mathcal{P}_{\rho_X} \Big(\frac{1}{N} \big( {\ba^{(\tau)}}^\intercal \bY \bb^{(\tau)} + \|\ba^{(\tau)}\| \|\bY \bb^{(\tau)}\|   \big) \Big)\\
         & \hspace{4cm} -  N \mathcal{P}_{\rho_X} \Big(\frac{1}{N} \big( {\ba^{(\tau)}}^\intercal \bY \bb^{(\tau)} - \|\ba^{(\tau)}\| \|\bY \bb^{(\tau)}\|   \big) \Big) \bigg] \Bigg\}
    \end{split}
    \label{Y-Gaussian Integral 2}
\end{equation}

Now, we introduce delta functions (for brevity we drop the limit term):
\begin{equation}
    \begin{split}
         \langle G_{i j}(z) \rangle = \int &\big( \prod_{k=1}^{N+M} \prod_{\tau =1}^n d \eta_k^{(\tau)} \big)  \big( \prod_{\tau =1}^n d p^{(\tau)}_1 \, d q^{(\tau)}_2 \, d q^{(\tau)}_3 \, d q^{(\tau)}_4 \big) \, \eta_i^{(1)} \eta_j^{(1)} \\ 
         &\times \prod_{\tau =1}^n \delta \big( q_1^{(\tau)} - \frac{1}{N} \| \ba^{(\tau)} \|^2 \big) \, \delta \big( q_2^{(\tau)} - \frac{1}{M} \| \bb^{(\tau)} \|^2 \big) \\
         &\hspace{3cm} \times \delta \big( q_3^{(\tau)} - \frac{1}{N} \| \bY \bb^{(\tau)} \|^2 \big) \, \delta \big( q_4^{(\tau)} - \frac{1}{N} {\ba^{(\tau)}}^\intercal \bY \bb^{(\tau)} \big) \\
         &\times  \exp \Big\{-\frac{1}{2} \sum_{\tau =1}^n z \|{\bbeta^{(\tau)}}\|^2  - N \mathcal{Q}_{\mu_W}^{(\alpha)} (q^{(\tau)}_1 q^{(\tau)}_2)\\
         &\hspace{2cm}-N \mathcal{P}_{\rho_X} \big(q_4^{(\tau)} + \sqrt{q_1^{(\tau)} q_3^{(\tau)}} \big) -N \mathcal{P}_{\rho_X} \big(q_4^{(\tau)} - \sqrt{q_1^{(\tau)} q_3^{(\tau)}} \big) \Big\}
    \end{split}
    \label{Y-Gaussian Integral delta}
\end{equation}
In the next step, we replace each delta with its Fourier transform. Note that for the parameters $q_1, q_2, q_3$ we use
$\delta \big(q_1^{\tau} - \frac{1}{N}\|\ba^{\tau}\|^2\big) \propto \int \, d \beta_1^{\tau}  \exp  \Big\{ -\frac{N}{2} \beta_1^{\tau} \big( q_1^{\tau} - \frac{1}{N}\|\ba^{\tau}\|^2 \big) \Big\}$, and for $q_4$ we use $\delta \big( q_4^{(\tau)} - \frac{1}{N} {\ba^{(\tau)}}^\intercal \bY \bb^{(\tau)} \big) \propto \int \, d \beta_1^{\tau}  \exp  \Big\{ -N \beta_1^{\tau} \big( q_4^{(\tau)} - \frac{1}{N} {\ba^{(\tau)}}^\intercal \bY \bb^{(\tau)} \big) \Big\}$. After rearranging, we find:
\begin{equation}
    \begin{split}
         \langle G_{i j}(z) \rangle &\propto \int \big( \prod_{\tau =1}^n d q^{(\tau)}_1 \, d q^{(\tau)}_2 \, d q^{(\tau)}_3 \, d q^{(\tau)}_4 \, d \beta^{(\tau)}_1 \, d \beta^{(\tau)}_2 \, d \beta^{(\tau)}_3 \, d \beta^{(\tau)}_4\big) \\
         &\times \exp \Big\{ \frac{N}{2} \sum_{\tau =1}^n  \mathcal{Q}_{\mu_W}^{(\alpha)} (q^{(\tau)}_1 q^{(\tau)}_2)  + \mathcal{P}_{\rho_X} \big(q_4^{(\tau)} + \sqrt{q_1^{(\tau)} q_3^{(\tau)}} \big) + \mathcal{P}_{\rho_X} \big(q_4^{(\tau)} - \sqrt{q_1^{(\tau)} q_3^{(\tau)}} \big) \\
         &\hspace{4cm} - \beta_1^{(\tau)} q_1^{(\tau)} - \frac{1}{\alpha} \beta_2^{(\tau)} q_2^{(\tau)} - \beta_3^{(\tau)} q_3^{(\tau)} - 2 \beta_4^{(\tau)} q_4^{(\tau)}\Big\} \\
         &\times \int \big( \prod_{k=1}^{N+M} \prod_{\tau =1}^n d \eta_k^{(\tau)} \big) \, \eta_i^{(1)} \eta_j^{(1)} \exp \Big\{-\frac{1}{2} \sum_{\tau =1}^n z \| {\bbeta^{(\tau)}}\| - \beta_1^{(\tau)} \| \ba^{(\tau)} \|^2 - \beta_2^{(\tau)} \| \bb^{(\tau)} \|^2 \\
         &\hspace{5cm} - \beta_3^{(\tau)} \| \bY \bb^{(\tau)} \|^2 - 2 \beta_4^{(\tau)} {\ba^{(\tau)}}^\intercal \bY \bb^{(\tau)} \Big\}
    \end{split}
    \label{Y-Gaussian Integral delta Fourier}
\end{equation}

The inner integral is a Gaussian integral, and can be written as:
\begin{equation}
\begin{split}
        \int \big( \prod_{k=1}^{N+M} &\prod_{\tau =1}^n d \eta_k^{(\tau)} \big) \, \eta_i^{(1)} \eta_j^{(1)} \\
        &\times \exp \Big\{ \sum_{\tau =1}^n -\frac{1}{2} {\bbeta^{(\tau)}}^\intercal \left[
\begin{array}{cc}
(z - \beta_1^{(\tau)}) \bI_N & -\beta_4^{(\tau)} \bY  \\
-\beta_4^{(\tau)} \bY^\intercal & (z-\beta_2^{(\tau)}) \bI_M - \beta_3^{(\tau)} \bY^\intercal \bY  
\end{array}
\right] \bbeta^{(\tau)} \Big\}
\end{split}
\label{Y-Gaussian}
\end{equation}

Denote the matrix in the exponent by $\bC_Y^{(\tau)}$. Using the formula for determinant of block matrices (see proposition 2.8.4 in \cite{bernstein2009matrix}), we have:: 
\begin{equation*}
\begin{split}
    \det \bC_Y^{(\tau)} &= \det \Big[ (z - \beta_1^{(\tau)}) \bI_N - {\beta_4^{(\tau)}}^2 \bY \big( (z-\beta_2^{(\tau)}) \bI_M - \beta_3^{(\tau)} \bY^\intercal \bY \big)^{-1} \bY^\intercal \Big] \\
    &\hspace{7cm} \times \det \big[(z-\beta_2^{(\tau)}) \bI_M - \beta_3^{(\tau)} \bY^\intercal \bY \big] \\
    &= \prod_{k=1}^N \big[ z - \beta_1^{(\tau)} - {\beta_4^{(\tau)}}^2 \frac{\sigma_k^2}{z-\beta_2^{(\tau)} - \beta_3^{(\tau)} \sigma_k^2} \big] \prod_{k=1}^N \big(z-\beta_2^{(\tau)} - \beta_3^{(\tau)} \sigma_k^2 \big) \, \big( z-\beta_2^{(\tau)} \big)^{M-N} \\
    &=  \big( z-\beta_2^{(\tau)} \big)^{M-N} \, \prod_{k=1}^N \big[ (z - \beta_1^{(\tau)})(z-\beta_2^{(\tau)} - \beta_3^{(\tau)} \sigma_k^2 ) - {\beta_4^{(\tau)}}^2 \sigma_k^2 \big] \\
    &= \big( z-\beta_2^{(\tau)} \big)^{M-N} \, \prod_{k=1}^N \Big[ (z - \beta_1^{(\tau)})(z-\beta_2^{(\tau)})  - \big( {\beta_4^{(\tau)}}^2 +  \beta_3^{(\tau)} (z - \beta_1^{(\tau)}) \big) \sigma_k^2 \Big]
\end{split}
\end{equation*}
where $\sigma_k$'s are the singular values of $\bY$. So computing the Gaussian integrals, \eqref{Y-Gaussian Integral delta Fourier} can be written as:
\begin{equation}
    \begin{split}
         \langle G_{i j}(z) \rangle \propto \int \big( \prod_{\tau =1}^n d q^{(\tau)}_1 \, d q^{(\tau)}_2 \, d q^{(\tau)}_3 \, d q^{(\tau)}_4 &\, d \beta^{(\tau)}_1 \, d \beta^{(\tau)}_2 \, d \beta^{(\tau)}_3 \, d \beta^{(\tau)}_4\big) \big( {\bC_Y^{(1)}}^{-1} \big)_{i j} \\
         &\times \exp \big\{  -\frac{N n}{2} F_0^Y ( \bm{q}_1, \bm{q}_2, \bm{q}_3, \bm{q}_4, \bm{\beta}_1, \bm{\beta}_2, \bm{\beta}_3, \bm{\beta}_4 ) \big\}
    \end{split}
    \label{Y-Saddle Point Integral}
\end{equation}
with
\begin{equation}
    \begin{split}
        F_0^Y ( \bm{q}_1, \bm{q}_2, \bm{q}_3, \bm{q}_4, & \bm{\beta}_1, \bm{\beta}_2, \bm{\beta}_3, \bm{\beta}_4 ) = \frac{1}{n} \sum_{\tau =1}^n \bigg[   \big( \frac{1}{\alpha} -1 \big) \ln ( z-\beta_2^{(\tau)} )\\
        & + \frac{1}{N} \sum_{k=1}^N \ln \Big( (z - \beta_1^{(\tau)})(z-\beta_2^{(\tau)})  - \big( {\beta_4^{(\tau)}}^2 +  \beta_3^{(\tau)} (z - \beta_1^{(\tau)}) \big) \sigma_k^2 \Big) \\
        &-\mathcal{Q}_{\mu_W}^{(\alpha)} (q^{(\tau)}_1 q^{(\tau)}_2)  - \mathcal{P}_{\rho_X} \big(q_4^{(\tau)} + \sqrt{q_1^{(\tau)} q_3^{(\tau)}} \big) - \mathcal{P}_{\rho_X} \big(q_4^{(\tau)} - \sqrt{q_1^{(\tau)} q_3^{(\tau)}} \big) \\
        &+  \beta_1^{(\tau)} q_1^{(\tau)} + \frac{1}{\alpha} \beta_2^{(\tau)} q_2^{(\tau)} + \beta_3^{(\tau)} q_3^{(\tau)} + 2 \beta_4^{(\tau)} q_4^{(\tau)} \bigg]
    \end{split}
    \label{Y-Free energy}
\end{equation}

We will evaluate the integral \eqref{Y-Gaussian Integral delta Fourier} using saddle-points of the function $F_0^Y$. From the \textit{replica symmetric ansatz} at the saddle-point we have:
\begin{equation*}
    \forall \tau \in \{1, \cdots, n\}: \quad \begin{cases}
        q_1^{\tau} = q_1, \quad q_2^{\tau} = q_2, \quad q_3^{\tau} = q_3, \quad q_4^{\tau} = q_4 \\
        \beta_1^{\tau} = \beta_1, \quad \beta_2^{\tau} = \beta_2, \quad \beta_3^{\tau} = \beta_3, \quad \beta_4^{\tau} = \beta_4
    \end{cases}
\end{equation*}

Finally, we find the solution to be:
\begin{equation}
    \begin{cases}
    \beta_1^* = \frac{\mathcal{C}_{\mu_W}^{(\alpha)}(q_1^* q_2^*)}{q_1^*} + \frac{1}{2} \sqrt{\frac{q_3^*}{q_1^*}} \Big( \mathcal{R}_{\rho_X} \big( q_4^* + \sqrt{q_1^* q_3^*} \big) - \mathcal{R}_{\rho_X} \big( q_4^* - \sqrt{q_1^* q_3^*} \big) \Big)\\
    \beta_2^* = \alpha \frac{\mathcal{C}_{\mu_W}^{(\alpha)}(q_1^* q_2^*)}{q_2^*}\\
    \beta_3^* = \frac{1}{2} \sqrt{\frac{q_1^*}{q_3^*}} \Big( \mathcal{R}_{\rho_X} \big( q_4^* + \sqrt{q_1^* q_3^*} \big) - \mathcal{R}_{\rho_X} \big( q_4^* - \sqrt{q_1^* q_3^*} \big) \Big)\\
    \beta_4^* = \frac{1}{2}  \Big( \mathcal{R}_{\rho_X} \big( q_4^* + \sqrt{q_1^* q_3^*} \big) + \mathcal{R}_{\rho_X} \big( q_4^* - \sqrt{q_1^* q_3^*} \big) \Big)\\
    q_1^* = \frac{(z-\beta_2^*) {\beta_4^*}^2}{Z_2(z)^2} \mathcal{G}_{\rho_Y} \big( \frac{Z_1(z)}{Z_2(z)} \big) + \frac{\beta_3^*}{Z_2(z)} \\
    q_2^* = \alpha \frac{z-\beta_1^*}{Z_2(z)} \mathcal{G}_{\rho_Y} \big( \frac{Z_1(z)}{Z_2(z)} \big) + \frac{1 - \alpha}{z - \beta_2^*} \\
    q_3^* = \frac{(z-\beta_1^*) Z_1(z)}{Z_2(z)^2} \mathcal{G}_{\rho_Y} \big( \frac{Z_1(z)}{Z_2(z)} \big) - \frac{z-\beta_1^*}{Z_2(z)} \\
    q_4^* = \frac{ \beta_4^* Z_1(z)}{Z_2(z)^2} \mathcal{G}_{\rho_Y} \big( \frac{Z_1(z)}{Z_2(z)} \big) - \frac{\beta_4^*}{Z_2(z)} 
    \end{cases} \quad \hspace{-80pt} \text{with }
    \begin{cases}
        Z_1(z) = (z-\beta_1^*)(z-\beta_2^*) \\
        Z_2(z) = {\beta_4^*}^2 + \beta_3^*(z-\beta_1^*)
    \end{cases}
    \label{Y-sol-app}
\end{equation}
where $\rho_Y$ is the limiting eigenvalue distribution of $\bY \bY^\intercal$.

The relation \eqref{Y-Saddle Point Integral} and the solutions \eqref{Y-sol-app} hold for arbitrary indices $i,j$, so we can state the relation in the matrix form. Computing the inverse of ${\bC_Y^*}^{-1}$ (see section \ref{Matrix-tool}), we have:
\begin{equation}
\begin{split}
    \big\langle \bG_{\mathcal{S}}(z) \big\rangle_{\bO, \bU, \bV} &= \Bigg\langle \left[
\begin{array}{cc}
\frac{1}{z} \bI_N + \frac{1}{z} \bS \bG_{S^\intercal S}(z^2) \bS^\intercal & \bS \bG_{S^\intercal S}(z^2) \\
\bG_{S^\intercal S}(z^2) \bS^\intercal & z \bG_{S^\intercal S}(z^2)
\end{array}
\right] \Bigg\rangle \\
&\hspace{-2pt}= \left[
\begin{array}{cc}
\frac{1}{z-\beta_1^*} \bI_N + \frac{{\beta_4^*}^2}{(z-\beta_1^*) Z_2(z)} \bY \bG_{Y^\intercal Y} \big(\frac{Z_1(z)}{Z_2(z)} \big) \bY^\intercal & \frac{\beta_4^*}{Z_2(z)} \bY  \bG_{Y^\intercal Y} \big(\frac{Z_1(z)}{Z_2(z)} \big)  \\
 \frac{\beta_4^*}{Z_2(z)}  \bG_{Y^\intercal Y} \big(\frac{Z_1(z)}{Z_2(z)} \big) \bY^\intercal & \frac{z - \beta_1^*}{Z_2(z)} \bG_{Y^\intercal Y} \big(\frac{Z_1(z)}{Z_2(z)} \big) 
\end{array}
\right]
\end{split}
\label{Y-resolvent relation}
\end{equation}

With this relation, we can further simplify the solution \eqref{Y-sol-app}. 

We start with comparing the trace of upper-left block in \eqref{Y-resolvent relation}. The normalized trace of the first block in $\big\langle \bG_{\mathcal{S}}(z) \big\rangle_{\bO, \bU, \bV}$ is computed in \eqref{trace-GS-X-first} to be $\mathcal{G}_{\bar{\mu}_S} (z)$. The normalized trace of the upper-left block in ${\bC_Y^*}^{-1}$ is:
\begin{equation}
    \begin{split}
        \frac{1}{N} \Tr \, \Big[  (z-\beta_1^*)^{-1} \bI_N +& \frac{{\beta_4^*}^2}{(z-\beta_1^*) Z_2(z)} \bY \bG_{Y^\intercal Y} \big(\frac{Z_1(z)}{Z_2(z)} \big) \bY^\intercal \Big] \\
        &= \frac{1}{N} \frac{1}{z-\beta_1^*} \sum_{k =1}^N \big[ 1 + \frac{{\beta_4^*}^2}{Z_2(z)} \frac{\sigma_k^2}{\frac{Z_1(z)}{Z_2(z)} - \sigma_k^2} \big] \\
        &= \frac{1}{N} \frac{1}{z-\beta_1^*} \sum_{k =1}^N \big[ \frac{{\beta_4^*}^2 Z_1(z)}{Z^2_2(z)} \frac{1}{\frac{Z_1(z)}{Z_2(z)} - \sigma_k^2} + 1- \frac{{\beta_4^*}^2}{Z_2(z)}\big] \\
        &= \frac{1}{N} \frac{1}{z-\beta_1^*}\frac{{\beta_4^*}^2 Z_1(z)}{Z^2_2(z)} \sum_{k =1}^N   \frac{1}{\frac{Z_1(z)}{Z_2(z)} - \sigma_k^2}  + \frac{1}{z-\beta_1^*} \frac{\beta_3^*(z-\beta_1^*)}{Z_2(z)} \\
        &= \frac{(z-\beta_2^*) {\beta_4^*}^2}{Z_2(z)^2} \mathcal{G}_{\rho_Y} \big( \frac{Z_1(z)}{Z_2(z)} \big) + \frac{\beta_3^*}{Z_2(z)} \\
        &= q_1^*
    \end{split}
    \label{q1-star}
\end{equation}
Thus, $q_1^* = \mathcal{G}_{\bar{\mu}_S}(z)$.  

The normalized trace of the lower-right block of $\big\langle \bG_{\mathcal{S}}(z) \big\rangle_{\bO, \bU, \bV}$ is $\alpha \mathcal{G}_{\bar{\mu}_S}(z) + (1- \alpha ) \frac{1}{z}$ (see \eqref{trace-GS-X-last}). The normalized trace of the lower-right block in ${\bC_Y^*}^{-1}$ is:
\begin{equation}
    \begin{split}
        \frac{1}{M} \Tr \, \Big[ \frac{z - \beta_1^*}{Z_2(z)} \bG_{Y^\intercal Y} \big(\frac{Z_1(z)}{Z_2(z)} \big) \Big] &= \frac{1}{M} \frac{z - \beta_1^*}{Z_2(z)} \sum_{k =1}^N   \frac{1}{\frac{Z_1(z)}{Z_2(z)} - \sigma_k^2} + \frac{M-N}{M} \frac{z - \beta_1^*}{Z_2(z)} \frac{Z_2(z)}{Z_1(z)} \\
        &= \frac{N}{M} \frac{1}{N} \frac{z - \beta_1^*}{Z_2(z)} \sum_{k =1}^N   \frac{1}{\frac{Z_1(z)}{Z_2(z)} - \sigma_k^2} + \frac{M-N}{M} \frac{z - \beta_1^*}{Z_1(z)}  \\
        &= \alpha \frac{z-\beta_1^*}{Z_2(z)} \mathcal{G}_{\rho_Y} \big( \frac{Z_1(z)}{Z_2(z)} \big) + \frac{1 - \alpha}{z - \beta_2^*} \\
        &= q_2^*
    \end{split}
    \label{q2-star}
\end{equation}
So, $q_2^* =  \alpha \mathcal{G}_{\bar{\mu}_S}(z) + (1- \alpha ) \frac{1}{z}$. 

With a bit of algebra, we can express the parameters $q^*_3, q^*_4$ in terms of $q^*_1, \beta^*_1, \beta^*_4$:
\begin{equation}
    q_3^* = \frac{(z - \beta_1^*)^2}{{\beta_4^*}^2} q_1^* - \frac{z - \beta_1^*}{{\beta_4^*}^2}, \quad q_4^* = \frac{z - \beta_1^*}{\beta_4^*} q_1^* - \frac{1}{\beta_4^*}
    \label{q-simplifications}
\end{equation}
Therefore, the solution can be written without involving $\mathcal{G}_{\rho_Y}$, as:
\begin{equation}
    \begin{cases}
    \beta_1^* = \frac{\mathcal{C}_{\mu_W}^{(\alpha)}(q_1^* q_2^*)}{q_1^*} + \frac{1}{2} \sqrt{\frac{q_3^*}{q_1^*}} \Big( \mathcal{R}_{\rho_X} \big( q_4^* + \sqrt{q_1^* q_3^*} \big) - \mathcal{R}_{\rho_X} \big( q_4^* - \sqrt{q_1^* q_3^*} \big) \Big)\\
    \beta_2^* = \alpha \frac{\mathcal{C}_{\mu_W}^{(\alpha)}(q_1^* q_2^*)}{q_2^*}\\
    \beta_3^* = \frac{1}{2} \sqrt{\frac{q_1^*}{q_3^*}} \Big( \mathcal{R}_{\rho_X} \big( q_4^* + \sqrt{q_1^* q_3^*} \big) - \mathcal{R}_{\rho_X} \big( q_4^* - \sqrt{q_1^* q_3^*} \big) \Big)\\
    \beta_4^* = \frac{1}{2}  \Big( \mathcal{R}_{\rho_X} \big( q_4^* + \sqrt{q_1^* q_3^*} \big) + \mathcal{R}_{\rho_X} \big( q_4^* - \sqrt{q_1^* q_3^*} \big) \Big)\\
    q_1^* = \mathcal{G}_{\bar{\mu}_S}(z) \\
    q_2^* = \alpha \mathcal{G}_{\bar{\mu}_S}(z) + (1- \alpha ) \frac{1}{z} \\
    q_3^* = \frac{(z - \beta_1^*)^2}{{\beta_4^*}^2} \mathcal{G}_{\bar{\mu}_S}(z) - \frac{z - \beta_1^*}{{\beta_4^*}^2} \\
    q_4^* = \frac{z - \beta_1^*}{\beta_4^*} \mathcal{G}_{\bar{\mu}_S}(z) - \frac{1}{\beta_4^*}
    \end{cases} 
    \label{Y-sol-app-simple}
\end{equation}
\begin{remark}\label{symmetric-rho-X-Y-RIE}
The simplifications in \eqref{q-simplifications} are derived with the assumption that $\beta_4^* \neq 0$. However, in the initial set of equations \eqref{Y-sol-app}, if $\rho_X$ is symmetric measure then $\beta^*_4 = q_4^* = 0$ is a solution. If $\rho_X$ is symmetric, then $\mathcal{R}_{\rho_X}(-z) = -\mathcal{R}_{\rho_X}(z)$, and plugging $q_4^* = 0$ in the expression for $\beta_4^*$ in \eqref{Y-sol-app}, we find that $\beta_4^* = 0$.
\end{remark}

\subsection{Overlaps and the optimal singular values}
From \eqref{Y-resolvent-overlap}, \eqref{Y-resolvent relation}, we find:
\begin{equation}
\begin{split}
    O_Y(\gamma, \sigma_i) &\approx \frac{1}{\pi \bar{\mu}_{S}(\gamma)} \,  {\rm Im} \, \lim_{z \to \gamma - \ci 0^+} \, \frac{\beta_4^*}{Z_2(z)}  {\by^{(r)}_i}^\intercal \bG_{Y^\intercal Y} \big(\frac{Z_1(z)}{Z_2(z)} \big) \bY^\intercal \by^{(l)}_i \\
    &= \frac{1}{\pi \bar{\mu}_{S}(\gamma)} \, {\rm Im} \, \lim_{z \to \gamma - \ci 0^+} \,   \beta_4^* \frac{\sigma_i}{ Z_1(z)  - Z_2(z) \sigma_i^2}
\end{split}
\label{Y-overlap-eq}
\end{equation}

From the overlap, we can compute the optimal singular values:
\begin{equation}
\begin{split}
    \widehat{\xi_y^*}_i &\approx \frac{1}{N} \sum_{j=1}^N \sigma_j O_Y(\gamma_i, \sigma_j) \\
    &\approx \frac{1}{\pi \bar{\mu}_{S}(\gamma_i)} \, {\rm Im} \, \lim_{z \to \gamma_i - \ci 0^+} \, \frac{1}{N} \sum_{j=1}^N  \beta_4^* \frac{\sigma_j^2}{ Z_1(z)  - Z_2(z) \sigma_j^2} \\
    &= \frac{1}{\pi \bar{\mu}_{S}(\gamma_i)} \, {\rm Im} \, \lim_{z \to \gamma_i - \ci 0^+} \, \frac{1}{N} \frac{\beta_4^*}{Z_2(z)} \sum_{j=1}^N   \frac{\sigma_j^2}{\frac{Z_1(z)}{Z_2(z)}  -  \sigma_j^2} \\
    &= \frac{1}{\pi \bar{\mu}_{S}(\gamma_i)} \, {\rm Im} \, \lim_{z \to \gamma_i - \ci 0^+} \, \frac{1}{N} \frac{\beta_4^*}{Z_2(z)} \sum_{j=1}^N  \bigg[\frac{\frac{Z_1(z)}{Z_2(z)}}{\frac{Z_1(z)}{Z_2(z)}  -  \sigma_j^2} - 1 \bigg] \\
    &\approx \frac{1}{\pi \bar{\mu}_{S}(\gamma_i)} \, {\rm Im} \, \lim_{z \to \gamma_i - \ci 0^+} \, \frac{ \beta_4^* Z_1(z)}{Z_2(z)^2} \mathcal{G}_{\rho_Y} \big( \frac{Z_1(z)}{Z_2(z)} \big) - \frac{\beta_4^*}{Z_2(z)}  \\
    &= \frac{1}{\pi \bar{\mu}_{S}(\gamma_i)} \, {\rm Im} \, \lim_{z \to \gamma_i - \ci 0^+} q_4^*
    \label{Y-optimal-sv-app}
\end{split}
\end{equation}
where in the last equality we used the solution we have found in \eqref{Y-sol-app}. Note that, based on \eqref{Y-sol-app-simple}, we do not need to have any knowledge about $\rho_Y$ to compute $q_4^*$. In the end, we need to divide the estimator by $\sqrt{\kappa}$ as we have absorbed it into $\bY$.

\subsubsection{Recovering the rectangular RIE for a denoising problem}\label{rect-RIE}
Note that if in the model \eqref{Y-model}, we put $\bX = \bI$ the model reduces to the additive denoising of $\bY$, and we recover the estimator recently proposed in \cite{pourkamali2023rectangular} for the rectangular case.

For $\bX= \bI$, $\mathcal{R}_{\rho_X}(z) = 1$, so \eqref{Y-sol-app-simple} reduces to:
\begin{equation}
    \begin{cases}
    \beta_1^* = \frac{\mathcal{C}_{\mu_W}^{(\alpha)}(q_1^* q_2^*)}{q_1^*}, \quad \beta_2^* = \alpha \frac{\mathcal{C}_{\mu_W}^{(\alpha)}(q_1^* q_2^*)}{q_2^*}, \quad \beta_3^* = 0, \quad \beta_4^* = 1\\
    q_1^* = \mathcal{G}_{\bar{\mu}_S}(z), \quad q_2^* = \alpha \mathcal{G}_{\bar{\mu}_S}(z) + (1- \alpha ) \frac{1}{z} \\
    q_3^* = (z - \beta_1^*)^2 \mathcal{G}_{\bar{\mu}_S}(z) - (z - \beta_1^*), \quad q_4^* = (z - \beta_1^*) \mathcal{G}_{\bar{\mu}_S}(z) - 1
    \end{cases} 
    \label{Y-denoising-sol}
\end{equation}
From \eqref{Y-optimal-sv-app}, we have:
\begin{equation}
\begin{split}
    \widehat{\xi_y^*}_i &= \frac{1}{\pi \bar{\mu}_{S}(\gamma_i)} \, {\rm Im} \, \lim_{z \to \gamma_i - \ci 0^+} q_4^* \\
    &= \frac{1}{\pi \bar{\mu}_{S}(\gamma_i)} \, {\rm Im} \, \lim_{z \to \gamma_i - \ci 0^+} z \mathcal{G}_{\bar{\mu}_S}(z) - \beta_1^* \mathcal{G}_{\bar{\mu}_S}(z) -1 \\
    &= \frac{1}{\pi \bar{\mu}_{S}(\gamma_i)} \, {\rm Im} \, \lim_{z \to \gamma_i - \ci 0^+} z \mathcal{G}_{\bar{\mu}_S}(z) - \frac{\mathcal{C}_{\mu_W}^{(\alpha)}(q_1^* q_2^*)}{q_1^*} \mathcal{G}_{\bar{\mu}_S}(z) -1 \\
    &=  \frac{1}{\pi \bar{\mu}_{S}(\gamma_i)} \, {\rm Im} \, \lim_{z \to \gamma_i - \ci 0^+} z \mathcal{G}_{\bar{\mu}_S}(z) - \mathcal{C}_{\mu_W}^{(\alpha)}(q_1^* q_2^*) -1 \\
    &=  \frac{1}{\pi \bar{\mu}_{S}(\gamma_i)} \, {\rm Im} \, \lim_{z \to \gamma_i - \ci 0^+} z \mathcal{G}_{\bar{\mu}_S}(z) - \mathcal{C}_{\mu_W}^{(\alpha)} \Big( \mathcal{G}_{\bar{\mu}_S}(z) \big( \alpha \mathcal{G}_{\bar{\mu}_S}(z) + (1- \alpha ) \frac{1}{z} \big) \Big) -1 \\
    &\stackrel{\text{(a)}}{=} \frac{1}{\pi \bar{\mu}_{S}(\gamma_i)} {\rm Im} \,\Bigg[  \gamma_i \mathcal{G}_{\bar{\mu}_S}(\gamma_i - \ci 0^+) - \mathcal{C}^{(\alpha)}_{\mu_W}\bigg(\frac{1}{\gamma_i}  \mathcal{G}_{\bar{\mu}_S}(\gamma_i - \ci 0^+) \Big(1 - \alpha + \alpha \gamma_i  \mathcal{G}_{\bar{\mu}_S}(\gamma_i - \ci 0^+) \Big) \bigg) \Bigg]  \\
    &\stackrel{\text{(b)}}{=} \gamma_i - \frac{1}{\pi \bar{\mu}_{S}(\gamma_i)} {\rm Im} \, \mathcal{C}^{(\alpha)}_{\mu_W}\bigg( \frac{1- \alpha}{\gamma_i} \pi \sH [\bar{\mu}_{S}](\gamma_i) + \alpha \big(\pi \sH [\bar{\mu}_{S}](\gamma_i)\big)^2  - \alpha  \big( \pi \bar{\mu}_{S}(\gamma_i)\big)^2 \\
    &\hspace{7cm}+ \ci  \pi \bar{\mu}_{S}(\gamma_i) \big(\frac{1-\alpha}{\gamma_i} + 2 \alpha \pi \sH [\bar{\mu}_{S}](\gamma_i) \big) \bigg)
    \label{Y-denosing-optimal-sv-app}
\end{split}
\end{equation}
where in (a) we used the analyticity of rectangular R-transform \cite{benaych2009rectangular}, and in (b), we used Plemelj formula \eqref{Plemelj formulae}. Note that, the final estimator should be divided by the $\sqrt{\kappa}$. 


\subsection{Examples}\label{Y-examples}
Throughout the numerical experiments, we consider the matrix $\bW$ to have i.i.d. Gaussian entries with variance $\nicefrac{1}{N}$, so $\mathcal{C}_{\mu_W}^{(\alpha)}(z) = \frac{1}{\alpha} z$. And, $\bX = \bF + c \bI$ where $\bF = \bF^\intercal \in \bR^{N \times N}$ has i.i.d. entries with variance $\nicefrac{1}{N}$, and $c \neq 0$ is a real number, so $\mathcal{R}_{\rho_X}(z) = z +c$. With these choices, the solution \eqref{Y-sol-app-simple} simplifies to:
\begin{equation}
    \begin{cases}
    \beta_1^* = \frac{1}{\alpha} q_2^* + q_3^*, \quad \beta_2^* = q_1^*, \quad \beta_3^* = q_1^*, \quad \beta_4^* =  q_4^* + c\\
    q_1^* = \mathcal{G}_{\bar{\mu}_S}(z), \quad q_2^* = \alpha \mathcal{G}_{\bar{\mu}_S}(z) + (1- \alpha ) \frac{1}{z} \\
    q_3^* = \frac{(z - \beta_1^*)^2}{{\beta_4^*}^2} \mathcal{G}_{\bar{\mu}_S}(z) - \frac{z - \beta_1^*}{{\beta_4^*}^2}, \quad q_4^* = \frac{z - \beta_1^*}{\beta_4^*} \mathcal{G}_{\bar{\mu}_S}(z) - \frac{1}{\beta_4^*}
    \end{cases} 
    \label{Y-sol-app-X,W-G}
\end{equation}
Note that in \eqref{Y-sol-app-X,W-G}, $q_1^*, q_2^*$ are given in terms of the observation, so to find the solution we only need to find the parameters $q_3^*, q_4^*$. In \eqref{Y-sol-app-X,W-G}, one can see that we have the relation $q_3^* = \frac{z - \beta_1^*}{\beta_4^*} q_4^*$. Writing the parameters $\beta_1^*, \beta_4^*$ in terms of $q_2^*, q_3^*, q_4^*$, after a bit of algebra we have the following relation:
\begin{equation}
    q_3^* = \frac{z - \frac{1}{\alpha}q_2^*}{2 q_4^* + c} q_4^*
    \label{q_3-q_4 rel}
\end{equation}
In the expression for $q_4^*$ in \eqref{Y-sol-app-X,W-G}, using \eqref{q_3-q_4 rel} we can rewrite $\beta_1^*, \beta_4^*$ in terms of $q_2^*, q_4^*$. After some manipulations we find that $q_4^*$ is the solution to the following cubic equation:
\begin{equation}
    2 x^3 + 3 c\, x^2 + \Big[ c^2 + 2 - \big(z - \mathcal{G}_{\bar{\mu}_S}(z) - \frac{1- \alpha}{\alpha} \frac{1}{z}\big) \mathcal{G}_{\bar{\mu}_S}(z) \Big]\, x - c \Big[ \big(z - \mathcal{G}_{\bar{\mu}_S}(z) - \frac{1- \alpha}{\alpha} \frac{1}{z}\big) \mathcal{G}_{\bar{\mu}_S}(z) - 1 \Big] = 0
    \label{q_4-eq}
\end{equation}
Based on our numerical simulations, we pick the following root for $q_4^*$:
\begin{equation}
    q_4^* = - \frac{c}{2} - \frac{12 - 3 c^2 + 6 A}{3 \sqrt[3]{B}} + \frac{\sqrt[3]{B}}{12}
    \label{q4-analyt}
\end{equation}
with
\begin{equation*}
\begin{split}
    A &=  {\mathcal{G}_{\bar{\mu}_S}(z)}^2 -  \frac{\mathcal{G}_{\bar{\mu}_S}(z)}{z} \big(1 - \frac{1}{\alpha} \big) -  \mathcal{G}_{\bar{\mu}_S}(z) z \\
    B &= - 216 c A + 4 \sqrt{ 4 \big( 12 - 3 c^2 + 6 A \big)^3 + 54^2 c^2 A^2} 
\end{split}
\end{equation*}
Once we have $q_4^*$, we can find $q_3^*$ using \eqref{q_3-q_4 rel}. In the end, $\beta_1^*, \cdots, \beta_4^*$ can be evaluated. Note that, for the RIE, only $q_4^*$ is required. Other parameters are used to evaluate the resolvent relation \eqref{Y-resolvent relation} and the overlap \eqref{Y-overlap-eq}. 
\subsubsection{Resolvent relation}
We take $\kappa =1$. In model \eqref{Y-model}, without loss of generality we can consider $\bY$ to be diagonal.

In figure \ref{Y-resolvent relation-Gaussian Y}, $\bY$ is the diagonal matrix obtained from the singular values of a  Gaussian matrix with i.i.d. entries of variance $\nicefrac{1}{N}$. In figure \ref{Y-resolvent relation-Uniform Y}, the non-zero entries (on main diagonal) of $\bY$ are uniformly distributed in $[1,3]$. As in previous cases, $\mu_S, \mathcal{G}_{\bar{\mu}_S}(z)$ are estimated numerically using Cauchy kernel, from which the parameters $\beta_1^*, \cdots, \beta_4^*$ are computed.

\begin{figure}
\begin{subfigure}[t]{\textwidth}
  \begin{subfigure}[t]{.4\textwidth}
    \centering
    \input{Figures/App/Y-est/Resolvent_Relation/Gaussian_real_entry1}
  \end{subfigure}
  \hfil
  \begin{subfigure}[t]{.4\textwidth}
  \centering
      \input{Figures/App/Y-est/Resolvent_Relation/Gaussian_im_entry1}
  \end{subfigure}
  \caption{{\small Entry $ i = j = 1$, first diagonal entry in the upper-left block}}
  \end{subfigure}
  \\
  \begin{subfigure}[t]{\textwidth}
  \begin{subfigure}[t]{.4\textwidth}
    \centering
    \input{Figures/App/Y-est/Resolvent_Relation/Gaussian_real_entry2}
  \end{subfigure}
  \hfil
  \begin{subfigure}[t]{.4\textwidth}
  \centering
      \input{Figures/App/Y-est/Resolvent_Relation/Gaussian_im_entry2}
  \end{subfigure}
  \caption{{\small Entry $ i = 1, j = 2$, non-diagonal entry in the upper-left block}}
  \end{subfigure}
  \\
  \begin{subfigure}[t]{\textwidth}
  \begin{subfigure}[t]{.4\textwidth}
    \centering
    \input{Figures/App/Y-est/Resolvent_Relation/Gaussian_real_entry3}
  \end{subfigure}
  \hfil
  \begin{subfigure}[t]{.4\textwidth}
  \centering
      \input{Figures/App/Y-est/Resolvent_Relation/Gaussian_im_entry3}
  \end{subfigure}
  \caption{{\small Entry $ i = j = 2001$, first diagonal entry in the lower-right block}}
  \end{subfigure}
  \\
  \begin{subfigure}[t]{\textwidth}
  \begin{subfigure}[t]{.4\textwidth}
    \centering
    \input{Figures/App/Y-est/Resolvent_Relation/Gaussian_real_entry4}
  \end{subfigure}
  \hfil
  \begin{subfigure}[t]{.4\textwidth}
  \centering
      \input{Figures/App/Y-est/Resolvent_Relation/Gaussian_im_entry4}
  \end{subfigure}
  \caption{{\small Entry $ i = 2001, j = 2002$, non-diagonal entry in the lower-right block}}
  \end{subfigure}
  \\
  \begin{subfigure}[t]{\textwidth}
  \begin{subfigure}[t]{.4\textwidth}
    \centering
    \input{Figures/App/Y-est/Resolvent_Relation/Gaussian_real_entry5}
  \end{subfigure}
  \hfil
  \begin{subfigure}[t]{.4\textwidth}
  \centering
      \input{Figures/App/Y-est/Resolvent_Relation/Gaussian_im_entry5}
  \end{subfigure}
  \caption{{\small Entry $ i = 1, j = 2001$, first entry in the upper-right block}}
  \end{subfigure}
    \\
  \begin{subfigure}[t]{\textwidth}
  \begin{subfigure}[t]{.4\textwidth}
    \centering
    \input{Figures/App/Y-est/Resolvent_Relation/Gaussian_real_entry6}
  \end{subfigure}
  \hfil
  \begin{subfigure}[t]{.4\textwidth}
  \centering
      \input{Figures/App/Y-est/Resolvent_Relation/Gaussian_im_entry6}
  \end{subfigure}
  \caption{{\small Entry $ i = 1, j = 2002$, second entry in the upper-right block}}
  \end{subfigure}
    \caption{\small  Illustration of \eqref{Y-resolvent relation}. $\bY \in \bR^{N \times M}$ is a diagonal matrix obtained from the singular values of a $N \times M$ matrix with i.i.d. entries of variance $\nicefrac{1}{N}$, $\bX = \bX^\intercal$ is shifted Wigner matrix with $c = 3$, and $\bZ$ is a Gaussian matrices with. The empirical estimate of $\bG_{\mathcal{S}}(z)$ (dashed blue line) is computed for $ z = \gamma_i - \ci \sqrt{\frac{1}{2N}}$ for $1 \leq i \leq N$, for $ N =2000, M =4000$. Theoretical one (solid orange line) is computed from the rhs of \eqref{Y-resolvent relation} with parameters computed from the generated matrix. Note that, the theoretical one has also fluctuations because the parameters $\beta^*_1,\cdots \beta^*_4$ are computed from the numerical estimate of $\mathcal{G}_{\bar{\mu}_S}(z)$.
    }
    \label{Y-resolvent relation-Gaussian Y}
\end{figure}

\begin{figure}
\begin{subfigure}[t]{\textwidth}
  \begin{subfigure}[t]{.4\textwidth}
    \centering
    \input{Figures/App/Y-est/Resolvent_Relation/Uniform_real_entry1}
  \end{subfigure}
  \hfil
  \begin{subfigure}[t]{.4\textwidth}
  \centering
      \input{Figures/App/Y-est/Resolvent_Relation/Uniform_im_entry1}
  \end{subfigure}
  \caption{{\small Entry $ i = j = 1$, first diagonal entry in the upper-left block}}
  \end{subfigure}
  \\
  \begin{subfigure}[t]{\textwidth}
  \begin{subfigure}[t]{.4\textwidth}
    \centering
    \input{Figures/App/Y-est/Resolvent_Relation/Uniform_real_entry2}
  \end{subfigure}
  \hfil
  \begin{subfigure}[t]{.4\textwidth}
  \centering
      \input{Figures/App/Y-est/Resolvent_Relation/Uniform_im_entry2}
  \end{subfigure}
  \caption{{\small Entry $ i = 1, j = 2$, non-diagonal entry in the upper-left block}}
  \end{subfigure}
  \\
  \begin{subfigure}[t]{\textwidth}
  \begin{subfigure}[t]{.4\textwidth}
    \centering
    \input{Figures/App/Y-est/Resolvent_Relation/Uniform_real_entry3}
  \end{subfigure}
  \hfil
  \begin{subfigure}[t]{.4\textwidth}
  \centering
      \input{Figures/App/Y-est/Resolvent_Relation/Uniform_im_entry3}
  \end{subfigure}
  \caption{{\small Entry $ i = j = 2001$, first diagonal entry in the lower-right block}}
  \end{subfigure}
  \\
  \begin{subfigure}[t]{\textwidth}
  \begin{subfigure}[t]{.4\textwidth}
    \centering
    \input{Figures/App/Y-est/Resolvent_Relation/Uniform_real_entry4}
  \end{subfigure}
  \hfil
  \begin{subfigure}[t]{.4\textwidth}
  \centering
      \input{Figures/App/Y-est/Resolvent_Relation/Uniform_im_entry4}
  \end{subfigure}
  \caption{{\small Entry $ i = 2001, j = 2002$, non-diagonal entry in the lower-right block}}
  \end{subfigure}
  \\
  \begin{subfigure}[t]{\textwidth}
  \begin{subfigure}[t]{.4\textwidth}
    \centering
    \input{Figures/App/Y-est/Resolvent_Relation/Uniform_real_entry5}
  \end{subfigure}
  \hfil
  \begin{subfigure}[t]{.4\textwidth}
  \centering
      \input{Figures/App/Y-est/Resolvent_Relation/Uniform_im_entry5}
  \end{subfigure}
  \caption{{\small Entry $ i = 1, j = 2001$, first entry in the upper-right block}}
  \end{subfigure}
    \\
  \begin{subfigure}[t]{\textwidth}
  \begin{subfigure}[t]{.4\textwidth}
    \centering
    \input{Figures/App/Y-est/Resolvent_Relation/Uniform_real_entry6}
  \end{subfigure}
  \hfil
  \begin{subfigure}[t]{.4\textwidth}
  \centering
      \input{Figures/App/Y-est/Resolvent_Relation/Uniform_im_entry6}
  \end{subfigure}
  \caption{{\small Entry $ i = 1, j = 2002$, second entry in the upper-right block}}
  \end{subfigure}
    \caption{\small  Illustration of \eqref{Y-resolvent relation}. $\bY \in \bR^{N \times M}$ is a diagonal matrix with (main) diagonal entries uniformly distributed in $[1,3]$, $\bX = \bX^\intercal$ is shifted Wigner matrix with $c = 3$, and $\bZ$ is a Gaussian matrices with. The empirical estimate of $\bG_{\mathcal{S}}(z)$ (dashed blue line) is computed for $ z = \gamma_i - \ci \sqrt{\frac{1}{2N}}$ for $1 \leq i \leq N$, for $ N =2000, M =4000$. Theoretical one (solid orange line) is computed from the rhs of \eqref{Y-resolvent relation} with parameters computed from the generated matrix. Note that, the theoretical one has also fluctuations because the parameters $\beta^*_1,\cdots \beta^*_4$ are computed from the numerical estimate of $\mathcal{G}_{\bar{\mu}_S}(z)$.}
    \label{Y-resolvent relation-Uniform Y}
\end{figure}

\subsubsection{Overlap}
To illustrate the formula for the overlap \eqref{Y-overlap-eq}, we fix the matrix $\bY$ and run experiments over various realization of the model \eqref{Y-model}. For each experiment, we record the overlap of $k$-th singular vectors left and right) of $\bS$ and singular vectors of $\bY$. To compute the theoretical prediction, we evaluate the parameters $\beta^*_1, \beta_2^*, \beta^*_3, \beta^*_4$, for $z = \bar{\gamma}_k - \ci 0^+$ where $\bar{\gamma}_k$ is the average of $k$-th singular value of $\bS$ in the experiments.

\begin{figure}
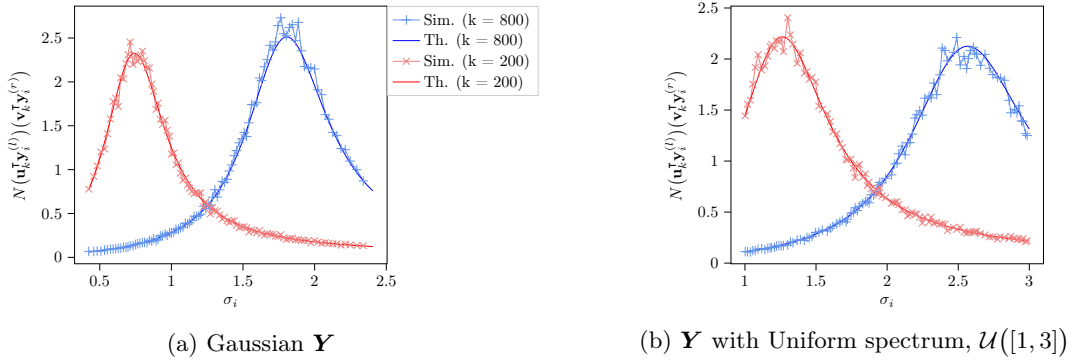

    \centering
\begin{subfigure}[t]{.4\textwidth}
    \centering
    \input{Figures/App/Y-est/Overlap/Overlap_GaussianY_XWGauss}
    \vspace{-11pt}
    \caption{\small Gaussian $\bY$}
     \label{fig:Y-Overlap-Gaussian Y-app}
\end{subfigure}
\hfil \hspace{5pt}
\begin{subfigure}[t]{.4\textwidth}
    \centering
    \input{Figures/App/Y-est/Overlap/Overlap_UniformY_XWGauss}
    \caption{\small $\bY$ with Uniform spectrum, $\mathcal{U}\big([1,3] \big)$}
    \label{fig:Y-Overlap-Uniform Y-app}
\end{subfigure}
\caption{\small Computation of the rescaled overlap. $\bX$ is a shifted Wigner matrix with $c=3$, and $\bW$ has i.i.d. Gaussian entries of variance $1/N$, and $N/M = 1/2$. The simulation results are average of 1000 experiments with fixed $\bY$, and $N = 1000, M =2000$. Some of the simulation points are dropped for clarity.}
\end{figure}

In figure \ref{fig:Y-Overlap-Gaussian Y-app}, the overlap is shown for $\bY$ with i.i.d. Gaussian entries of variance $\frac{1}{N}$, so $\mu_Y$ is the Marchenko-Pastur law with aspect-ratio $\alpha$. In figure \ref{fig:Y-Overlap-Uniform Y-app}, matrix $\bY$ is constructed as $\bY = \bU_Y \bSig \bV_Y^\intercal$, where $\bU_Y \in \bR^{N \times N}, \bV_Y \in \bR^{M \times M}$ are Haar distributed orthogonal matrices, and singular values $\sigma_1, \cdots, \sigma_N$ are chosen independently uniformly from $[1,3]$, so $\mu_Y = \mathcal{U}\big([1,3] \big)$.


\subsubsection{RIE performance}\label{App-Y-RIE}
In this section, we investigate the performance of our proposed estimators for $\bY$. To construct the RIE for $\bY$, we only need $q^*_4$ which we use \eqref{q4-analyt}. We compare performances of the optimal RIE \eqref{Y-optimal-sv-app} with the one of oracle estimator \eqref{Y-oracle-estimator}.

In figures \ref{Gaussian-Uniform-Y},\ref{Y-RIE-Bernoulli}, the MSE of RIE and the oracle estimator is plotted for three cases of priors: $\bY$ with Gaussian entries, $\bY$ with uniform spectral density, and $\bY$ with Bernoulli spectral density. In all cases, observe that the RIE has the same performance as the oracle estimator.

\begin{figure}
    \centering
\begin{subfigure}[t]{.4\textwidth}
    \centering
\begin{tikzpicture}[scale = 0.6]

\definecolor{darkgray176}{RGB}{176,176,176}
\definecolor{lightcoral}{RGB}{240,128,128}
\definecolor{lightslategray}{RGB}{119,136,153}
\definecolor{royalblue}{RGB}{65,105,225}

\begin{axis}[
legend cell align={left},
legend style={fill opacity=0.8, draw opacity=1, text opacity=1, draw=white!80!black},
tick align=outside,
tick pos=left,
x grid style={darkgray176},
xlabel={$\kappa$},
xmin=-0.145, xmax=5.25,
xtick style={color=black},
y grid style={darkgray176},
ylabel={${\rm MSE}$},
scaled y ticks=false,
yticklabel style={
  /pgf/number format/precision=3,
  /pgf/number format/fixed},
ymin=0.064786805742015, ymax=0.573597120391406,
ytick style={color=black}
]
\path [draw=lightcoral, semithick]
(axis cs:0.1,0.549661082272082)
--(axis cs:0.1,0.550125520168613);

\path [draw=lightcoral, semithick]
(axis cs:0.3,0.321989672579266)
--(axis cs:0.3,0.322421886897764);

\path [draw=lightcoral, semithick]
(axis cs:0.6,0.218784249584983)
--(axis cs:0.6,0.219232264927173);

\path [draw=lightcoral, semithick]
(axis cs:1,0.16604121311557)
--(axis cs:1,0.166316661960242);

\path [draw=lightcoral, semithick]
(axis cs:2,0.119527057415692)
--(axis cs:2,0.119752132841484);

\path [draw=lightcoral, semithick]
(axis cs:3,0.102412188285689)
--(axis cs:3,0.102626738208753);

\path [draw=lightcoral, semithick]
(axis cs:4,0.0934347110080447)
--(axis cs:4,0.0935938983668105);

\path [draw=lightcoral, semithick]
(axis cs:5,0.0879145473169873)
--(axis cs:5,0.0880634842521682);

\path [draw=lightslategray, semithick]
(axis cs:0.1,0.550036974837565)
--(axis cs:0.1,0.550469378816434);

\path [draw=lightslategray, semithick]
(axis cs:0.3,0.322134280811544)
--(axis cs:0.3,0.322564203050586);

\path [draw=lightslategray, semithick]
(axis cs:0.6,0.218896855577694)
--(axis cs:0.6,0.21934855486424);

\path [draw=lightslategray, semithick]
(axis cs:1,0.16616363784447)
--(axis cs:1,0.166437557272511);

\path [draw=lightslategray, semithick]
(axis cs:2,0.119681521964011)
--(axis cs:2,0.119899328531154);

\path [draw=lightslategray, semithick]
(axis cs:3,0.102600079218673)
--(axis cs:3,0.102806396908418);

\path [draw=lightslategray, semithick]
(axis cs:4,0.0936446888651997)
--(axis cs:4,0.0938227941177117);

\path [draw=lightslategray, semithick]
(axis cs:5,0.0881784818306752)
--(axis cs:5,0.0883263573376689);

\addplot [semithick, red, mark=triangle*, mark size=3, mark options={solid,rotate=180}, only marks]
table {%
0.1 0.549893301220348
0.3 0.322205779738515
0.6 0.219008257256078
1 0.166178937537906
2 0.119639595128588
3 0.102519463247221
4 0.0935143046874276
5 0.0879890157845777
};
\addlegendentry{Oracle estimator, ${\bm{\Xi}_Y^*}(\bS)$}
\addplot [semithick, royalblue, mark=triangle*, mark size=3, mark options={solid}, only marks]
table {%
0.1 0.550253176826999
0.3 0.322349241931065
0.6 0.219122705220967
1 0.166300597558491
2 0.119790425247583
3 0.102703238063546
4 0.0937337414914557
5 0.088252419584172
};
\addlegendentry{RIE, $\widehat{{\bm{\Xi}_Y^*}}(\bS)$}
\draw (axis cs:0.2,0.549893301220348) node[
  scale=0.6,
  anchor=base west,
  text=black,
  rotate=0.0
]{0.07\%};
\draw (axis cs:0.15,0.342205779738515) node[
  scale=0.6,
  anchor=base west,
  text=black,
  rotate=0.0
]{0.04\%};
\draw (axis cs:0.45,0.239008257256078) node[
  scale=0.6,
  anchor=base west,
  text=black,
  rotate=0.0
]{0.05\%};
\draw (axis cs:0.85,0.186178937537906) node[
  scale=0.6,
  anchor=base west,
  text=black,
  rotate=0.0
]{0.07\%};
\draw (axis cs:1.85,0.139639595128588) node[
  scale=0.6,
  anchor=base west,
  text=black,
  rotate=0.0
]{0.13\%};
\draw (axis cs:2.85,0.122519463247221) node[
  scale=0.6,
  anchor=base west,
  text=black,
  rotate=0.0
]{0.18\%};
\draw (axis cs:3.85,0.113514304687428) node[
  scale=0.6,
  anchor=base west,
  text=black,
  rotate=0.0
]{0.23\%};
\draw (axis cs:4.85,0.107989015784578) node[
  scale=0.6,
  anchor=base west,
  text=black,
  rotate=0.0
]{0.3\%};
\end{axis}

\end{tikzpicture}
    \caption{Gaussian}
\end{subfigure}
\hfil
\begin{subfigure}[t]{.4\textwidth}
    \centering
\begin{tikzpicture}[scale = 0.6]

\definecolor{darkgray176}{RGB}{176,176,176}
\definecolor{lightcoral}{RGB}{240,128,128}
\definecolor{lightslategray}{RGB}{119,136,153}
\definecolor{royalblue}{RGB}{65,105,225}

\begin{axis}[
legend cell align={left},
legend style={fill opacity=0.8, draw opacity=1, text opacity=1, draw=white!80!black},
tick align=outside,
tick pos=left,
x grid style={darkgray176},
xlabel={$\kappa$},
xmin=-0.145, xmax=5.35,
xtick style={color=black},
y grid style={darkgray176},
ylabel={${\rm MSE}$},
scaled y ticks=false,
yticklabel style={
  /pgf/number format/precision=3,
  /pgf/number format/fixed},
ymin=0.0432528431531933, ymax=0.39457092906823,
ytick style={color=black}
]
\path [draw=lightcoral, semithick]
(axis cs:0.1,0.373512909726238)
--(axis cs:0.1,0.378437895486202);

\path [draw=lightcoral, semithick]
(axis cs:0.3,0.195568222689053)
--(axis cs:0.3,0.198455746381194);

\path [draw=lightcoral, semithick]
(axis cs:0.6,0.129719083279093)
--(axis cs:0.6,0.130959578976959);

\path [draw=lightcoral, semithick]
(axis cs:1,0.0999373844438707)
--(axis cs:1,0.101667574902742);

\path [draw=lightcoral, semithick]
(axis cs:2,0.0749482719476051)
--(axis cs:2,0.0762414479090259);

\path [draw=lightcoral, semithick]
(axis cs:3,0.0662698130081426)
--(axis cs:3,0.0682014561186748);

\path [draw=lightcoral, semithick]
(axis cs:4,0.062168909249415)
--(axis cs:4,0.0630242993297389);

\path [draw=lightcoral, semithick]
(axis cs:5,0.0592218470584223)
--(axis cs:5,0.0604937979516368);

\path [draw=lightslategray, semithick]
(axis cs:0.1,0.373668314553029)
--(axis cs:0.1,0.378601925163001);

\path [draw=lightslategray, semithick]
(axis cs:0.3,0.195673773538471)
--(axis cs:0.3,0.198554358375894);

\path [draw=lightslategray, semithick]
(axis cs:0.6,0.129836791528522)
--(axis cs:0.6,0.131071783500439);

\path [draw=lightslategray, semithick]
(axis cs:1,0.10008161988262)
--(axis cs:1,0.101814811169172);

\path [draw=lightslategray, semithick]
(axis cs:2,0.0751747889504974)
--(axis cs:2,0.0764695398504073);

\path [draw=lightslategray, semithick]
(axis cs:3,0.0665717243180533)
--(axis cs:3,0.0685106206536402);

\path [draw=lightslategray, semithick]
(axis cs:4,0.0625469464427706)
--(axis cs:4,0.0634091398549601);

\path [draw=lightslategray, semithick]
(axis cs:5,0.0596922287297291)
--(axis cs:5,0.0609637801713104);

\addplot [semithick, red, mark=triangle*, mark size=3, mark options={solid,rotate=180}, only marks]
table {%
0.1 0.37597540260622
0.3 0.197011984535123
0.6 0.130339331128026
1 0.100802479673306
2 0.0755948599283155
3 0.0672356345634087
4 0.062596604289577
5 0.0598578225050295
};
\addlegendentry{Oracle estimator, ${\bm{\Xi}_Y^*}(\bS)$}
\addplot [semithick, royalblue, mark=triangle*, mark size=3, mark options={solid}, only marks]
table {%
0.1 0.376135119858015
0.3 0.197114065957182
0.6 0.13045428751448
1 0.100948215525896
2 0.0758221644004523
3 0.0675411724858467
4 0.0629780431488653
5 0.0603280044505198
};
\addlegendentry{RIE, $\widehat{{\bm{\Xi}_Y^*}}(\bS)$}
\draw (axis cs:0.2,0.37597540260622) node[
  scale=0.6,
  anchor=base west,
  text=black,
  rotate=0.0
]{0.04\%};
\draw (axis cs:0.15,0.217011984535123) node[
  scale=0.6,
  anchor=base west,
  text=black,
  rotate=0.0
]{0.05\%};
\draw (axis cs:0.45,0.150339331128026) node[
  scale=0.6,
  anchor=base west,
  text=black,
  rotate=0.0
]{0.09\%};
\draw (axis cs:0.85,0.120802479673306) node[
  scale=0.6,
  anchor=base west,
  text=black,
  rotate=0.0
]{0.14\%};
\draw (axis cs:1.85,0.0955948599283155) node[
  scale=0.6,
  anchor=base west,
  text=black,
  rotate=0.0
]{0.3\%};
\draw (axis cs:2.85,0.0872356345634087) node[
  scale=0.6,
  anchor=base west,
  text=black,
  rotate=0.0
]{0.45\%};
\draw (axis cs:3.85,0.082596604289577) node[
  scale=0.6,
  anchor=base west,
  text=black,
  rotate=0.0
]{0.61\%};
\draw (axis cs:4.85,0.0798578225050295) node[
  scale=0.6,
  anchor=base west,
  text=black,
  rotate=0.0
]{0.79\%};
\end{axis}

\end{tikzpicture}
    \caption{Uniform spectral, $\mathcal{U}\big([1,3] \big)$}
\end{subfigure}
\caption{\small Estimating $\bY$. MSE is normalized by the norm of the signal, $\| \bY \|_{\rm F}^2$. $\bX$ is a shifted Wigner matrix with $c=3$, and $\bW$ has i.i.d. Gaussian entries of variance $1/N$, and $N/M = 1/2$. The RIE is applied to $N=2000, M =4000$, and the results are averaged over 10 runs (error bars are invisible). Average relative error between RIE $\widehat{{\bm{\Xi}_Y^*}}(\bS)$ and Oracle estimator is also reported.}
\label{Gaussian-Uniform-Y}
\end{figure}
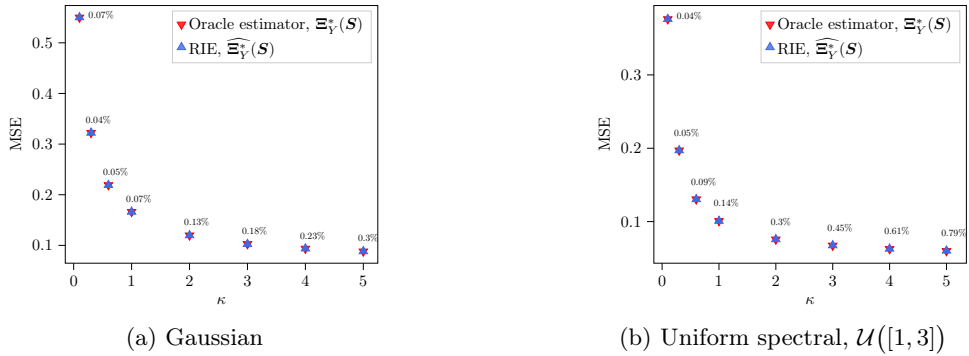
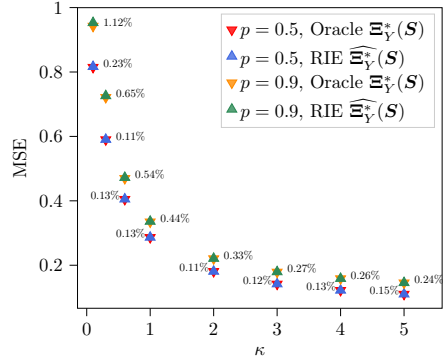
\begin{figure}
    \centering
\begin{tikzpicture}[scale = 0.7]

\definecolor{darkgray176}{RGB}{176,176,176}
\definecolor{darkorange}{RGB}{255,140,0}
\definecolor{lightcoral}{RGB}{240,128,128}
\definecolor{lightgreen}{RGB}{144,238,144}
\definecolor{lightslategray}{RGB}{119,136,153}
\definecolor{orange}{RGB}{255,165,0}
\definecolor{royalblue}{RGB}{65,105,225}
\definecolor{seagreen}{RGB}{46,139,87}

\begin{axis}[
legend cell align={left},
legend style={fill opacity=0.8, draw opacity=1, text opacity=1, draw=white!80!black},
tick align=outside,
tick pos=left,
x grid style={darkgray176},
xlabel={$\kappa$},
xmin=-0.145, xmax=5.6,
xtick style={color=black},
y grid style={darkgray176},
ylabel={${\rm MSE}$},
scaled y ticks=false,
yticklabel style={
  /pgf/number format/precision=3,
  /pgf/number format/fixed},
ymin=0.0664117960560262, ymax=0.99973729237639,
ytick style={color=black}
]
\path [draw=lightcoral, semithick]
(axis cs:0.1,0.812566666180997)
--(axis cs:0.1,0.816845876447459);

\path [draw=lightcoral, semithick]
(axis cs:0.3,0.586470886512217)
--(axis cs:0.3,0.59313479938925);

\path [draw=lightcoral, semithick]
(axis cs:0.6,0.402668609673547)
--(axis cs:0.6,0.406712329293766);

\path [draw=lightcoral, semithick]
(axis cs:1,0.284494703682938)
--(axis cs:1,0.288096618937717);

\path [draw=lightcoral, semithick]
(axis cs:2,0.179456733414115)
--(axis cs:2,0.182012119921784);

\path [draw=lightcoral, semithick]
(axis cs:3,0.140699557517812)
--(axis cs:3,0.14254426926014);

\path [draw=lightcoral, semithick]
(axis cs:4,0.121364457268204)
--(axis cs:4,0.122789194172475);

\path [draw=lightcoral, semithick]
(axis cs:5,0.108835682252406)
--(axis cs:5,0.111397134376699);

\path [draw=lightslategray, semithick]
(axis cs:0.1,0.814436035507314)
--(axis cs:0.1,0.818695188936649);

\path [draw=lightslategray, semithick]
(axis cs:0.3,0.587114136523164)
--(axis cs:0.3,0.593845598320586);

\path [draw=lightslategray, semithick]
(axis cs:0.6,0.403205998708727)
--(axis cs:0.6,0.407257437811785);

\path [draw=lightslategray, semithick]
(axis cs:1,0.284851674102815)
--(axis cs:1,0.288481320675235);

\path [draw=lightslategray, semithick]
(axis cs:2,0.179653030829168)
--(axis cs:2,0.182206338914283);

\path [draw=lightslategray, semithick]
(axis cs:3,0.14086870967582)
--(axis cs:3,0.142707409594779);

\path [draw=lightslategray, semithick]
(axis cs:4,0.121520893007553)
--(axis cs:4,0.122946694567711);

\path [draw=lightslategray, semithick]
(axis cs:5,0.10899656503901)
--(axis cs:5,0.111562193461588);

\path [draw=orange, semithick]
(axis cs:0.1,0.938860741069166)
--(axis cs:0.1,0.945885525926177);

\path [draw=orange, semithick]
(axis cs:0.3,0.718784173249564)
--(axis cs:0.3,0.72374028871986);

\path [draw=orange, semithick]
(axis cs:0.6,0.468353816404558)
--(axis cs:0.6,0.471003302494996);

\path [draw=orange, semithick]
(axis cs:1,0.333929873106405)
--(axis cs:1,0.335382328354642);

\path [draw=orange, semithick]
(axis cs:2,0.219420663047858)
--(axis cs:2,0.221211232957778);

\path [draw=orange, semithick]
(axis cs:3,0.178325385983598)
--(axis cs:3,0.179418847603011);

\path [draw=orange, semithick]
(axis cs:4,0.157637970024636)
--(axis cs:4,0.158826557075442);

\path [draw=orange, semithick]
(axis cs:5,0.144737747336776)
--(axis cs:5,0.14585523605734);

\path [draw=lightgreen, semithick]
(axis cs:0.1,0.948638969522396)
--(axis cs:0.1,0.95731340618001);

\path [draw=lightgreen, semithick]
(axis cs:0.3,0.723209713502785)
--(axis cs:0.3,0.728754075443292);

\path [draw=lightgreen, semithick]
(axis cs:0.6,0.470763917695929)
--(axis cs:0.6,0.473666312321464);

\path [draw=lightgreen, semithick]
(axis cs:1,0.335401963822167)
--(axis cs:1,0.336835757193538);

\path [draw=lightgreen, semithick]
(axis cs:2,0.220125533866838)
--(axis cs:2,0.221964116919503);

\path [draw=lightgreen, semithick]
(axis cs:3,0.178792679702436)
--(axis cs:3,0.179931739911604);

\path [draw=lightgreen, semithick]
(axis cs:4,0.158030593702699)
--(axis cs:4,0.159264906631313);

\path [draw=lightgreen, semithick]
(axis cs:5,0.145057974949517)
--(axis cs:5,0.146242635111564);

\addplot [semithick, red, mark=triangle*, mark size=3, mark options={solid,rotate=180}, only marks]
table {%
0.1 0.814706271314228
0.3 0.589802842950734
0.6 0.404690469483657
1 0.286295661310327
2 0.180734426667949
3 0.141621913388976
4 0.122076825720339
5 0.110116408314553
};
\addlegendentry{$p=0.5$, Oracle  ${\bm{\Xi}_Y^*}(\bS)$}
\addplot [semithick, royalblue, mark=triangle*, mark size=3, mark options={solid}, only marks]
table {%
0.1 0.816565612221982
0.3 0.590479867421875
0.6 0.405231718260256
1 0.286666497389025
2 0.180929684871726
3 0.1417880596353
4 0.122233793787632
5 0.110279379250299
};
\addlegendentry{$p=0.5$, RIE $\widehat{{\bm{\Xi}_Y^*}}(\bS)$}
\addplot [semithick, darkorange, mark=triangle*, mark size=3, mark options={solid,rotate=180}, only marks]
table {%
0.1 0.942373133497671
0.3 0.721262230984712
0.6 0.469678559449777
1 0.334656100730524
2 0.220315948002818
3 0.178872116793304
4 0.158232263550039
5 0.145296491697058
};
\addlegendentry{$p=0.9$, Oracle  ${\bm{\Xi}_Y^*}(\bS)$}
\addplot [semithick, seagreen, mark=triangle*, mark size=3, mark options={solid}, only marks]
table {%
0.1 0.952976187851203
0.3 0.725981894473039
0.6 0.472215115008696
1 0.336118860507853
2 0.22104482539317
3 0.17936220980702
4 0.158647750167006
5 0.14565030503054
};
\addlegendentry{$p=0.9$, RIE $\widehat{{\bm{\Xi}_Y^*}}(\bS)$}
\draw (axis cs:0.2,0.814706271314228) node[
  scale=0.6,
  anchor=base west,
  text=black,
  rotate=0.0
]{0.23\%};
\draw (axis cs:0.4,0.589802842950734) node[
  scale=0.6,
  anchor=base west,
  text=black,
  rotate=0.0
]{0.11\%};
\draw (axis cs:0,0.404690469483657) node[
  scale=0.6,
  anchor=base west,
  text=black,
  rotate=0.0
]{0.13\%};
\draw (axis cs:0.4,0.286295661310327) node[
  scale=0.6,
  anchor=base west,
  text=black,
  rotate=0.0
]{0.13\%};
\draw (axis cs:1.4,0.180734426667949) node[
  scale=0.6,
  anchor=base west,
  text=black,
  rotate=0.0
]{0.11\%};
\draw (axis cs:2.4,0.141621913388976) node[
  scale=0.6,
  anchor=base west,
  text=black,
  rotate=0.0
]{0.12\%};
\draw (axis cs:3.4,0.122076825720339) node[
  scale=0.6,
  anchor=base west,
  text=black,
  rotate=0.0
]{0.13\%};
\draw (axis cs:4.4,0.110116408314553) node[
  scale=0.6,
  anchor=base west,
  text=black,
  rotate=0.0
]{0.15\%};
\draw (axis cs:0.2,0.942373133497671) node[
  scale=0.6,
  anchor=base west,
  text=black,
  rotate=0.0
]{1.12\%};
\draw (axis cs:0.4,0.721262230984712) node[
  scale=0.6,
  anchor=base west,
  text=black,
  rotate=0.0
]{0.65\%};
\draw (axis cs:0.7,0.469678559449777) node[
  scale=0.6,
  anchor=base west,
  text=black,
  rotate=0.0
]{0.54\%};
\draw (axis cs:1.1,0.334656100730524) node[
  scale=0.6,
  anchor=base west,
  text=black,
  rotate=0.0
]{0.44\%};
\draw (axis cs:2.1,0.220315948002818) node[
  scale=0.6,
  anchor=base west,
  text=black,
  rotate=0.0
]{0.33\%};
\draw (axis cs:3.1,0.178872116793304) node[
  scale=0.6,
  anchor=base west,
  text=black,
  rotate=0.0
]{0.27\%};
\draw (axis cs:4.1,0.158232263550039) node[
  scale=0.6,
  anchor=base west,
  text=black,
  rotate=0.0
]{0.26\%};
\draw (axis cs:5.1,0.145296491697058) node[
  scale=0.6,
  anchor=base west,
  text=black,
  rotate=0.0
]{0.24\%};
\end{axis}

\end{tikzpicture}
    \caption{\small Estimating $\bY$ with Bernoulli spectral prior. MSE is normalized by the norm of the signal, $\| \bY \|_{\rm F}^2$. $\bY$ has Bernoulli spectral distribution with parameter $p$. $\bX$ is a shifted Wigner matrix with $c=3$, and $\bW$ has i.i.d. Gaussian entries of variance $1/N$, and $N/M = 1/2$. The RIE is applied to $N=2000, M =4000$, and the results are averaged over 10 runs (error bars are invisible).Average relative error between RIE $\widehat{{\bm{\Xi}_Y^*}}(\bS)$ and Oracle estimator is also reported. }
    \label{Y-RIE-Bernoulli}
\end{figure}

\begin{figure}
    \centering
\begin{tikzpicture}[scale = 0.7]

\definecolor{darkgray176}{RGB}{176,176,176}
\definecolor{lightcoral}{RGB}{240,128,128}
\definecolor{lightslategray}{RGB}{119,136,153}
\definecolor{royalblue}{RGB}{65,105,225}

\begin{axis}[
legend cell align={left},
legend style={
  fill opacity=0.8,
  draw opacity=1,
  text opacity=1,
  at={(1,1)},
  anchor=north east,
  draw=white!80!black
},
tick align=outside,
tick pos=left,
x dir=reverse,
x grid style={darkgray176},
xlabel={$\alpha$},
xmin=0.2125, xmax=1.0375,
xtick style={color=black},
y grid style={darkgray176},
ylabel={${\rm MSE}$},
scaled y ticks=false,
yticklabel style={
  /pgf/number format/precision=3,
  /pgf/number format/fixed},
ymin=0.0689457440046494, ymax=0.103604653916272,
ytick style={color=black}
]
\path [draw=lightcoral, semithick]
(axis cs:1,0.101605301848394)
--(axis cs:1,0.101814381680697);

\path [draw=lightcoral, semithick]
(axis cs:0.666666666666667,0.0943341146541025)
--(axis cs:0.666666666666667,0.0944994744558913);

\path [draw=lightcoral, semithick]
(axis cs:0.5,0.0879080493935161)
--(axis cs:0.5,0.0880321390797517);

\path [draw=lightcoral, semithick]
(axis cs:0.4,0.0823237261266984)
--(axis cs:0.4,0.0825017797010767);

\path [draw=lightcoral, semithick]
(axis cs:0.333333333333333,0.0776983360040941)
--(axis cs:0.333333333333333,0.0778588957836681);

\path [draw=lightcoral, semithick]
(axis cs:0.285714285714286,0.0737940830744484)
--(axis cs:0.285714285714286,0.0739039183200334);

\path [draw=lightcoral, semithick]
(axis cs:0.25,0.0705211490006323)
--(axis cs:0.25,0.0706353732048547);

\path [draw=lightslategray, semithick]
(axis cs:1,0.101813026923096)
--(axis cs:1,0.102029248920289);

\path [draw=lightslategray, semithick]
(axis cs:0.666666666666667,0.0945651752580815)
--(axis cs:0.666666666666667,0.094734534880508);

\path [draw=lightslategray, semithick]
(axis cs:0.5,0.0881622403960948)
--(axis cs:0.5,0.0882936907126535);

\path [draw=lightslategray, semithick]
(axis cs:0.4,0.0826136605487134)
--(axis cs:0.4,0.0828007001095422);

\path [draw=lightslategray, semithick]
(axis cs:0.333333333333333,0.0780266367782406)
--(axis cs:0.333333333333333,0.078198126821867);

\path [draw=lightslategray, semithick]
(axis cs:0.285714285714286,0.0741623420219247)
--(axis cs:0.285714285714286,0.0742780724630596);

\path [draw=lightslategray, semithick]
(axis cs:0.25,0.07093446396058)
--(axis cs:0.25,0.0710506672161462);

\addplot [semithick, red, mark=triangle*, mark size=3, mark options={solid,rotate=180}, only marks]
table {%
1 0.101709841764546
0.666666666666667 0.0944167945549969
0.5 0.0879700942366339
0.4 0.0824127529138875
0.333333333333333 0.0777786158938811
0.285714285714286 0.0738490006972409
0.25 0.0705782611027435
};
\addlegendentry{Oracle estimator, ${\bm{\Xi}_Y^*}(\bS)$}
\addplot [semithick, royalblue, mark=triangle*, mark size=3, mark options={solid}, only marks]
table {%
1 0.101921137921693
0.666666666666667 0.0946498550692948
0.5 0.0882279655543742
0.4 0.0827071803291278
0.333333333333333 0.0781123818000538
0.285714285714286 0.0742202072424921
0.25 0.0709925655883631
};
\addlegendentry{RIE, $\widehat{{\bm{\Xi}_Y^*}}(\bS)$}
\draw (axis cs:0.95,0.101709841764546) node[
  scale=0.6,
  anchor=base west,
  text=black,
  rotate=0.0
]{0.21\%};
\draw (axis cs:0.766666666666667,0.0944167945549969) node[
  scale=0.6,
  anchor=base west,
  text=black,
  rotate=0.0
]{0.25\%};
\draw (axis cs:0.6,0.0879700942366339) node[
  scale=0.6,
  anchor=base west,
  text=black,
  rotate=0.0
]{0.29\%};
\draw (axis cs:0.5,0.0824127529138875) node[
  scale=0.6,
  anchor=base west,
  text=black,
  rotate=0.0
]{0.36\%};
\draw (axis cs:0.433333333333333,0.0777786158938811) node[
  scale=0.6,
  anchor=base west,
  text=black,
  rotate=0.0
]{0.43\%};
\draw (axis cs:0.385714285714286,0.0738490006972409) node[
  scale=0.6,
  anchor=base west,
  text=black,
  rotate=0.0
]{0.5\%};
\draw (axis cs:0.35,0.0705782611027435) node[
  scale=0.6,
  anchor=base west,
  text=black,
  rotate=0.0
]{0.59\%};
\end{axis}

\end{tikzpicture}
    \caption{\small MSE of estimating $\bY$ as a function of aspect-ratio $\alpha$, $\bY$ has Gaussain entries of variance $\nicefrac{1}{N}$, and $\kappa = 5$. MSE is normalized by the norm of the signal, $\| \bY \|_{\rm F}^2$. $\bX$ is a shifted Wigner matrix with $c=3$, and $\bW$ has i.i.d. Gaussian entries of variance $1/N$. The RIE is applied to $N=2000, M = \nicefrac{1}{\alpha} N$, and the results are averaged over 10 runs (error bars are invisible). Average relative error between RIE $\widehat{{\bm{\Xi}_Y^*}}(\bS)$ and Oracle estimator is also reported.}
    \label{Y-ar}
\end{figure}
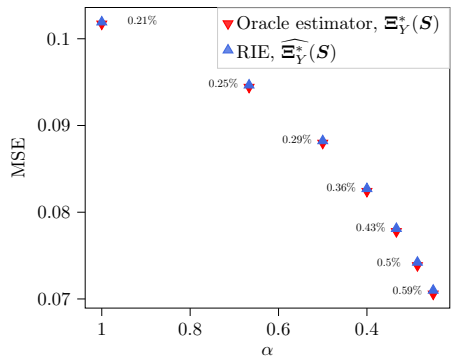

\paragraph{Effect of aspect-ratio $\alpha$.} In Figure \ref{Y-ar}, we take $\bY$ to have Gaussian entries (with variance $\frac{1}{N}$), and the MSE is depicted for various values of the aspect-ratio $\alpha$. We see that as $M$ increases ($\alpha$ decreases) the estimation error (of $\bY$) decreases.

\begin{figure}
    \centering
\begin{subfigure}[t]{.4\textwidth}
    \centering
\begin{tikzpicture}[scale = 0.6]

\definecolor{darkgray176}{RGB}{176,176,176}
\definecolor{lightcoral}{RGB}{240,128,128}
\definecolor{lightgreen}{RGB}{144,238,144}
\definecolor{lightslategray}{RGB}{119,136,153}
\definecolor{royalblue}{RGB}{65,105,225}
\definecolor{seagreen}{RGB}{46,139,87}

\begin{axis}[
legend cell align={left},
legend style={fill opacity=0.8, draw opacity=1, text opacity=1, draw=white!80!black},
tick align=outside,
tick pos=left,
x grid style={darkgray176},
xlabel={$\kappa$},
xmin=-0.145, xmax=5.245,
xtick style={color=black},
y grid style={darkgray176},
ylabel={${\rm MSE}$},
scaled y ticks=false,
yticklabel style={
  /pgf/number format/precision=3,
  /pgf/number format/fixed},
ymin=0.0640057793258399, ymax=1.06337486483025,
ytick style={color=black}
]
\path [draw=lightcoral, semithick]
(axis cs:0.1,0.699131126715838)
--(axis cs:0.1,0.699779917575325);

\path [draw=lightcoral, semithick]
(axis cs:0.3,0.459389815641978)
--(axis cs:0.3,0.460150458706304);

\path [draw=lightcoral, semithick]
(axis cs:0.6,0.321924184916511)
--(axis cs:0.6,0.32240043027887);

\path [draw=lightcoral, semithick]
(axis cs:1,0.242305026563169)
--(axis cs:1,0.242736682664744);

\path [draw=lightcoral, semithick]
(axis cs:2,0.166062700772127)
--(axis cs:2,0.166246715272348);

\path [draw=lightcoral, semithick]
(axis cs:3,0.135735751221648)
--(axis cs:3,0.136106014409718);

\path [draw=lightcoral, semithick]
(axis cs:4,0.119609382554311)
--(axis cs:4,0.119756334729497);

\path [draw=lightcoral, semithick]
(axis cs:5,0.109431646848768)
--(axis cs:5,0.109534531824536);

\path [draw=lightslategray, semithick]
(axis cs:0.1,0.699911736129243)
--(axis cs:0.1,0.700607387774203);

\path [draw=lightslategray, semithick]
(axis cs:0.3,0.459633840106745)
--(axis cs:0.3,0.460380188092452);

\path [draw=lightslategray, semithick]
(axis cs:0.6,0.322066502437168)
--(axis cs:0.6,0.322547756458355);

\path [draw=lightslategray, semithick]
(axis cs:1,0.242423181140006)
--(axis cs:1,0.242857429419913);

\path [draw=lightslategray, semithick]
(axis cs:2,0.166178134530113)
--(axis cs:2,0.166369986159607);

\path [draw=lightslategray, semithick]
(axis cs:3,0.135869274960594)
--(axis cs:3,0.136244503072035);

\path [draw=lightslategray, semithick]
(axis cs:4,0.119753444167918)
--(axis cs:4,0.119906386426967);

\path [draw=lightslategray, semithick]
(axis cs:5,0.109615783267211)
--(axis cs:5,0.109705976349919);

\path [draw=lightgreen, semithick]
(axis cs:0.1,1.01725334566236)
--(axis cs:0.1,1.01794899730732);

\path [draw=lightgreen, semithick]
(axis cs:0.3,0.720603507418589)
--(axis cs:0.3,0.721349855404296);

\path [draw=lightgreen, semithick]
(axis cs:0.6,0.554374638336762)
--(axis cs:0.6,0.554855892357949);

\path [draw=lightgreen, semithick]
(axis cs:1,0.457288459956666)
--(axis cs:1,0.457722708236573);

\path [draw=lightgreen, semithick]
(axis cs:2,0.355027078665358)
--(axis cs:2,0.355218930294852);

\path [draw=lightgreen, semithick]
(axis cs:3,0.307548145684649)
--(axis cs:3,0.30792337379609);

\path [draw=lightgreen, semithick]
(axis cs:4,0.278846365689276)
--(axis cs:4,0.278999307948325);

\path [draw=lightgreen, semithick]
(axis cs:5,0.259027747581651)
--(axis cs:5,0.259117940664359);

\addplot [semithick, red, mark=triangle*, mark size=3, mark options={solid,rotate=180}, only marks]
table {%
0.1 0.699455522145582
0.3 0.459770137174141
0.6 0.32216230759769
1 0.242520854613957
2 0.166154708022238
3 0.135920882815683
4 0.119682858641904
5 0.109483089336652
};
\addlegendentry{Oracle estimator, ${\bm{\Xi}_Y^*}(\bS)$}
\addplot [semithick, royalblue, mark=triangle*, mark size=3, mark options={solid}, only marks]
table {%
0.1 0.700259561951723
0.3 0.460007014099599
0.6 0.322307129447762
1 0.24264030527996
2 0.16627406034486
3 0.136056889016315
4 0.119829915297443
5 0.109660879808565
};
\addlegendentry{RIE, $\widehat{{\bm{\Xi}_Y^*}}(\bS)$}
\addplot [semithick, seagreen, mark=x, mark size=3, mark options={solid}, only marks]
table {%
0.1 1.01760117148484
0.3 0.720976681411443
0.6 0.554615265347356
1 0.457505584096619
2 0.355123004480105
3 0.30773575974037
4 0.278922836818801
5 0.259072844123005
};
\addlegendentry{RIE + $f_p(x)$}
\end{axis}

\end{tikzpicture}
    \caption{$ p = 0.5$}
\end{subfigure}
\hfil
\begin{subfigure}[t]{.4\textwidth}
    \centering
\begin{tikzpicture}[scale = 0.6]

\definecolor{darkgray176}{RGB}{176,176,176}
\definecolor{lightcoral}{RGB}{240,128,128}
\definecolor{lightgreen}{RGB}{144,238,144}
\definecolor{lightslategray}{RGB}{119,136,153}
\definecolor{royalblue}{RGB}{65,105,225}
\definecolor{seagreen}{RGB}{46,139,87}

\begin{axis}[
legend cell align={left},
legend style={fill opacity=0.8, draw opacity=1, text opacity=1, draw=white!80!black},
tick align=outside,
tick pos=left,
x grid style={darkgray176},
xlabel={$\kappa$},
xmin=-0.145, xmax=5.245,
xtick style={color=black},
y grid style={darkgray176},
ylabel={${\rm MSE}$},
scaled y ticks=false,
yticklabel style={
  /pgf/number format/precision=3,
  /pgf/number format/fixed},
ymin=-0.0291445647502558, ymax=1.04950111790214,
ytick style={color=black}
]
\path [draw=lightcoral, semithick]
(axis cs:0.1,0.917747523917354)
--(axis cs:0.1,0.918109705685828);

\path [draw=lightcoral, semithick]
(axis cs:0.3,0.791326573746199)
--(axis cs:0.3,0.792133219202896);

\path [draw=lightcoral, semithick]
(axis cs:0.6,0.661645409752188)
--(axis cs:0.6,0.662350283580213);

\path [draw=lightcoral, semithick]
(axis cs:1,0.549527387769611)
--(axis cs:1,0.550198776456851);

\path [draw=lightcoral, semithick]
(axis cs:2,0.398933415002584)
--(axis cs:2,0.399533466473634);

\path [draw=lightcoral, semithick]
(axis cs:3,0.322046268874099)
--(axis cs:3,0.322582814931751);

\path [draw=lightcoral, semithick]
(axis cs:4,0.274556671507888)
--(axis cs:4,0.275232986731542);

\path [draw=lightcoral, semithick]
(axis cs:5,0.242474029304169)
--(axis cs:5,0.242876349306416);

\path [draw=lightslategray, semithick]
(axis cs:0.1,0.922541862287869)
--(axis cs:0.1,0.923573645470489);

\path [draw=lightslategray, semithick]
(axis cs:0.3,0.792746599766245)
--(axis cs:0.3,0.793582637213448);

\path [draw=lightslategray, semithick]
(axis cs:0.6,0.662282275786786)
--(axis cs:0.6,0.662999550229876);

\path [draw=lightslategray, semithick]
(axis cs:1,0.549881826029265)
--(axis cs:1,0.550554957450395);

\path [draw=lightslategray, semithick]
(axis cs:2,0.399117032658234)
--(axis cs:2,0.399730404205225);

\path [draw=lightslategray, semithick]
(axis cs:3,0.322188490242287)
--(axis cs:3,0.322724479772883);

\path [draw=lightslategray, semithick]
(axis cs:4,0.274684109976373)
--(axis cs:4,0.275361149301729);

\path [draw=lightslategray, semithick]
(axis cs:5,0.242596501370718)
--(axis cs:5,0.243003706942305);

\path [draw=lightgreen, semithick]
(axis cs:0.1,0.99943998550805)
--(axis cs:0.1,1.00047176869067);

\path [draw=lightgreen, semithick]
(axis cs:0.3,0.970516404885996)
--(axis cs:0.3,0.971352442333199);

\path [draw=lightgreen, semithick]
(axis cs:0.6,0.792371460137561)
--(axis cs:0.6,0.793088734580652);

\path [draw=lightgreen, semithick]
(axis cs:1,0.536251427816851)
--(axis cs:1,0.536924559237981);

\path [draw=lightgreen, semithick]
(axis cs:2,0.198812577643737)
--(axis cs:2,0.199425949190727);

\path [draw=lightgreen, semithick]
(axis cs:3,0.0830102190336211)
--(axis cs:3,0.0835462085642174);

\path [draw=lightgreen, semithick]
(axis cs:4,0.0387071145201993)
--(axis cs:4,0.0393841538455548);

\path [draw=lightgreen, semithick]
(axis cs:5,0.0198847844612168)
--(axis cs:5,0.0202919900328034);

\addplot [semithick, red, mark=triangle*, mark size=3, mark options={solid,rotate=180}, only marks]
table {%
0.1 0.917928614801591
0.3 0.791729896474547
0.6 0.661997846666201
1 0.549863082113231
2 0.399233440738109
3 0.322314541902925
4 0.274894829119715
5 0.242675189305293
};
\addlegendentry{Oracle estimator, ${\bm{\Xi}_Y^*}(\bS)$}
\addplot [semithick, royalblue, mark=triangle*, mark size=3, mark options={solid}, only marks]
table {%
0.1 0.923057753879179
0.3 0.793164618489846
0.6 0.662640913008331
1 0.55021839173983
2 0.39942371843173
3 0.322456485007585
4 0.275022629639051
5 0.242800104156512
};
\addlegendentry{RIE, $\widehat{{\bm{\Xi}_Y^*}}(\bS)$}
\addplot [semithick, seagreen, mark=x, mark size=3, mark options={solid}, only marks]
table {%
0.1 0.99995587709936
0.3 0.970934423609597
0.6 0.792730097359106
1 0.536587993527416
2 0.199119263417232
3 0.0832782137989192
4 0.039045634182877
5 0.0200883872470101
};
\addlegendentry{RIE + $f_p(x)$}
\end{axis}

\end{tikzpicture}
    \caption{$p = 0.9$}
\end{subfigure}
\caption{\small Estimating $\bY$ with Bernoulli-Rademacher \textit{entries}. MSE is normalized by the norm of the signal, $\| \bY \|_{\rm F}^2$. $\bX$ is a shifted Wigner matrix with $c=3$, and $\bW$ has i.i.d. Gaussian entries of variance $1/N$, and $N/M = 1/2$. The RIE is applied to $N=2000, M =4000$, and the results are averaged over 10 runs (error bars are invisible).}
\label{Sparse-Y}
\end{figure}
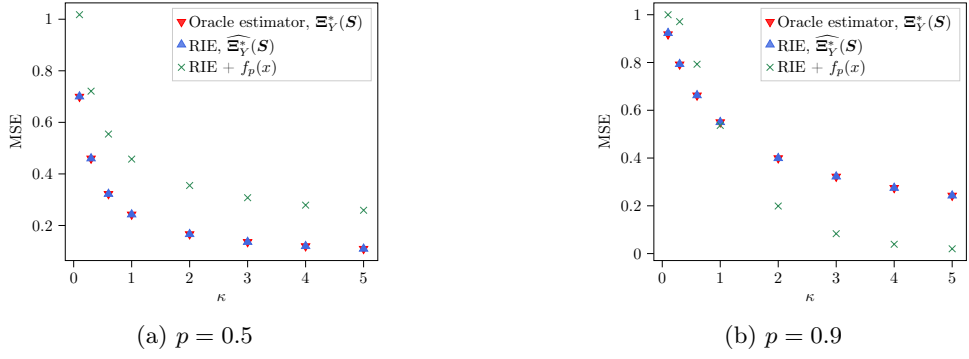

\paragraph{Sparse $\bY$: a non-rotation invariant example.} We consider $\bY$ to have i.i.d. entries from the Bernoulli-Rademacher distribution,
\begin{equation*}
    Y_{i,j} = \begin{cases}
        +\frac{1}{\sqrt{N}} &\text{with probability } \frac{1-p}{2} \\
        0 &\text{with probability } p \\
        -\frac{1}{\sqrt{N}} &\text{with probability } \frac{1-p}{2} 
    \end{cases}, \quad \quad \forall \quad 1\leq i \leq N, \quad 1\leq j \leq M
\end{equation*}
With the normalization $\nicefrac{1}{\sqrt{N}}$, the spectrum of $\bY$ does not grow with the dimension and has a finite support, thus we can apply our estimator to reconstruct $\bY$. \textit{Note that the prior of $\bY$ is not rotationally invariant, and neither the oracle estimator nor the RIE are optimal}. Therefore, taking the prior into account, we apply a thresholding function on the entries of the matrix obtained from the RIE, $\widehat{{\bm{\Xi}_Y^*}}(\bS)$. We apply the following function on each entry of the estimator:
\begin{equation*}
    f_h(x) = \begin{cases}
        +\frac{1}{\sqrt{N}} &\text{if } x > \frac{h}{\sqrt{N}} \\
        0 &\text{if } |x| \leq  \frac{h}{\sqrt{N}} \\
        -\frac{1}{\sqrt{N}} &\text{if } x < - \frac{h}{\sqrt{N}}
    \end{cases}, \quad \quad \text{for } h \in [0,1]
\end{equation*}
In figure \ref{Sparse-Y}, the MSE of the oracle estimator, RIE, and RIE$+ f_p(x)$ (with $h = p$) is plotted. A few remarks on this figure are in order. First, RIEs are not limited to rotationally invariant priors and can give non-trivial estimates for non-rotationally invariant priors, although they are sub-optimal. The RIE's output can be refined, or used as a warmed-up initialization for other algorithms to get a better estimate.

In figure \ref{Sparse-optth-Y}, for one experiment, the MSE is plotted for RIE and RIE$+f(x)$ with the best $h$ among $\{0, 0.1, \cdots, 1\}$. We observe that for the particular case of Bernoulli-Rademacher prior, the thresholding stage can improve the MSE when SNR is greater than 1, however the parameter $h$ should be chosen properly.
\begin{figure}
    \centering
\begin{subfigure}[t]{.4\textwidth}
    \centering
\begin{tikzpicture}[scale = 0.6]

\definecolor{darkgray176}{RGB}{176,176,176}
\definecolor{royalblue}{RGB}{65,105,225}
\definecolor{seagreen}{RGB}{46,139,87}

\begin{axis}[
legend cell align={left},
legend style={fill opacity=0.8, draw opacity=1, text opacity=1, draw=white!80!black},
tick align=outside,
tick pos=left,
x grid style={darkgray176},
xlabel={$\kappa$},
xmin=-0.145, xmax=5.245,
xtick style={color=black},
y grid style={darkgray176},
ylabel={${\rm MSE}$},
scaled y ticks=false,
yticklabel style={
  /pgf/number format/precision=3,
  /pgf/number format/fixed},
ymin=0.0214581370291006, ymax=0.914555268052649,
ytick style={color=black}
]
\addplot [semithick, royalblue, mark=triangle*, mark size=3, mark options={solid}, only marks]
table {%
0.1 0.699886126910039
0.3 0.459918077849467
0.6 0.321939706356816
1 0.242361147916157
2 0.166077919136607
3 0.135970067081875
4 0.119700284006126
5 0.109474157589461
};
\addlegendentry{RIE, $\widehat{{\bm{\Xi}_Y^*}}(\bS)$}
\addplot [semithick, seagreen, mark=x, mark size=3, mark options={solid}, only marks]
table {%
0.1 0.873959943915215
0.3 0.608162210060273
0.6 0.419221491647004
1 0.292432397561689
2 0.15950631159318
3 0.105636998717536
4 0.0781482419909289
5 0.0620534611665346
};
\addlegendentry{RIE + $f_{h^*}(x)$}
\end{axis}

\end{tikzpicture}
    \caption{$ p = 0.5$}
\end{subfigure}
\hfil
\begin{subfigure}[t]{.4\textwidth}
    \centering
\begin{tikzpicture}[scale = 0.6]

\definecolor{darkgray176}{RGB}{176,176,176}
\definecolor{royalblue}{RGB}{65,105,225}
\definecolor{seagreen}{RGB}{46,139,87}

\begin{axis}[
legend cell align={left},
legend style={fill opacity=0.8, draw opacity=1, text opacity=1, draw=white!80!black},
tick align=outside,
tick pos=left,
x grid style={darkgray176},
xlabel={$\kappa$},
xmin=-0.145, xmax=5.245,
xtick style={color=black},
y grid style={darkgray176},
ylabel={${\rm MSE}$},
scaled y ticks=false,
yticklabel style={
  /pgf/number format/precision=3,
  /pgf/number format/fixed},
ymin=-0.0290680820257207, ymax=1.04778774105391,
ytick style={color=black}
]
\addplot [semithick, royalblue, mark=triangle*, mark size=3, mark options={solid}, only marks]
table {%
0.1 0.917658188383288
0.3 0.791857982805171
0.6 0.661982324900161
1 0.549742282038492
2 0.399673604264371
3 0.322738626541779
4 0.27447443776136
5 0.24248625994927
};
\addlegendentry{RIE, $\widehat{{\bm{\Xi}_Y^*}}(\bS)$}
\addplot [semithick, seagreen, mark=x, mark size=3, mark options={solid}, only marks]
table {%
0.1 0.998839749095741
0.3 0.923507147971258
0.6 0.705540450924008
1 0.47486221980095
2 0.186775458355369
3 0.0833265426116476
4 0.0386944014881592
5 0.0198799099324442
};
\addlegendentry{RIE + $f_{h^*}(x)$}
\end{axis}

\end{tikzpicture}
    \caption{$p = 0.9$}
\end{subfigure}
\caption{\small Estimating $\bY$ with Bernoulli-Rademacher \textit{entries}. MSE is normalized by the norm of the signal, $\| \bY \|_{\rm F}^2$. $\bX$ is a shifted Wigner matrix with $c=3$, and $\bW$ has i.i.d. Gaussian entries of variance $1/N$, and $N/M = 1/2$. The RIE is applied to $N=2000, M =4000$, and thresholding function is applied with the best $h$ among $\{0, 0.1, \cdots, 1\}$. Results are averaged over 10 runs (error bars are invisible).}
\label{Sparse-optth-Y}
\end{figure}
\clearpage
\section{Comparison of RIEs for MF and denoising}
For estimating $\bX$, we have derived the estimator \eqref{X-optimal-ev-app} for general priors $\rho_X, \mu_Y, \mu_W$. This estimator simplifies greatly, with parameters in \eqref{X-Y,W-Gauss}, when both $\mu_Y, \mu_W$ are Marchenko-Pastur distribution, i.e. both $\bY, \bW$ having i.i.d. Gaussian entries of variance $\nicefrac{1}{N}$. Similarly, although the RIE for $\bY$ in \eqref{Y-optimal-sv-app} is derived for the general priors, it reduces to a rather simple estimator if $\rho_X$, $\mu_W$  are taken to be shifted Wigner, and Marchenko-Pastur distribution, respectively. Therefore, in our numerical examples on factorization problem, we consider $\bX$ to be a shifted Wigner matrix, and $\bY, \bW$ to be Gaussian matrices.

In each experiment, the factors $\bX$, $\bY$ are estimated simultaneously using RIE from the observation matrix $\bS$. In addition to the MSE of estimating each factor, we also compute the MSE of estimating the product $\bX \bY$. We compare the MSE of the product with the MSE of the oracle estimator and the RIE introduced in \cite{pourkamali2023rectangular} for the denoising problem. The oracle estimator for the denoising is constructed as:
\begin{equation}
\bm{\Xi}^*_{XY}(\bS) = \sum_{i = 1}^N {\xi_{xy}^*}_i \, \bu_i \bv_i^\intercal, \quad
    {\xi_{xy}^*}_i = \bu_i^\intercal \bX \bY \bv_i
    \label{XY-oracle-estimator}
\end{equation}
where $\bu_i, \bv_i$'s are left/right singular vectors of $\bS$. In the RIE proposed in \cite{pourkamali2023rectangular}, the singular values are estimated by (see section \ref{rect-RIE})
\begin{equation}
\begin{split}
    \widehat{\xi_{xy}^*}_i &= \frac{1}{\sqrt{\kappa}} \Bigg[ \gamma_i - \frac{1}{\pi \bar{\mu}_{S}(\gamma_i)} {\rm Im} \, \mathcal{C}^{(\alpha)}_{\mu_W}\bigg( \frac{1- \alpha}{\gamma_i} \pi \sH [\bar{\mu}_{S}](\gamma_i) + \alpha \big(\pi \sH [\bar{\mu}_{S}](\gamma_i)\big)^2  - \alpha  \big( \pi \bar{\mu}_{S}(\gamma_i)\big)^2 \\
    &\hspace{7cm}+ \ci  \pi \bar{\mu}_{S}(\gamma_i) \big(\frac{1-\alpha}{\gamma_i} + 2 \alpha \pi \sH [\bar{\mu}_{S}](\gamma_i) \big) \bigg) \Bigg]
    \label{XY-denosing-optimal-sv}
\end{split}
\end{equation}
Note that, in general the MSE of the denoising RIE $\widehat{{\bm{\Xi}_{XY}^*}}(\bS)$, is less than the MSE of the prdouct of the estimated factors $\widehat{{\bm{\Xi}_X^*}}(\bS) \widehat{{\bm{\Xi}_Y^*}}(\bS)$.

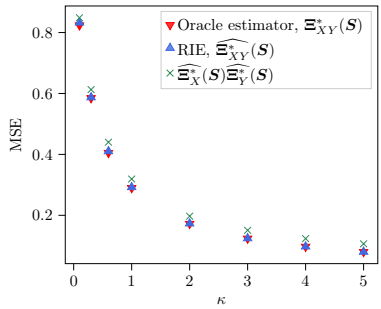
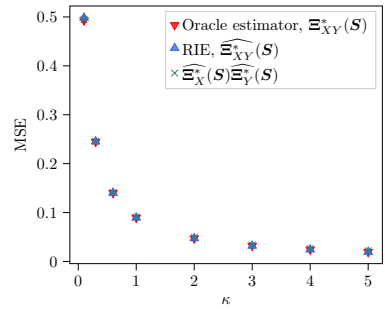
\begin{figure}
    \centering
\begin{subfigure}[t]{.4\textwidth}
    \centering
\begin{tikzpicture}[scale = 0.6]

\definecolor{darkgray176}{RGB}{176,176,176}
\definecolor{lightcoral}{RGB}{240,128,128}
\definecolor{lightgreen}{RGB}{144,238,144}
\definecolor{lightslategray}{RGB}{119,136,153}
\definecolor{royalblue}{RGB}{65,105,225}
\definecolor{seagreen}{RGB}{46,139,87}

\begin{axis}[
legend cell align={left},
legend style={fill opacity=0.8, draw opacity=1, text opacity=1, draw=white!80!black},
tick align=outside,
tick pos=left,
x grid style={darkgray176},
xlabel={$\kappa$},
xmin=-0.145, xmax=5.245,
xtick style={color=black},
y grid style={darkgray176},
ylabel={${\rm MSE}$},
scaled y ticks=false,
yticklabel style={
  /pgf/number format/precision=3,
  /pgf/number format/fixed},
ymin=0.0413329418062274, ymax=0.887241708163143,
ytick style={color=black}
]
\path [draw=lightcoral, semithick]
(axis cs:0.1,0.823478241414999)
--(axis cs:0.1,0.823971553047559);

\path [draw=lightcoral, semithick]
(axis cs:0.3,0.584527545396763)
--(axis cs:0.3,0.585374439580424);

\path [draw=lightcoral, semithick]
(axis cs:0.6,0.404711933839447)
--(axis cs:0.6,0.405368118382134);

\path [draw=lightcoral, semithick]
(axis cs:1,0.289423188364281)
--(axis cs:1,0.29004931760945);

\path [draw=lightcoral, semithick]
(axis cs:2,0.171959678819973)
--(axis cs:2,0.172393703279081);

\path [draw=lightcoral, semithick]
(axis cs:3,0.123565211654094)
--(axis cs:3,0.123784104869595);

\path [draw=lightcoral, semithick]
(axis cs:4,0.0968680454232269)
--(axis cs:4,0.0970951684035276);

\path [draw=lightcoral, semithick]
(axis cs:5,0.0797833402769963)
--(axis cs:5,0.0799880080148496);

\path [draw=lightslategray, semithick]
(axis cs:0.1,0.832197639498174)
--(axis cs:0.1,0.832809204799342);

\path [draw=lightslategray, semithick]
(axis cs:0.3,0.586633426361618)
--(axis cs:0.3,0.58747005186858);

\path [draw=lightslategray, semithick]
(axis cs:0.6,0.409743854988)
--(axis cs:0.6,0.410495071384984);

\path [draw=lightslategray, semithick]
(axis cs:1,0.291763049203933)
--(axis cs:1,0.292429209784479);

\path [draw=lightslategray, semithick]
(axis cs:2,0.172775315694239)
--(axis cs:2,0.173210051621457);

\path [draw=lightslategray, semithick]
(axis cs:3,0.124210994710361)
--(axis cs:3,0.124434455425909);

\path [draw=lightslategray, semithick]
(axis cs:4,0.0972604797836797)
--(axis cs:4,0.0974891503771267);

\path [draw=lightslategray, semithick]
(axis cs:5,0.0800523030002821)
--(axis cs:5,0.080257916903967);

\path [draw=lightgreen, semithick]
(axis cs:0.1,0.84834581452111)
--(axis cs:0.1,0.848791309692374);

\path [draw=lightgreen, semithick]
(axis cs:0.3,0.612406464049448)
--(axis cs:0.3,0.613114342032113);

\path [draw=lightgreen, semithick]
(axis cs:0.6,0.439774046638864)
--(axis cs:0.6,0.440276251534087);

\path [draw=lightgreen, semithick]
(axis cs:1,0.318881692133742)
--(axis cs:1,0.319320058520465);

\path [draw=lightgreen, semithick]
(axis cs:2,0.196918530756298)
--(axis cs:2,0.197286537645992);

\path [draw=lightgreen, semithick]
(axis cs:3,0.150676219889456)
--(axis cs:3,0.151002017652665);

\path [draw=lightgreen, semithick]
(axis cs:4,0.123368470669134)
--(axis cs:4,0.123651472373388);

\path [draw=lightgreen, semithick]
(axis cs:5,0.106303616789717)
--(axis cs:5,0.106649453332115);

\addplot [semithick, red, mark=triangle*, mark size=3, mark options={solid,rotate=180}, only marks]
table {%
0.1 0.823724897231279
0.3 0.584950992488593
0.6 0.405040026110791
1 0.289736252986866
2 0.172176691049527
3 0.123674658261845
4 0.0969816069133772
5 0.079885674145923
};
\addlegendentry{Oracle estimator, ${\bm{\Xi}_{XY}^*}(\bS)$}
\addplot [semithick, royalblue, mark=triangle*, mark size=3, mark options={solid}, only marks]
table {%
0.1 0.832503422148758
0.3 0.587051739115099
0.6 0.410119463186492
1 0.292096129494206
2 0.172992683657848
3 0.124322725068135
4 0.0973748150804032
5 0.0801551099521246
};
\addlegendentry{RIE, $\widehat{{\bm{\Xi}_{XY}^*}}(\bS)$}
\addplot [semithick, seagreen, mark=x, mark size=3, mark options={solid}, only marks]
table {%
0.1 0.848568562106742
0.3 0.61276040304078
0.6 0.440025149086475
1 0.319100875327104
2 0.197102534201145
3 0.15083911877106
4 0.123509971521261
5 0.106476535060916
};
\addlegendentry{$\widehat{{\bm{\Xi}_X^*}}(\bS) \widehat{{\bm{\Xi}_Y^*}}(\bS)$}
\end{axis}

\end{tikzpicture}
    \caption{$c = 1$}
\end{subfigure}
\hfil
\begin{subfigure}[t]{.4\textwidth}
    \centering
\begin{tikzpicture}[scale = 0.6]

\definecolor{darkgray176}{RGB}{176,176,176}
\definecolor{lightcoral}{RGB}{240,128,128}
\definecolor{lightgreen}{RGB}{144,238,144}
\definecolor{lightslategray}{RGB}{119,136,153}
\definecolor{royalblue}{RGB}{65,105,225}
\definecolor{seagreen}{RGB}{46,139,87}

\begin{axis}[
legend cell align={left},
legend style={fill opacity=0.8, draw opacity=1, text opacity=1, draw=white!80!black},
tick align=outside,
tick pos=left,
x grid style={darkgray176},
xlabel={$\kappa$},
xmin=-0.145, xmax=5.245,
xtick style={color=black},
y grid style={darkgray176},
ylabel={${\rm MSE}$},
scaled y ticks=false,
yticklabel style={
  /pgf/number format/precision=3,
  /pgf/number format/fixed},
ymin=-0.00444403826887389, ymax=0.522421968969946,
ytick style={color=black}
]
\path [draw=lightcoral, semithick]
(axis cs:0.1,0.492892007269047)
--(axis cs:0.1,0.493265390248217);

\path [draw=lightcoral, semithick]
(axis cs:0.3,0.244624420437456)
--(axis cs:0.3,0.244964320377858);

\path [draw=lightcoral, semithick]
(axis cs:0.6,0.140196278853719)
--(axis cs:0.6,0.140366957323577);

\path [draw=lightcoral, semithick]
(axis cs:1,0.0895179520819202)
--(axis cs:1,0.0896450010478514);

\path [draw=lightcoral, semithick]
(axis cs:2,0.0471425691393406)
--(axis cs:2,0.047248780769465);

\path [draw=lightcoral, semithick]
(axis cs:3,0.0320344303816539)
--(axis cs:3,0.0320885898484053);

\path [draw=lightcoral, semithick]
(axis cs:4,0.0242386353813771)
--(axis cs:4,0.0243044456628581);

\path [draw=lightcoral, semithick]
(axis cs:5,0.0195044166056179)
--(axis cs:5,0.0195341566503052);

\path [draw=lightslategray, semithick]
(axis cs:0.1,0.498082351344284)
--(axis cs:0.1,0.498473514095454);

\path [draw=lightslategray, semithick]
(axis cs:0.3,0.245236610343702)
--(axis cs:0.3,0.245597594839824);

\path [draw=lightslategray, semithick]
(axis cs:0.6,0.140342823593346)
--(axis cs:0.6,0.140515221498633);

\path [draw=lightslategray, semithick]
(axis cs:1,0.0895726174392163)
--(axis cs:1,0.0896985580522907);

\path [draw=lightslategray, semithick]
(axis cs:2,0.0471594878888491)
--(axis cs:2,0.0472656000443664);

\path [draw=lightslategray, semithick]
(axis cs:3,0.0320453577315453)
--(axis cs:3,0.0320994712905457);

\path [draw=lightslategray, semithick]
(axis cs:4,0.0242460264397597)
--(axis cs:4,0.0243117088964962);

\path [draw=lightslategray, semithick]
(axis cs:5,0.0195098699967647)
--(axis cs:5,0.0195397007987366);

\path [draw=lightgreen, semithick]
(axis cs:0.1,0.494942144522653)
--(axis cs:0.1,0.49524286213741);

\path [draw=lightgreen, semithick]
(axis cs:0.3,0.245597598923706)
--(axis cs:0.3,0.24591629856205);

\path [draw=lightgreen, semithick]
(axis cs:0.6,0.141207244009918)
--(axis cs:0.6,0.141360736085495);

\path [draw=lightgreen, semithick]
(axis cs:1,0.0908091145972185)
--(axis cs:1,0.090995777142181);

\path [draw=lightgreen, semithick]
(axis cs:2,0.0489802129034483)
--(axis cs:2,0.0490918503494514);

\path [draw=lightgreen, semithick]
(axis cs:3,0.0337924773021265)
--(axis cs:3,0.0339217814304475);

\path [draw=lightgreen, semithick]
(axis cs:4,0.0261977327235399)
--(axis cs:4,0.0263232726258618);

\path [draw=lightgreen, semithick]
(axis cs:5,0.021654237048952)
--(axis cs:5,0.0217768028844965);

\addplot [semithick, red, mark=triangle*, mark size=3, mark options={solid,rotate=180}, only marks]
table {%
0.1 0.493078698758632
0.3 0.244794370407657
0.6 0.140281618088648
1 0.0895814765648858
2 0.0471956749544028
3 0.0320615101150296
4 0.0242715405221176
5 0.0195192866279616
};
\addlegendentry{Oracle estimator, ${\bm{\Xi}_{XY}^*}(\bS)$}
\addplot [semithick, royalblue, mark=triangle*, mark size=3, mark options={solid}, only marks]
table {%
0.1 0.498277932719869
0.3 0.245417102591763
0.6 0.140429022545989
1 0.0896355877457535
2 0.0472125439666078
3 0.0320724145110455
4 0.024278867668128
5 0.0195247853977506
};
\addlegendentry{RIE, $\widehat{{\bm{\Xi}_{XY}^*}}(\bS)$}
\addplot [semithick, seagreen, mark=x, mark size=3, mark options={solid}, only marks]
table {%
0.1 0.495092503330032
0.3 0.245756948742878
0.6 0.141283990047707
1 0.0909024458696998
2 0.0490360316264499
3 0.033857129366287
4 0.0262605026747008
5 0.0217155199667242
};
\addlegendentry{$\widehat{{\bm{\Xi}_X^*}}(\bS) \widehat{{\bm{\Xi}_Y^*}}(\bS)$}
\end{axis}

\end{tikzpicture}
    \caption{$c = 3$}
\end{subfigure}
\caption{\small MSE of the product of the factors. MSE is normalized by the norm of the signal $\| \bX \bY\|_{\rm F}^2$. $\bX$ is a shifted Wigner matrix with $c=1, c=3$, and both $\bY$ and $\bW$ are $N \times M$ matrices with i.i.d. Gaussian entries of variance $1/N$, and $N/M = 1/2$. The RIE is applied to $N=2000, M =4000$. Results are averaged over 10 runs (error bars are invisible).}
\label{MF-prod}
\end{figure}

In figures \ref{MF-c=1},\ref{MF-c=3}, the MSE of estimating the factors is illustrated for $c=1$ and $c=3$ respectively. The MSE of estimating the product is shown in figure \ref{MF-prod}. 
\begin{figure}
    \centering
\begin{subfigure}[t]{.4\textwidth}
    \centering
\begin{tikzpicture}[scale = 0.6]

\definecolor{darkgray176}{RGB}{176,176,176}
\definecolor{lightcoral}{RGB}{240,128,128}
\definecolor{lightslategray}{RGB}{119,136,153}
\definecolor{royalblue}{RGB}{65,105,225}

\begin{axis}[
legend cell align={left},
legend style={fill opacity=0.8, draw opacity=1, text opacity=1, draw=white!80!black},
tick align=outside,
tick pos=left,
x grid style={darkgray176},
xlabel={$\kappa$},
xmin=-0.145, xmax=5.35,
xtick style={color=black},
y grid style={darkgray176},
ylabel={${\rm MSE}$},
scaled y ticks=false,
yticklabel style={
  /pgf/number format/precision=3,
  /pgf/number format/fixed},
ymin=0.158989648894357, ymax=0.489776450653492,
ytick style={color=black}
]
\path [draw=lightcoral, semithick]
(axis cs:0.1,0.472679862890143)
--(axis cs:0.1,0.47369940142939);

\path [draw=lightcoral, semithick]
(axis cs:0.3,0.394383432704544)
--(axis cs:0.3,0.395189230169388);

\path [draw=lightcoral, semithick]
(axis cs:0.6,0.322359971106662)
--(axis cs:0.6,0.323545953558721);

\path [draw=lightcoral, semithick]
(axis cs:1,0.271617606398936)
--(axis cs:1,0.272463997304963);

\path [draw=lightcoral, semithick]
(axis cs:2,0.217437387501637)
--(axis cs:2,0.21811034000889);

\path [draw=lightcoral, semithick]
(axis cs:3,0.194396584030318)
--(axis cs:3,0.195341349525913);

\path [draw=lightcoral, semithick]
(axis cs:4,0.182018974565411)
--(axis cs:4,0.18282015642648);

\path [draw=lightcoral, semithick]
(axis cs:5,0.174025412610681)
--(axis cs:5,0.174762125551437);

\path [draw=lightslategray, semithick]
(axis cs:0.1,0.47377559294999)
--(axis cs:0.1,0.474740686937167);

\path [draw=lightslategray, semithick]
(axis cs:0.3,0.395400823592722)
--(axis cs:0.3,0.396203440825488);

\path [draw=lightslategray, semithick]
(axis cs:0.6,0.32478470995362)
--(axis cs:0.6,0.32586208415824);

\path [draw=lightslategray, semithick]
(axis cs:1,0.273584125988031)
--(axis cs:1,0.274333828104676);

\path [draw=lightslategray, semithick]
(axis cs:2,0.218852589334871)
--(axis cs:2,0.219525971584205);

\path [draw=lightslategray, semithick]
(axis cs:3,0.196200360549737)
--(axis cs:3,0.197066901589207);

\path [draw=lightslategray, semithick]
(axis cs:4,0.183443623804713)
--(axis cs:4,0.184262054496322);

\path [draw=lightslategray, semithick]
(axis cs:5,0.175366459606412)
--(axis cs:5,0.176170039823591);

\addplot [semithick, red, mark=triangle*, mark size=3, mark options={solid,rotate=180}, only marks]
table {%
0.1 0.473189632159767
0.3 0.394786331436966
0.6 0.322952962332692
1 0.272040801851949
2 0.217773863755263
3 0.194868966778116
4 0.182419565495946
5 0.174393769081059
};
\addlegendentry{Oracle estimator, ${\bm{\Xi}_X^*}(\bS)$}
\addplot [semithick, royalblue, mark=triangle*, mark size=3, mark options={solid}, only marks]
table {%
0.1 0.474258139943579
0.3 0.395802132209105
0.6 0.32532339705593
1 0.273958977046354
2 0.219189280459538
3 0.196633631069472
4 0.183852839150517
5 0.175768249715002
};
\addlegendentry{RIE, $\widehat{{\bm{\Xi}_X^*}}(\bS)$}
\draw (axis cs:0.2,0.473189632159767) node[
  scale=0.6,
  anchor=base west,
  text=black,
  rotate=0.0
]{0.23\%};
\draw (axis cs:0.15,0.404786331436966) node[
  scale=0.6,
  anchor=base west,
  text=black,
  rotate=0.0
]{0.26\%};
\draw (axis cs:0.45,0.332952962332692) node[
  scale=0.6,
  anchor=base west,
  text=black,
  rotate=0.0
]{0.73\%};
\draw (axis cs:0.85,0.282040801851949) node[
  scale=0.6,
  anchor=base west,
  text=black,
  rotate=0.0
]{0.71\%};
\draw (axis cs:1.85,0.227773863755263) node[
  scale=0.6,
  anchor=base west,
  text=black,
  rotate=0.0
]{0.65\%};
\draw (axis cs:2.85,0.204868966778116) node[
  scale=0.6,
  anchor=base west,
  text=black,
  rotate=0.0
]{0.91\%};
\draw (axis cs:3.85,0.192419565495946) node[
  scale=0.6,
  anchor=base west,
  text=black,
  rotate=0.0
]{0.79\%};
\draw (axis cs:4.85,0.184393769081059) node[
  scale=0.6,
  anchor=base west,
  text=black,
  rotate=0.0
]{0.79\%};
\end{axis}

\end{tikzpicture}
    \caption{Estimating $\bX$}
\end{subfigure}
\hfil
\begin{subfigure}[t]{.4\textwidth}
    \centering
\begin{tikzpicture}[scale = 0.6]

\definecolor{darkgray176}{RGB}{176,176,176}
\definecolor{lightcoral}{RGB}{240,128,128}
\definecolor{lightslategray}{RGB}{119,136,153}
\definecolor{royalblue}{RGB}{65,105,225}

\begin{axis}[
legend cell align={left},
legend style={fill opacity=0.8, draw opacity=1, text opacity=1, draw=white!80!black},
tick align=outside,
tick pos=left,
x grid style={darkgray176},
xlabel={$\kappa$},
xmin=-0.145, xmax=5.35,
xtick style={color=black},
y grid style={darkgray176},
ylabel={${\rm MSE}$},
scaled y ticks=false,
yticklabel style={
  /pgf/number format/precision=3,
  /pgf/number format/fixed},
ymin=0.511630570465254, ymax=0.934756886300313,
ytick style={color=black}
]
\path [draw=lightcoral, semithick]
(axis cs:0.1,0.913534806416141)
--(axis cs:0.1,0.914054528076912);

\path [draw=lightcoral, semithick]
(axis cs:0.3,0.801744489184537)
--(axis cs:0.3,0.802333687334179);

\path [draw=lightcoral, semithick]
(axis cs:0.6,0.714432593344145)
--(axis cs:0.6,0.71524679242528);

\path [draw=lightcoral, semithick]
(axis cs:1,0.653954871657974)
--(axis cs:1,0.654674672043728);

\path [draw=lightcoral, semithick]
(axis cs:2,0.587440039833787)
--(axis cs:2,0.588400318916892);

\path [draw=lightcoral, semithick]
(axis cs:3,0.557954986361824)
--(axis cs:3,0.559011283611603);

\path [draw=lightcoral, semithick]
(axis cs:4,0.541548775105845)
--(axis cs:4,0.542902957038355);

\path [draw=lightcoral, semithick]
(axis cs:5,0.530863584821393)
--(axis cs:5,0.531884750800902);

\path [draw=lightslategray, semithick]
(axis cs:0.1,0.915126009006564)
--(axis cs:0.1,0.915523871944174);

\path [draw=lightslategray, semithick]
(axis cs:0.3,0.802308699980285)
--(axis cs:0.3,0.802887691176402);

\path [draw=lightslategray, semithick]
(axis cs:0.6,0.715408417207449)
--(axis cs:0.6,0.716211006714881);

\path [draw=lightslategray, semithick]
(axis cs:1,0.654855088789837)
--(axis cs:1,0.655562701787716);

\path [draw=lightslategray, semithick]
(axis cs:2,0.588214340720012)
--(axis cs:2,0.58915826130104);

\path [draw=lightslategray, semithick]
(axis cs:3,0.559067528213006)
--(axis cs:3,0.560077743866405);

\path [draw=lightslategray, semithick]
(axis cs:4,0.542554740628416)
--(axis cs:4,0.543893438153047);

\path [draw=lightslategray, semithick]
(axis cs:5,0.531831453314418)
--(axis cs:5,0.53284388897055);

\addplot [semithick, red, mark=triangle*, mark size=3, mark options={solid,rotate=180}, only marks]
table {%
0.1 0.913794667246526
0.3 0.802039088259358
0.6 0.714839692884712
1 0.654314771850851
2 0.587920179375339
3 0.558483134986713
4 0.5422258660721
5 0.531374167811148
};
\addlegendentry{Oracle estimator, ${\bm{\Xi}_Y^*}(\bS)$}
\addplot [semithick, royalblue, mark=triangle*, mark size=3, mark options={solid}, only marks]
table {%
0.1 0.915324940475369
0.3 0.802598195578344
0.6 0.715809711961165
1 0.655208895288777
2 0.588686301010526
3 0.559572636039705
4 0.543224089390731
5 0.532337671142484
};
\addlegendentry{RIE, $\widehat{{\bm{\Xi}_Y^*}}(\bS)$}
\draw (axis cs:0.2,0.913794667246526) node[
  scale=0.6,
  anchor=base west,
  text=black,
  rotate=0.0
]{0.17\%};
\draw (axis cs:0.15,0.812039088259358) node[
  scale=0.6,
  anchor=base west,
  text=black,
  rotate=0.0
]{0.07\%};
\draw (axis cs:0.45,0.724839692884713) node[
  scale=0.6,
  anchor=base west,
  text=black,
  rotate=0.0
]{0.14\%};
\draw (axis cs:0.85,0.664314771850851) node[
  scale=0.6,
  anchor=base west,
  text=black,
  rotate=0.0
]{0.14\%};
\draw (axis cs:1.85,0.597920179375339) node[
  scale=0.6,
  anchor=base west,
  text=black,
  rotate=0.0
]{0.13\%};
\draw (axis cs:2.85,0.568483134986713) node[
  scale=0.6,
  anchor=base west,
  text=black,
  rotate=0.0
]{0.2\%};
\draw (axis cs:3.85,0.5522258660721) node[
  scale=0.6,
  anchor=base west,
  text=black,
  rotate=0.0
]{0.18\%};
\draw (axis cs:4.85,0.541374167811148) node[
  scale=0.6,
  anchor=base west,
  text=black,
  rotate=0.0
]{0.18\%};
\end{axis}

\end{tikzpicture}
    \caption{Estimating $\bY$}
\end{subfigure}
\caption{\small MSE of factorization problem. MSE is normalized by the norm of the signal. $\bX$ is a shifted Wigner matrix with $c=1$, and both $\bY$ and $\bW$ are $N \times M$ matrices with i.i.d. Gaussian entries of variance $1/N$, and $N/M = 1/2$. The RIE is applied to $N=2000, M =4000$. In each run, the observation matrix $\bS$ is generated according to \eqref{MF-model}, and the factors $\bX$, $\bY$ are estimated simultaneously from $\bS$. Results are averaged over 10 runs (error bars are invisible). Average relative error between RIEs and Oracle estimators is also reported.}
\label{MF-c=1}
\end{figure}

\begin{figure}
    \centering
\begin{subfigure}[t]{.4\textwidth}
    \centering
\begin{tikzpicture}[scale = 0.6]

\definecolor{darkgray176}{RGB}{176,176,176}
\definecolor{lightcoral}{RGB}{240,128,128}
\definecolor{lightslategray}{RGB}{119,136,153}
\definecolor{royalblue}{RGB}{65,105,225}

\begin{axis}[
legend cell align={left},
legend style={fill opacity=0.8, draw opacity=1, text opacity=1, draw=white!80!black},
tick align=outside,
tick pos=left,
x grid style={darkgray176},
xlabel={$\kappa$},
xmin=-0.145, xmax=5.35,
xtick style={color=black},
y grid style={darkgray176},
ylabel={${\rm MSE}$},
scaled y ticks=false,
yticklabel style={
  /pgf/number format/precision=3,
  /pgf/number format/fixed},
ymin=0.051004703654956, ymax=0.0870318364024802,
ytick style={color=black}
]
\path [draw=lightcoral, semithick]
(axis cs:0.1,0.0845223337073945)
--(axis cs:0.1,0.0847650669307829);

\path [draw=lightcoral, semithick]
(axis cs:0.3,0.0695045792085804)
--(axis cs:0.3,0.0695789430909292);

\path [draw=lightcoral, semithick]
(axis cs:0.6,0.0621021052803107)
--(axis cs:0.6,0.0622225947678647);

\path [draw=lightcoral, semithick]
(axis cs:1,0.0582500053145135)
--(axis cs:1,0.0583380871902252);

\path [draw=lightcoral, semithick]
(axis cs:2,0.0549332420892584)
--(axis cs:2,0.0550216159124154);

\path [draw=lightcoral, semithick]
(axis cs:3,0.0536634859156279)
--(axis cs:3,0.0537741575108468);

\path [draw=lightcoral, semithick]
(axis cs:4,0.053014139373991)
--(axis cs:4,0.0531597338012953);

\path [draw=lightcoral, semithick]
(axis cs:5,0.0526423005980253)
--(axis cs:5,0.0527091126095964);

\path [draw=lightslategray, semithick]
(axis cs:0.1,0.0851679437706787)
--(axis cs:0.1,0.0853942394594109);

\path [draw=lightslategray, semithick]
(axis cs:0.3,0.069933961494274)
--(axis cs:0.3,0.0700339675618426);

\path [draw=lightslategray, semithick]
(axis cs:0.6,0.0624333362236129)
--(axis cs:0.6,0.0625649787217132);

\path [draw=lightslategray, semithick]
(axis cs:1,0.0585547287504115)
--(axis cs:1,0.0587089014225088);

\path [draw=lightslategray, semithick]
(axis cs:2,0.0552572318077674)
--(axis cs:2,0.0553898033415283);

\path [draw=lightslategray, semithick]
(axis cs:3,0.0539883982727616)
--(axis cs:3,0.0540976318997757);

\path [draw=lightslategray, semithick]
(axis cs:4,0.0533780373607833)
--(axis cs:4,0.053492338038451);

\path [draw=lightslategray, semithick]
(axis cs:5,0.0530262646696547)
--(axis cs:5,0.0531418929943894);

\addplot [semithick, red, mark=triangle*, mark size=3, mark options={solid,rotate=180}, only marks]
table {%
0.1 0.0846437003190887
0.3 0.0695417611497548
0.6 0.0621623500240877
1 0.0582940462523693
2 0.0549774290008369
3 0.0537188217132373
4 0.0530869365876432
5 0.0526757066038108
};
\addlegendentry{Oracle estimator, ${\bm{\Xi}_X^*}(\bS)$}
\addplot [semithick, royalblue, mark=triangle*, mark size=3, mark options={solid}, only marks]
table {%
0.1 0.0852810916150448
0.3 0.0699839645280583
0.6 0.062499157472663
1 0.0586318150864602
2 0.0553235175746479
3 0.0540430150862687
4 0.0534351876996172
5 0.0530840788320221
};
\addlegendentry{RIE, $\widehat{{\bm{\Xi}_X^*}}(\bS)$}
\draw (axis cs:0.2,0.0846437003190887) node[
  scale=0.6,
  anchor=base west,
  text=black,
  rotate=0.0
]{0.75\%};
\draw (axis cs:0.15,0.0715417611497548) node[
  scale=0.6,
  anchor=base west,
  text=black,
  rotate=0.0
]{0.64\%};
\draw (axis cs:0.45,0.0641623500240877) node[
  scale=0.6,
  anchor=base west,
  text=black,
  rotate=0.0
]{0.54\%};
\draw (axis cs:0.85,0.0602940462523694) node[
  scale=0.6,
  anchor=base west,
  text=black,
  rotate=0.0
]{0.58\%};
\draw (axis cs:1.85,0.0569774290008369) node[
  scale=0.6,
  anchor=base west,
  text=black,
  rotate=0.0
]{0.63\%};
\draw (axis cs:2.85,0.0557188217132373) node[
  scale=0.6,
  anchor=base west,
  text=black,
  rotate=0.0
]{0.6\%};
\draw (axis cs:3.85,0.0550869365876432) node[
  scale=0.6,
  anchor=base west,
  text=black,
  rotate=0.0
]{0.66\%};
\draw (axis cs:4.85,0.0546757066038108) node[
  scale=0.6,
  anchor=base west,
  text=black,
  rotate=0.0
]{0.78\%};

\end{axis}

\end{tikzpicture}
    \caption{Estimating $\bX$}
\end{subfigure}
\hfil
\begin{subfigure}[t]{.4\textwidth}
    \centering
\begin{tikzpicture}[scale = 0.6]

\definecolor{darkgray176}{RGB}{176,176,176}
\definecolor{lightcoral}{RGB}{240,128,128}
\definecolor{lightslategray}{RGB}{119,136,153}
\definecolor{royalblue}{RGB}{65,105,225}

\begin{axis}[
legend cell align={left},
legend style={fill opacity=0.8, draw opacity=1, text opacity=1, draw=white!80!black},
tick align=outside,
tick pos=left,
x grid style={darkgray176},
xlabel={$\kappa$},
xmin=-0.145, xmax=5.35,
xtick style={color=black},
y grid style={darkgray176},
ylabel={${\rm MSE}$},
scaled y ticks=false,
yticklabel style={
  /pgf/number format/precision=3,
  /pgf/number format/fixed},
ymin=0.0646435909226282, ymax=0.573697668514425,
ytick style={color=black}
]
\path [draw=lightcoral, semithick]
(axis cs:0.1,0.549576280471562)
--(axis cs:0.1,0.549987162887662);

\path [draw=lightcoral, semithick]
(axis cs:0.3,0.321891239380813)
--(axis cs:0.3,0.322306600898997);

\path [draw=lightcoral, semithick]
(axis cs:0.6,0.21879431381576)
--(axis cs:0.6,0.219058982288363);

\path [draw=lightcoral, semithick]
(axis cs:1,0.165897025331504)
--(axis cs:1,0.166093501403753);

\path [draw=lightcoral, semithick]
(axis cs:2,0.119571899325948)
--(axis cs:2,0.119746750849501);

\path [draw=lightcoral, semithick]
(axis cs:3,0.102332798931918)
--(axis cs:3,0.102494346389936);

\path [draw=lightcoral, semithick]
(axis cs:4,0.0933102516473477)
--(axis cs:4,0.0935141635911479);

\path [draw=lightcoral, semithick]
(axis cs:5,0.0877824126313462)
--(axis cs:5,0.0878693040875659);

\path [draw=lightslategray, semithick]
(axis cs:0.1,0.550164725514011)
--(axis cs:0.1,0.550558846805707);

\path [draw=lightslategray, semithick]
(axis cs:0.3,0.322318823531444)
--(axis cs:0.3,0.32272237202712);

\path [draw=lightslategray, semithick]
(axis cs:0.6,0.219078210827559)
--(axis cs:0.6,0.219344011402768);

\path [draw=lightslategray, semithick]
(axis cs:1,0.166100669270005)
--(axis cs:1,0.166298768188349);

\path [draw=lightslategray, semithick]
(axis cs:2,0.119709746176119)
--(axis cs:2,0.119887070213836);

\path [draw=lightslategray, semithick]
(axis cs:3,0.102503888156776)
--(axis cs:3,0.102668461388967);

\path [draw=lightslategray, semithick]
(axis cs:4,0.0934547147884752)
--(axis cs:4,0.0936628602902976);

\path [draw=lightslategray, semithick]
(axis cs:5,0.0879082036062515)
--(axis cs:5,0.0879984404071738);

\addplot [semithick, red, mark=triangle*, mark size=3, mark options={solid,rotate=180}, only marks]
table {%
0.1 0.549781721679612
0.3 0.322098920139905
0.6 0.218926648052062
1 0.165995263367629
2 0.119659325087725
3 0.102413572660927
4 0.0934122076192478
5 0.0878258583594561
};
\addlegendentry{Oracle estimator, ${\bm{\Xi}_Y^*}(\bS)$}
\addplot [semithick, royalblue, mark=triangle*, mark size=3, mark options={solid}, only marks]
table {%
0.1 0.550361786159859
0.3 0.322520597779282
0.6 0.219211111115164
1 0.166199718729177
2 0.119798408194977
3 0.102586174772871
4 0.0935587875393864
5 0.0879533220067126
};
\addlegendentry{RIE, $\widehat{{\bm{\Xi}_Y^*}}(\bS)$}
\draw (axis cs:0.2,0.549781721679612) node[
  scale=0.6,
  anchor=base west,
  text=black,
  rotate=0.0
]{0.11\%};
\draw (axis cs:0.15,0.332098920139905) node[
  scale=0.6,
  anchor=base west,
  text=black,
  rotate=0.0
]{0.13\%};
\draw (axis cs:0.45,0.228926648052062) node[
  scale=0.6,
  anchor=base west,
  text=black,
  rotate=0.0
]{0.13\%};
\draw (axis cs:0.85,0.175995263367629) node[
  scale=0.6,
  anchor=base west,
  text=black,
  rotate=0.0
]{0.12\%};
\draw (axis cs:1.85,0.129659325087725) node[
  scale=0.6,
  anchor=base west,
  text=black,
  rotate=0.0
]{0.12\%};
\draw (axis cs:2.85,0.112413572660927) node[
  scale=0.6,
  anchor=base west,
  text=black,
  rotate=0.0
]{0.17\%};
\draw (axis cs:3.85,0.103412207619248) node[
  scale=0.6,
  anchor=base west,
  text=black,
  rotate=0.0
]{0.16\%};
\draw (axis cs:4.85,0.0978258583594561) node[
  scale=0.6,
  anchor=base west,
  text=black,
  rotate=0.0
]{0.15\%};
\end{axis}
\end{tikzpicture}
    \caption{Estimating $\bY$}
\end{subfigure}
\caption{\small MSE of factorization problem. MSE is normalized by the norm of the signal. $\bX$ is a shifted Wigner matrix with $c=3$, and both $\bY$ and $\bW$ are $N \times M$ matrices with i.i.d. Gaussian entries of variance $1/N$, and $N/M = 1/2$. The RIE is applied to $N=2000, M =4000$. In each run, the observation matrix $\bS$ is generated according to \eqref{MF-model}, and the factors $\bX$, $\bY$ are estimated simultaneously from $\bS$. Results are averaged over 10 runs (error bars are invisible). Average relative error between RIEs and Oracle estimators is also reported.}
\label{MF-c=3}
\end{figure}

\clearpage
\section{Case of $\alpha \geq 1$}\label{alpha>1}
In this section we consider the case where $M \leq N$ and $N/M \to \alpha \geq 1$ as $N \to \infty$. Throughout this section $\bm{\Gamma} \in \bR^{N \times M}$ is a (tall) matrix with $\bm{\Gamma}_M$ in its upper $M \times M$ block, and the rest zero entries. $\bm{\Gamma}_M$ is diagonal matrix constructed from $\bm{\gamma} \in \bR^M$ which are the singular values of $\bS$.

Similar to the case of $\alpha \leq 1$, resolvent of the matrix $\cS \in \bR^{(N+M) \times (N+M)}$ plays a central role in deriving the RIEs. For the case of $M \geq N$, with $\bS = \bU_S \bm{\Gamma} \bV_S^{\intercal}$, the matrix $\cS$ has the following eigen-decomposition:
\begin{equation}
  \cS  = \left[
\begin{array}{ccc}
\hat{\bU}_S^{(1)} & -\hat{\bU}_S^{(1)} &   \bU_S^{(2)} \\
\hat{\bV}_S & -\hat{\bV}_S &  \mathbf{0}
\end{array}
\right] \left[
\begin{array}{ccc}
\bGam_M & \mathbf{0} & \mathbf{0}\\
\mathbf{0} & -\bGam_M &  \mathbf{0}\\
\mathbf{0} & \mathbf{0} & \mathbf{0} 
\end{array}
\right]  \left[
\begin{array}{ccc}
\hat{\bU}_S^{(1)} & -\hat{\bU}_S^{(1)} &   \bU_S^{(2)} \\
\hat{\bV}_S & -\hat{\bV}_S &  \mathbf{0}
\end{array}
\right]^\intercal
\label{EVD of S-alphag1}
\end{equation}
with $\bU_S = \left[
\begin{array}{cc}
\bU_S^{(1)} & \bU_S^{(2)}
\end{array}
\right]$ in which $\bU_S^{(1)} \in \bR^{N \times M}$. And, $\hat{\bU}_S^{(1)} = \frac{1}{\sqrt{2}} \bU_S^{(1)}$, $\hat{\bV}_S= \frac{1}{\sqrt{2}} \bV_S$. The resolvent of $\cS$ can be written as:
\begin{equation*}
    \bG_{\mathcal{S}}(x- \ci \epsilon) = \sum_{k=1}^{2M} \frac{x + \ci \epsilon }{(x - \tilde{\gamma}_k)^2+\epsilon^2} \bs_k \bs_k^\intercal +  \frac{x + \ci \epsilon }{x^2+\epsilon^2} \sum_{k=2M+1}^{M+N} \bs_k \bs_k^\intercal
\end{equation*}
where $\tilde{\gamma}_k$ are the eigenvalues of $\cS$, which are in fact the (signed) singular values of $\bS$, $\tilde{\gamma}_1 = \gamma_1, \hdots, \tilde{\gamma}_M=\gamma_M, \tilde{\gamma}_{M+1}= - \gamma_1, \hdots, \tilde{\gamma}_{2M}=-\gamma_M$.

\subsection{Estimating $\bX$}
The RIE for $\bX$ is constructed in the same way as in the case of $\alpha \leq 1$, \eqref{X-RIE-class}. However, in the present case the observation matrix $\bS$ has $M$ (non-trivially zero) singular values and we need to estimate $N$ eigenvalues for the RIE. As it will be clear, the $N-M$ eigenvalues are chosen to be equal.

\subsubsection{Relation between overlap and the resolvent}
Define the vectors $\tilde{\bx}_i = [ \bx_i^\intercal ,  \mathbf{0}_{M} ]^\intercal$ for $\bx_i$ eigenvectors of $\bX$. We have
\begin{equation}
\begin{split}
    \tilde{\bx}_i^\intercal \big( {\rm Im}\, \bG_{\mathcal{S}}(x - \ci \epsilon) \big) \tilde{\bx}_i =  \sum_{k=1}^{2M} \frac{ \epsilon }{(x - \tilde{\gamma}_k)^2+\epsilon^2} \big( \tilde{\bx}_i^\intercal \bs_k \big)^2  +  \frac{\epsilon }{x^2+\epsilon^2} \sum_{k=2M+1}^{M+N} \big( \tilde{\bx}_i^\intercal \bs_k \big)^2 
\end{split}
\end{equation}
Given the structure of $\bs_k$'s in \eqref{EVD of S-alphag1}, we have:
\begin{equation*}
  \big( \tilde{\bx}_i^\intercal \bs_k \big)^2 = \begin{cases}
      \frac{1}{2}  \big( \bx_i^\intercal \bu_k \big)^2  & {\rm for } \hspace{5pt} 1 \leq k \leq M \\
      \frac{1}{2}  \big( \bx_i^\intercal \bu_{k-M} \big)^2  & {\rm for } \hspace{5pt} M + 1 \leq k \leq 2 M \\
      \big( \bx_i^\intercal \bu_{k-M} \big)^2  & {\rm for } \hspace{5pt} 2 M + 1 \leq k \leq M + N 
  \end{cases}
\end{equation*}

We assume that in the limit of large N this quantity concentrates on $O_X(\gamma_j, \lambda_i)$ and depends only on the singular values and eigenvalue pairs $(\gamma_j, \lambda_i)$. This assumption implies that the singular vectors associated with $0$ singular values ($\bu_j$ for $M+1 \leq j \leq N$) all have the same overlap with the eigenvectors of $\bX$, $O_X(0, \lambda_i)$.  We thus have:
\begin{equation}
    \tilde{\bx}_i^\intercal \big( {\rm Im}\, \bG_{\mathcal{S}}(x - \ci \epsilon) \big) \tilde{\bx}_i \xrightarrow[]{N \to \infty} \frac{1}{\alpha} \int_\bR \frac{\epsilon}{(x - t)^2+ \epsilon^2} O_X(t,\lambda_i) \bar{\mu}_{S}(t) \, dt + \big( 1 - \frac{1}{\alpha} \big) \frac{\epsilon }{x^2+\epsilon^2} O_X(0, \lambda_i)
\end{equation}
where the overlap function $O_X(t,\lambda_i) $ is extended (continuously) to arbitrary values within the support of $\bar{\mu}_{S}$ (the symmetrized limiting singular value distribution of $\bS$) with the property that $O_X(t,\lambda_i) = O_X(-t,\lambda_i)$ for $ t \in {\rm supp} (\mu_S)$ . Sending $\epsilon \to 0$, we find 
\begin{equation}
    \tilde{\bx}_i^\intercal \big( {\rm Im}\, \bG_{\mathcal{S}}(x - \ci \epsilon) \big)\tilde{\bx}_i \to \frac{1}{\alpha} \pi \bar{\mu}_{S}(x) O_X(x, \lambda_i) + \big( 1 - \frac{1}{\alpha} \big) \pi \delta(x) O_X(x, \lambda_i)
    \label{X-resolvent-overlap-alphag1}
\end{equation}

\subsubsection{Resolvent relation}
We derive the resolvent relation for the same model as in \eqref{X-model}. The derivation is similar to the procedure explained in section \ref{X-resolvent-rel}, and we omit here. The final resolvent relation is the same as \eqref{X-resolvent-relation-app}, with parameters satisfying:
\begin{equation}
    \begin{cases}
    \zeta_1^* = \frac{1}{\alpha} \frac{\mathcal{C}_{\mu_W}^{(\nicefrac{1}{\alpha})}(p_1^* p_2^*)}{p_1^*}, \quad \zeta_2^* = \frac{1}{p_2^*} \big( \mathcal{C}_{\mu_W}^{(\nicefrac{1}{\alpha})}(p_1^* p_2^*) + \mathcal{C}_{\mu_Y}^{(\nicefrac{1}{\alpha})}(p_2^* p_3^*) \big), \quad \zeta_3^* = \frac{1}{\alpha} \frac{\mathcal{C}_{\mu_Y}^{(\nicefrac{1}{\alpha})}(p_2^* p_3^*)}{p_3^*} \\
    \\
    p_1^* = \frac{1}{\zeta_3^*} \mathcal{G}_{\rho_{X^2}} \big( \frac{z - \zeta_1^* }{\zeta_3^*} \big), \quad p_2^* = \frac{1}{z - \zeta_2^*}, \quad p_3^* = \frac{z - \zeta_1^*}{{\zeta_3^*}^2} \mathcal{G}_{\rho_{X^2}} \big( \frac{z - \zeta_1^* }{\zeta_3^*} \big) - \frac{1}{\zeta_3^*}
    \end{cases}
    \label{X-sol-alphag1}
\end{equation}
Again, with the same procedure as \eqref{trace-GS-X-first},\eqref{trace-GS-X-last}, the saddle point equations \eqref{X-sol-alphag1} can be rewritten in a simplified form, which does not involve $\rho_{X^2}$,  as:
\begin{equation}
    \begin{cases}
    \zeta_1^* = \frac{1}{\alpha} \frac{\mathcal{C}_{\mu_W}^{(\nicefrac{1}{\alpha})}(p_1^* p_2^*)}{p_1^*}, \quad \zeta_2^* = z - \frac{1}{\mathcal{G}_{\bar{\mu}_S} (z)}, \quad \zeta_3^* = \frac{1}{\alpha} \frac{\mathcal{C}_{\mu_Y}^{(\nicefrac{1}{\alpha})}(p_2^* p_3^*)}{p_3^*} \\
    \\
    p_1^* = \frac{1}{\alpha} \mathcal{G}_{\bar{\mu}_S} (z) + \big(1 - \frac{1}{\alpha} \big) \frac{1}{z}, \quad p_2^* = \mathcal{G}_{\bar{\mu}_S} (z), \quad p_3^* = \frac{z - \zeta_1^*}{\alpha \zeta_3^*} \mathcal{G}_{\bar{\mu}_S} (z) + \frac{z - \zeta_1^*}{ \zeta_3^*} \big(1 - \frac{1}{\alpha} \big) \frac{1}{z} - \frac{1}{\zeta_3^*} 
    \end{cases}
    \label{X-sol-alphag1-sec}
\end{equation}
with $\bar{\mu}_S$ the limiting ESD of non-trivial singular values of $\bS$. Note that $\zeta_1^*, \zeta_2^*$ can be computed from the observation matrix, and we only need to find $\zeta_3^*$ satisfying the following equation:
\begin{equation}
    (z - \zeta_1^*) \big[  \frac{1}{\alpha} \mathcal{G}_{\bar{\mu}_S} (z) + \big(1 - \frac{1}{\alpha} \big) \frac{1}{z} \big] - 1 = \frac{1}{\alpha} \mathcal{C}_{\mu_Y}^{(\nicefrac{1}{\alpha})}\Big( \frac{1}{\zeta_3^*} \mathcal{G}_{\bar{\mu}_S} (z) (z - \zeta_1^*) \big[  \frac{1}{\alpha} \mathcal{G}_{\bar{\mu}_S} (z) + \big(1 - \frac{1}{\alpha} \big) \frac{1}{z} \big] \Big)
    \label{zeta3-X-est-alphag1}
\end{equation}

Note that both sets of equations \eqref{X-resolvent-overlap-alphag1}, \eqref{X-sol-alphag1-sec} and \eqref{X-resolvent-overlap-app}, \eqref{X-sol-app-sec} match for $\alpha = 1$.

\subsubsection{Overlaps and optimal eigenvalues}
From \eqref{X-resolvent-overlap-alphag1}, \eqref{X-resolvent-relation-app}, for $\gamma$ a non-trivially zero singular value of $\bS$ we find:
\begin{equation}
\begin{split}
    O_X(\gamma, \lambda_i) &\approx \frac{\alpha}{\pi \bar{\mu}_{S}(\gamma)} \, {\rm Im} \, \lim_{z \to \gamma - \ci 0^+} \, \bx_i^\intercal \, {\zeta_3^*}^{-1} \bG_{X^2} \big( \frac{z - \zeta^*_1 }{\zeta^*_3} \big) \, \bx_i \\
    &= \frac{\alpha}{\pi \bar{\mu}_{S}(\gamma)} \, {\rm Im} \, \lim_{z \to \gamma - \ci 0^+} \,  \frac{1}{ z - \zeta^*_1  - \zeta^*_3 \lambda_i^2}
\end{split}
\label{X-overlap-eq-alphag1-nontr}
\end{equation}
And, in the case of $M>N$, for zero singular values we have:
\begin{equation}
\begin{split}
    O_X(0, \lambda_i) &\approx \frac{\alpha}{(\alpha - 1) \pi } \, {\rm Im} \, \lim_{z \to  - \ci 0^+} \, \bx_i^\intercal \, {\zeta_3^*}^{-1} \bG_{X^2} \big( \frac{z - \zeta^*_1 }{\zeta^*_3} \big) \, \bx_i \\
    &= \frac{\alpha}{(\alpha - 1) \pi } \, {\rm Im} \, \lim_{z \to  - \ci 0^+} \,  \frac{1}{ z - \zeta^*_1  - \zeta^*_3 \lambda_i^2}
\end{split}
\label{X-overlap-eq-alphag1-triv}
\end{equation}
Finally, the optimal eigenvalues can be derived in the same way as in \eqref{X-optimal-ev-app}. For $1 \leq i \leq M$, we have:
\begin{equation}
\begin{split}
    \widehat{\xi_x^*}_i = \frac{\alpha}{2 \kappa \pi \bar{\mu}_{S}(\gamma_i) } \, {\rm Im}  \, \lim_{z \to \gamma_i - \ci 0^+}  \,  \bigg\{  \frac{1}{\zeta_3^*}  \Big[\mathcal{G}_{\rho_X}\Big(\sqrt{\frac{z-\zeta_1^*}{\kappa \zeta_3^*}}\Big) +  \mathcal{G}_{\rho_{X}}\Big(-\sqrt{\frac{z-\zeta_1^*}{\kappa \zeta_3^*}}\Big) \Big] \bigg\}
\end{split}
\label{X-optimal-ev-alphag1-nontr}
\end{equation}
And, for all $ M+1 \leq i \leq N$ :
\begin{equation}
\begin{split}
    \widehat{\xi_x^*}_i = \frac{\alpha}{2 \kappa (\alpha - 1) \pi } \, {\rm Im}  \, \lim_{z \to  - \ci 0^+}  \,  \bigg\{  \frac{1}{\zeta_3^*}  \Big[\mathcal{G}_{\rho_X}\Big(\sqrt{\frac{z-\zeta_1^*}{\kappa \zeta_3^*}}\Big) +  \mathcal{G}_{\rho_{X}}\Big(-\sqrt{\frac{z-\zeta_1^*}{\kappa \zeta_3^*}}\Big) \Big] \bigg\}
\end{split}
\label{X-optimal-ev-alphag1-triv}
\end{equation}


\subsubsection{Numerical Examples}
For matrices $\bY, \bW \in \bR^{N \times M}$ with i.i.d. Gaussian entries of variance $\nicefrac{1}{N}$ and $ M > N$, we have that $\mathcal{C}^{(\nicefrac{1}{\alpha})}_{\mu_Y}(z) = \mathcal{C}^{(\nicefrac{1}{\alpha})}_{\mu_W}(z) =  z$ which leads to a simplification of equations \eqref{X-sol-alphag1-sec}: 
\begin{equation}
    \begin{cases}
    \zeta_1^* = \frac{1}{\alpha} p_2^*, \quad \zeta_2^* = z - \frac{1}{\mathcal{G}_{\bar{\mu}_S} (z)}, \quad \zeta_3^* = \frac{1}{\alpha}p_2^*  \\
    \\
    p_1^* = \frac{1}{\alpha} \mathcal{G}_{\bar{\mu}_S} (z) + \big(1 - \frac{1}{\alpha} \big) \frac{1}{z}, \quad p_2^* = \mathcal{G}_{\bar{\mu}_S} (z), \quad p_3^* = \frac{z - \zeta_1^*}{\alpha \zeta_3^*} \mathcal{G}_{\bar{\mu}_S} (z) + \frac{z - \zeta_1^*}{ \zeta_3^*} \big(1 - \frac{1}{\alpha} \big) \frac{1}{z} - \frac{1}{\zeta_3^*} 
    \end{cases}
    \label{X-Y,W-Gauss-alphag1}
\end{equation}
Therefore, $\zeta_1^* = \zeta_3^* = \frac{1}{\alpha} \mathcal{G}_{\bar{\mu}_S} (z)$.

\begin{figure}
    \centering
\begin{subfigure}[t]{.4\textwidth}
    \centering
\begin{tikzpicture}[scale = 0.6]

\definecolor{darkgray176}{RGB}{176,176,176}
\definecolor{lightcoral}{RGB}{240,128,128}
\definecolor{lightslategray}{RGB}{119,136,153}
\definecolor{royalblue}{RGB}{65,105,225}

\begin{axis}[
legend cell align={left},
legend style={fill opacity=0.8, draw opacity=1, text opacity=1, draw=white!80!black},
tick align=outside,
tick pos=left,
x grid style={darkgray176},
xlabel={$\kappa$},
xmin=-0.145, xmax=5.7,
xtick style={color=black},
y grid style={darkgray176},
ylabel={${\rm MSE}$},
yticklabel style={
  /pgf/number format/precision=3,
  /pgf/number format/fixed},
ymin=0.0832227579219618, ymax=0.0995406468450475,
ytick style={color=black}
]
\path [draw=lightcoral, semithick]
(axis cs:0.1,0.0955091398880763)
--(axis cs:0.1,0.0956785267080883);

\path [draw=lightcoral, semithick]
(axis cs:0.3,0.0905023151498059)
--(axis cs:0.3,0.0906534747970803);

\path [draw=lightcoral, semithick]
(axis cs:0.6,0.0876858649996227)
--(axis cs:0.6,0.0878797641770603);

\path [draw=lightcoral, semithick]
(axis cs:1,0.0861893082974154)
--(axis cs:1,0.0864400625926959);

\path [draw=lightcoral, semithick]
(axis cs:2,0.0848520124688365)
--(axis cs:2,0.0850426215414082);

\path [draw=lightcoral, semithick]
(axis cs:3,0.0843690552202474)
--(axis cs:3,0.0845096748799747);

\path [draw=lightcoral, semithick]
(axis cs:4,0.0841400874651255)
--(axis cs:4,0.0842791876463864);

\path [draw=lightcoral, semithick]
(axis cs:5,0.0839644801457385)
--(axis cs:5,0.0841352298479381);

\path [draw=lightslategray, semithick]
(axis cs:0.1,0.0985882211762003)
--(axis cs:0.1,0.0987989246212709);

\path [draw=lightslategray, semithick]
(axis cs:0.3,0.091879270040749)
--(axis cs:0.3,0.0920344140030081);

\path [draw=lightslategray, semithick]
(axis cs:0.6,0.0885444176050747)
--(axis cs:0.6,0.0887582642153025);

\path [draw=lightslategray, semithick]
(axis cs:1,0.0868590479356263)
--(axis cs:1,0.0871603262078542);

\path [draw=lightslategray, semithick]
(axis cs:2,0.0855194732071358)
--(axis cs:2,0.0857209643013465);

\path [draw=lightslategray, semithick]
(axis cs:3,0.0850815934922544)
--(axis cs:3,0.0852736154785304);

\path [draw=lightslategray, semithick]
(axis cs:4,0.0848705580179885)
--(axis cs:4,0.0851677417507626);

\path [draw=lightslategray, semithick]
(axis cs:5,0.0847201851128233)
--(axis cs:5,0.0849826886768351);

\addplot [semithick, red, mark=triangle*, mark size=3, mark options={solid,rotate=180}, only marks]
table {%
0.1 0.0955938332980823
0.3 0.0905778949734431
0.6 0.0877828145883415
1 0.0863146854450556
2 0.0849473170051223
3 0.0844393650501111
4 0.0842096375557559
5 0.0840498549968383
};
\addlegendentry{Oracle estimator, ${\bm{\Xi}_X^*}(\bS)$}
\addplot [semithick, royalblue, mark=triangle*, mark size=3, mark options={solid}, only marks]
table {%
0.1 0.0986935728987356
0.3 0.0919568420218786
0.6 0.0886513409101886
1 0.0870096870717403
2 0.0856202187542411
3 0.0851776044853924
4 0.0850191498843755
5 0.0848514368948292
};
\addlegendentry{RIE, $\widehat{{\bm{\Xi}_X^*}}(\bS)$}
\draw (axis cs:0.2,0.0955938332980823) node[
  scale=0.6,
  anchor=base west,
  text=black,
  rotate=0.0
]{3.24\%};
\draw (axis cs:0.4,0.0905778949734431) node[
  scale=0.6,
  anchor=base west,
  text=black,
  rotate=0.0
]{1.52\%};
\draw (axis cs:0.7,0.0877828145883415) node[
  scale=0.6,
  anchor=base west,
  text=black,
  rotate=0.0
]{0.99\%};
\draw (axis cs:1.1,0.0863146854450556) node[
  scale=0.6,
  anchor=base west,
  text=black,
  rotate=0.0
]{0.81\%};
\draw (axis cs:2.1,0.0849473170051223) node[
  scale=0.6,
  anchor=base west,
  text=black,
  rotate=0.0
]{0.79\%};
\draw (axis cs:3.1,0.0844393650501111) node[
  scale=0.6,
  anchor=base west,
  text=black,
  rotate=0.0
]{0.87\%};
\draw (axis cs:4.1,0.0842096375557559) node[
  scale=0.6,
  anchor=base west,
  text=black,
  rotate=0.0
]{0.96\%};
\draw (axis cs:5.1,0.0840498549968383) node[
  scale=0.6,
  anchor=base west,
  text=black,
  rotate=0.0
]{0.95\%};
\end{axis}

\end{tikzpicture}
    \caption{Shifted Wigner, $c=3$}
\end{subfigure}
\hfil
\begin{subfigure}[t]{.4\textwidth}
    \centering
\begin{tikzpicture}[scale = 0.6]

\definecolor{darkgray176}{RGB}{176,176,176}
\definecolor{lightcoral}{RGB}{240,128,128}
\definecolor{lightslategray}{RGB}{119,136,153}
\definecolor{royalblue}{RGB}{65,105,225}

\begin{axis}[
legend cell align={left},
legend style={fill opacity=0.8, draw opacity=1, text opacity=1, draw=white!80!black},
tick align=outside,
tick pos=left,
x grid style={darkgray176},
xlabel={$\kappa$},
xmin=-0.145, xmax=5.7,
xtick style={color=black},
y grid style={darkgray176},
ylabel={${\rm MSE}$},
scaled y ticks=false,
yticklabel style={
  /pgf/number format/precision=3,
  /pgf/number format/fixed},
ymin=0.135555306836279, ymax=0.173282677917095,
ytick style={color=black}
]
\path [draw=lightcoral, semithick]
(axis cs:0.1,0.16914301432887)
--(axis cs:0.1,0.169607882939905);

\path [draw=lightcoral, semithick]
(axis cs:0.3,0.152155346113923)
--(axis cs:0.3,0.152375825997895);

\path [draw=lightcoral, semithick]
(axis cs:0.6,0.145215612001563)
--(axis cs:0.6,0.145484501978514);

\path [draw=lightcoral, semithick]
(axis cs:1,0.141885628492589)
--(axis cs:1,0.142090767916165);

\path [draw=lightcoral, semithick]
(axis cs:2,0.139074056336078)
--(axis cs:2,0.139315034879198);

\path [draw=lightcoral, semithick]
(axis cs:3,0.138110532276817)
--(axis cs:3,0.138406165522007);

\path [draw=lightcoral, semithick]
(axis cs:4,0.137585833244901)
--(axis cs:4,0.137871706492);

\path [draw=lightcoral, semithick]
(axis cs:5,0.137270187339952)
--(axis cs:5,0.137658833032843);

\path [draw=lightslategray, semithick]
(axis cs:0.1,0.171160918138194)
--(axis cs:0.1,0.171567797413421);

\path [draw=lightslategray, semithick]
(axis cs:0.3,0.153056482213234)
--(axis cs:0.3,0.153374278342356);

\path [draw=lightslategray, semithick]
(axis cs:0.6,0.145958262702928)
--(axis cs:0.6,0.146203634106491);

\path [draw=lightslategray, semithick]
(axis cs:1,0.142585901105339)
--(axis cs:1,0.142880730750559);

\path [draw=lightslategray, semithick]
(axis cs:2,0.139864928846079)
--(axis cs:2,0.140169988426055);

\path [draw=lightslategray, semithick]
(axis cs:3,0.139036299481753)
--(axis cs:3,0.139283043660949);

\path [draw=lightslategray, semithick]
(axis cs:4,0.138520703969679)
--(axis cs:4,0.138877447431376);

\path [draw=lightslategray, semithick]
(axis cs:5,0.138257810712016)
--(axis cs:5,0.138753681862458);

\addplot [semithick, red, mark=triangle*, mark size=3, mark options={solid,rotate=180}, only marks]
table {%
0.1 0.169375448634388
0.3 0.152265586055909
0.6 0.145350056990039
1 0.141988198204377
2 0.139194545607638
3 0.138258348899412
4 0.13772876986845
5 0.137464510186398
};
\addlegendentry{Oracle estimator, ${\bm{\Xi}_X^*}(\bS)$}
\addplot [semithick, royalblue, mark=triangle*, mark size=3, mark options={solid}, only marks]
table {%
0.1 0.171364357775807
0.3 0.153215380277795
0.6 0.14608094840471
1 0.142733315927949
2 0.140017458636067
3 0.139159671571351
4 0.138699075700527
5 0.138505746287237
};
\addlegendentry{RIE, $\widehat{{\bm{\Xi}_X^*}}(\bS)$}
\draw (axis cs:0.2,0.169375448634388) node[
  scale=0.6,
  anchor=base west,
  text=black,
  rotate=0.0
]{1.17\%};
\draw (axis cs:0.4,0.152265586055909) node[
  scale=0.6,
  anchor=base west,
  text=black,
  rotate=0.0
]{0.62\%};
\draw (axis cs:0.7,0.145350056990039) node[
  scale=0.6,
  anchor=base west,
  text=black,
  rotate=0.0
]{0.5\%};
\draw (axis cs:1.1,0.141988198204377) node[
  scale=0.6,
  anchor=base west,
  text=black,
  rotate=0.0
]{0.52\%};
\draw (axis cs:2.1,0.139194545607638) node[
  scale=0.6,
  anchor=base west,
  text=black,
  rotate=0.0
]{0.59\%};
\draw (axis cs:3.1,0.138258348899412) node[
  scale=0.6,
  anchor=base west,
  text=black,
  rotate=0.0
]{0.65\%};
\draw (axis cs:4.1,0.13772876986845) node[
  scale=0.6,
  anchor=base west,
  text=black,
  rotate=0.0
]{0.7\%};
\draw (axis cs:5.1,0.137464510186398) node[
  scale=0.6,
  anchor=base west,
  text=black,
  rotate=0.0
]{0.76\%};
\end{axis}

\end{tikzpicture}
    \caption{Wishart, $\alpha' = \nicefrac{1}{4}$}
\end{subfigure}
\caption{\small Estimating $\bX$. The MSE is normalized by the norm of the signal, $\| \bX \|_{\rm F}^2$. Both $\bY$ and $\bW$ are $N \times M$ matrices with i.i.d. Gaussian entries of variance $1/N$, and aspect ratio $N/M = 2$. The RIE is applied to $N=2000, M =1000$, and the results are averaged over 10 runs (error bars are invisible). Average relative error between RIE $\widehat{{\bm{\Xi}_X^*}}(\bS)$ and Oracle estimator is also reported.}
\label{Wigner-Wishart-X-alphag1}
\end{figure}
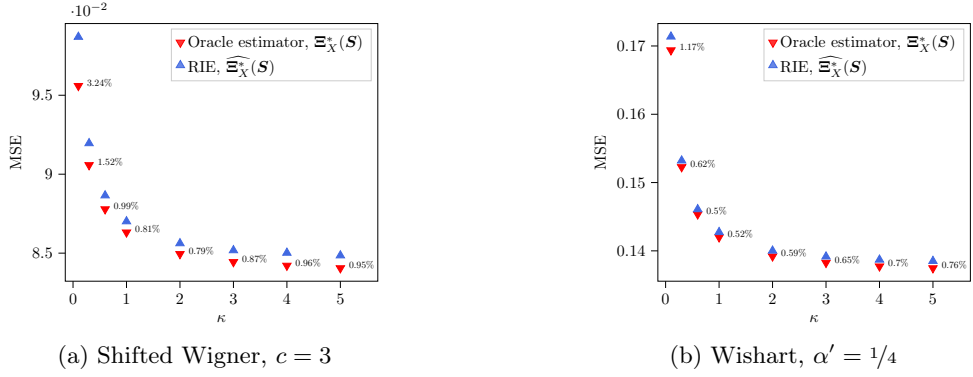

In Figure \ref{Wigner-Wishart-X-alphag1}, the MSE of the Oracle estimator and the RIE \eqref{X-optimal-ev-alphag1-nontr}, \eqref{X-optimal-ev-alphag1-triv} is illustrated for shifted Wigner $\bX$ with $c=3$, and Wishart with aspect-ratio $\alpha' = \nicefrac{1}{4}$.

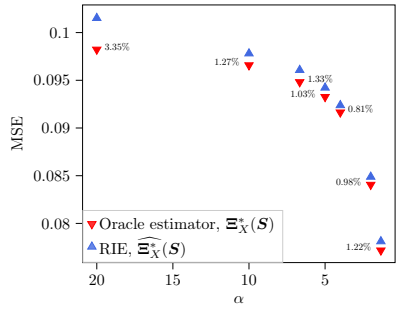
\begin{figure}
    \centering
\begin{tikzpicture}[scale = 0.6]

\definecolor{darkgray176}{RGB}{176,176,176}
\definecolor{lightcoral}{RGB}{240,128,128}
\definecolor{lightslategray}{RGB}{119,136,153}
\definecolor{royalblue}{RGB}{65,105,225}

\begin{axis}[
legend cell align={left},
legend style={
  fill opacity=0.8,
  draw opacity=1,
  text opacity=1,
  at={(0,0)},
  anchor=south west,
  draw=white!80!black
},
tick align=outside,
tick pos=left,
x dir=reverse,
x grid style={darkgray176},
xlabel={$\alpha$},
xmin=0.4, xmax=20.9333333333333,
xtick style={color=black},
y grid style={darkgray176},
ylabel={${\rm MSE}$},
yticklabel style={
  /pgf/number format/precision=3,
  /pgf/number format/fixed},
ymin=0.0758941436001279, ymax=0.10289741549371,
ytick style={color=black}
]
\path [draw=lightcoral, semithick]
(axis cs:20,0.0981399445168636)
--(axis cs:20,0.098306162651221);

\path [draw=lightcoral, semithick]
(axis cs:10,0.0964893210575839)
--(axis cs:10,0.0966741584433507);

\path [draw=lightcoral, semithick]
(axis cs:6.66666666666667,0.0947420632643503)
--(axis cs:6.66666666666667,0.0948804760329659);

\path [draw=lightcoral, semithick]
(axis cs:5,0.0931538203338095)
--(axis cs:5,0.0933702003956913);

\path [draw=lightcoral, semithick]
(axis cs:4,0.0915500667572145)
--(axis cs:4,0.0917224284019286);

\path [draw=lightcoral, semithick]
(axis cs:2,0.083979348574348)
--(axis cs:2,0.0841337235618452);

\path [draw=lightcoral, semithick]
(axis cs:1.33333333333333,0.0771215650498362)
--(axis cs:1.33333333333333,0.0772321391515741);

\path [draw=lightslategray, semithick]
(axis cs:20,0.101363486174665)
--(axis cs:20,0.101669994044002);

\path [draw=lightslategray, semithick]
(axis cs:10,0.0976793207715828)
--(axis cs:10,0.0979356482314223);

\path [draw=lightslategray, semithick]
(axis cs:6.66666666666667,0.0959602135455994)
--(axis cs:6.66666666666667,0.0961825953420452);

\path [draw=lightslategray, semithick]
(axis cs:5,0.0940966252044989)
--(axis cs:5,0.0943491129036909);

\path [draw=lightslategray, semithick]
(axis cs:4,0.092293341093632)
--(axis cs:4,0.0924663249148742);

\path [draw=lightslategray, semithick]
(axis cs:2,0.084757120379403)
--(axis cs:2,0.0850081901041899);

\path [draw=lightslategray, semithick]
(axis cs:1.33333333333333,0.0780057340930656)
--(axis cs:1.33333333333333,0.0782261788609194);

\addplot [semithick, red, mark=triangle*, mark size=3, mark options={solid,rotate=180}, only marks]
table {%
20 0.0982230535840423
10 0.0965817397504673
6.66666666666667 0.0948112696486581
5 0.0932620103647504
4 0.0916362475795715
2 0.0840565360680966
1.33333333333333 0.0771723148803089
};
\addlegendentry{Oracle estimator, ${\bm{\Xi}_X^*}(\bS)$}
\addplot [semithick, royalblue, mark=triangle*, mark size=3, mark options={solid}, only marks]
table {%
20 0.101516740109333
10 0.0978074845015026
6.66666666666667 0.0960714044438223
5 0.0942228690540949
4 0.0923798330042531
2 0.0848826552417965
1.33333333333333 0.0781159564769925
};
\addlegendentry{RIE, $\widehat{{\bm{\Xi}_X^*}}(\bS)$}
\draw (axis cs:19.7,0.0982230535840423) node[
  scale=0.6,
  anchor=base west,
  text=black,
  rotate=0.0
]{3.35\%};
\draw (axis cs:12.5,0.0965817397504673) node[
  scale=0.6,
  anchor=base west,
  text=black,
  rotate=0.0
]{1.27\%};
\draw (axis cs:6.36666666666667,0.0948112696486581) node[
  scale=0.6,
  anchor=base west,
  text=black,
  rotate=0.0
]{1.33\%};
\draw (axis cs:7.5,0.0932620103647504) node[
  scale=0.6,
  anchor=base west,
  text=black,
  rotate=0.0
]{1.03\%};
\draw (axis cs:3.7,0.0916362475795715) node[
  scale=0.6,
  anchor=base west,
  text=black,
  rotate=0.0
]{0.81\%};
\draw (axis cs:4.5,0.0840565360680966) node[
  scale=0.6,
  anchor=base west,
  text=black,
  rotate=0.0
]{0.98\%};
\draw (axis cs:3.83333333333333,0.0771723148803089) node[
  scale=0.6,
  anchor=base west,
  text=black,
  rotate=0.0
]{1.22\%};
\end{axis}

\end{tikzpicture}
    \caption{\small MSE of estimating $\bX$ as a function of aspect-ratio $\alpha > 1$, prior on $\bX$ is shifted Wigner with $c =3$, and $\kappa = 5$. MSE is normalized by the norm of the signal, $\| \bX \|_{\rm F}^2$. Both $\bY$ and $\bW$ are $N \times M$ matrices with i.i.d. Gaussian entries of variance $1/N$. The RIE is applied to $N=2000, M = \nicefrac{1}{\alpha} N$, and the results are averaged over 10 runs (error bars are invisible). Average relative error between RIE $\widehat{{\bm{\Xi}_X^*}}(\bS)$ and Oracle estimator is also reported.}
    \label{X-ar-alphag1}
\end{figure}

\paragraph{Effect of aspect-ratio $\alpha$.} In Figure \ref{X-ar-alphag1}, we take $\bX$ to be a shifted Wigner matrix with $c=3$, and the MSE is depicted for various values of the aspect-ratio $\alpha > 1$. We see that as $M$ decreases ($\alpha$ increases) the estimation error (of $\bY$) increases.

\subsection{Estimating $\bY$}
\subsubsection{Relation between overlap and the resolvent}
For the vectors  $\br_i = \left[
\begin{array}{c}
\mathbf{0}_N \\
\by_i^{(r)}
\end{array}
\right]$, $\bl_i =\left[
\begin{array}{c}
\by_i^{(l)} \\
\mathbf{0}_M
\end{array}
\right]$ with $\by_i^{(r)}, \by_i^{(l)}$ right/ left singular vectors of $\bY$, we have
\begin{equation}
\begin{split}
    \br_i^\intercal \big( {\rm Im}\, \bG_{\mathcal{S}}(x - \ci \epsilon) \big) \bl_i = \sum_{k=1}^{2 M} \frac{\epsilon }{(x - \tilde{\gamma}_k)^2+\epsilon^2} \big( \br_i^\intercal \bs_k \big) \big( \bl_i^\intercal \bs_k \big) + \frac{\epsilon}{x^2 + \epsilon^2} \sum_{k=2M+1}^{M+N}  \big( \br_i^\intercal \bs_k \big) \big( \bl_i^\intercal \bs_k \big)
\end{split}
\end{equation}
Given the structure of $\bs_k$'s in \eqref{EVD of S-alphag1},  we have:
\begin{equation*}
    \big( \br_i^\intercal \bs_k \big) \big( \bl_i^\intercal \bs_k \big) = \begin{cases}
      \frac{1}{2}  \big(  \bu_k^\intercal \by_i^{(l)}\big) \big(  \bv_k^\intercal \by_i^{(r)} \big) & {\rm for } \hspace{5pt} 1 \leq k \leq M \\
      - \frac{1}{2}  \big(  \bu_{k-M}^\intercal \by_i^{(l)}\big) \big(  \bv_{k-M}^\intercal \by_i^{(r)} \big)  & {\rm for }  \hspace{5pt} M+1 \leq k \leq 2M \\
      0 & {\rm for }  \hspace{5pt} 2M+1 \leq k \leq N+M
  \end{cases}
\end{equation*}
Therefore, in the limit $N \to \infty$, we have:
\begin{equation}
     \br_i^\intercal \big( {\rm Im}\, \bG_{\mathcal{S}}(x - \ci \epsilon) \big) \bl_i \xrightarrow[]{N \to \infty} \frac{1}{\alpha} \int_\bR \frac{\epsilon}{(x - t)^2+ \epsilon^2} O_Y(t,\sigma_i) \bar{\mu}_{S}(t) \, dt 
\end{equation}
where the overlap function $O_Y(t,\lambda_i) $ is extended (continuously) to arbitrary values within the support of $\bar{\mu}_{S}$ with the property that $O_Y(-t,\lambda_i) = - O_Y(t,\lambda_i)$ for $ t \in {\rm supp} (\mu_S)$ . Sending $\epsilon \to 0$, we find 
\begin{equation}
    \br_i^\intercal \big( {\rm Im}\, \bG_{\mathcal{S}}(x - \ci \epsilon) \big) \bl_i \approx \frac{1}{\alpha} \pi \bar{\mu}_{S}(x) O_Y(x, \sigma_i)
    \label{Y-resolvent-overlap-alphag1}
\end{equation}

\subsubsection{Resolvent relation}
The resolvent relation for the model \eqref{Y-model} with $M<N$ is the same as in \eqref{Y-resolvent relation} with parameters satisfying:
\begin{equation}
        \begin{cases}
    \beta_1^* = \frac{1}{\alpha} \frac{\mathcal{C}_{\mu_W}^{(\alpha)}(q_1^* q_2^*)}{q_1^*} + \frac{1}{2} \sqrt{\frac{q_3^*}{q_1^*}} \Big( \mathcal{R}_{\rho_X} \big( q_4^* + \sqrt{q_1^* q_3^*} \big) - \mathcal{R}_{\rho_X} \big( q_4^* - \sqrt{q_1^* q_3^*} \big) \Big)\\
    \beta_2^* = \frac{\mathcal{C}_{\mu_W}^{(\alpha)}(q_1^* q_2^*)}{q_2^*}\\
    \beta_3^* = \frac{1}{2} \sqrt{\frac{q_1^*}{q_3^*}} \Big( \mathcal{R}_{\rho_X} \big( q_4^* + \sqrt{q_1^* q_3^*} \big) - \mathcal{R}_{\rho_X} \big( q_4^* - \sqrt{q_1^* q_3^*} \big) \Big)\\
    \beta_4^* = \frac{1}{2}  \Big( \mathcal{R}_{\rho_X} \big( q_4^* + \sqrt{q_1^* q_3^*} \big) + \mathcal{R}_{\rho_X} \big( q_4^* - \sqrt{q_1^* q_3^*} \big) \Big)\\
    q_1^* = \frac{1}{\alpha} \frac{(z-\beta_2^*) {\beta_4^*}^2}{Z_2(z)^2} \mathcal{G}_{\rho_Y} \big( \frac{Z_1(z)}{Z_2(z)} \big) + \frac{1}{\alpha}\frac{\beta_3^*}{Z_2(z)} + \frac{\alpha - 1}{\alpha} \frac{1}{z - \beta_1^*}\\
    q_2^* = \frac{z-\beta_1^*}{Z_2(z)} \mathcal{G}_{\rho_Y} \big( \frac{Z_1(z)}{Z_2(z)} \big)  \\
    q_3^* = \frac{1}{\alpha} \frac{(z-\beta_1^*) Z_1(z)}{Z_2(z)^2} \mathcal{G}_{\rho_Y} \big( \frac{Z_1(z)}{Z_2(z)} \big) - \frac{1}{\alpha} \frac{z-\beta_1^*}{Z_2(z)} \\
    q_4^* = \frac{1}{\alpha} \frac{ \beta_4^* Z_1(z)}{Z_2(z)^2} \mathcal{G}_{\rho_Y} \big( \frac{Z_1(z)}{Z_2(z)} \big) - \frac{1}{\alpha} \frac{\beta_4^*}{Z_2(z)} 
    \end{cases} \quad \hspace{-80pt} \text{with }
    \begin{cases}
        Z_1(z) = (z-\beta_1^*)(z-\beta_2^*) \\
        Z_2(z) = {\beta_4^*}^2 + \beta_3^*(z-\beta_1^*)
    \end{cases}
    \label{Y-sol-alphag1}
\end{equation}
With the same procedure as \eqref{q1-star},\eqref{q2-star}, the saddle point equations \eqref{Y-sol-alphag1} can be rewritten in a simplified form:
\begin{equation}
    \begin{cases}
    \beta_1^* = \frac{1}{\alpha} \frac{\mathcal{C}_{\mu_W}^{(\alpha)}(q_1^* q_2^*)}{q_1^*} + \frac{1}{2} \sqrt{\frac{q_3^*}{q_1^*}} \Big( \mathcal{R}_{\rho_X} \big( q_4^* + \sqrt{q_1^* q_3^*} \big) - \mathcal{R}_{\rho_X} \big( q_4^* - \sqrt{q_1^* q_3^*} \big) \Big)\\
    \beta_2^* = \frac{\mathcal{C}_{\mu_W}^{(\alpha)}(q_1^* q_2^*)}{q_2^*}\\
    \beta_3^* = \frac{1}{2} \sqrt{\frac{q_1^*}{q_3^*}} \Big( \mathcal{R}_{\rho_X} \big( q_4^* + \sqrt{q_1^* q_3^*} \big) - \mathcal{R}_{\rho_X} \big( q_4^* - \sqrt{q_1^* q_3^*} \big) \Big)\\
    \beta_4^* = \frac{1}{2}  \Big( \mathcal{R}_{\rho_X} \big( q_4^* + \sqrt{q_1^* q_3^*} \big) + \mathcal{R}_{\rho_X} \big( q_4^* - \sqrt{q_1^* q_3^*} \big) \Big)\\
    q_1^* = \frac{1}{\alpha} \mathcal{G}_{\bar{\mu}_S} (z) + \big(1 - \frac{1}{\alpha} \big) \frac{1}{z} \\
    q_2^* = \mathcal{G}_{\bar{\mu}_S}(z) \\
    q_3^* = \frac{(z - \beta_1^*)^2}{{\beta_4^*}^2} q_1^* - \frac{z - \beta_1^*}{{\beta_4^*}^2} \\
    q_4^* = \frac{z - \beta_1^*}{\beta_4^*} q_1^* - \frac{1}{\beta_4^*}
    \end{cases} 
    \label{Y-sol-alphag1-simple}
\end{equation}

Note that both sets of equations \eqref{Y-resolvent-overlap-alphag1}, \eqref{Y-sol-alphag1-simple} and \eqref{Y-resolvent-overlap}, \eqref{Y-sol-app-simple} match for $\alpha = 1$.

\subsubsection{Overlaps and optimal singular values}
From \eqref{Y-resolvent relation}, \eqref{Y-resolvent-overlap-alphag1}, we have:
\begin{equation}
\begin{split}
    O_Y(\gamma, \sigma_i) &\approx \frac{\alpha}{\pi \bar{\mu}_{S}(\gamma)} \,  {\rm Im} \, \lim_{z \to \gamma - \ci 0^+} \, \frac{\beta_4^*}{Z_2(z)}  {\by^{(r)}_i}^\intercal \bG_{Y^\intercal Y} \big(\frac{Z_1(z)}{Z_2(z)} \big) \bY^\intercal \by^{(l)}_i \\
    &= \frac{\alpha}{\pi \bar{\mu}_{S}(\gamma)} \, {\rm Im} \, \lim_{z \to \gamma - \ci 0^+} \,   \beta_4^* \frac{\sigma_i}{ Z_1(z)  - Z_2(z) \sigma_i^2}
\end{split}
\label{Y-overlap-eq-alphag1}
\end{equation}
Similar to \eqref{Y-optimal-sv-app}, we can compute the optimal singular values to be:
\begin{equation}
\begin{split}
    \widehat{\xi_y^*}_i = \frac{\alpha}{\pi \bar{\mu}_{S}(\gamma_i)} \, {\rm Im} \, \lim_{z \to \gamma_i - \ci 0^+} q_4^*
    \label{Y-optimal-sv-alphag1}
\end{split}
\end{equation}

\subsubsection{Numerical examples}
We consider the matrix $\bW$ to have i.i.d. Gaussian entries with variance $\nicefrac{1}{N}$, so $\mathcal{C}_{\mu_W}^{(\nicefrac{1}{\alpha})}(z) =  z$. And, $\bX = \bF + c \bI$ where $\bF = \bF^\intercal \in \bR^{N \times N}$ has i.i.d. entries with variance $\nicefrac{1}{N}$, and $c \neq 0$ is a real number, so $\mathcal{R}_{\rho_X}(z) = z +c$. With these choices, the solution \eqref{Y-sol-alphag1-simple} simplifies to:
\begin{equation}
    \begin{cases}
    \beta_1^* = \frac{1}{\alpha} q_2^* + q_3^*, \quad \beta_2^* = q_1^*, \quad \beta_3^* = q_1^*, \quad \beta_4^* =  q_4^* + c\\
    q_1^* = \frac{1}{\alpha} \mathcal{G}_{\bar{\mu}_S} (z) + \big(1 - \frac{1}{\alpha} \big) \frac{1}{z} , \quad q_2^* = \mathcal{G}_{\bar{\mu}_S}(z) \\
    q_3^* = \frac{(z - \beta_1^*)^2}{{\beta_4^*}^2} q_1^* - \frac{z - \beta_1^*}{{\beta_4^*}^2}, \quad q_4^* = \frac{z - \beta_1^*}{\beta_4^*} q_1^* - \frac{1}{\beta_4^*}
    \end{cases} 
    \label{Y-sol-alphag1-X,W-G}
\end{equation}
After a bit of algebra, we find that $q_4^*$ is the solution to the following qubic equation:
\begin{equation}
\begin{split}
    2 x^3 + 3 c\, x^2 + \Big[ c^2 + 2 & - \big(z - \frac{1}{\alpha} \mathcal{G}_{\bar{\mu}_S}(z) \big) \big( \frac{1}{\alpha} \mathcal{G}_{\bar{\mu}_S}(z) + \frac{\alpha -1}{\alpha z} \big) \Big]\, x \\
    &- c \Big[ \big(z - \frac{1}{\alpha} \mathcal{G}_{\bar{\mu}_S}(z) \big) \big( \frac{1}{\alpha} \mathcal{G}_{\bar{\mu}_S}(z) + \frac{\alpha -1}{\alpha z} \big) - 1 \Big] = 0
\end{split}
    \label{q_4-eq-alphag1}
\end{equation}

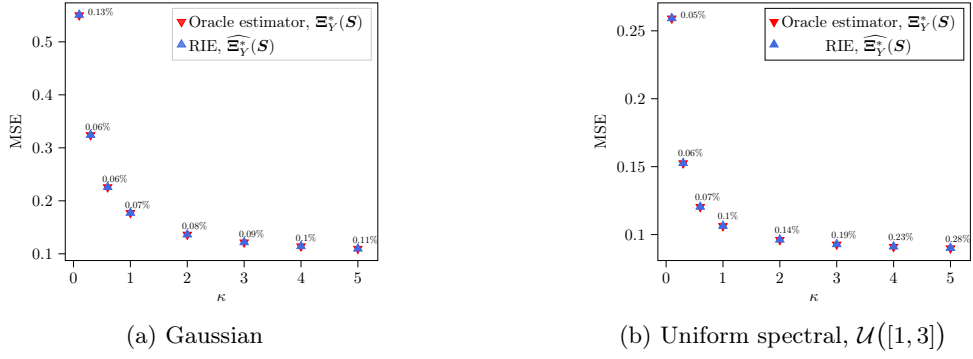
\begin{figure}
    \centering
\begin{subfigure}[t]{.4\textwidth}
    \centering
\begin{tikzpicture}[scale = 0.6]

\definecolor{darkgray176}{RGB}{176,176,176}
\definecolor{lightcoral}{RGB}{240,128,128}
\definecolor{lightslategray}{RGB}{119,136,153}
\definecolor{royalblue}{RGB}{65,105,225}

\begin{axis}[
legend cell align={left},
legend style={fill opacity=0.8, draw opacity=1, text opacity=1, draw=white!80!black},
tick align=outside,
tick pos=left,
x grid style={darkgray176},
xlabel={$\kappa$},
xmin=-0.145, xmax=5.35,
xtick style={color=black},
y grid style={darkgray176},
ylabel={${\rm MSE}$},
scaled y ticks=false,
yticklabel style={
  /pgf/number format/precision=3,
  /pgf/number format/fixed},
ymin=0.0873224132178036, ymax=0.573513400797319,
ytick style={color=black}
]
\path [draw=lightcoral, semithick]
(axis cs:0.1,0.549688108009911)
--(axis cs:0.1,0.550712127236451);

\path [draw=lightcoral, semithick]
(axis cs:0.3,0.323741184301787)
--(axis cs:0.3,0.324362725679531);

\path [draw=lightcoral, semithick]
(axis cs:0.6,0.225443812969368)
--(axis cs:0.6,0.226039455243397);

\path [draw=lightcoral, semithick]
(axis cs:1,0.176933563565292)
--(axis cs:1,0.177440839836802);

\path [draw=lightcoral, semithick]
(axis cs:2,0.136071133689125)
--(axis cs:2,0.136449039919983);

\path [draw=lightcoral, semithick]
(axis cs:3,0.121457979739386)
--(axis cs:3,0.121824827884159);

\path [draw=lightcoral, semithick]
(axis cs:4,0.11406173711567)
--(axis cs:4,0.114354683211531);

\path [draw=lightcoral, semithick]
(axis cs:5,0.109422003562327)
--(axis cs:5,0.109729127562647);

\path [draw=lightslategray, semithick]
(axis cs:0.1,0.550385268844045)
--(axis cs:0.1,0.551413810452796);

\path [draw=lightslategray, semithick]
(axis cs:0.3,0.323953109197546)
--(axis cs:0.3,0.324565046791671);

\path [draw=lightslategray, semithick]
(axis cs:0.6,0.225573940148503)
--(axis cs:0.6,0.226174229861782);

\path [draw=lightslategray, semithick]
(axis cs:1,0.177061217625596)
--(axis cs:1,0.177560845578233);

\path [draw=lightslategray, semithick]
(axis cs:2,0.136172357630392)
--(axis cs:2,0.136558095820947);

\path [draw=lightslategray, semithick]
(axis cs:3,0.121564581547344)
--(axis cs:3,0.12193227832868);

\path [draw=lightslategray, semithick]
(axis cs:4,0.114176854358612)
--(axis cs:4,0.114474316532998);

\path [draw=lightslategray, semithick]
(axis cs:5,0.109537919599659)
--(axis cs:5,0.109860479556356);

\addplot [semithick, red, mark=triangle*, mark size=3, mark options={solid,rotate=180}, only marks]
table {%
0.1 0.550200117623181
0.3 0.324051954990659
0.6 0.225741634106382
1 0.177187201701047
2 0.136260086804554
3 0.121641403811773
4 0.1142082101636
5 0.109575565562487
};
\addlegendentry{Oracle estimator, ${\bm{\Xi}_Y^*}(\bS)$}
\addplot [semithick, royalblue, mark=triangle*, mark size=3, mark options={solid}, only marks]
table {%
0.1 0.550899539648421
0.3 0.324259077994609
0.6 0.225874085005142
1 0.177311031601914
2 0.136365226725669
3 0.121748429938012
4 0.114325585445805
5 0.109699199578008
};
\addlegendentry{RIE, $\widehat{{\bm{\Xi}_Y^*}}(\bS)$}
\draw (axis cs:0.2,0.550200117623181) node[
  scale=0.6,
  anchor=base west,
  text=black,
  rotate=0.0
]{0.13\%};
\draw (axis cs:0.15,0.334051954990659) node[
  scale=0.6,
  anchor=base west,
  text=black,
  rotate=0.0
]{0.06\%};
\draw (axis cs:0.45,0.235741634106382) node[
  scale=0.6,
  anchor=base west,
  text=black,
  rotate=0.0
]{0.06\%};
\draw (axis cs:0.85,0.187187201701047) node[
  scale=0.6,
  anchor=base west,
  text=black,
  rotate=0.0
]{0.07\%};
\draw (axis cs:1.85,0.146260086804554) node[
  scale=0.6,
  anchor=base west,
  text=black,
  rotate=0.0
]{0.08\%};
\draw (axis cs:2.85,0.131641403811773) node[
  scale=0.6,
  anchor=base west,
  text=black,
  rotate=0.0
]{0.09\%};
\draw (axis cs:3.85,0.1242082101636) node[
  scale=0.6,
  anchor=base west,
  text=black,
  rotate=0.0
]{0.1\%};
\draw (axis cs:4.85,0.119575565562487) node[
  scale=0.6,
  anchor=base west,
  text=black,
  rotate=0.0
]{0.11\%};
\end{axis}

\end{tikzpicture}
    \caption{Gaussian}
\end{subfigure}
\hfil
\begin{subfigure}[t]{.4\textwidth}
    \centering
\begin{tikzpicture}[scale = 0.6]

\definecolor{darkgray176}{RGB}{176,176,176}
\definecolor{lightcoral}{RGB}{240,128,128}
\definecolor{lightslategray}{RGB}{119,136,153}
\definecolor{royalblue}{RGB}{65,105,225}

\begin{axis}[
tick align=outside,
tick pos=left,
x grid style={darkgray176},
xlabel={$\kappa$},
xmin=-0.145, xmax=5.35,
xtick style={color=black},
y grid style={darkgray176},
ylabel={${\rm MSE}$},
scaled y ticks=false,
yticklabel style={
  /pgf/number format/precision=3,
  /pgf/number format/fixed},
ymin=0.0808876462376157, ymax=0.270567390958171,
ytick style={color=black}
]
\path [draw=lightcoral, semithick]
(axis cs:0.1,0.25641674013027)
--(axis cs:0.1,0.261813001386205);

\path [draw=lightcoral, semithick]
(axis cs:0.3,0.150846932383893)
--(axis cs:0.3,0.154219572074255);

\path [draw=lightcoral, semithick]
(axis cs:0.6,0.119582853034795)
--(axis cs:0.6,0.121123150971919);

\path [draw=lightcoral, semithick]
(axis cs:1,0.105678187638245)
--(axis cs:1,0.106828056329156);

\path [draw=lightcoral, semithick]
(axis cs:2,0.0957254147226043)
--(axis cs:2,0.0965079474229338);

\path [draw=lightcoral, semithick]
(axis cs:3,0.0921692862934061)
--(axis cs:3,0.0932220065772616);

\path [draw=lightcoral, semithick]
(axis cs:4,0.0904779077268622)
--(axis cs:4,0.091441334699417);

\path [draw=lightcoral, semithick]
(axis cs:5,0.0895094528158227)
--(axis cs:5,0.0903014741740077);

\path [draw=lightslategray, semithick]
(axis cs:0.1,0.256548369669406)
--(axis cs:0.1,0.261945584379964);

\path [draw=lightslategray, semithick]
(axis cs:0.3,0.15094046605923)
--(axis cs:0.3,0.154309681239261);

\path [draw=lightslategray, semithick]
(axis cs:0.6,0.119673666075984)
--(axis cs:0.6,0.121208026609527);

\path [draw=lightslategray, semithick]
(axis cs:1,0.105780007457477)
--(axis cs:1,0.106928653959593);

\path [draw=lightslategray, semithick]
(axis cs:2,0.0958562266014766)
--(axis cs:2,0.0966497574516814);

\path [draw=lightslategray, semithick]
(axis cs:3,0.0923469297914053)
--(axis cs:3,0.0933956079395494);

\path [draw=lightslategray, semithick]
(axis cs:4,0.0906893256330594)
--(axis cs:4,0.0916518998298158);

\path [draw=lightslategray, semithick]
(axis cs:5,0.0897635023994584)
--(axis cs:5,0.0905563681910522);

\addplot [semithick, red, mark=triangle*, mark size=3, mark options={solid,rotate=180}, only marks]
table {%
0.1 0.259114870758237
0.3 0.152533252229074
0.6 0.120353002003357
1 0.1062531219837
2 0.096116681072769
3 0.0926956464353338
4 0.0909596212131396
5 0.0899054634949152
};
\addlegendentry{Oracle estimator, ${\bm{\Xi}_Y^*}(\bS)$}
\addplot [semithick, royalblue, mark=triangle*, mark size=3, mark options={solid}, only marks]
table {%
0.1 0.259246977024685
0.3 0.152625073649246
0.6 0.120440846342755
1 0.106354330708535
2 0.096252992026579
3 0.0928712688654773
4 0.0911706127314376
5 0.0901599352952553
};
\addlegendentry{RIE, $\widehat{{\bm{\Xi}_Y^*}}(\bS)$}
\draw (axis cs:0.2,0.259114870758237) node[
  scale=0.6,
  anchor=base west,
  text=black,
  rotate=0.0
]{0.05\%};
\draw (axis cs:0.15,0.157533252229074) node[
  scale=0.6,
  anchor=base west,
  text=black,
  rotate=0.0
]{0.06\%};
\draw (axis cs:0.45,0.125353002003357) node[
  scale=0.6,
  anchor=base west,
  text=black,
  rotate=0.0
]{0.07\%};
\draw (axis cs:0.85,0.1112531219837) node[
  scale=0.6,
  anchor=base west,
  text=black,
  rotate=0.0
]{0.1\%};
\draw (axis cs:1.85,0.101116681072769) node[
  scale=0.6,
  anchor=base west,
  text=black,
  rotate=0.0
]{0.14\%};
\draw (axis cs:2.85,0.0976956464353338) node[
  scale=0.6,
  anchor=base west,
  text=black,
  rotate=0.0
]{0.19\%};
\draw (axis cs:3.85,0.0959596212131396) node[
  scale=0.6,
  anchor=base west,
  text=black,
  rotate=0.0
]{0.23\%};
\draw (axis cs:4.85,0.0949054634949152) node[
  scale=0.6,
  anchor=base west,
  text=black,
  rotate=0.0
]{0.28\%};
\end{axis}

\end{tikzpicture}
    \caption{Uniform spectral, $\mathcal{U}\big([1,3] \big)$}
\end{subfigure}
\caption{\small Estimating $\bY$. MSE is normalized by the norm of the signal, $\| \bY \|_{\rm F}^2$. $\bX$ is a shifted Wigner matrix with $c=3$, and $\bW$ has i.i.d. Gaussian entries of variance $1/N$, and $N/M = 2$. The RIE is applied to $N=2000, M =1000$, and the results are averaged over 10 runs (error bars are invisible).}
\label{Gaussian-Uniform-Y-alphag1}
\end{figure}

In figure \ref{Gaussian-Uniform-Y-alphag1} the MSE of RIE and the oracle estimator is plotted for two cases of priors: $\bY$ with Gaussian entries and $\bY$ with uniform spectral density.

\begin{figure}
    \centering
\begin{tikzpicture}[scale = 0.7]

\definecolor{darkgray176}{RGB}{176,176,176}
\definecolor{lightcoral}{RGB}{240,128,128}
\definecolor{lightslategray}{RGB}{119,136,153}
\definecolor{royalblue}{RGB}{65,105,225}

\begin{axis}[
legend cell align={left},
legend style={
  fill opacity=0.8,
  draw opacity=1,
  text opacity=1,
  at={(0,0)},
  anchor=south west,
  draw=white!80!black
},
tick align=outside,
tick pos=left,
x dir=reverse,
x grid style={darkgray176},
xlabel={$\alpha$},
xmin=0.4, xmax=20.9333333333333,
xtick style={color=black},
y grid style={darkgray176},
ylabel={${\rm MSE}$},
yticklabel style={
  /pgf/number format/precision=3,
  /pgf/number format/fixed},
ymin=0.104762490621842, ymax=0.117977453065667,
ytick style={color=black}
]
\path [draw=lightcoral, semithick]
(axis cs:20,0.116400132719598)
--(axis cs:20,0.117305416488343);

\path [draw=lightcoral, semithick]
(axis cs:10,0.116155980448036)
--(axis cs:10,0.116407035410802);

\path [draw=lightcoral, semithick]
(axis cs:6.66666666666667,0.115100931772049)
--(axis cs:6.66666666666667,0.115644708476546);

\path [draw=lightcoral, semithick]
(axis cs:5,0.114340522483055)
--(axis cs:5,0.114746359242298);

\path [draw=lightcoral, semithick]
(axis cs:4,0.113440343000258)
--(axis cs:4,0.113922367453468);

\path [draw=lightcoral, semithick]
(axis cs:2,0.109530299422505)
--(axis cs:2,0.10973404711094);

\path [draw=lightcoral, semithick]
(axis cs:1.33333333333333,0.105363170732925)
--(axis cs:1.33333333333333,0.10561354781082);

\path [draw=lightslategray, semithick]
(axis cs:20,0.116473274855184)
--(axis cs:20,0.117376772954584);

\path [draw=lightslategray, semithick]
(axis cs:10,0.116231648776338)
--(axis cs:10,0.116490081376043);

\path [draw=lightslategray, semithick]
(axis cs:6.66666666666667,0.115184919308687)
--(axis cs:6.66666666666667,0.115725397705931);

\path [draw=lightslategray, semithick]
(axis cs:5,0.114424965064127)
--(axis cs:5,0.114831827762254);

\path [draw=lightslategray, semithick]
(axis cs:4,0.113535465277463)
--(axis cs:4,0.114021388495366);

\path [draw=lightslategray, semithick]
(axis cs:2,0.109651600932245)
--(axis cs:2,0.10986339305337);

\path [draw=lightslategray, semithick]
(axis cs:1.33333333333333,0.105523124544209)
--(axis cs:1.33333333333333,0.105771138511803);

\addplot [semithick, red, mark=triangle*, mark size=3, mark options={solid,rotate=180}, only marks]
table {%
20 0.11685277460397
10 0.116281507929419
6.66666666666667 0.115372820124298
5 0.114543440862677
4 0.113681355226863
2 0.109632173266722
1.33333333333333 0.105488359271872
};
\addlegendentry{Oracle estimator, ${\bm{\Xi}_Y^*}(\bS)$}
\addplot [semithick, royalblue, mark=triangle*, mark size=3, mark options={solid}, only marks]
table {%
20 0.116925023904884
10 0.116360865076191
6.66666666666667 0.115455158507309
5 0.114628396413191
4 0.113778426886414
2 0.109757496992807
1.33333333333333 0.105647131528006
};
\addlegendentry{RIE, $\widehat{{\bm{\Xi}_Y^*}}(\bS)$}
\draw (axis cs:19.7,0.11685277460397) node[
  scale=0.6,
  anchor=base west,
  text=black,
  rotate=0.0
]{0.06\%};
\draw (axis cs:12.5,0.116281507929419) node[
  scale=0.6,
  anchor=base west,
  text=black,
  rotate=0.0
]{0.07\%};
\draw (axis cs:6.36666666666667,0.115372820124298) node[
  scale=0.6,
  anchor=base west,
  text=black,
  rotate=0.0
]{0.07\%};
\draw (axis cs:7.5,0.114543440862677) node[
  scale=0.6,
  anchor=base west,
  text=black,
  rotate=0.0
]{0.07\%};
\draw (axis cs:3.7,0.113681355226863) node[
  scale=0.6,
  anchor=base west,
  text=black,
  rotate=0.0
]{0.09\%};
\draw (axis cs:4.5,0.109632173266722) node[
  scale=0.6,
  anchor=base west,
  text=black,
  rotate=0.0
]{0.11\%};
\draw (axis cs:3.83333333333333,0.105488359271872) node[
  scale=0.6,
  anchor=base west,
  text=black,
  rotate=0.0
]{0.15\%};
\end{axis}
\end{tikzpicture}
    \caption{\small MSE of estimating $\bY$ as a function of aspect-ratio $\alpha > 1$, $\bY$ has Gaussain entries of variance $\nicefrac{1}{N}$, and $\kappa = 5$. MSE is normalized by the norm of the signal, $\| \bY \|_{\rm F}^2$. $\bX$ is a shifted Wigner matrix with $c=3$, and $\bW$ has i.i.d. Gaussian entries of variance $1/N$. The RIE is applied to $N=2000, M = \nicefrac{1}{\alpha} N$, and the results are averaged over 10 runs (error bars are invisible). Average relative error between RIE $\widehat{{\bm{\Xi}_Y^*}}(\bS)$ and Oracle estimator is also reported.}
    \label{Y-ar-alphag1}
\end{figure}
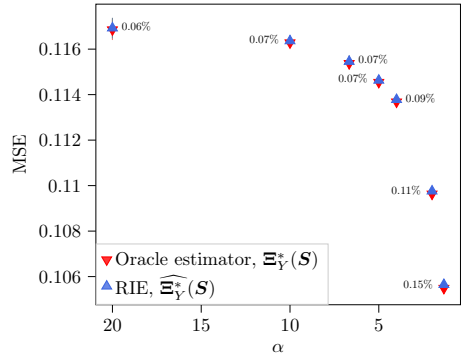

\paragraph{Effect of aspect-ratio $\alpha$.} In Figure \ref{Y-ar-alphag1}, we take $\bY$ to have Gaussian entries (with variance $\frac{1}{N}$), and the MSE is depicted for various values of the aspect-ratio $\alpha > 1$. We see that as $M$ decreases ($\alpha$ increases) the estimation error (of $\bY$) increases.

\clearpage
\section{Spherical integrals and matrix lemmas}

\subsection{Spherical Integrals}\label{spherical integral app}
For two symmetric matrices $\bA, \bB \in \bR^{N \times N}$, the \textit{spherical integral} is defined as:
\begin{equation*}
    \mathcal{I}_N (\bA, \bB) = \Big\langle \exp \big\{ \frac{N}{2} \Tr \bA \bU \bB \bU^\intercal \big\} \Big\rangle_{\bU}
\end{equation*}
where the average is w.r.t. the \textit{Haar} measure over the group of (real) orthogonal $N\times N$ matrices. The spherical integrals can also be defined w.r.t. the unitary or symplectic group. These integrals are often referred to as \textit{Harish Chandra-Itzykson-Zuber} (HCIZ) integrals in mathematical physics literature. The study of these objects dates back to the work of mathematician Harish Chandra \cite{harish1957differential} and they have since been extensively studied and developed in both physics and mathematics. In particular, \cite{guionnet2005fourier} studied the limit of the integral in the case where one of the matrices, say $\bA$,  has finite rank. 

\begin{theorem}[\textbf{Rank-one  spherical integral}, Guionnet and Ma{\"\i}da \cite{guionnet2005fourier}]\label{sphericla-integral} Let $\theta$ be the only non-zero eigenvalue of $\bA$ (so it is rank one), and the empirical eigenvalue distribution of $\bB$ converge weakly towards $\rho_B$. Then, for $\theta$ sufficiently small (see details in Theorem 2 in \cite{guionnet2005fourier}), we have:
\begin{equation}
    \lim_{N \to \infty} \frac{1}{N} \ln \mathcal{I}_N (\bA, \bB) = \frac{1}{2} \int_0^{\theta} \mathcal{R}_{\rho_B} (t) \, dt \equiv \frac{1}{2} \mathcal{P}_{\rho_B}(\theta)
    \label{spherical-limit}
\end{equation}
\end{theorem}
When $\bA$ has higher (but finite) rank, theorem 7 in \cite{guionnet2005fourier} states that the limit is the sum over eigenvalues of the expression on the rhs of \eqref{spherical-limit}. 

\paragraph{Non-symmetric case.} In the non-symmetric case the \textit{rectangular spherical integral} is defined, for the matrices $\bA \in \bR^{M \times N}, \bB \in \bR^{N \times M}$, as:
\begin{equation*}
    \mathcal{J}_N (\bA, \bB) = \Big\langle \exp \big\{ \sqrt{N M} \Tr \bA \bU \bB \bV \big\} \Big\rangle_{\bU,\bV}
\end{equation*}
where $\bU \in \bR^{ N \times N}, \bV \in \bR^{M \times M}$, and the expectation is w.r.t. the \textit{Haar} measure over orthogonal matrices of size $N\times N$ and $M\times M$. 
\begin{theorem}[\textbf{Rank-one  rectangular spherical integral}, Benaych-Georges \cite{benaych2011rectangular}]\label{rect-sphericla-integral} Let $N/M \to \alpha \in (0,1]$, and $\theta$ be the only non-zero singular value of $\bA$, and the empirical singular value distribution of $\bB$ converges weakly towards $\mu_B$. Then, for $\theta$ sufficiently small (see details in Theorem 2.2 in \cite{benaych2011rectangular}), we have:
\begin{equation}
    \lim_{N \to \infty} \frac{1}{N} \ln \mathcal{J}_N (\bA, \bB) =  \int_0^{\theta} \frac{\mathcal{C}_{\mu_B}^{(\alpha)}(t^2)} {t} \, dt = \frac{1}{2} \int_0^{\theta^2} \frac{\mathcal{C}_{\mu_B}^{(\alpha)}(t)}{t} \, dt \equiv \frac{1}{2} \mathcal{Q}_{\mu_B}^{(\alpha)}(\theta^2)
    \label{rect-spherical-limit}
\end{equation}
\end{theorem}
In our derivation, we use a generalization of this formula, namely when $\bA$ has higher (but fixed) rank, the limit is the sum over singular values of the expression on the rhs of \eqref{rect-spherical-limit}. Although we are not aware if this generalization has been proved, we believe that the ideas found in \cite{guionnet2022asymptotics} can be applied to show it holds.

\begin{remark}
It is known that additional terms may be present on the rhs of \eqref{spherical-limit} and \eqref{rect-spherical-limit} when the parameter $\theta$ is "large". This has been rigorously proved at least in the case of symmetric $\bA$ and $\bB$ (see theorem 6 in \cite{guionnet2005fourier}). In the replica calculation the order of magnitude of this parameter is determined by the solutions of the saddle point equations, but  it is difficult to fully control its order of magnitude. However the numerics show very good agreement between our explicit RIEs and the Oracle estimator, which strongly suggests it is sound to use \eqref{spherical-limit} and \eqref{rect-spherical-limit}.
\end{remark}

\subsection{Matrix analysis tools}\label{Matrix-tool}
\begin{proposition}[\textbf{Inverse of a block matrix}, Bernstein \cite{bernstein2009matrix}] \label{inverse-block-matrix}
For a block matrix $\bF = \left[
\begin{array}{cc}
\bA& \bB  \\
\bC & \bD
\end{array}
\right]$ with $\bA \in \bR^{N \times N}, \bB \in \bR^{N \times M}, \bC \in \bR^{M \times N}, \bD \in \bR^{M \times M}$, if $\bA$ and $\bD - \bC \bA^{-1} \bB $, are non-singular, then,
\begin{equation*}
    \bF^{-1} = \left[
\begin{array}{cc}
\bA^{-1} + \bA^{-1} \bB (\bD - \bC \bA^{-1} \bB)^{-1} \bC \bA^{-1}& -\bA^{-1} \bB (\bD - \bC \bA^{-1} \bB)^{-1}  \\
-(\bD - \bC \bA^{-1} \bB)^{-1} \bC \bA^{-1}  & (\bD - \bC \bA^{-1} \bB)^{-1}
\end{array}
\right]
\end{equation*}
\end{proposition}

\paragraph{Block structure of $\bG_\mathcal{S}(z)$} 
The matrix $\bG_\mathcal{S}(z)$ is:
\begin{equation*}
        \bG_{\mathcal{S}}(z) = \big(z \bI - \cS \big)^{-1} = \left[
\begin{array}{cc}
z \bI_N & -\bS \\
-\bS^\intercal & z \bI_M
\end{array}
\right]^{-1} 
\end{equation*}
Using Proposition \ref{inverse-block-matrix}, first we need to compute the inverse matrix $\big(z \bI_M - (-\bS^{\intercal}) (z \bI_N)^{-1} (- \bS) \big)^{-1}$ which simply reads:
\begin{equation*}
        \big(z \bI_M - \frac{1}{z} \bS^{\intercal} \bS \big)^{-1} = z \big(z^2 \bI_M - \bS^{\intercal} \bS \big)^{-1} = z \bG_{S^\intercal S} (z^2)
\end{equation*}
Consequently, we find:
\begin{equation}
        \bG_{\mathcal{S}}(z) = \left[
\begin{array}{cc}
\frac{1}{z} \bI_N + \frac{1}{z} \bS \bG_{S^\intercal S}(z^2) \bS^\intercal & \bS \bG_{S^\intercal S}(z^2) \\
\bG_{S^\intercal S}(z^2) \bS^\intercal & z \bG_{S^\intercal S}(z^2)
\end{array}
\right]
    \label{resolvent_S}
\end{equation}

\paragraph{Inverse of $\bC_X^*$}  For $\bC_X^*$ since the blocks $\bB, \bC$ are zero, the inverse is simply:
\begin{equation}
\begin{split}
    {\bC_X^*}^{-1} &= \left[
\begin{array}{cc}
\big[ (z - \zeta_1^*) \bI_N - \zeta_3^* \bX^2 \big]^{-1} & \mathbf{0}  \\
\mathbf{0} & \big[(z-\zeta_2^*) \bI_M \big]^{-1}
\end{array}
\right] \\
&= \left[
\begin{array}{cc}
 \frac{1}{\zeta_3^*}\big[\frac{z - \zeta_1^*}{\zeta_3^*} \bI_N -  \bX^2 \big]^{-1} & \mathbf{0}  \\
\mathbf{0} & \frac{1}{z-\zeta_2^*} \bI_M 
\end{array}
\right] \\
&= \left[
\begin{array}{cc}
 \frac{1}{\zeta_3^*}\bG_{X^2} \big( \frac{z - \zeta_1^*}{\zeta_3^*} \big) & \mathbf{0}  \\
\mathbf{0} & \frac{1}{z-\zeta_2^*} \bI_M 
\end{array}
\right]
\end{split}
\label{CX-inv}
\end{equation}

\paragraph{Inverse of $\bC_Y^*$}  Let the block structure of $\bC_Y^*$ be as in Proposition \ref{inverse-block-matrix}, then
\begin{equation*}
    \begin{split}
        (\bD - \bC \bA^{-1} \bB)^{-1} &= \Big( (z-\beta_2^*) \bI_M - \beta_3^* \bY^\intercal \bY  - \frac{{\beta_4^*}^2}{z - \beta_1^*} \bY^\intercal \bY \Big)^{-1} \\
        &= \Big( (z-\beta_2^*) \bI_M - \big( \beta_3^* + \frac{{\beta_4^*}^2}{z - \beta_1^*} \big) \bY^\intercal \bY \Big)^{-1} \\
        &= (z - \beta_1^*) \Big( Z_1(z) \bI_M - Z_2(z) \bY^\intercal \bY \Big)^{-1} \\
        &= \frac{z - \beta_1^*}{Z_2(z)} \Big( \frac{Z_1(z)}{Z_2(z)} \bI_M -  \bY^\intercal \bY \Big)^{-1} \\
        &= \frac{z - \beta_1^*}{Z_2(z)} \bG_{Y^\intercal Y} \big(\frac{Z_1(z)}{Z_2(z)} \big) 
    \end{split}
\end{equation*}
where $\bG_{Y^\intercal Y}$ is the resolvent of the matrix $\bY^\intercal \bY$. So, we have
\begin{equation*}
    {\bC_Y^*}^{-1} = \left[
\begin{array}{cc}
(z-\beta_1^*)^{-1} \bI_N + \frac{{\beta_4^*}^2}{(z-\beta_1^*) Z_2(z)} \bY \bG_{Y^\intercal Y} \big(\frac{Z_1(z)}{Z_2(z)} \big) \bY^\intercal & \frac{\beta_4^*}{Z_2(z)} \bY  \bG_{Y^\intercal Y} \big(\frac{Z_1(z)}{Z_2(z)} \big)  \\
\frac{\beta_4^*}{Z_2(z)}  \bG_{Y^\intercal Y} \big(\frac{Z_1(z)}{Z_2(z)} \big) \bY^\intercal & \frac{z - \beta_1^*}{Z_2(z)} \bG_{Y^\intercal Y} \big(\frac{Z_1(z)}{Z_2(z)} \big) 
\end{array}
\right]
\end{equation*}

\begin{lemma}\label{eigenvalue of xy^T + y^Tx}
Consider two vectors $\bx, \by \in \bR^N$. The symmetric matrix $\bx \by^\intercal + \by \bx^\intercal$ has rank at most two with non-zero eigenvalues $\bx^\intercal \by \pm \|\bx\| \|\by \|$.
\end{lemma}
\begin{proof}
Construct the matrices $\bA \in \bR^{2 \times N}, \bB \in \bR^{N \times 2}$ as follows:
\begin{equation*}
    \bA = \left[
\begin{array}{c}
 \bx^\intercal \\
\by^\intercal 
\end{array}
\right], \quad \bB =  \left[
\begin{array}{cc}
 \by & \bx
\end{array}
\right]
\end{equation*}
Then, we have that $\bx \by^\intercal + \by \bx^\intercal = \bB \bA$. Using the lemma \ref{Sylvester}, we have that:
\begin{equation*}
    z^2 \det \big( z \bI_N - \bB \bA \big) = z^N \det \big( z \bI_2 - \bA \bB \big)
\end{equation*}
So, the characteristic polynomial of $\bx \by^\intercal + \by \bx^\intercal$ is $z^{N-2} \det \big( z \bI_2 - \bA \bB \big)$, which implies that the $\bx \by^\intercal + \by \bx^\intercal$ has eigenvalue 0 with multiplicity $N-2$, plus the eigenvalues of the $2\times 2$ matrix $\bA \bB$. The matrix $\bA \bB$ is:
\begin{equation*}
    \bA \bB = \left[
\begin{array}{cc}
 \bx^\intercal \by & \| \bx \|^2 \\
 \| \by \|^2 & \bx^\intercal \by
\end{array}
\right]
\end{equation*}
which has two eigenvalues  $\bx^\intercal \by \pm \|\bx\| \|\by \|$.
\end{proof}

\begin{lemma}\label{Sylvester}
    For matrices $\bA \in \bR^{M \times N}, \bB \in \bR^{N \times M}$, we have:
    \begin{equation*}
    z^M \det \big( z \bI_N - \bB \bA \big) = z^N \det \big( z \bI_M - \bA \bB \big)
\end{equation*}
\end{lemma}
\begin{proof}
    Construct the matrices $\bC , \bD \in \bR^{(M+N) \times (M+N)}$ as follows:
    \begin{equation*}
        \bC = \left[
\begin{array}{cc}
 z \bI_M & \bA \\
 \bB & \bI_N
\end{array}
\right], \quad \bD = \left[
\begin{array}{cc}
  \bI_M & \bm{0}_{M \times N} \\
 -\bB & z \bI_N
\end{array}
\right]
    \end{equation*}
    We have:
    \begin{equation*}
        \det \bC \bD = z^N \det \big( z \bI_M - \bA \bB), \quad \det \bD \bC = z^M \det \big( z \bI_N - \bB \bA \big)
    \end{equation*}
    The result follows from the fact that $\det \bC \bD = \det \bD \bC$.
\end{proof}

\end{document}